\documentclass[preprint,superscriptaddress,secnumarabic,amssymb, nobibnotes, aps, pre]{revtex4-2}

\usepackage{bm}
\usepackage{setspace}
\usepackage{tikz}
\usetikzlibrary{automata,patterns,positioning}
\usepackage{graphicx}
\usepackage{epstopdf, epsfig}
\usepackage{makecell}
\usepackage{bm}
\usepackage{mathrsfs}
\usepackage{amsmath}
\usepackage{array}
\usepackage{xcolor}

\definecolor{matchred}{RGB}{227,26,28}

\newcommand{\bnabla}{\boldsymbol{\nabla}}
\newcommand{\bcdot}{\boldsymbol{\cdot}}

\newcommand\qtwod{quasi-two-di\-men\-sion\-al}

\newcommand\TS{TS}
\newcommand\TSL{TS-like}
\newcommand\eg{e.g.}  

\newcommand{\vect}[1]{\bm{#1}}

\newcommand{\pde}[2]{\frac{\partial #1}{\partial #2}}
\newcommand{\pdesqr}[2]{\frac{\partial^2 #1}{\partial #2^2}}

\newcommand\Fig{Figure} 
\newcommand\fig{Fig.}  
\newcommand\figs{Figs.}  
\newcommand\tbl{Table}  

\newcommand\alphaCrit{\alpha_\mathrm{crit}}

\newcommand\dUP{\mathrm{d}}

\newcommand\Ezero{E_0}
\newcommand\Euv{E}
\newcommand\Ev{E_\mathrm{v}}

\newcommand\Np{N_\mathrm{p}}

\newcommand\Rey{\mathrm{Re}}  
\newcommand\Har{\mathrm{Ha}}  
\newcommand\Pra{\mathrm{Pr}}  
\newcommand\Sr{\mathrm{Sr}}
\newcommand\Wo{\mathrm{Wo}}
\newcommand\Rez{\mathrm{Re}}
\newcommand\Imz{\mathrm{Im}}

\newcommand\ReyCrit{\Rey_{\mathrm{crit}}}
\newcommand\meanfoco{\bar{c}_\kappa}
\newcommand\Nf{N_\mathrm{f}}
\newcommand\Ny{N_\mathrm{y}}
\newcommand\phased{\psi_\mathrm{d}}
\newcommand{\Nc}{N_\mathrm{c}}
\newcommand\UoneB{U_{1,\mathrm{B}}}
\newcommand\UtwoB{U_{2,\mathrm{B}}}
\newcommand\UoneBv{\vect{U}_{1,\mathrm{B}}}
\newcommand\UtwoBv{\vect{U}_{2,\mathrm{B}}}
\newcommand\rrs{r_\mathrm{s}}
\newcommand\als{\alpha_\mathrm{s}}

\newcommand{\tpe}{t_\mathrm{P}}
\newcommand{\EU}{E_\mathrm{U}}
\newcommand{\twonv}{\lVert \tilde{v} \rVert_2}
\newcommand{\tmr}[1]{\textcolor{matchred}{#1}}
\newcommand\vsk{\hat{v}_{\mathrm{s},j}}
\newcommand\usk{\hat{u}_{\mathrm{s},j}}
\newcommand\csk{\hat{\omega}_{\mathrm{s},j}}
\newcommand\cvb{\check{\vect{b}}}
\newcommand\cvn{\check{\vect{\nabla}}}
\newcommand\cvu{\check{\vect{u}}}

\graphicspath{ {Figures/} }

\begin{document}
\title{Stability of pulsatile quasi-two-dimensional duct flows under a transverse magnetic field}%

\author{Christopher J. Camobreco}%
\email{christopher.camobreco@monash.edu}
\affiliation{Department of Mechanical and Aerospace Engineering, Monash University, Victoria 3800, Australia}
\author{Alban Poth{\'e}rat}%
\email{alban.potherat@coventry.ac.uk}
\affiliation{Fluid and Complex Systems Research Centre, Coventry University, Coventry CV15FB, United Kingdom}
\author{Gregory J. Sheard}%
\email{greg.sheard@monash.edu}
\affiliation{Department of Mechanical and Aerospace Engineering, Monash University, Victoria 3800, Australia}
\date{\today}%
\begin{abstract}
This manuscript has been accepted for publication in Physical Review Fluids, see https://journals.aps.org/prfluids/accepted/53075Se8O0b1b109b1cc0061b280aaa122f0f92dc. The stability of a pulsatile quasi-two-dimensional duct flow was numerically investigated. Flow was driven, in concert, by a constant pressure gradient and by the synchronous oscillation of the lateral walls. This prototypical setup serves to aid understanding of unsteady magnetohydrodynamic flows in liquid metal coolant ducts subjected to transverse magnetic fields, motivated by the conditions expected in magnetic confinement fusion reactors. A wide range of wall oscillation frequencies and amplitudes, relative to the constant pressure gradient, were simulated. Focus was placed on the driving pulsation optimized for the greatest reduction in the critical Reynolds number, for a range of friction parameters $H$ (proportional to magnetic field strength). An almost $70$\% reduction in the critical Reynolds number, relative to that for the steady base flow, was obtained toward the hydrodynamic limit ($H=10^{-7}$), while just over a $90$\% reduction was obtained by $H=10$. For all oscillation amplitudes, increasing $H$ consistently led to an increasing percentage reduction in the critical Reynolds number. This is a promising result, given fusion relevant conditions of $H \geq 10^4$. These reductions were obtained by selecting a frequency that both ensures prominent inflection points are maintained in the base flow, and a growth in perturbation energy in phase with the deceleration of the base flow. Nonlinear simulations of perturbations driven at the optimized frequency and amplitude still satisfied the no net growth condition at the greatly reduced critical Reynolds numbers. However, two complications were introduced by nonlinearity. First, although the linear mode undergoes a symmetry breaking process, turbulence was not triggered. Second, a streamwise invariant sheet of negative velocity formed, able to arrest the linear decay of the perturbation. Although the nonlinearly modulated base flow maintained a higher time-averaged energy, it also stabilized the flow, with exponential growth not observed at supercritical Reynolds numbers. 
\end{abstract}
\maketitle
\section{Introduction}\label{sec:introduction}
The aim of this paper is to assess the generation and promotion of turbulence in oscillatory magnetohydrodynamic (MHD) duct flows. Motivation stems from proposed designs of dual purpose tritium breeder/coolant ducts in magnetic confinement fusion reactors \cite{Abdou2015blanket}. These coolant ducts are plasma facing, hence subjected to both high temperatures and a strong pervading transverse magnetic field \citep{Smolentsev2008characterization}. At the same time, obtaining turbulent heat transfer rates is crucial to the long term operation of self-cooled duct designs \citep{Moreau2010flow}. This can be achieved by keeping the flow turbulent. Various strategies to promote turbulence in MHD flows include: the placement of physical obstacles of various cross section \citep{Cassels2016heat, Hussam2011dynamics, Hussam2012enhancing}, inhomogeneity in electrical boundary conditions \citep{Buhler1996instabilities}, electrode stimulation \citep{Hamid2016combining, Hamid2016heat} and localized magnetic obstacles \citep{Cuevas2006flow}. The approach to promote turbulence taken in this work is to superimpose a time periodic flow, of specified frequency and amplitude, onto an underlying steady flow. The benchmark used, particularly in the linear analysis, is the critical Reynolds number for the steady flow. The goal is to obtain the greatest reduction in the critical Reynolds number (considered as the degree of destabilization) with the addition of a time periodic flow component, of optimized frequency and amplitude. Ultimately, this approach seeks an estimate of the lowest Reynolds number at which turbulence may be incited and sustained by the addition of a pulsatile component to the base flow.

In MHD flows, the predominant action of the Lorentz force on the electrically conducting fluid is to diffuse momentum along magnetic field lines \citep{Davidson2001introduction, Sommeria1982why}. When the Lorentz force dominates both diffusive and inertial forces, the flow becomes quasi-two-dimensional (Q2D) \citep{Potherat2010direct, Thess2007transition, Zikanov1998direct}. In the limit of quasi-static Q2D MHD, the magnetic field is imposed, and the Lorentz force dominates all other forces far from walls normal to the field. Three dimensionality only remains when asymptotically small in amplitude, or in regions of asymptotically small thickness. The boundary layers remain intrinsically three dimensional. Hartmann boundary layers form on walls perpendicular to magnetic field lines, with a thickness scaling as $\Har^{-1}$ \citep{Potherat2015decay, Sommeria1982why}, while the thickness of parallel wall Shercliff boundary layers scales as $\Har^{-1/2}$ \citep{Potherat2007quasi}. The Hartmann number $\Har = aB(\sigma/\rho\nu)^{1/2}$ represents the square root of the ratio of electromagnetic to viscous forces, where $a$ is the distance between Hartmann walls, $B$ the imposed magnetic field strength, and $\sigma$, $\rho$ and $\nu$ the incompressible Newtonian fluid's electrical conductivity, density and kinematic viscosity, respectively. Nevertheless, although not asymptotically small, three dimensionality in Shercliff layers remains small enough for Q2D models to represent them with high accuracy \cite{Potherat2000effective}. The remaining core flow is uniform and well two-dimensionalized, in fusion relevant regimes \citep{Smolentsev2008characterization}. A Q2D model proposed by Ref.~\citep{Sommeria1982why} (hereafter the SM82 model) is applied, which governs flow quantities averaged along the magnetic field direction. In the Q2D setup, the Hartmann walls are accounted for with the addition of linear friction acting on the bulk flow, valid for laminar Hartmann layers \citep{Sommeria1982why}. Shercliff layers still remain in the averaged velocity field, even in the quasi-static limit of a dominant Lorentz force, of thickness scaling as $H^{-1/2}$ \citep{Potherat2007quasi}, where $H = 2(L/a)^2\Har$ is the friction parameter, and $L$ the characteristic wall-normal length. The accuracy of the SM82 model is well established for the duct problem \citep{Cassels2019from3D, Kanaris2013numerical,Muck2000flows}, with less than $10$\% error between the quasi-two-dimensional and the three-dimensional laminar boundary layer profiles \citep{Potherat2000effective}. 

The linear stability of steady Q2D duct flow was first analysed by Ref.~\citep{Potherat2007quasi}. As the magnetic field is strongly stabilizing, the critical Reynolds number for a steady base flow, beyond which modal instabilities grow, scales as $\Rey_\mathrm{crit,s} =  4.835\times10^4 H^{1/2}$ for $H \gtrsim 1000$ \citep{Camobreco2020transition, Potherat2007quasi, Vo2017linear}. The Reynolds number $\Rey = U_0 L/\nu$ represents the ratio of inertial to viscous forces. In this work, both transient and steady inertial forces will be encapsulated in $U_0$, a characteristic velocity based on both the steady and oscillating flow components. Instability occurs via Tollmien--Schlichting (\TS) waves originating in the Shercliff layers. The instabilities become isolated at the  duct walls with increasing magnetic field strength \citep{Camobreco2020transition, Potherat2007quasi}, eventually behaving as per an instability in an isolated exponential boundary layer \citep{Camobreco2020role, Camobreco2020transition, Potherat2007quasi}. To the authors' knowledge, oscillatory or pulsatile Q2D flows are yet to be analysed under a transverse magnetic field. Weak in-plane fields have been analysed for oscillatory flows, although pulsatility was not considered \citep{Thomas2010linear, Thomas2013global}.

The destabilization of hydrodynamic plane channel flows with the imposition of an oscillating flow component was first convincingly assessed by Ref.~\citep{Kerczek1982instability}. Using series expansions to evaluate Floquet exponents, the range of frequencies that induce destabilization was determined.  Womersly numbers $1\leq \Wo \lesssim 13$ were destabilizing and $\Wo \geq 14$ stabilizing, for low Reynolds numbers and pulsation amplitudes, where the Womersly number $\Wo=\omega L^2/\nu$ characterizes the square root of transient inertial to viscous forces, and where $\omega$ is the pulsation frequency. The problem was revisited with advanced computational power and techniques \citep{Thomas2011linear, Pier2017linear}. However, even large-scale Floquet matrix problems struggled to adequately resolve larger-amplitude pulsations  \citep{Thomas2011linear, Pier2017linear}, as the required number of Fourier modes rapidly increases with increasing pulsation amplitude. Instead, direct forward evolution of the linearized Navier--Stokes equations is required. Improved bounds for destabilizing frequencies of $5\leq \Wo < 13$ were determined \citep{Pier2017linear}, with the optimum frequency for destabilization at $\Wo=7$. The optimized amplitude ratio for the pulsation was also found to be near unity (steady and oscillatory velocity maximums of equal amplitude) at lower frequencies \citep{Thomas2011linear}. In addition, a small destabilization was observed at very high frequencies, for small pulsation amplitudes. Although Ref.~\citep{Thomas2011linear} did not focus on obtaining the maximum destabilization, an approximately $33$\% reduction in the critical Reynolds number (relative to the steady result) was observed at the lowest frequency tested, near an amplitude ratio of unity. Further improvement, with an approximately $57$\% reduction in the critical Reynolds number \cite{Thomas2015linear}, was attained by the imposition of an oscillation with two modes of different frequencies. Given the size of the parameter space, there remains significant potential to further destabilize both hydrodynamic and MHD flows, with single-frequency optimized pulsations. 

At lower frequencies the perturbation energy varies over several orders of magnitude within a single period of evolution \citep{Pier2017linear, Singer1989numerical}. This intracylcic growth and decay predominantly occurs during the deceleration and acceleration phases of the base flow, respectively. The intracylcic growth increases exponentially with increasing pulsation amplitude \citep{Pier2017linear}. At smaller pulsation amplitudes, a `cruising' regime \citep{Pier2017linear} has been identified, where the perturbation energy remains of similar nonlinear magnitude throughout the entire cycle. At larger pulsation amplitudes, and at smaller frequencies, a `ballistic' regime \citep{Pier2017linear} was identified, where the perturbation energy varies by many orders of magnitude over the cycle, and is propelled from a linear to nonlinear regime through this growth. However, in full nonlinear simulations of Stokes boundary layers, an incomplete decay of the perturbation over one cycle is observed \citep{Ozdemir2014direct}. This has little effect on growth in the next cycle, thereby leading to either an intermittent or sustained turbulent state \citep{Ozdemir2014direct}. Thus, ballistic regimes form an enticing means to sustain turbulence under fusion relevant conditions. To assess the effectiveness of this strategy, we must understand the conditions of transition to turbulence in a duct flow pervaded by a strong enough magnetic field to assume quasi-two-dimensionality. Specifically, this paper seeks to answer the following questions:

\begin{itemize}
\item{Will superimposing an oscillatory flow onto an underlying steady base flow still be effective at reducing the critical Reynolds number in high $H$, fusion relevant, regimes?}
\item{What pulsation frequencies and amplitudes are most effective at destabilizing the flow, both hydrodynamically, and toward fusion relevant regimes?}
\item{Are the parameters at which reductions in $\ReyCrit$ are observed viable for both SM82 modelling, and fusion relevant applications?}
\item{Are reductions in $\ReyCrit$ sufficient to observe turbulence at correspondingly lower $\Rey$?}
\end{itemize}

This paper proceeds as follows: In Sec.~\ref{sec:prob_set}, the problem is nondimensionalized, and the base flow for the duct problem derived in the SM82 framework. Particular focus is placed on the dependence of the base flow on all four nondimensional parameters. Pressure- and wall-driven flows are compared, before determining the bounds for validity of the SM82 approximation for pulsatile flows. In Sec.~\ref{sec:lin_for}, the linear problem is formulated, and both the Floquet and timestepper methods are introduced. The long-term stability behavior is considered in Sec.~\ref{sec:lin_res1}, with particular focus on the optimal conditions for destabilization. Intracyclic growth and the linear mode structure are analysed in more detail in Sec.~\ref{sec:lin_res2}. Sec.~\ref{sec:nlin_all} focuses on targeted direct numerical simulations (DNS) of the optimized pulsations. Emphasis is placed on comparing linear and nonlinear evolutions, and symmetry breaking induced by nonlinearity.

\section{Problem setup}\label{sec:prob_set}
\subsection{Geometry and base flows} \label{sec:geom}

\begin{figure}[h!]
\centering
\begin{tikzpicture}
\node [inner sep=0pt] (img)  {\includegraphics[width=0.435\columnwidth]{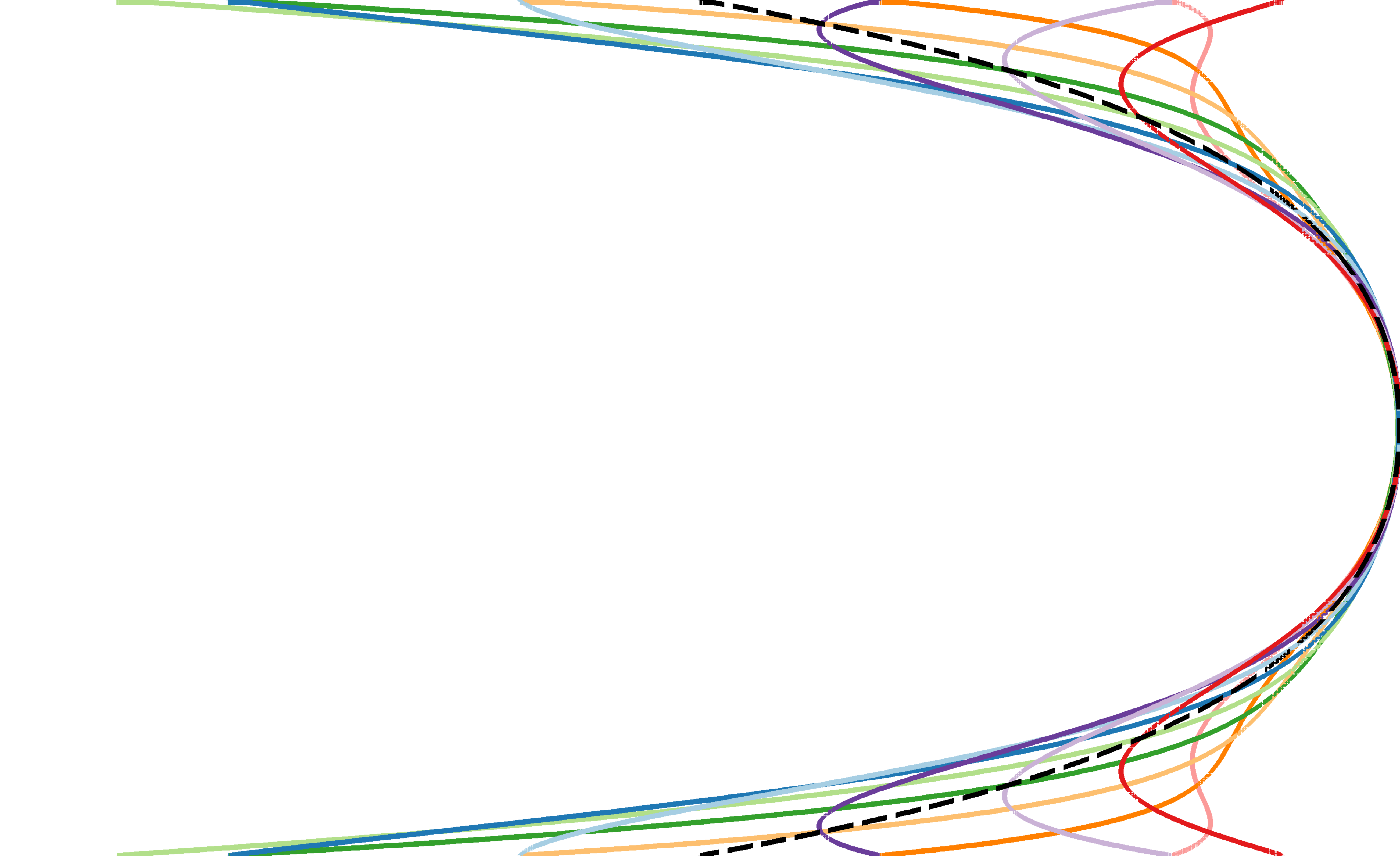}};

\draw[dashed,line width = 0.2mm, dash pattern=on 3pt off 5pt] (-3.6,-2.2) -- (-3.6,2.2); 
\draw[line width = 0.4mm] (-3.6,-2.18) -- (3.6,-2.18); 
\draw[dashed,line width = 0.2mm, dash pattern=on 3pt off 5pt] (3.6,-2.2) -- (3.6,2.2); 
\draw[line width = 0.4mm] (-3.6,2.18) -- (3.6,2.18);  

\node at (-2.5,1.2) [circle,draw=black,fill=black,inner sep=0.4mm] {};
\node at (-2.5,1.2) [circle,draw=black,inner sep=2mm] {};
\node at (-2,1.2) {$\check{\vect{B}}$};

\node at (0,-2.55) {$\check{u}= U_2\cos(\omega \,\check{t})$, $\check{v}=0$, $\hat{\vect{u}}=0$};
\node at (0,2.55) {$\check{u}= U_2\cos(\omega \,\check{t})$, $\check{v}=0$, $\hat{\vect{u}}=0$};

\fill[pattern=north east lines, pattern color=black] (-3.62,2.3) rectangle (3.62,2.2);
\fill[pattern=north east lines, pattern color=black] (-3.62,-2.3) rectangle (3.62,-2.2);

\draw[line width = 0.2mm,<->] (-3.6,-2.8) -- (3.6,-2.8);
\draw[line width = 0.2mm,<->] (3.8,-2.2) -- (3.8,2.2);
\node at (4.2,0) {$2L$};
\node at (0,-3.1) {$2\pi/\alpha$};

\draw[line width = 0.4mm,->] (-3.6,0) -- (-2.8,0) node[anchor=north west] {};
\draw[line width = 0.4mm,->] (-3.6,0) -- (-3.6,0.8) node[anchor= west] {$y$};
\node at (-3.6,0) [circle,draw=black,inner sep=0.6mm,line width = 0.4mm] {};
\node at (-2.8,-0.2) {$x$};
\node at (-3.75,-0.2) {$z$};
\end{tikzpicture}
\caption{A schematic representation of the system under investigation. Solid lines denote the oscillating, impermeable, no-slip walls. Short dashed lines indicate the streamwise extent of the periodic domain, defined by streamwise wave number $\alpha$. Examples of the steady base flow component ($\UoneB(y)$; dashed line) and the normalized total pulsatile base flow ($(1+1/\Gamma)U(y,t)$; 11 colored lines over the full period, $2\pi$) are overlaid,  at $H=10$, $\Gamma=10$, $\Sr = 5\times10^{-3}$ and $\Rey = 1.5\times10^4$.}
\label{fig:prob}
\end{figure}
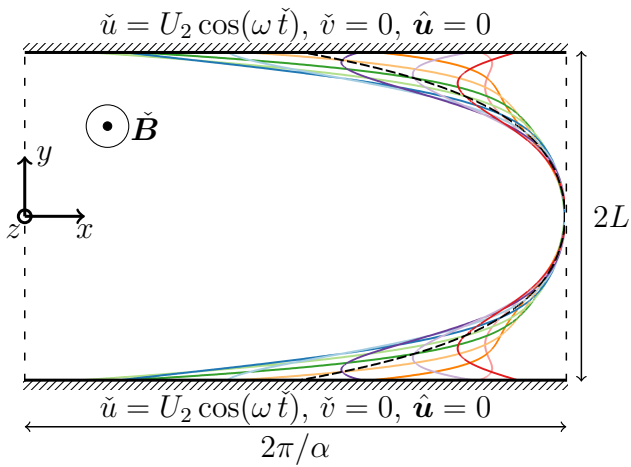

This study considers a duct with rectangular cross-section of wall-normal height $2L$ ($y$ direction) and transverse width $a$ ($z$ direction), subjected to a uniform magnetic field $B\vect{e_z}$. The duct is uniform and of infinite streamwise extent ($x$ direction). A steady base flow component is driven by a constant pressure gradient, producing a maximum undisturbed dimensional velocity $U_1$. An oscillatory base flow component is driven by synchronous oscillation of both lateral walls at velocity $U_2\cos(\omega \check{t})$, with maximum dimensional velocity $U_2$. The pulsatile flow, the sum of the steady and oscillatory components, has a maximum velocity over the cycle of $U_0$. In the limits $\Har = aB(\sigma/\rho\nu)^{1/2} \gg 1$ and $N = aB^2\sigma/\rho U_0 \gg 1$, the flow is Q2D and can be modelled by the SM82 model \citep{Sommeria1982why,Potherat2000effective}. A more detailed assessment of the the validity of the SM82 model follows in Sec.~\ref{sec:vali}. Normalizing lengths by $L$, velocity by $U_0$, time by $1/\omega$ and pressure by $\rho U_0^2$, the governing momentum and mass conservation equations become
\begin{equation} \label{eq:non_dim_m}
\Sr\pde{\vect{u}}{t} = -(\vect{u}\vect{\cdot}\vect{\nabla}_\perp)\vect{u} - \vect{\nabla}_\perp p + \frac{1}{\Rey}\nabla_\perp^2\vect{u} - \frac{H}{\Rey}\vect{u},
\end{equation}
\begin{equation} \label{eq:non_dim_c}
\vect{\nabla_\perp} \vect{\cdot} \vect{u} = 0,
\end{equation}
where $\vect{u}=(u,v)$ is the \qtwod\ velocity vector, representing the $z$-averaged field, and $\vect{\nabla}_\perp=(\partial_x,\partial_y)$ is the two-dimensional gradient operator. Four nondimensional parameters govern this problem: the Reynolds number $\Rey = U_0 L/\nu$, the Strouhal number $\Sr = \omega L /U_0$, the Hartman friction parameter $H=2B(L^2/a)(\sigma/\rho\nu)^{1/2}$ and the amplitude ratio $\Gamma = U_1/U_2$. $\Gamma=0$ represents a flow purely driven by oscillating walls (no pressure gradient) and $\Gamma \rightarrow \infty$ a pressure driven flow (no wall motion). The Womersly number $\Wo^2 = \Sr\Rey$ is sometimes used instead of $\Sr$ as a dimensionless frequency. 

The nondimensional pulsatile base flow is $U(y,t) = \gamma_1\UoneB(y) + \gamma_2\UtwoB(y,t)$, where $\gamma_1=\Gamma/(\Gamma+1)$ and $\gamma_2 = 1/(\Gamma+1)$, following Ref. \citep{Thomas2011linear}, with steady component $\UoneB(y)$ and oscillating component $\UtwoB(y,t)$. This work considers $1 \leq \Gamma<\infty$. Thus, the magnitude of the steady component of the base flow is never smaller than that of the oscillating component, ensuring net transfer of tritium/heat is dominant. The nondimensional wall oscillation is $\cos(t)/\Gamma$, and the maximum velocity over the cycle $U_0 = \max_{\{y,t\}}(U)=1/(1+1/\Gamma)$ for $\Gamma \geq 1$ (henceforth, $\Gamma \geq 1$). The normalized time $\tpe = t/2\pi$ is also defined. To assess the degree of destabilization, the Reynolds number ratio $\rrs = [\Rey/(1+1/\Gamma)]/\mathit{Re}_\mathrm{crit,s}$ is defined, comparing the Reynolds number in this problem to the critical Reynolds number for a purely steady base flow \citep{Camobreco2020transition,Potherat2007quasi, Vo2017linear}. The wave number is similarly rescaled, as $\als = \alpha/\alpha_\mathrm{crit,s}$. 

\begin{figure}
\begin{center}
\addtolength{\extrarowheight}{-10pt}
\addtolength{\tabcolsep}{-2pt}
\begin{tabular}{ llll }
\footnotesize{(a)} & \footnotesize{\hspace{4mm} $\Sr=5\times10^{-2}$, $\Rey=5\times10^3$, $H=1$}  &  &   \\
\makecell{\vspace{26mm} \\  \vspace{34mm} \rotatebox{90}{\footnotesize{$y$}}} & \makecell{\includegraphics[width=0.458\textwidth]{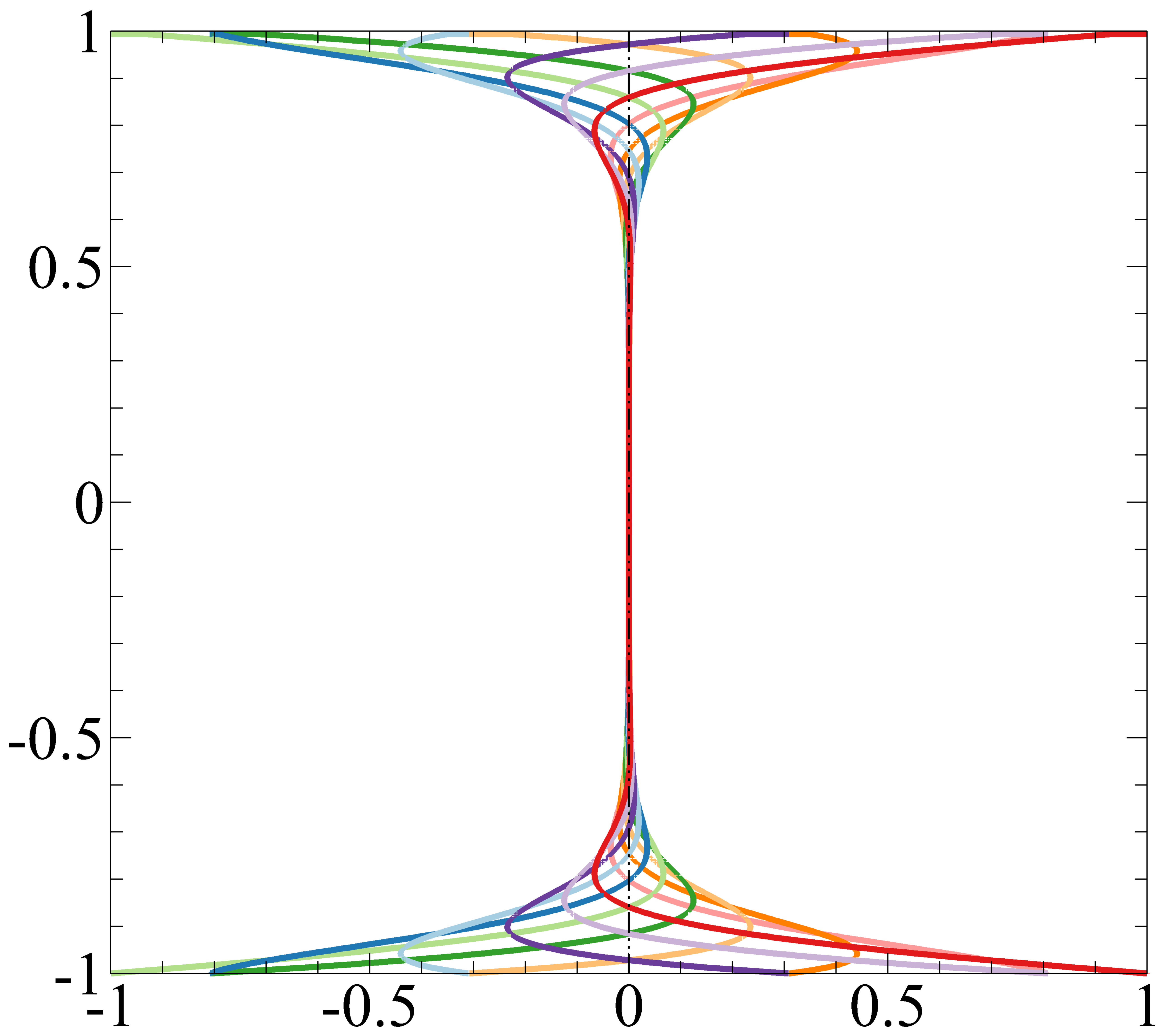}} &
\makecell{\vspace{26mm} \\  \vspace{34mm} \rotatebox{90}{\footnotesize{$y$}}} &
\makecell{\includegraphics[width=0.458\textwidth]{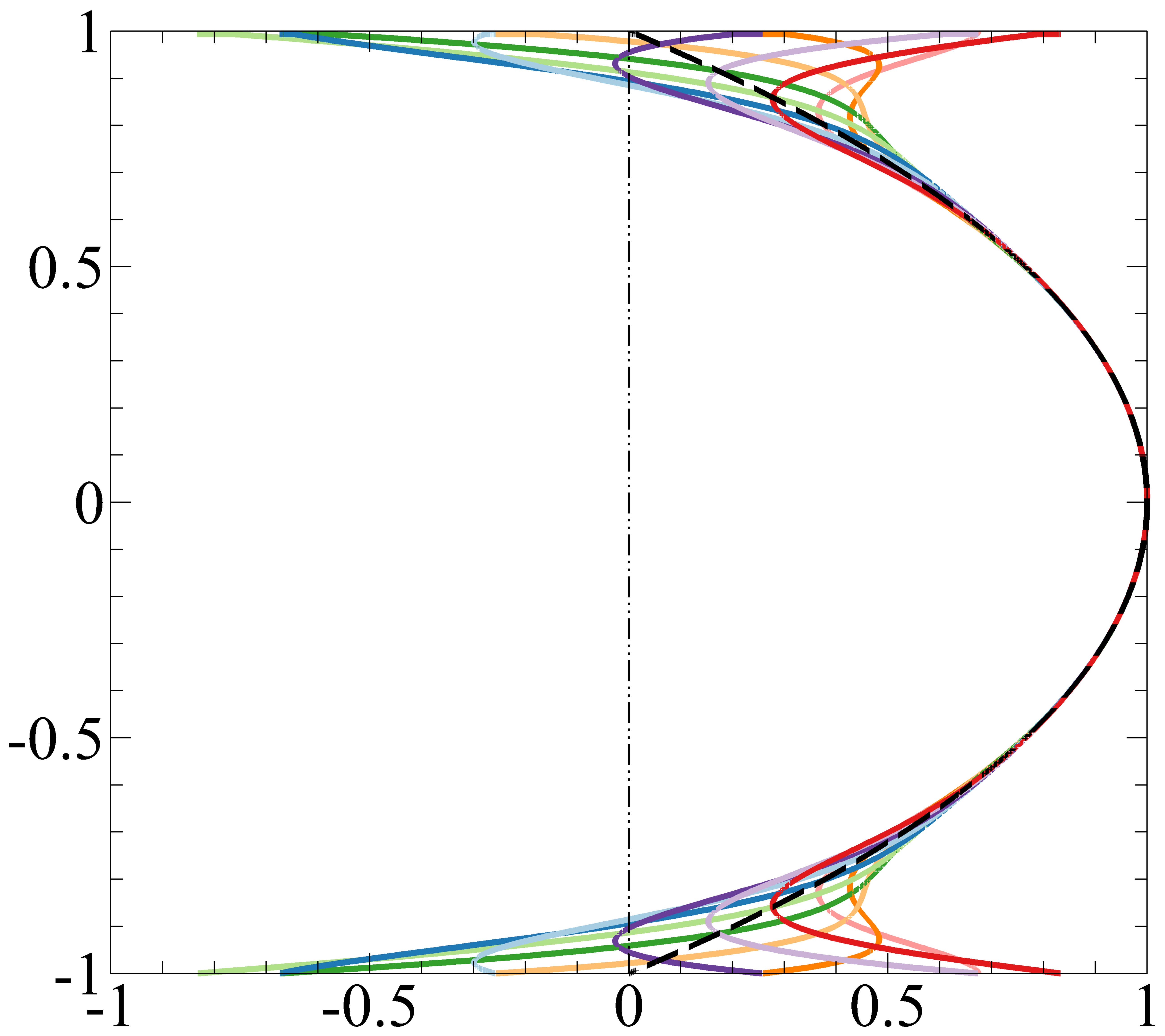}} \\
 \footnotesize{(b)} & \footnotesize{\hspace{4mm} $\Sr=5\times10^{-3}$, $\Rey=1.5\times10^4$, $H=10$}  &  &   \\
\makecell{\vspace{26mm} \\  \vspace{34mm} \rotatebox{90}{\footnotesize{$y$}}} & \makecell{\includegraphics[width=0.458\textwidth]{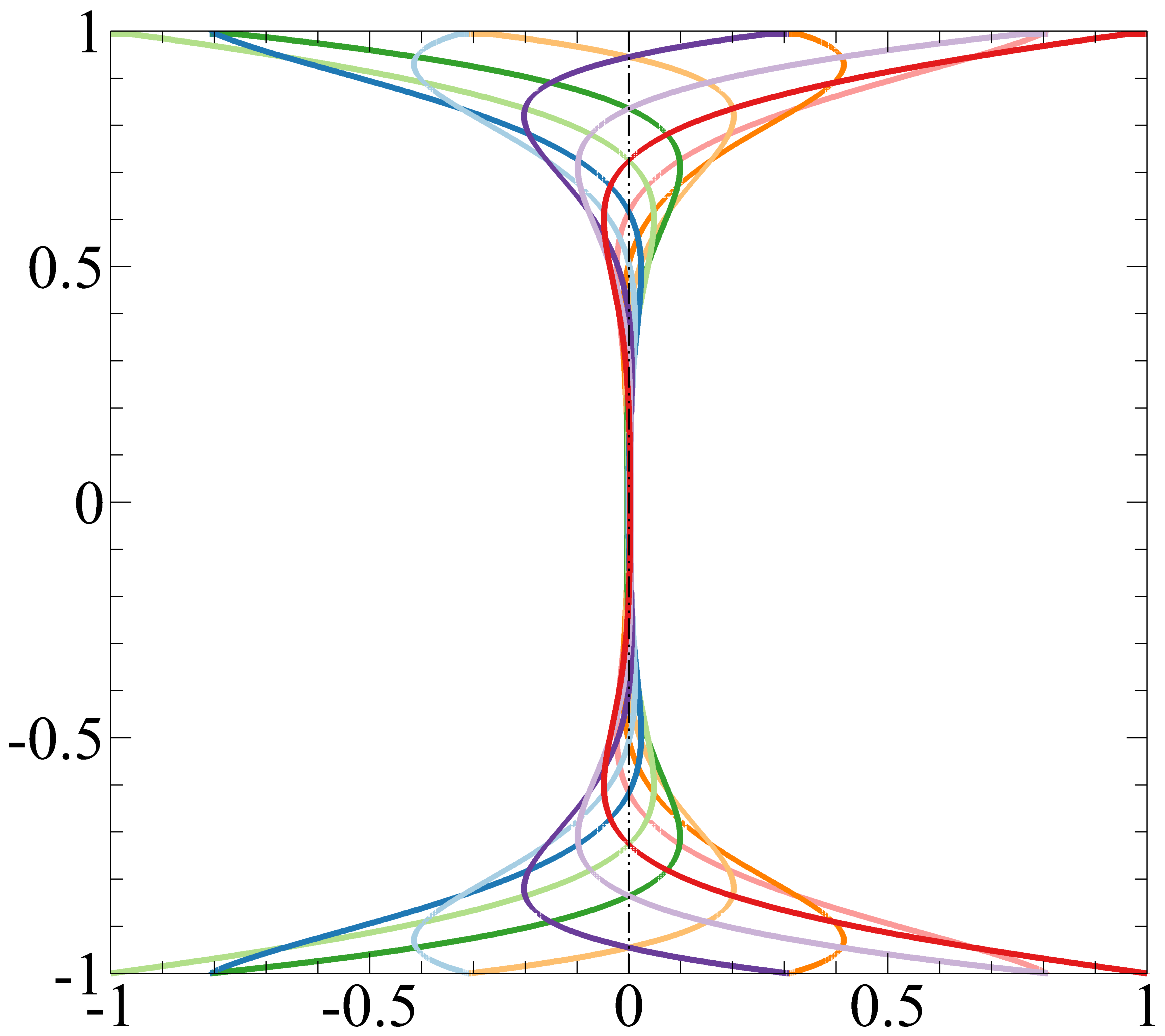}} &
\makecell{\vspace{26mm} \\  \vspace{34mm} \rotatebox{90}{\footnotesize{$y$}}} &
\makecell{\includegraphics[width=0.458\textwidth]{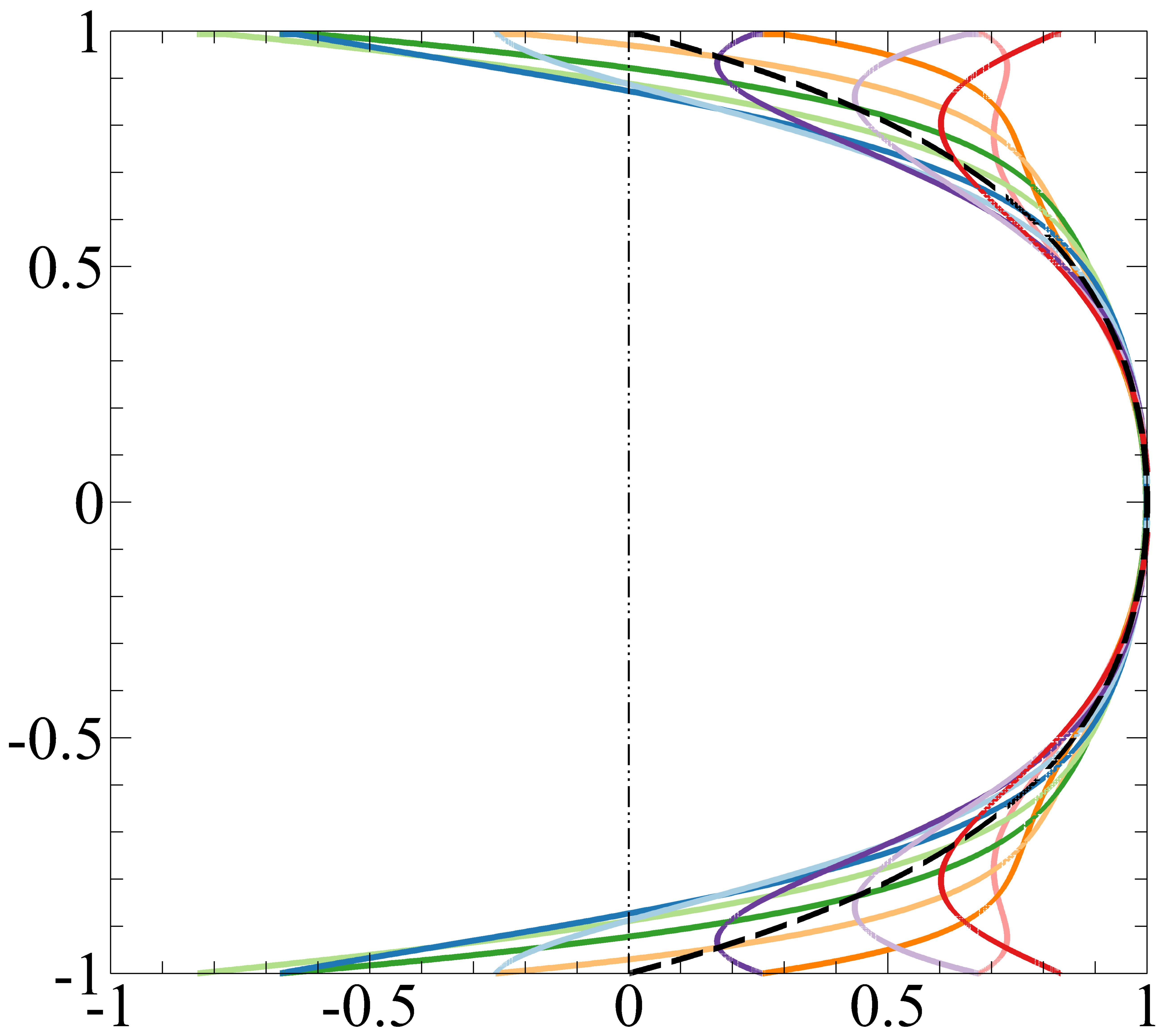}} \\
 \footnotesize{(c)} & \footnotesize{\hspace{4mm} $\Sr=5\times10^{-4}$, $\Rey=4.5\times10^4$, $H=100$}  &  &   \\
\makecell{\vspace{26mm} \\  \vspace{34mm} \rotatebox{90}{\footnotesize{$y$}}} & \makecell{\includegraphics[width=0.458\textwidth]{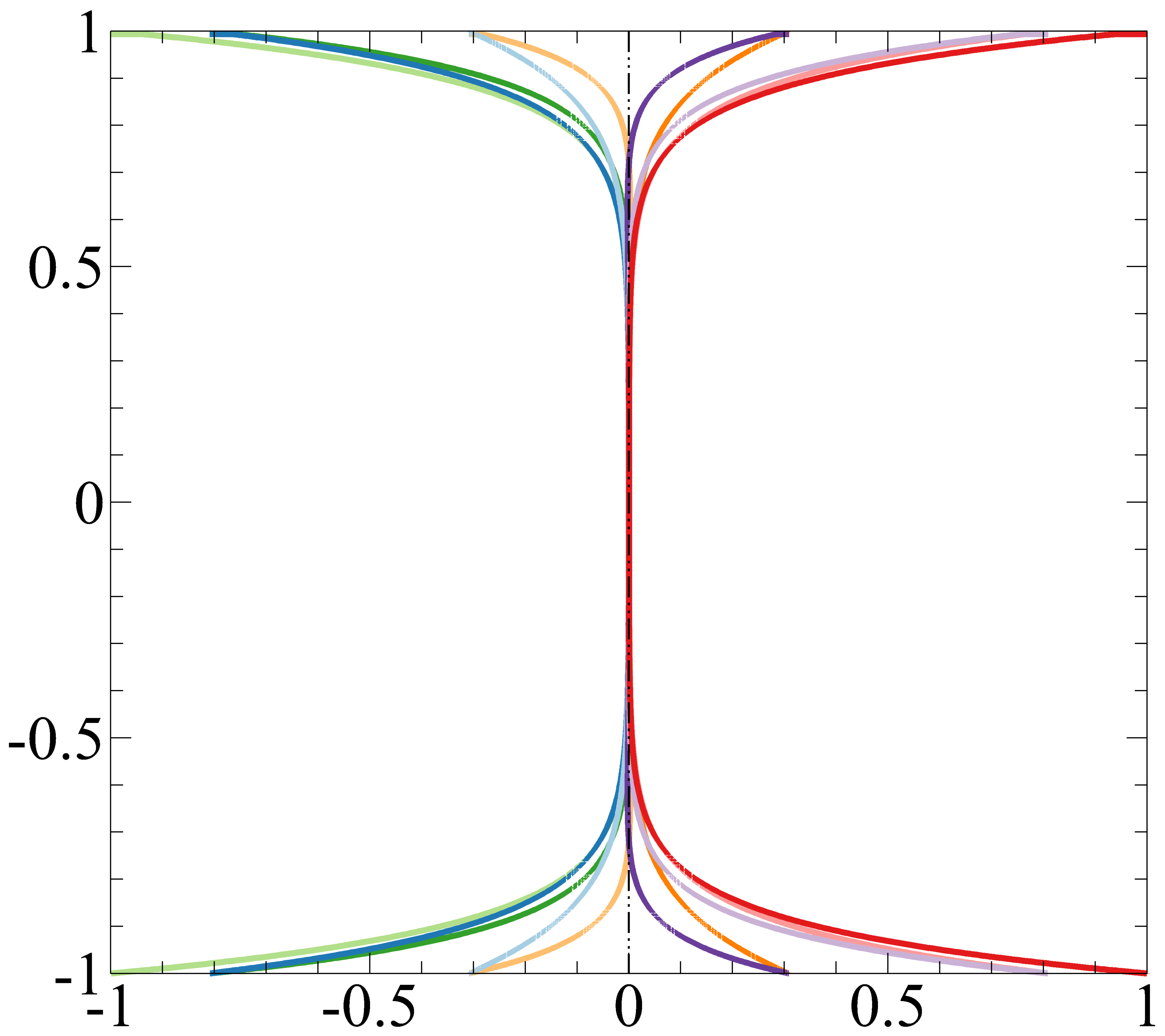}} &
\makecell{\vspace{26mm}  \\  \vspace{34mm} \rotatebox{90}{\footnotesize{$y$}}} & 
\makecell{\includegraphics[width=0.458\textwidth]{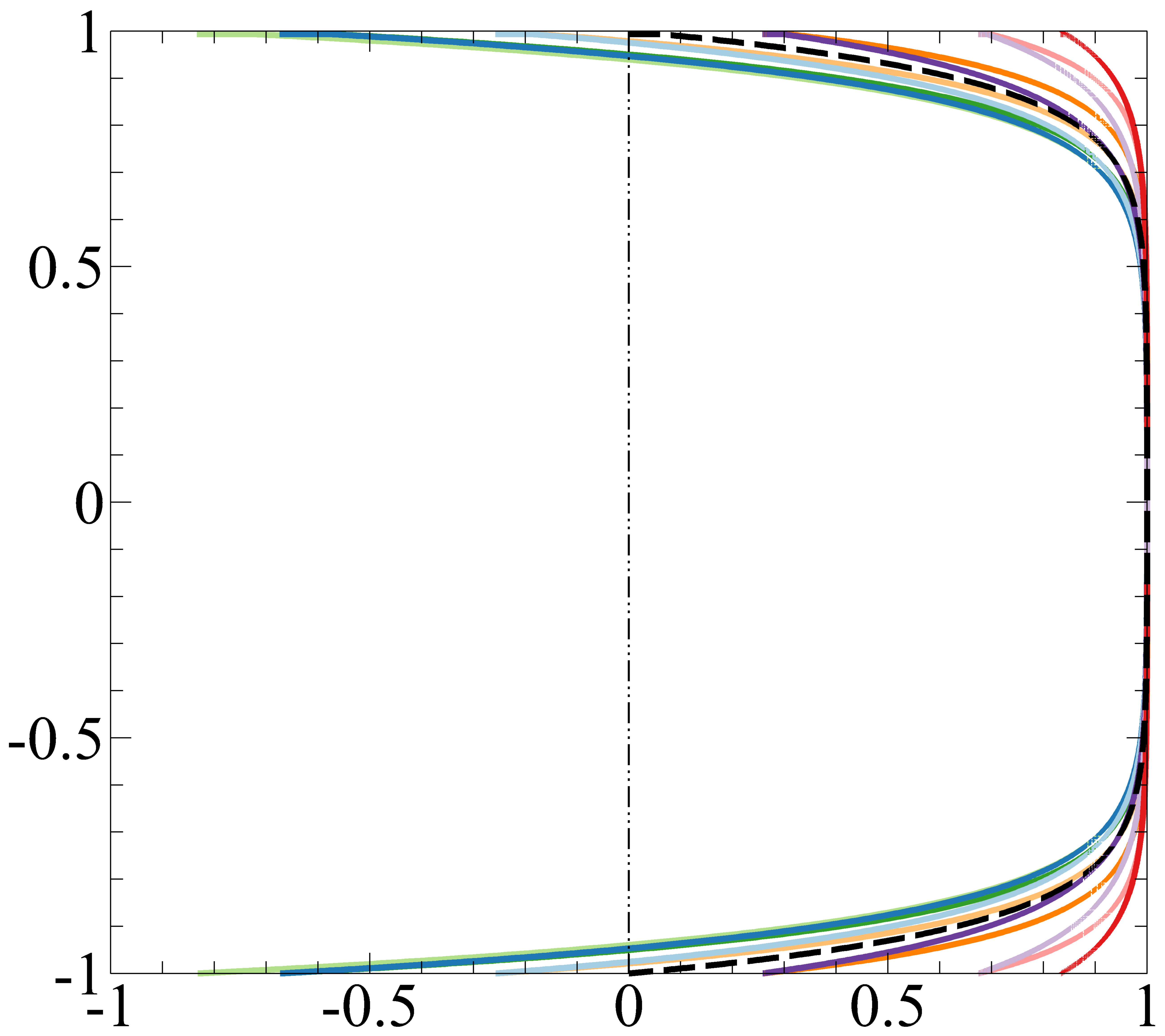}} \\
 & \hspace{35mm} \footnotesize{$\UtwoB$} & & \hspace{26mm} \footnotesize{$(1+1/\Gamma)\,U(y,t)$} \\
\end{tabular}
\addtolength{\tabcolsep}{+2pt}
\addtolength{\extrarowheight}{+10pt}
\end{center}
    \caption{Base flow profiles at $\Gamma=1.2$. Equispaced over one period: oscillating component (left),  $(1+1/\Gamma)$ rescaled pulsatile base flow (right).  A black dashed line denotes the steady component, $\UoneB$.}
    \label{fig:base_flows}
\end{figure}

Instantaneous variables $(\vect{u},p)$ are decomposed into base $(\vect{U},P)$ and perturbation $(\hat{\vect{u}},\hat{p})$ components via small parameter $\epsilon$, as $\vect{u} = \vect{U} + \epsilon \hat{\vect{u}}$; $p = P + \epsilon \hat{p}$. The fully developed, steady, parallel flow $\UoneBv=\UoneB(y)\vect{e_x}$, with boundary conditions $\UoneB(y \pm 1)=0$, and a constant driving pressure gradient scaled to achieve a unit maximum velocity is \citep{Potherat2007quasi},
%
\begin{equation}\label{eq:U1B}
\UoneB = \frac{\cosh(H^{1/2})}{\cosh(H^{1/2})-1}\left(1-\frac{\cosh(H^{1/2}y)}{\cosh(H^{1/2})}\right).
\end{equation}
The fully developed, time periodic, parallel flow $\UtwoBv=\UtwoB(t,y)\vect{e_x} = \UtwoB(t+2\pi,y)\vect{e_x}$, with boundary conditions $\UtwoB(y \pm 1) = \cos(t)$, $\partial \UtwoB/ \partial t |_{y \pm 1}= -\sin(t)$ expresses as
%
\begin{equation}\label{eq:U2B}
\UtwoB  = \Rez\left(\frac{\cosh[(r+si)y]}{\cosh(r+si)}e^{it}\right) = b(y)e^{it} + b^*(y)e^{-it},
\end{equation}
where the inverse boundary layer thickness and the wave number of the wall-normal oscillations are represented by
\begin{eqnarray}\label{eq:rands}
r&=&[(\Sr\Rey)^2+H^2]^{1/4} \cos([\tan^{-1}(\Sr\Rey/H)]/2), \\ \nonumber
s&=&[(\Sr\Rey)^2+H^2]^{1/4} \sin([\tan^{-1}(\Sr\Rey/H)]/2),
\end{eqnarray}
%
respectively, $i=(-1)^{1/2}$ and $*$ represents the complex conjugate. In the hydrodynamic limit of $H \rightarrow 0$, $r=s=(\Sr\Rey/2)^{1/2}$. In the limit of $H \rightarrow \infty$, at constant $\Rey$ and $\Sr$, $r \sim  H^{1/2}$ and $s \rightarrow 0$. If $\Rey$ is also varied, it must vary at a rate $H^p$, with $p \geq 1$, for the limiting cases to differ. Note that the oscillating component of the base flow depends only on two parameters ($\Sr\Rey=\Wo^2$ and $H$). Although these choices mean the base flow is $\Rey$ dependent, they allow $\ReyCrit$ to be found at a constant frequency (constant $\Sr$), as a constant $\Wo$ instead represents a constant oscillating boundary layer thickness. Examples of the base flow at $\Gamma=1.2$ are illustrated in \fig\ \ref{fig:base_flows}, with the total pulsatile profile plotted as $(1+1/\Gamma)\,U(y,t)$ to show oscillation about the steady component $\UoneB$. 

Both dominant transient inertial forces (large $\Sr$) or dominant frictional forces (large $H$) are capable of flattening the central region of the oscillating flow component. In \fig\ \ref{fig:base_flows}(a), the oscillating component is flattened by large transient inertial forces, while the steady flow still exhibits a curved Poiseuille-like profile as $H$ is small. Whereas, in \fig\ \ref{fig:base_flows}(c), it is the large $H$ value that is flattening both the steady and oscillating flow components. However, inflection points, which are important for intracyclic growth, are no longer present in \fig\ \ref{fig:base_flows}(c), as $H$ is large, but can be observed in the boundary layers of \figs\ \ref{fig:base_flows}(a) and \ref{fig:base_flows}(b), as $\Sr$ is large.

It is instructive to consider the velocity profile for the simpler problem of the SM82 equivalent of an isolated Stokes layer, $U(y,t) = e^{-ry}\cos(sy-t)$, where $r$ and $s$ remain as defined in Eq.~(\ref{eq:rands}), except scaled by $H^{-1/2}$ to account for the isolated boundary layer nondimensionalization. This highlights the effects of $r$ and $s$ on the boundary layer, as the base flow becomes akin to a damped harmonic oscillator. Increasing either $H$ or $\Sr\Rey$ increases $r$ in turn, and reduces the boundary layer thickness. However, increasing $H$ reduces $s$. Thus, inflection points are eliminated with increasing $H$, and the boundary layer just appears as shifted exponential profiles, as is observed in \fig\ \ref{fig:base_flows}(c). Decreasing $\Sr\Rey$ reduces $s$, and also eliminates inflection points, whereas increasing $\Sr\Rey$ increases $s$, promoting inflection points, but containing them within a thinner oscillating boundary layer.

It is also worth considering the pulsatile base flow in a broader context, as past literature is divided on the method of oscillation. Among many others, Refs.~\citep{Pier2017linear,Straatman2002hydrodynamic} impose an oscillatory pressure gradient, while Refs.~\citep{Blenn2006linear, Thomas2011linear} impose oscillating walls. For the unbounded, oscillatory Stokes flow, the eigenvalues of the linear operator, with either imposed oscillation, have been proven identical \citep{Blenn2002linear}. Furthermore, it has also been shown that (transient) energy growth is also identical between the two methods of oscillation \citep{Biau2016transient}. However, the full linear and nonlinear problems can be shown to be identical. Defining a motionless frame $G$, and a frame $\bar{G}$ in motion with arbitrary, time varying velocity $\vect{V}(t)$, the two frames are related through:
\begin{equation} \label{eq:gal}
\bar{\vect{x}} = \vect{x} - \int \vect{V} \mathrm{d}t,\,\,\, \bar{t}=t,\,\,\, \bar{\vect{u}} = \vect{u} - \vect{V}.
\end{equation}
Under extended Galilean invariance, $\partial \bar{\vect{u}}/\partial \bar{\vect{x}} = \partial \vect{u}/\partial \vect{x}$ and $\Sr\partial\bar{\vect{u}}/\partial \bar{t} + (\bar{\vect{u}} \bcdot \bar{\bnabla}_\perp)\bar{\vect{u}} = \Sr(\partial \vect{u}/\partial t - \partial \vect{V}/\partial t) + (\vect{u} \bcdot \bnabla_\perp)\vect{u}$ \citep{Pope2000turbulent}. In the frame $G$, a constant driving pressure gradient, and oscillatory wall motion $U(y\pm1,t)=\UtwoB(y\pm1,t)/\Gamma$, are imposed. $\vect{V}(t) = (\UtwoB(y\pm1,t)/\Gamma,0)$ is selected so the walls appear stationary, $\bar{U}(y\pm1,t)=0$, in the moving frame $\bar{G}$ . Substituting the relations in Eq.~(\ref{eq:gal}) into Eqs.~(\ref{eq:non_dim_m}) and (\ref{eq:non_dim_c}), the governing equations in the moving frame become
\begin{equation} \label{eq:non_dim_m_move}
\Sr\bigg(\pde{\bar{\vect{u}}}{\bar{t}} + \pde{\vect{V}}{t}\bigg) = -(\bar{\vect{u}}\vect{\cdot}\bar{\vect{\nabla}}_\perp)\bar{\vect{u}} - \bar{\vect{\nabla}}_\perp p + \frac{1}{\Rey}\bar{\nabla}_\perp^2\bar{\vect{u}} - \frac{H}{\Rey}(\bar{\vect{u}}+\vect{V}),
\end{equation}
\begin{equation} \label{eq:non_dim_c_move}
\bar{\vect{\nabla}}_\perp \vect{\cdot} \bar{\vect{u}} = 0.
\end{equation}
As the pressure does not have a conversion relation, the driving pressure in the moving frame can be freely chosen as
%
\begin{equation}\label{eq:pres}
\bar{p}(t) = p+\frac{x}{\Gamma}\bigg(\Sr\pde{\UtwoB(y\pm1,t)}{t} + \frac{H}{\Rey}\UtwoB(y\pm1,t)\bigg).
\end{equation}
Substituting Eq.~(\ref{eq:pres}) into Eq.~(\ref{eq:non_dim_m_move}) and cancelling yields
\begin{equation} \label{eq:non_dim_m_can}
\Sr\pde{\bar{\vect{u}}}{\bar{t}} = -(\bar{\vect{u}}\vect{\cdot}\bar{\vect{\nabla}}_\perp)\bar{\vect{u}} - \bar{\vect{\nabla}}_\perp \bar{p} + \frac{1}{\Rey}\bar{\nabla}_\perp^2\bar{\vect{u}} - \frac{H}{\Rey}\bar{\vect{u}},
\end{equation}
\begin{equation} \label{eq:non_dim_c_can}
\bar{\vect{\nabla}}_\perp \vect{\cdot} \bar{\vect{u}} = 0.
\end{equation}
Thus, in the frame $\bar{G}$, the governing equations, Eqs.~(\ref{eq:non_dim_m_can}) and (\ref{eq:non_dim_c_can}), are identical to the governing equations in $G$, Eqs.~(\ref{eq:non_dim_m}) and (\ref{eq:non_dim_c}). However, in $\bar{G}$ the walls are stationary, and the pressure forcing $\bar{p}$ is the sum of a steady and oscillatory component. Thus, the linear and nonlinear dynamics when the flow is driven by oscillatory wall motion ($G$), or an oscillatory pressure gradient ($\bar{G}$), are identical in all respects, as they are both the same problem viewed in different frames of reference. These arguments do not hold if $H=0$ in the steady limit ($\Gamma \rightarrow \infty$, $\UtwoB=0$), or if the oscillation of both walls is not synchronous. Note that the constant pressure gradient in the fixed frame could also be considered as a constant wall motion, for non-zero $H$. If so, the oscillations would be about a finite wall velocity, rather than about zero.

\subsection{Validity of SM82 for pulsatile flows} \label{sec:vali}

With the pulsatile base flow established, the realm of validity of the SM82 model is assessed. The dimensional equation governing the induced magnetic field $\cvb$ is \cite{Muller2001magnetofluiddynamics}, 
\begin{equation} \label{eq:mag_ind}
\pde{\cvb}{\check{t}} = B_0(\vect{e}_z \vect{\cdot} \cvn )\cvu + (\cvb \vect{\cdot} \cvn )\cvu - (\cvu \vect{\cdot} \cvn) \cvb + \frac{1}{\mu_0\sigma} \check{\nabla}^2 \cvb,
\end{equation}
where a background uniform steady field $B_0\vect{e}_z$ is imposed. The aim is to show that the induced magnetic field diffuses $R_\mathrm{m}$ times faster than it locally varies, where the magnetic Reynolds number $R_\mathrm{m}=\mu_0 \sigma U_1 L$ and where $\mu_0$ is the permeability of free space. The low-$R_\mathrm{m}$ approximation assumes that one of the bilinear terms is much smaller than the diffusive term, $|(\cvu \vect{\cdot} \cvn) \cvb| \ll |(\mu_0\sigma)^{-1} \check{\nabla}^2 \cvb|$. Once non-dimensionalized by $U_1$ and $L$ this imposes an $R_\mathrm{m} \ll 1$ constraint. This is well satisfied for liquid metal duct flows, with $R_\mathrm{m}$ of the order of $10^{-2}$ \citep{Moreau1990magnetohydrodynamics,Knaepen2004magnetohydrodynamic}. Note that $|B_0(\vect{e}_z \vect{\cdot} \cvn )\cvu|$ remains of the same order as $|(\mu_0\sigma)^{-1} \check{\nabla}^2 \cvb|$ when the background magnetic field is imposed.

The quasi-static approximation assumes $|\partial \cvb/\partial t| \ll |(\mu_0\sigma)^{-1} \check{\nabla}^2 \cvb|$. Note that a low $R_\mathrm{m}$ does not necessarily imply that $|\partial \cvb/\partial \check{t}|$ is small. Based on a typical out-of-plane steady velocity scale of $a/U_1$,  $|\partial \cvb/\partial \check{t}|$ may be reasonably assumed to scale as $|(\cvu \vect{\cdot} \cvn) \cvb|$, and thereby be small if $R_\mathrm{m}$ were small. However, a pulsatile flow introduces an additional velocity timescale, based on the forcing frequency, to also compare against. Hence, non-dimensionalizing $|\partial \cvb/\partial \check{t}| \ll |(\mu_0\sigma)^{-1} \check{\nabla}^2 \cvb|$ based on a timescale of $1/\omega$ yields a constraint on the shielding parameter $R_\omega = \mu_0 \sigma \omega L^2 \ll 1$ \cite{Moreau1990magnetohydrodynamics}. This translates to $R_\mathrm{m} \Sr \ll 1$, or $\Sr \ll R_\mathrm{m}^{-1}$, to ensure that diffusion of the induced field is not contained to small boundary regions of the domain. Given $R_\mathrm{m}$ of $10^{-2}$ is typical of liquid metal duct flows at moderate Reynolds numbers \citep{Moreau1990magnetohydrodynamics,Knaepen2004magnetohydrodynamic}, since $R_\mathrm{m}=\Rey\Pra_m$ and the magnetic Prandtl number $\Pra_m=\nu\mu_0\sigma$ is of the order of $10^{-6}$ for liquid metals \cite{Potherat2015decay}, the shielding condition of $\Sr \ll R_\mathrm{m}^{-1}$ requires $\Sr \ll 100$.

Furthermore, for the induced magnetic field to be treated as steady, the induced magnetic field must vary rapidly relative to a slowly varying velocity field. This requires the Alfv\'{e}n timescale (time taken for the Alfv\'{e}n velocity to cross the duct width) be much smaller than the pulsation (transient inertial) timescale. The Alfv\'{e}n velocity $v_\mathrm{A} = B/(\mu_0 \rho)^{1/2} = (N_L/R_\mathrm{m})^{1/2}(U_1 L/a)$ is expressed in terms of the interaction parameter $N_L=a^2B^2\sigma/\rho U_1 L$. Thus the Alfv\'{e}n timescale is $\tau_\mathrm{A} = a/v_\mathrm{A} = (R_\mathrm{m}/N_L)^{1/2}(a^2/U_1 L)$, while the steady inertial timescale $\tau_\mathrm{I,L}=L/U_1$ and the pulsation timescale $\tau_\mathrm{P}=1/\omega$. Thus, $\tau_\mathrm{A}/\tau_\mathrm{I,L} =  (R_\mathrm{m}/N_L)^{1/2}(a^2/L^2)$ and $\tau_\mathrm{A}/\tau_\mathrm{P} =  (R_\mathrm{m}/N_L)^{1/2}Sr(U_0/U_1)(a^2/L^2)$. If $\Sr(U_0/U_1)<1$, or equally $\Sr(1+1/\Gamma)<1$, no SM82 assumptions are in question. This requires $\Sr < 1/2$ at $\Gamma=1$ (and $\Sr < 1$  for $\Gamma \rightarrow \infty$) at equivalent $N \gg 1$ and $R_\mathrm{m} \ll 1$ conditions as for a steady case. Recall that $\Sr \ll 100$ was required from the shielding constraint.

Finally, the quasi-static approximation is only valid if Alfv\'{e}n waves dissipate much faster than they propagate. This is ensured if $|\partial \cvb/\partial t| \ll |(\mu_0\sigma)^{-1} \check{\nabla}^2 \cvb|$ is satisfied when considering the last remaining characteristic timescale, the Alfv\'{e}n timescale $\tau_\mathrm{A} = a/v_\mathrm{A}$. This places a condition on the Lundquist number $S=(N_L R_\mathrm{m})^{1/2} = \Har \Pra_m^{1/2} \ll 1$. Given $\Pra_m$ of the order of $10^{-6}$ \cite{Potherat2015decay}, and with $R_\mathrm{m}$ of $10^{-2}$ \citep{Moreau1990magnetohydrodynamics,Knaepen2004magnetohydrodynamic}, this translates to conditions on the interaction parameter and Hartmann number of $N_L \lesssim 100$ and $\Har \lesssim 1000$, respectively.

An additional component of the SM82 model is the quasi-two-dimensional approximation, which requires the timescale for two-dimensionalization to occur via diffusion of momentum along magnetic field lines, $\tau_\mathrm{2D}=(\rho/\sigma B^2)(a^2/L^2)=(1/N_L)(a^4/U_1L^3)$ \citep{Potherat2007quasi}, be much smaller than the inertial and pulsation timescales. These ratios are $\tau_\mathrm{2D}/\tau_\mathrm{I,L}=(1/N_L)(a^4/L^4)$ and $\tau_\mathrm{2D}/\tau_\mathrm{P}=(\Sr/N_L)(U_0/U_1)(a^4/L^4)$. Thus, if $\Sr < 1/2$ for otherwise equivalent conditions as for a steady case, momentum is diffused across the duct more rapidly by the magnetic field than by steady or transient inertial forces. The SM82 approximation also assumes $1 \ll \Har \lesssim 1000$ and $N \gg 1$, $N_L \lesssim 100$. These constraints can be met with any $H$ if $a$ and $L$ are chosen appropriately, as discussed in Ref.~\citep{Vo2017linear}. 

The SM82 model is more generally applicable to flows which exhibit a linear friction and a strong tendency to two-dimensionalize. Axisymmetric quasigeostrophic flows, with frictional forces imparted by Ekman layers, and Hele-Shaw (shallow water) flows, with Rayleigh friction, both tend to two-dimensionality if the aspect ratio $L/a$ is small. In these flows a formally equivalent Q2D model can be derived \cite{Buhler1996instabilities,Vo2015effect} (with the addition of a term representing the Coriolis force in the quasigeostrophic case), although the physical meaning of the friction term differs, as do the bounds of validity \cite{Vo2017linear}.

\section{Linear stability analysis}\label{sec:lin_all}
\subsection{Formulation and validation}\label{sec:lin_for}

Linear stability is assessed via the exponential growth rate of disturbances, with unstable perturbations exhibiting net growth each period. The linearized evolution equations
\begin{equation} \label{eq:non_dim_lin_m}
\Sr\pde{\hat{\vect{u}}}{t} = -(\hat{\vect{u}}\vect{\cdot}\vect{\nabla}_\perp)\vect{U} -(\vect{U} \vect{\cdot}\vect{\nabla}_\perp)\hat{\vect{u}} - \vect{\nabla}_\perp \hat{p} + \frac{1}{\Rey}\nabla_\perp^2\hat{\vect{u}} - \frac{H}{\Rey}\hat{\vect{u}},
\end{equation}
\begin{equation} \label{eq:non_dim_lin_c}
\vect{\nabla_\perp} \vect{\cdot} \hat{\vect{u}} = 0.
\end{equation}
%
are obtained by neglecting terms of $O(\epsilon^2)$ in the decomposed Navier--Stokes equations. A single fourth-order equation governing the linearized evolution of the perturbation is obtained by taking twice the curl of Eq.~(\ref{eq:non_dim_lin_m}), and substituting Eq.~(\ref{eq:non_dim_lin_c}). By additionally decomposing perturbations into plane wave solutions of the form $\hat{v}(y,t)=e^{i\alpha x}\tilde{v}(y,t)$, by virtue of the streamwise invariant base flow $U(y,t)$, yields
\begin{equation} \label{eq:linearised_v}
\pde{\tilde{v}}{t}   = \mathscr{L}^{-1}\left[\frac{i\alpha}{\Sr}\pdesqr{U}{y} - \frac{Ui\alpha}{\Sr} \mathscr{L} + \frac{1}{\Sr\Rey}\mathscr{L}^2 - \frac{H}{\Sr\Rey}\mathscr{L} \right]\tilde{v},
\end{equation}
where $\mathscr{L}=(\partial^2/\partial y^2 - \alpha^2)$, and where the perturbation eigenvector $\tilde{v}(y,t)$ still contains both exponential and intracyclic time dependence. Integrating Eq.~(\ref{eq:linearised_v}) forward in time, with a third-order forward Adams--Bashforth scheme \cite{Hairer1993solving}, and with the renormalization $\twonv=1$ at the start of each period, forms the timestepper method. After sufficient forward evolution all but the fastest growing mode is washed away, providing the net growth of the leading eigenmode over one period. A Krylov subspace scheme \citep{Barkley2008direct} is also implemented to aid convergence and provide the leading few eigenvalues $\lambda_j$ with largest growth rate (real component). The domain $y \in [-1,1]$ is discretized with $\Nc+1$ Chebyshev nodes. The derivative operators, incorporating boundary conditions, are approximated with spectral derivative matrices \citep{Trefethen2000spectral}. The spatial resolution requirements are halved by incorporating a symmetry (resp.~antisymmetry) condition along the duct centreline, and resolving even (resp.~odd) perturbations separately. Even perturbations were consistently found to be less stable than odd perturbations.

The eigenvalues of the discretized forward evolution operator are also determined with a Floquet matrix approach \citep{Blenn2006linear,Thomas2011linear}. The exponential and time periodic growth components of the eigenvector are separated by defining
\begin{equation} \label{eq:floq_dec}
\tilde{v}(y,t) = e^{\mu_\mathrm{F} t} \sum_{n=-\infty}^{n=\infty} \tilde{v}_n(y) e^{int},
\end{equation}
with Floquet multiplier $\mu_\mathrm{F}$ and harmonic $n$. This sum is numerically truncated to $n \in [-T,T]$, to obtain a finite set of coupled equations 
\begin{eqnarray} \label{eq:floq_odes}
\mu \tilde{v}_n = -\frac{i\alpha}{\Sr} \bigg( M\tilde{v}_{n+1} &+&M^*\tilde{v}_{n-1}\bigg) \\ + \bigg\{\frac{1}{\Sr\Rey}\mathscr{L}^{-1}\mathscr{L}^2 &-& \frac{H}{\Sr\Rey} -in  -\frac{i\alpha\gamma_1}{\Sr}\bigg[\mathscr{L}^{-1}\bigg(\UoneB\mathscr{L} - \frac{\partial^2 \UoneB}{\partial y^2}\bigg)\bigg] \bigg\}\tilde{v}_n \nonumber,
\end{eqnarray}
after substituting Eq.~(\ref{eq:floq_dec}) into Eq.~(\ref{eq:linearised_v}), where $M=\gamma_2[\mathscr{L}^{-1}(b\mathscr{L}-\partial^2 b/\partial y^2)]$. This system of Chebyshev-discretized equations is set up as a block tridiagonal system, with the coefficients of $\tilde{v}_{n+1}$, $\tilde{v}_n$ and $\tilde{v}_{n-1}$ placed on super-, central- and sub-diagonals, respectively. Spectral derivative matrices are built as before. The MATLAB function \texttt{eigs} is used to find a subset of eigenvalues of the block tridiagonal system located near zero real component (neutral stability), with convergence tolerance $10^{-14}$. $\Rey$ and $\alpha$ are varied until only a single wave number, $\alphaCrit$, attains zero growth rate, at $\ReyCrit$ (for specified $\Sr$, $\Gamma$ and $H$). 

The numerical requirements for the Floquet and timestepper approaches are highly parameter dependent. Validation against the hydrodynamic oscillatory problem \citep{Blenn2006linear} is provided in \tbl\ \ref{tab:tab_1}. Further assurance of the validity of the numerical method is provided in the excellent agreement between pulsatile and steady $\ReyCrit$ values (e.g.~$\rrs \rightarrow 1$) at very small and large $\Sr$ in Sec.~\ref{sec:lin_res1}, and the agreement between the timestepper and Floquet growth rates shown in Sec.~\ref{sec:lin_res1}. Sporadic resolution testing, post determination of $\ReyCrit$, was also performed, with an example shown in \tbl\ \ref{tab:tab_2}.

\begin{table}
\begin{center}
\begin{tabular}{ cccccc } 
\hline
$\Nc$ ($T=300$)  &  $\Rez({\lambda_1})$ & $|$\% Error$|$ & $T$ ($\Nc=150$) & $\Rez({\lambda_1})$   & $|$\% Error$|$    \\
\hline
50  & 0.4719273115651  & 3.02$\times10^1$    & 200 & 0.9493815978240 & 4.04$\times10^1$   \\
100 & 0.6762032203289  & 6.39$\times10^{-3}$ & 250 & 0.6761968753200 & 5.45$\times10^{-3}$  \\
150 & 0.6761968755932  & 5.45$\times10^{-3}$ & 300 & 0.6761968755932 & 5.45$\times10^{-3}$  \\
Ref.~\citep{Blenn2006linear}, even & 0.67616   & 0 &  & 0.67616 & 0 \\
\hline
50  & 0.4689789806609 & $3.06\times10^1$    & 200 & 0.8329627125585 & $2.33\times10^1$   \\
100 & 0.6756830883343 & $6.38\times10^{-3}$ & 250 & 0.6756767389579 & $5.44\times10^{-3}$  \\
150 & 0.6756767389579 & $5.44\times10^{-3}$ & 300 & 0.6756767389579 & $5.44\times10^{-3}$  \\
Ref.~\citep{Blenn2006linear}, odd &  0.67564  & 0 &  & 0.67564 & 0 \\
\hline
\end{tabular}
\caption{$\Gamma=0$, $H=0$ cases validating and testing the resolution of the Floquet matrix method, considering the real part of even and odd modes separately. From Ref.~\citep{Blenn2006linear}, parameters convert as $\Sr=h_\mathrm{BB06}/\Rey_\mathrm{BB06}$ and $\Rey = 2h_\mathrm{BB06}\Rey_\mathrm{BB06}$, where $h_\mathrm{BB06}=16$ and $\Rey_\mathrm{BB06}=847.5$. $\Nc$ accounts for the entire domain.}
\label{tab:tab_1}
\end{center}
\end{table}

%
%
%

As a rough guide, for the Floquet method, $\Nc$ varies between $100$ and $400$ and $T$ between $100$ and $600$, with an eigenvalue subset size of around $200$. For the timestepper, $\Nc$ varies between $40$ to $240$, with $10^5$ to $4\times10^7$ time steps per period, and $6$ to $4000$ iterations. As discussed in Refs.~\citep{Thomas2011linear, Pier2017linear}, with increasing pulsation amplitude (decreasing $\Gamma$), decreasing $\Sr$ and increasing $\Rey$, the intracylcic growth can become stupendously large. The matrix method becomes problematic when the intracylcic growth exceeds four to six orders of magnitude, while the timestepper withstands approximately ten to fifteen orders of magnitude of intracylcic growth (the perturbation norm $\twonv$ does not cleanly converge thereafter). Very roughly, for $\Sr \lesssim 10^{-3}$ and/or $\Gamma \lesssim 2$ and/or $\Rey \gtrsim 10^5$ when $H \geq 10$ the intracyclic growth was greater than even the timestepper could handle. However, given the specific aims of this work, this does not obstruct too large a fraction of the parameter space we wish to explore. 

\begin{table}
\begin{center}
\begin{tabular}{ cccccc } 
\hline
$\Nc$  &  \makecell{Time steps \\ (per period)} & Iterations & \makecell{ $\twonv$ \\ (final iteration)}  & $\Rez(\lambda_1)$  & $\Imz(\lambda_1)$   \\
\hline
100 & $4\times10^5$ & 40 & 0.991293824970121 & -0.001391699032636 & 0.962888347220989   \\
140 & $4\times10^5$ & 20 & 1.000006449491397 &  0.000001028054446 & 0.955814791449918  \\
180 & $4\times10^5$ & 20 & 0.999993672187703 & -0.000001007773833 & 0.955795855565797  \\
220 & $7\times10^5$ & 20 & 0.999993546425103 & -0.000001027780526 & 0.955795848436100 \\
240 & $10^6$        & 10 & 0.999993662207549 & -0.000001011050606 & 0.955795855979765  \\
\hline
\end{tabular}
\caption{Resolution test at $H=10$, $\Gamma=10$ (at large $\Rey$, and small $\Sr=1.12\times10^{-2}$). The Floquet method was used to determine $\ReyCrit=8.1243\times10^5$ and $\alphaCrit=0.91137$, at $\Nc=200$ and $T=400$. This $\ReyCrit$ and $\alphaCrit$ were input into the timestepper to validate the timestepper, and the Floquet $\ReyCrit$ value (note the neutrally stable growth rate $\Rez(\lambda_1)\approx 0$). $\Nc$ accounts for the entire domain, with an even mode enforced.}
\label{tab:tab_2}
\end{center}
\end{table}

\subsection{Long-term behavior}\label{sec:lin_res1}

A neutrally stable perturbation exhibits no net growth or decay over each cycle. Neutral stability is first achieved at $\ReyCrit$ and $\alphaCrit$, as $\Rey$ is increased. However, such a definition conceals the intracylic dynamics, which strongly influence $\ReyCrit$, as is further discussed in Sec.~\ref{sec:lin_res2}. 

\begin{figure}
\begin{center}
\addtolength{\extrarowheight}{-10pt}
\addtolength{\tabcolsep}{-2pt}
\begin{tabular}{ llll }
\makecell{\vspace{15mm} \footnotesize{(a)} \\  \vspace{26mm} \rotatebox{90}{\footnotesize{$\ReyCrit/(1+1/\Gamma)$}}} & \makecell{\includegraphics[width=0.458\textwidth]{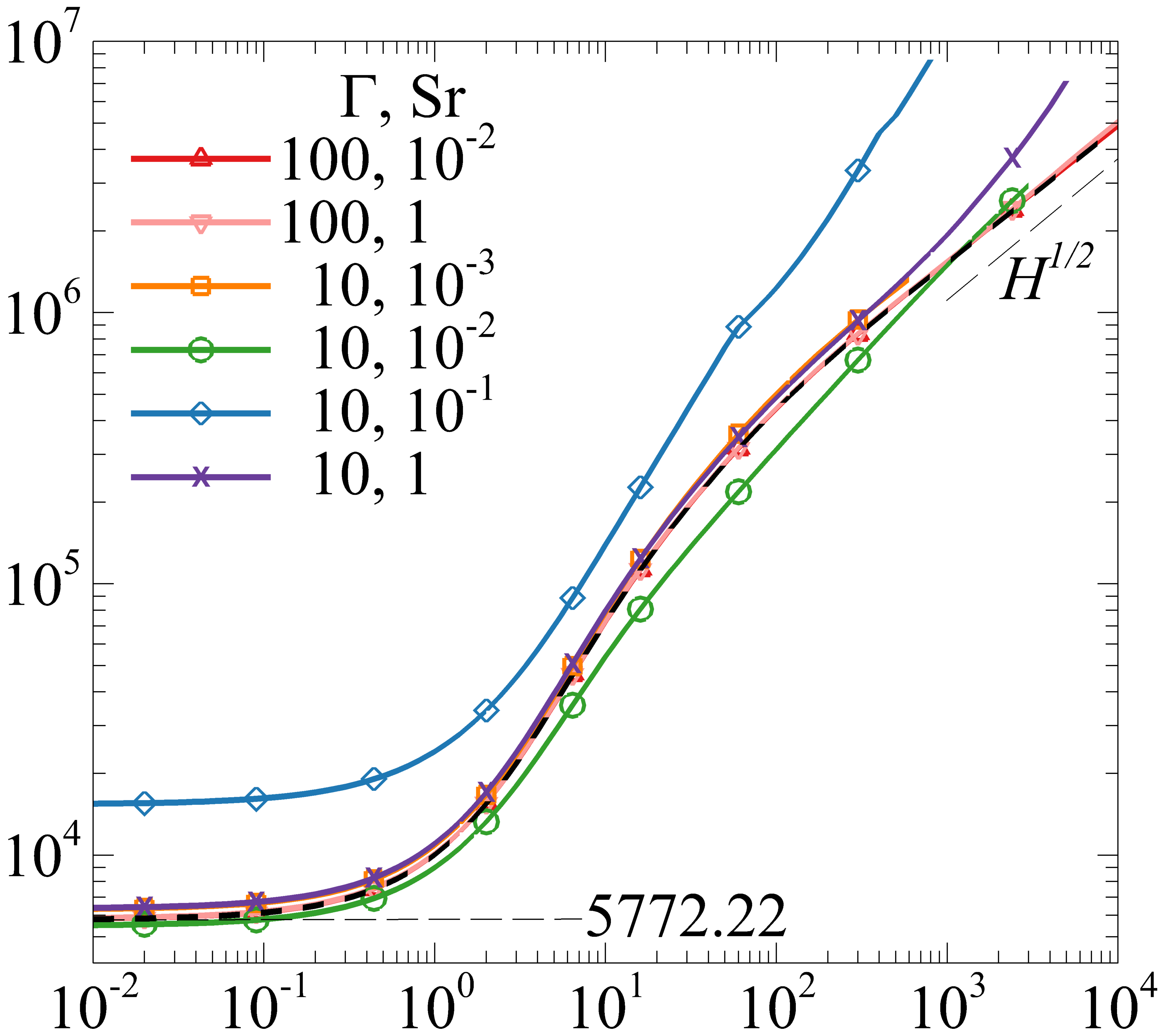}} &
\makecell{\vspace{24mm} \footnotesize{(b)} \\  \vspace{34mm} \rotatebox{90}{\footnotesize{$\alphaCrit$}}}
 & \makecell{\includegraphics[width=0.458\textwidth]{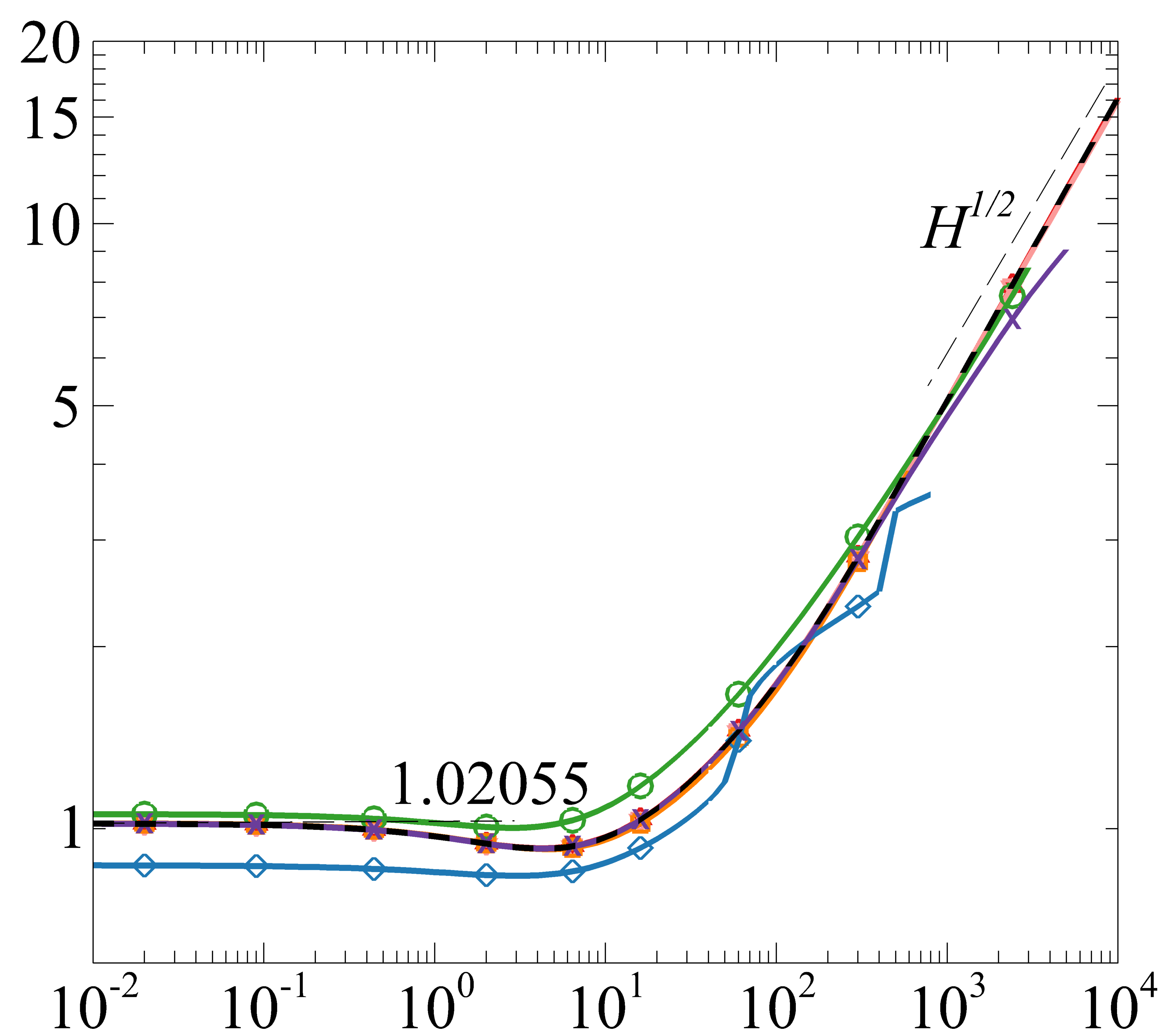}} \\
 & \hspace{36mm} \footnotesize{$H$} & & \hspace{36mm} \footnotesize{$H$} \\
 \makecell{\vspace{25mm} \footnotesize{(c)} \\  \vspace{35mm} \rotatebox{90}{\footnotesize{$\rrs$}}} & \makecell{\includegraphics[width=0.458\textwidth]{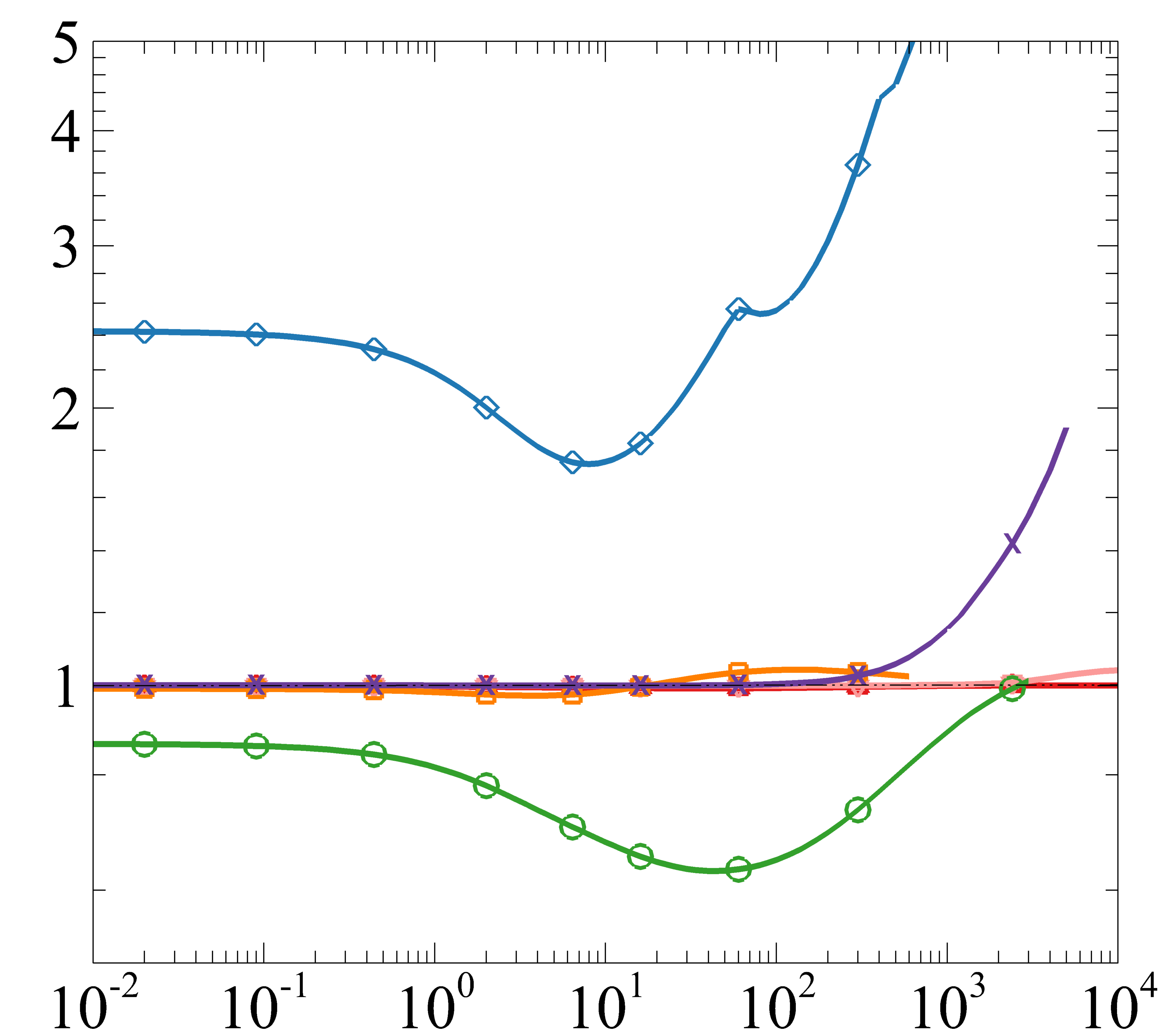}} &
\makecell{\vspace{25mm} \footnotesize{(d)} \\  \vspace{35mm} \rotatebox{90}{\footnotesize{$\als$}}}
 & \makecell{\includegraphics[width=0.458\textwidth]{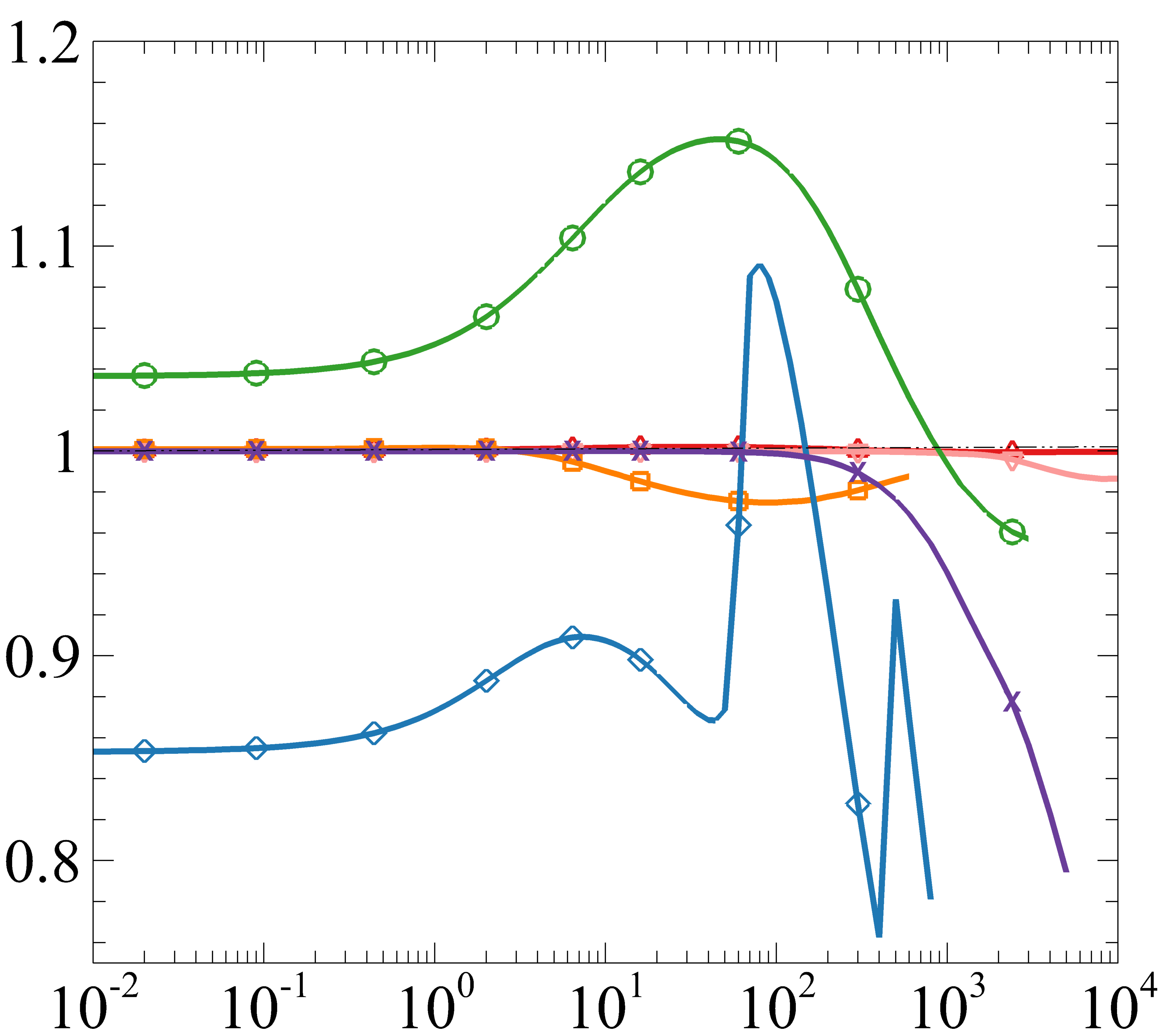}} \\
 & \hspace{36mm} \footnotesize{$H$} & & \hspace{36mm} \footnotesize{$H$} \\
\end{tabular}
\addtolength{\tabcolsep}{+2pt}
\addtolength{\extrarowheight}{+10pt}
\end{center}
    \caption{Rescaled $\ReyCrit$ and $\alphaCrit$ as a function of $H$ for $10^{-3} \leq \Sr \leq 1$ and $\Gamma \geq 10$. The steady ($\Gamma \rightarrow \infty$) results from Ref.~\citep{Camobreco2020transition} have been included for direct comparison in the top row (black dashed line), and are divided out to compute $\rrs$ and $\als$ in the bottom row.}
    \label{fig:vary_H}
\end{figure}
Two key results are shown in \fig\ \ref{fig:vary_H}, considering the effect of varying $H$ on $\ReyCrit$. First, at large $H$, $\ReyCrit$ for a purely steady base flow scales as $H^{1/2}$, while all pulsatile cases scale as $H^p$, with $1/2 \leq p < 1$. For large $H$, $r$ is dominated by $[(SrRe)^2+H^2]^{1/4}$, which is always greater than $H^{1/2}$. As the isolated boundary layer thickness is defined by $e^{-ry}$ (Sec.~\ref{sec:prob_set}), increasing $H$ stabilizes pulsatile base flows more rapidly than steady base flows. Thus, the thinner pulsatile boundary layers are always more stable than their thicker counterpart exhibited by steady base flows. Note that in the high $H$ regime, when the boundary layers are isolated for any frequency pulsation, the stability results are defined solely by the dynamics of an isolated boundary layer, as observed in steady MHD or Q2D studies \citep{Camobreco2020transition, Potherat2007quasi, Takashima1998stability, Vo2017linear}, and for high frequency oscillatory hydrodynamic flows \citep{Blenn2006linear}. Second, variations in the pulsation frequency and amplitude roughly act to translate the stability curves, without significantly changing the overall trends (a slight change, the local minimums in \fig\ \ref{fig:vary_H}(c), are explained when considering $\Sr$ variations at fixed $H$ shortly). At $\Gamma=100$, differences between pulsatile and steady results are not easily observed, confirming the accuracy of the Floquet solver. The $\Gamma=10$ curves overlay the steady trend at respective high and low frequencies of $\Sr = 1$ and $\Sr=10^{-3}$. At $\Sr=10^{-2}$, the flow is more unstable as $H\rightarrow 0$, with $\rrs \rightarrow 0.8651$. However, for $H \gtrsim 2400$ the additional stability conferred by thinner pulsatile boundary layers pushes $\rrs$ above unity. The pulsatile flow is then more stable than the steady counterpart. Note that so long as $\ReyCrit$ varies as $H^p$ with $p<1$ (as observed for all $H$ simulated), then $\Rey$ does not increase quickly enough to offset the eventual $s \rightarrow 0$ and $r \sim H^{1/2}$ trends as $H \rightarrow \infty$. Eventually, the exponent $p$ should settle to $1/2$, after which $\ReyCrit$ should vary as $H^{1/2}$ for very large $H>10^4$. At $\Sr=10^{-1}$, the flow is hydrodynamically more stable ($\rrs \rightarrow 2.4258$ as $H\rightarrow 0$), and is even more strongly stabilized at higher $H$. The  $\Sr=10^{-1}$ curve in \Fig\ \ref{fig:vary_H}(c) is not smooth as different least stable modes become dominant, as shown in the jumps in critical wave number, clearest in \fig\ \ref{fig:vary_H}(d). In steady Q2D flows \citep{Camobreco2020transition, Potherat2007quasi, Vo2017linear}, $\alphaCrit$ also scales with $H^{1/2}$ for high $H$, like $\ReyCrit$. However, perplexingly for the pulsatile cases, the $\alphaCrit$ trends are as $H^q$, with a lower exponent than the steady case, $q \leq 1/2$.

\begin{figure}
\begin{center}
\addtolength{\extrarowheight}{-10pt}
\addtolength{\tabcolsep}{-2pt}
\begin{tabular}{ llll }
\makecell{\vspace{25mm} \footnotesize{(a)} \\  \vspace{35mm} \rotatebox{90}{\footnotesize{$\rrs$}}} & \makecell{\includegraphics[width=0.458\textwidth]{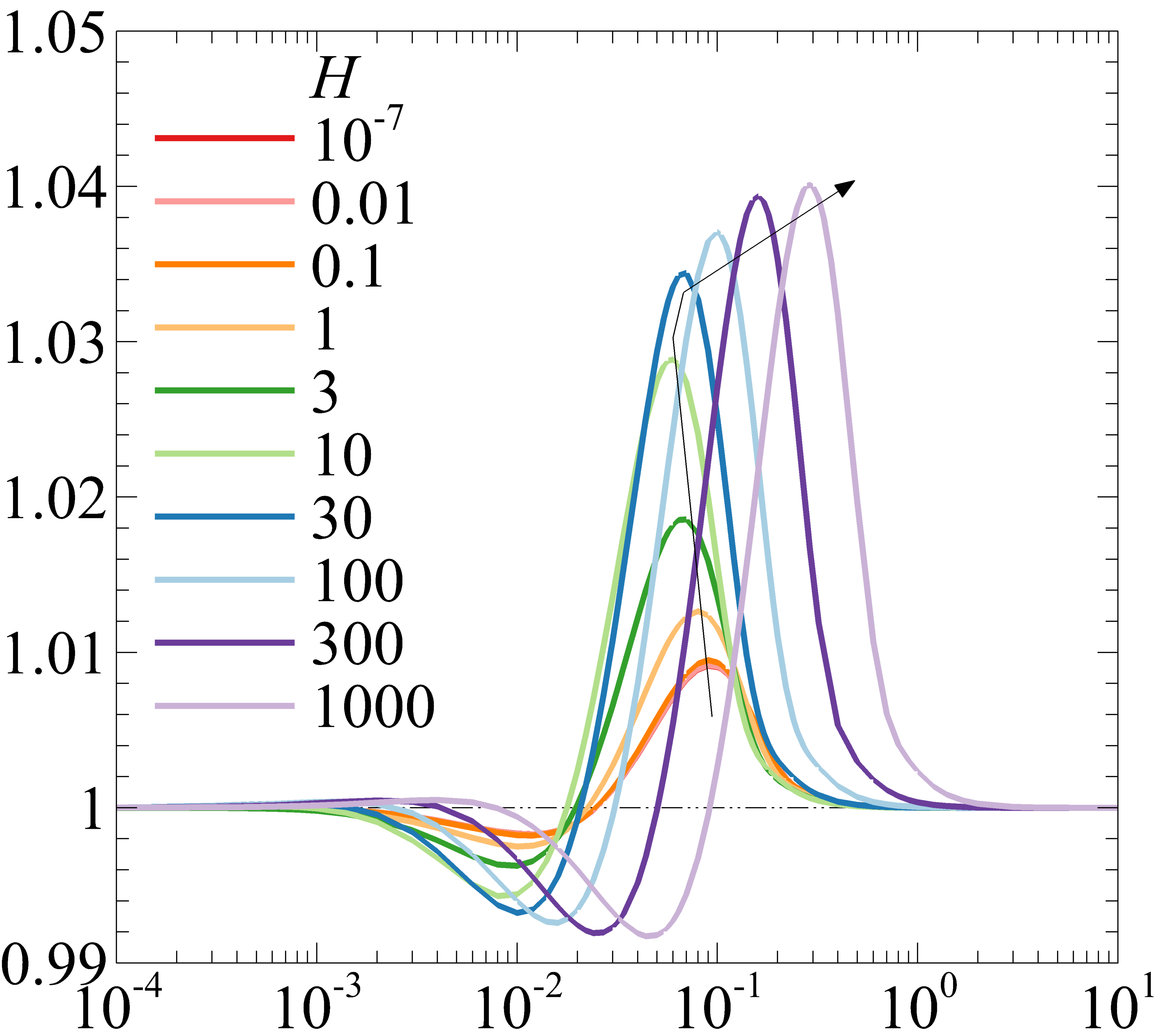}} &
\makecell{\vspace{25mm} \footnotesize{(b)} \\  \vspace{35mm} \rotatebox{90}{\footnotesize{$\als$}}}
 & \makecell{\includegraphics[width=0.458\textwidth]{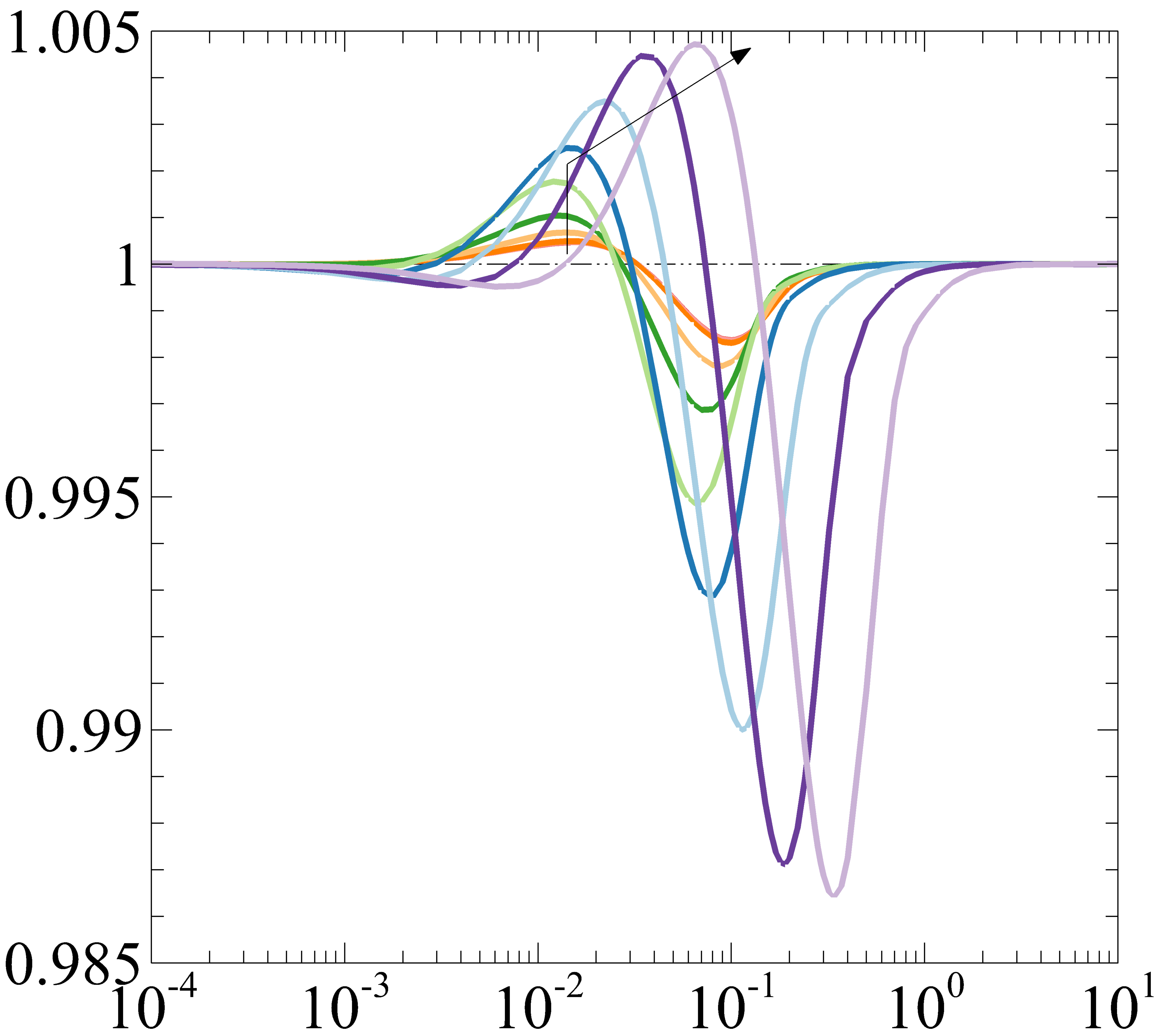}} \\
 & \hspace{36mm} \footnotesize{$\Sr$} & & \hspace{36mm} \footnotesize{$\Sr$} \\
\end{tabular}
\addtolength{\tabcolsep}{+2pt}
\addtolength{\extrarowheight}{+10pt}
\end{center}
    \caption{Variation in $\rrs$ and $\als$ as a function of $\Sr$ at $\Gamma=100$, curves of constant $H$ (arrows indicate increasing $H$). As $\Sr \rightarrow 0$ and $\Sr \rightarrow \infty$, the agreement with the steady result is further validation.}
    \label{fig:vary_Sr_G100}
\end{figure}

\begin{figure}
\begin{center}
\addtolength{\extrarowheight}{-10pt}
\addtolength{\tabcolsep}{-2pt}
\begin{tabular}{ llll }
\makecell{\vspace{25mm} \footnotesize{(a)} \\  \vspace{35mm} \rotatebox{90}{\footnotesize{$\rrs$}}} & \makecell{\includegraphics[width=0.458\textwidth]{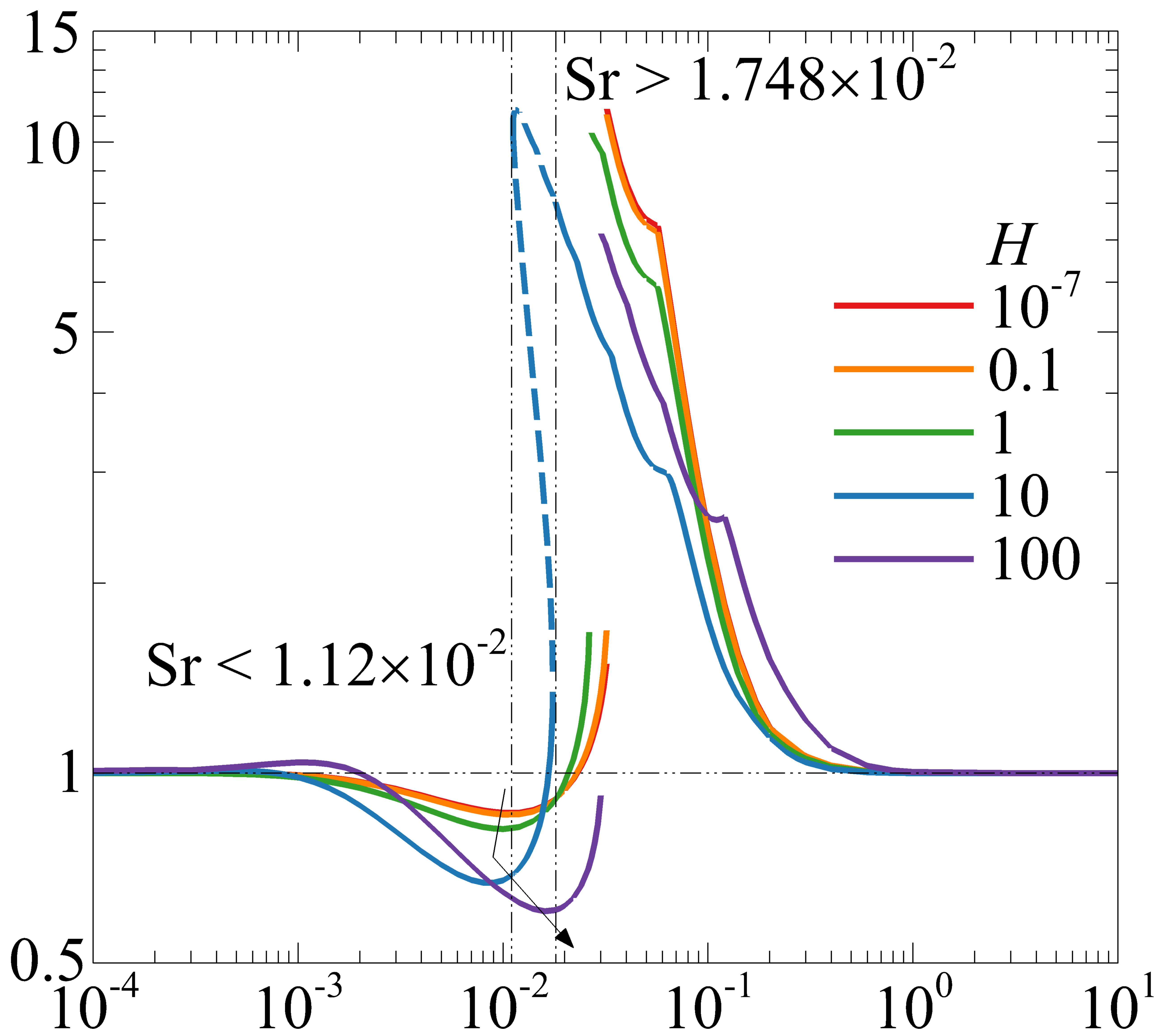}} &
\makecell{\vspace{25mm} \footnotesize{(b)} \\  \vspace{35mm} \rotatebox{90}{\footnotesize{$\als$}}}
 & \makecell{\includegraphics[width=0.458\textwidth]{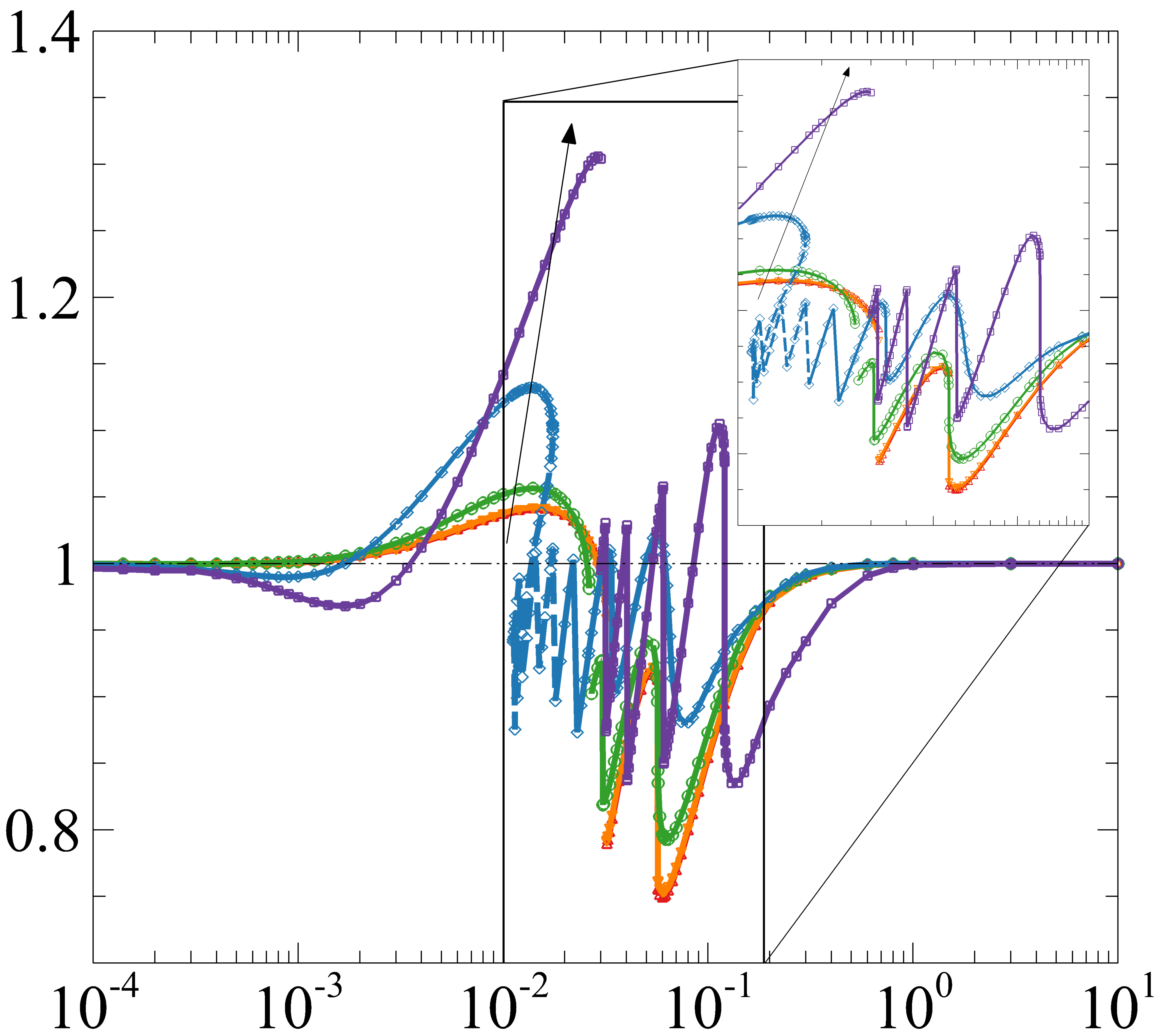}} \\
 & \hspace{36mm} \footnotesize{$\Sr$} & & \hspace{36mm} \footnotesize{$\Sr$} \\
\end{tabular}
\addtolength{\tabcolsep}{+2pt}
\addtolength{\extrarowheight}{+10pt}
\end{center}
    \caption{Variation in $\rrs$ and $\als$ as a function of $\Sr$ at $\Gamma=10$, curves of constant $H$ (arrows indicate increasing $H$). Dashed curve indicates restabilization and a second destabilization with increasing $\Rey > \ReyCrit$ at $H=10$. The stable region is below the continuous solid-dashed-solid curve.}
    \label{fig:vary_Sr_G10}
\end{figure}

Variations in $\rrs$ as a function of $\Sr$ are depicted for various $H$ under a weak pulsatility of $\Gamma=100$ in \fig\ \ref{fig:vary_Sr_G100}(a) and at $\Gamma=10$ in \fig\ \ref{fig:vary_Sr_G10}(a). The deviations from the steady $\ReyCrit$ are modest at $\Gamma=100$ (between $-1$\% and $+4$\%). However, it helps provide a clearer picture of the underlying dynamics. Considering the hydrodynamic case (approximated by $H=10^{-7}$) as an example, the steady $\ReyCrit$ is approached ($\rrs \rightarrow 1$) as $\Sr \rightarrow 0$. In this limit, transient inertial forces act so slowly that viscosity can smooth out all wall-normal oscillations in the velocity profile over the entire duct within a single oscillation period ($2\pi$). Although large intracylic growth occurs during the deceleration phase of the base flow (effectively due to an adverse pressure gradient), this is not augmented by additional growth as inflection points are absent. Therefore, the growth is entirely cancelled out by decay (due to an equivalent-magnitude favorable pressure gradient) in the acceleration phase. With increasing $\Sr$, inflection points are present over a greater fraction of the deceleration phase, in spite of the action of viscosity, and become more prominent, providing a reduction in $\rrs$. However, increasing $\Sr$ reduces the effective duration of the deceleration phase of the base flow, leaving less time for intracyclic growth. Thus, the local minimum in $\rrs$ occurs when the benefits of promoting and maintaining inflection points for a larger time (increasing $\Sr$) is counteracted by reducing the duration of the growth phase (decreasing $\Sr$). However, although increasing $\Sr$ promotes inflection points, these points also become increasingly isolated as the oscillating boundary layers become thinner. The thinner boundary layers reduce constructive interference between modes at each wall, stabilizing the flow \citep{Camobreco2020transition}. Eventually, the oscillating boundary layers become so thin that they are immaterial, and $\rrs$ drops to recover the steady value ($\Sr \rightarrow \infty$). 

The other friction parameters are now considered. For larger $H$, as $H$ is increased, the curves in figure \fig\ \ref{fig:vary_Sr_G100}(a) shift to larger $\Sr$. Increasing $H$ smooths inflection points within the pulsatile boundary layer. Recall that a pulsatile isolated SM82 boundary layer has the form $e^{-ry}\cos(sy-t)$, and increasing $H$ decreases $s$, thereby increasing the wavelength of wall-normal oscillations in the base flow. Larger $\Sr$ values are then required to offset the larger $H$ values, ensuring that inflection points remain within the boundary layer, and provide enough intracylic growth to reduce $\rrs$. Thus, the local minimum of $\rrs$ does not strongly depend on $H$, although the corresponding $\Sr$ value varies greatly. Importantly for fusion relevant regimes, the percentage reduction in $\ReyCrit$ appears to steadily improve with increasing $H$, although the shift to higher $\Sr$ may eventually invalidate the SM82 assumption requiring $\Sr < 1/2$ for $\Gamma \geq 1$. The pulsatile boundary layers also become increasingly isolated with increasing $H$, as $r$ increases with $H$, resulting in the steady increase in the maximum of $\rrs$. At $\Gamma=100$, the variations in $\ReyCrit$ are small, with the Reynolds number dependence of the base flow having little effect, relative to the $\Sr$ and $H$ variations (this is not the case at $\Gamma=10$). As a last note, for $\Gamma=100$, the smooth $\als$ curves in \fig\ \ref{fig:vary_Sr_G100}(b) also show that the variations in $\rrs$ represent the same instability mode for all $\Sr$ (henceforth the \TSL\ mode).

At the lower $\Gamma=10$, \fig\ \ref{fig:vary_Sr_G10}, the oscillating component plays a much greater role. The underlying behaviors discussed for $\Gamma=100$ still hold for smaller $\Sr$, including the region of minimum $\rrs$, and for much larger $\Sr$. Furthermore, the local minimum in $\rrs$ still becomes more pronounced with increasing $H$, with an approximately $33.0\%$ reduction in $\ReyCrit$, compared to the steady value, at $H=10$. $H=1000$ could not be computed over a wide range of $\Sr$ at $\Gamma=10$, but the partial data collected (not shown) demonstrated a further reduction in $\rrs$, of up to $42.4$\%. 

The degree of stabilization at $\Gamma=10$ is far more striking. The sudden jumps in $\als$, shown in the inset of \fig\ \ref{fig:vary_Sr_G10}(b), indicate different instability modes. These modes are increasingly stable, with much larger accompanying $\rrs$ values (the $H=10$ case peaks with an approximately $804\%$ increase over the steady $\ReyCrit$).  Because the Reynolds numbers are significantly far from the steady $\ReyCrit$ values, the change in Reynolds number has had a noticeable effect on the base flow profiles. At larger Reynolds numbers the oscillating boundary layers become much thinner, so inflection points are not positioned where  they could underpin sizeable intracyclic growth. 

\begin{figure}
\begin{center}
\addtolength{\extrarowheight}{-10pt}
\addtolength{\tabcolsep}{-2pt}
\begin{tabular}{ llll }
\footnotesize{(a)} & \footnotesize{\hspace{5mm} $\Sr=1.7\times10^{-2}$}  &  \footnotesize{(b)} & \footnotesize{\hspace{5mm} $\Sr=1.8\times10^{-2}$}  \\
\makecell{\\  \vspace{10mm} \rotatebox{90}{\footnotesize{$\Rez(\lambda_1)$}}} & \makecell{\includegraphics[width=0.458\textwidth]{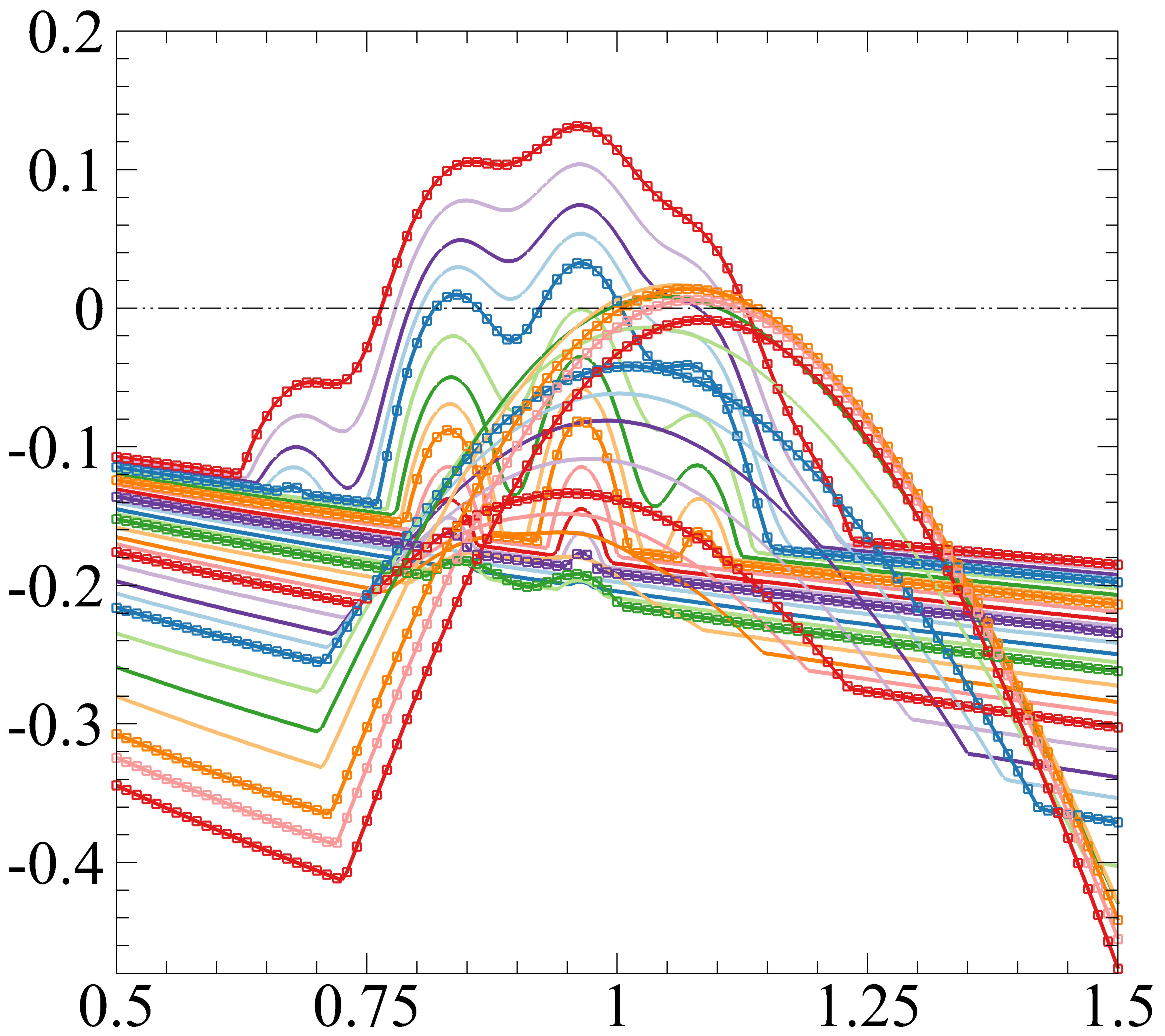}} &
\makecell{ \\  \vspace{10mm} \rotatebox{90}{\footnotesize{$\Rez(\lambda_1)$}}}
 & \makecell{\includegraphics[width=0.458\textwidth]{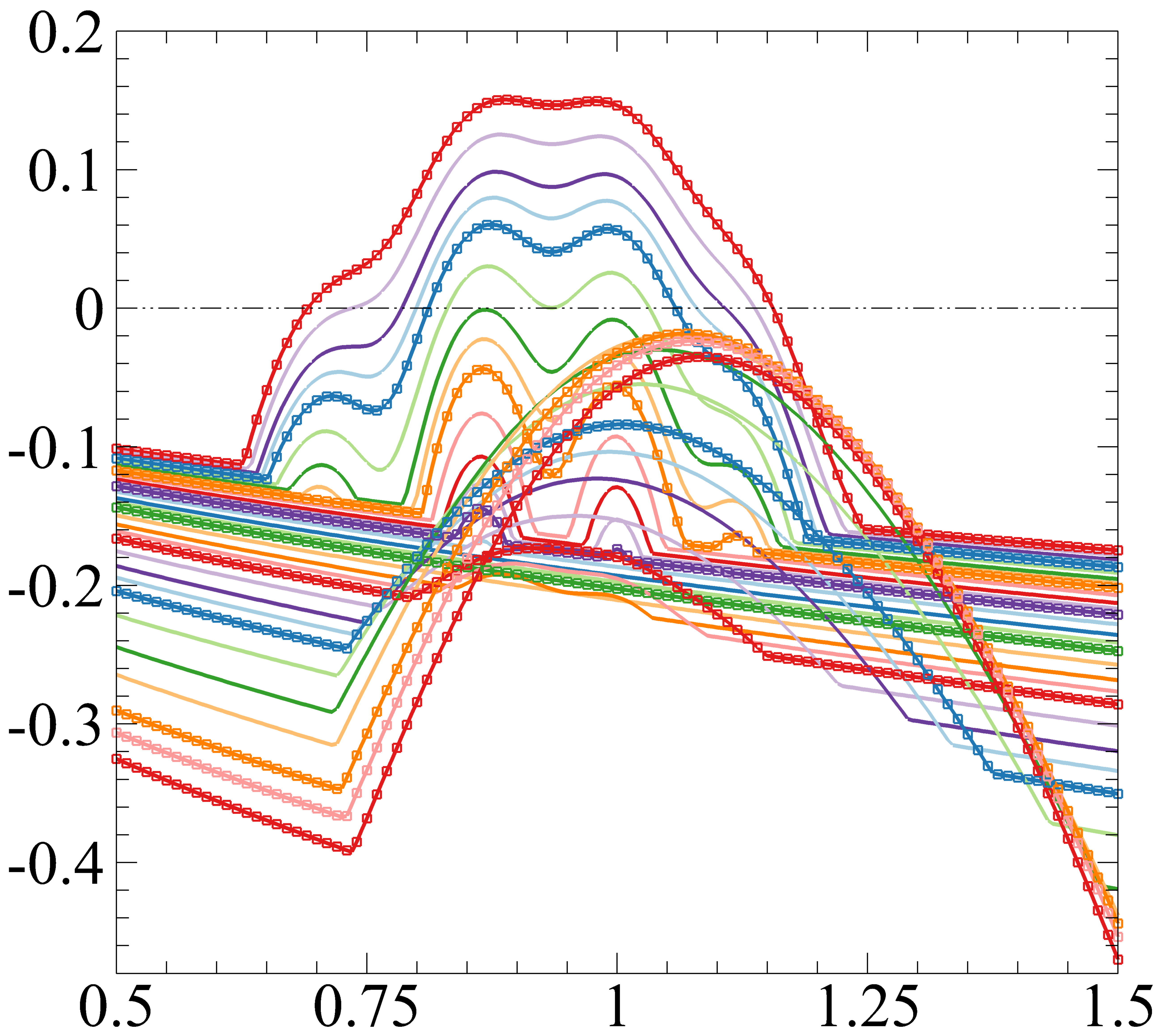}} \\
 & \hspace{37.5mm} \footnotesize{$\alpha$} & & \hspace{37.5mm} \footnotesize{$\alpha$} \\
\end{tabular}
\addtolength{\tabcolsep}{+2pt}
\addtolength{\extrarowheight}{+10pt}
\end{center}
    \caption{Exponential growth rate as a function of $\alpha$ with increasing $\Rey$ ($8\times10^4$ through $8\times10^5$) at $H=10$, $\Gamma=10$, comparing $\Sr$. At $\Sr=1.8\times10^{-2}$ the \TSL\ mode does not become unstable, thus $\ReyCrit=6.40840\times10^5$ is much larger than $\ReyCrit=8.50617\times10^4$ at $\Sr=1.7\times10^{-2}$. As additional validation, symbols (timestepper) show excellent agreement with curves (Floquet).}
    \label{fig:Sr1718}
\end{figure}

This explains the discontinuous change in $\rrs$ with a slight shift in $\Sr$. At fixed $\Sr$, at Reynolds numbers near the steady $\ReyCrit$ value, a \TSL\ mode is excited, but not necessarily unstable. The \TSL\ mode is based on the instability of the steady flow, i.e.~the \TS\ wave. For $\Rey>\ReyCrit$ the exponential growth rate increases with increasing Reynolds number. However, the same increase in $\Rey$ increasingly isolates and thins the boundary layers, thus reducing the exponential growth rate. The isolation of the boundary layers (the effect of $\Rey$ on the base flow) eventually overcomes any increases in exponential growth rate (the effect of $\Rey$ on the perturbation). At higher $\Sr$, when the oscillating boundary layers are naturally further apart, the increased isolation prevents the instability of the \TSL\ mode. This is shown at $\Sr=1.8\times10^{-2}$ in \fig\ \ref{fig:Sr1718}(b), or to the right of the discontinuity in $\rrs$ on \fig\ \ref{fig:vary_Sr_G10}(a). The sudden increase in $\rrs$ in \fig\ \ref{fig:vary_Sr_G10}(a) reflects the stabilization of the \TSL\ mode (another mode is destabilized at a much higher $\Rey$). At smaller $\Sr$, the effect of $\Rey$ on increasing the growth rate allows the \TSL\ mode to become unstable, if only briefly at $\Sr=1.7\times10^{-2}$ in \fig\ \ref{fig:Sr1718}(a). With further increasing $\Rey$ the isolation and thinning of the boundary layers leads to the \TSL\ mode becoming stable again; the stable region is bounded by the dashed curve in \fig\ \ref{fig:vary_Sr_G10}(a). At $\Sr=1.7\times10^{-2}$, a different mode becomes unstable at much higher $\Rey$, as is also shown in \fig\ \ref{fig:Sr1718}(a). This mode is a very similar to that  at $\Sr=1.8\times10^{-2}$, so the dashed curve in \fig\ \ref{fig:vary_Sr_G10}(a) follows the trend of increasing $\rrs$ from the right of the discontinuity. Eventually, for all $\Sr<1.12\times10^{-2}$ ($H=10$, $\Gamma=10$), with oscillating boundary layers that `start out' closer together, at least one mode is unstable for all $\Rey$.

\begin{figure}
\begin{center}
\addtolength{\extrarowheight}{-10pt}
\addtolength{\tabcolsep}{-2pt}
\begin{tabular}{ llll }
\makecell{\vspace{24mm} \footnotesize{(a)} \\  \vspace{34mm} \rotatebox{90}{\footnotesize{$\alpha$}}} & \makecell{\includegraphics[width=0.458\textwidth]{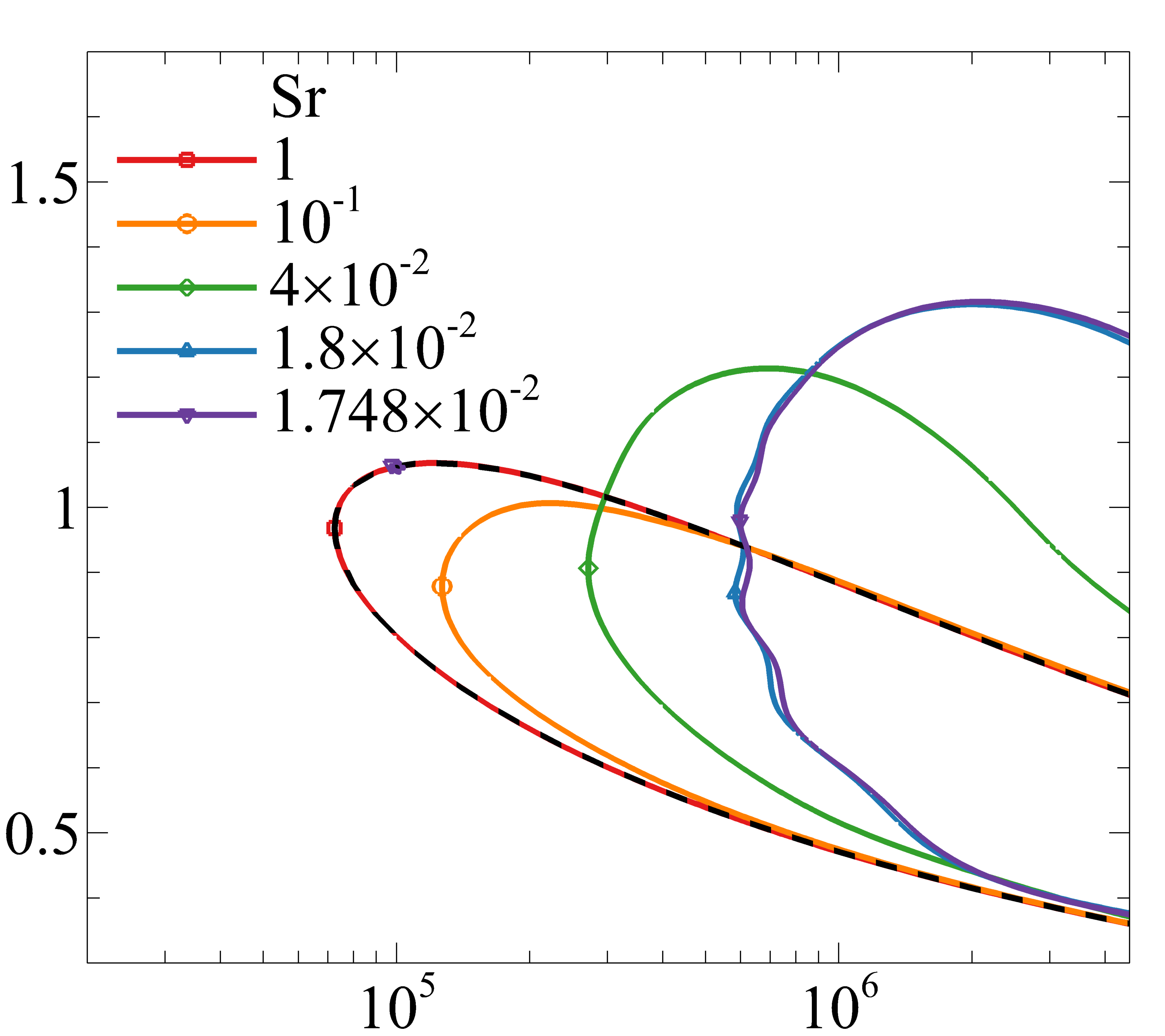}} &
\makecell{\vspace{24mm} \footnotesize{(b)} \\  \vspace{34mm} \rotatebox{90}{\footnotesize{$\alpha$}}}
 & \makecell{\includegraphics[width=0.458\textwidth]{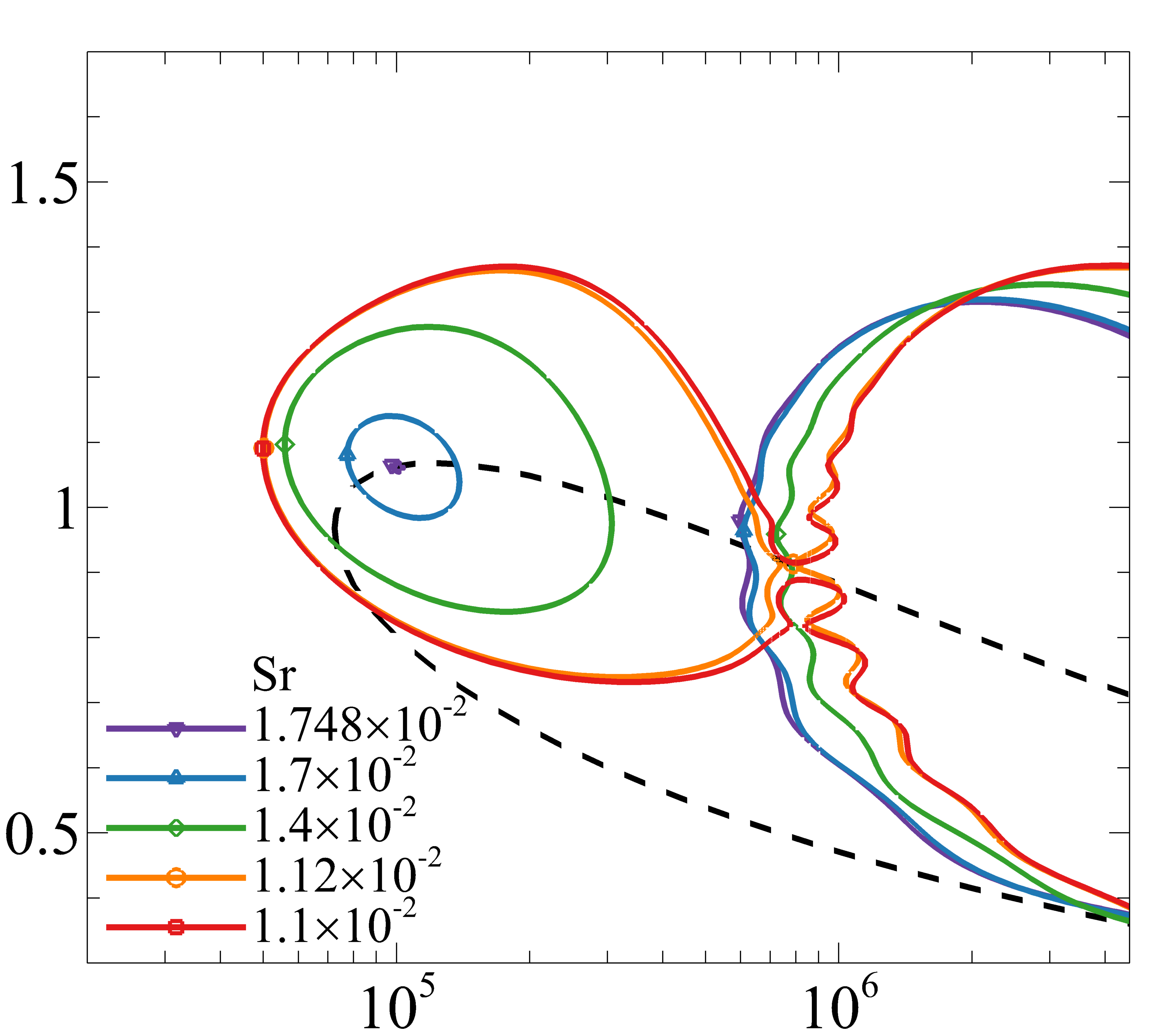}} \\
 & \hspace{30mm} \footnotesize{$\Rey/(1+1/\Gamma)$} & & \hspace{30mm} \footnotesize{$\Rey/(1+1/\Gamma)$} \\
\end{tabular}
\begin{tabular}{ ll }
\makecell{\vspace{24mm} \footnotesize{(c)} \\  \vspace{34mm} \rotatebox{90}{\footnotesize{$\alpha$}}}
 & \makecell{\includegraphics[width=0.458\textwidth]{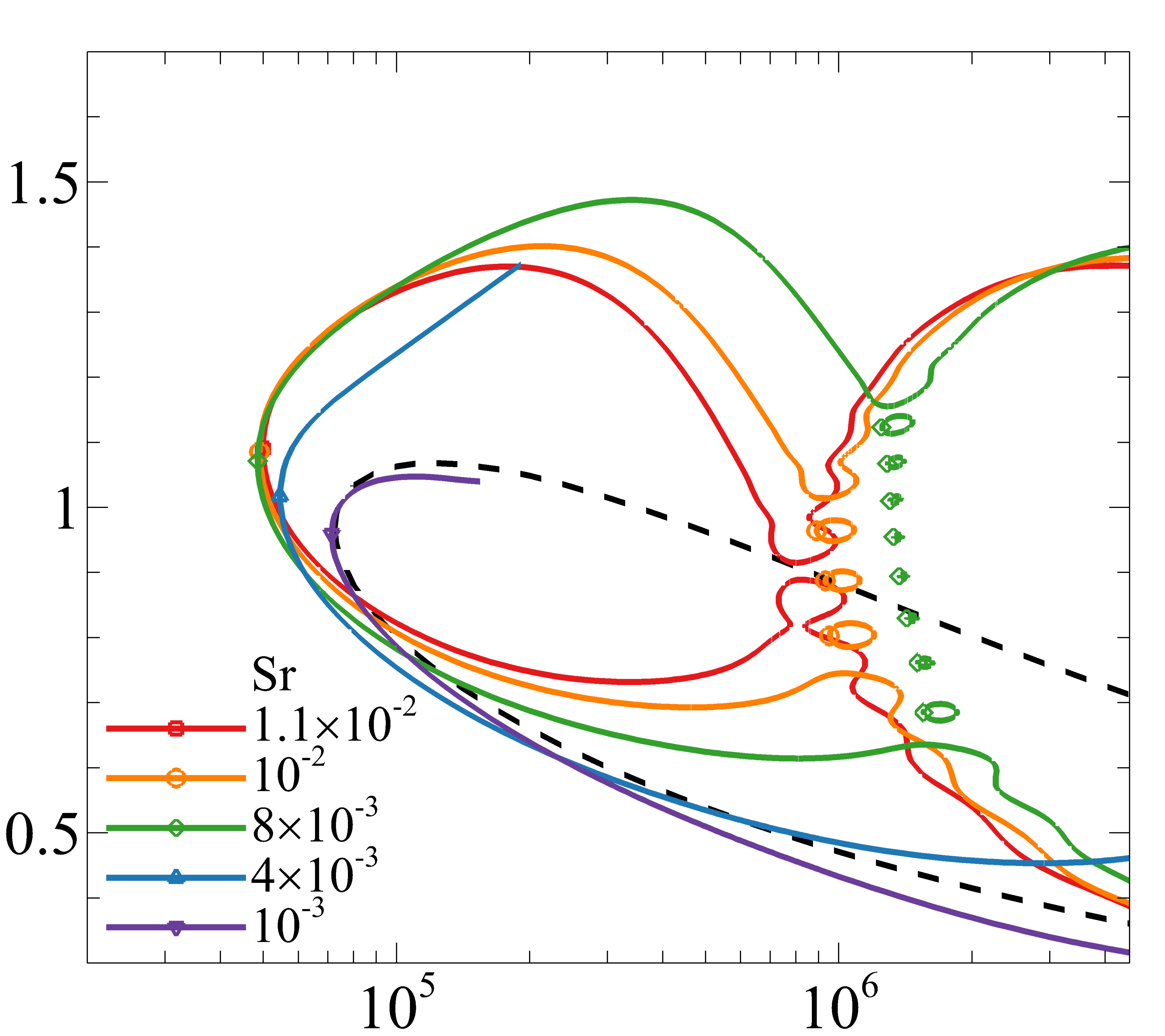}} \\
 & \hspace{30mm} \footnotesize{$\Rey/(1+1/\Gamma)$} \\
\end{tabular}
\addtolength{\tabcolsep}{+2pt}
\addtolength{\extrarowheight}{+10pt}
\end{center}
    \caption{Neutral curves for various $\Sr$, at $\Gamma=10$, $H=10$, with instability to the right of open curves. (a) $\Sr$ from the steady result, to the first destabilization of the \TSL\ mode (unstable pocket) at $\Sr \leq 1.748\times10^{-2}$. (b) Dominance of the \TSL\ mode, and eventual vanishing of the restabilization region for $\Sr \leq 1.1\times10^{-2}$. (c) Instability for all $\Rey>\ReyCrit$, including the local $\ReyCrit$ minimum (near $\Sr=9\times10^{-3}$). However, stable pockets form at higher $\Rey$. The dashed black curve corresponds to the steady base flow at $H=10$ \citep{Camobreco2020transition}.}
    \label{fig:neut_curves}
\end{figure}

\begin{figure}
\begin{center}
\addtolength{\extrarowheight}{-10pt}
\addtolength{\tabcolsep}{-2pt}
\begin{tabular}{ llll }
\footnotesize{(a)} & \footnotesize{\hspace{4mm} $\Sr=1.12\times10^{-2}$}  & \footnotesize{(b)} & \footnotesize{\hspace{4mm} $\Sr=8\times10^{-3}$}   \\
\makecell{\\  \vspace{10mm} } &
\makecell{\includegraphics[width=0.458\textwidth]{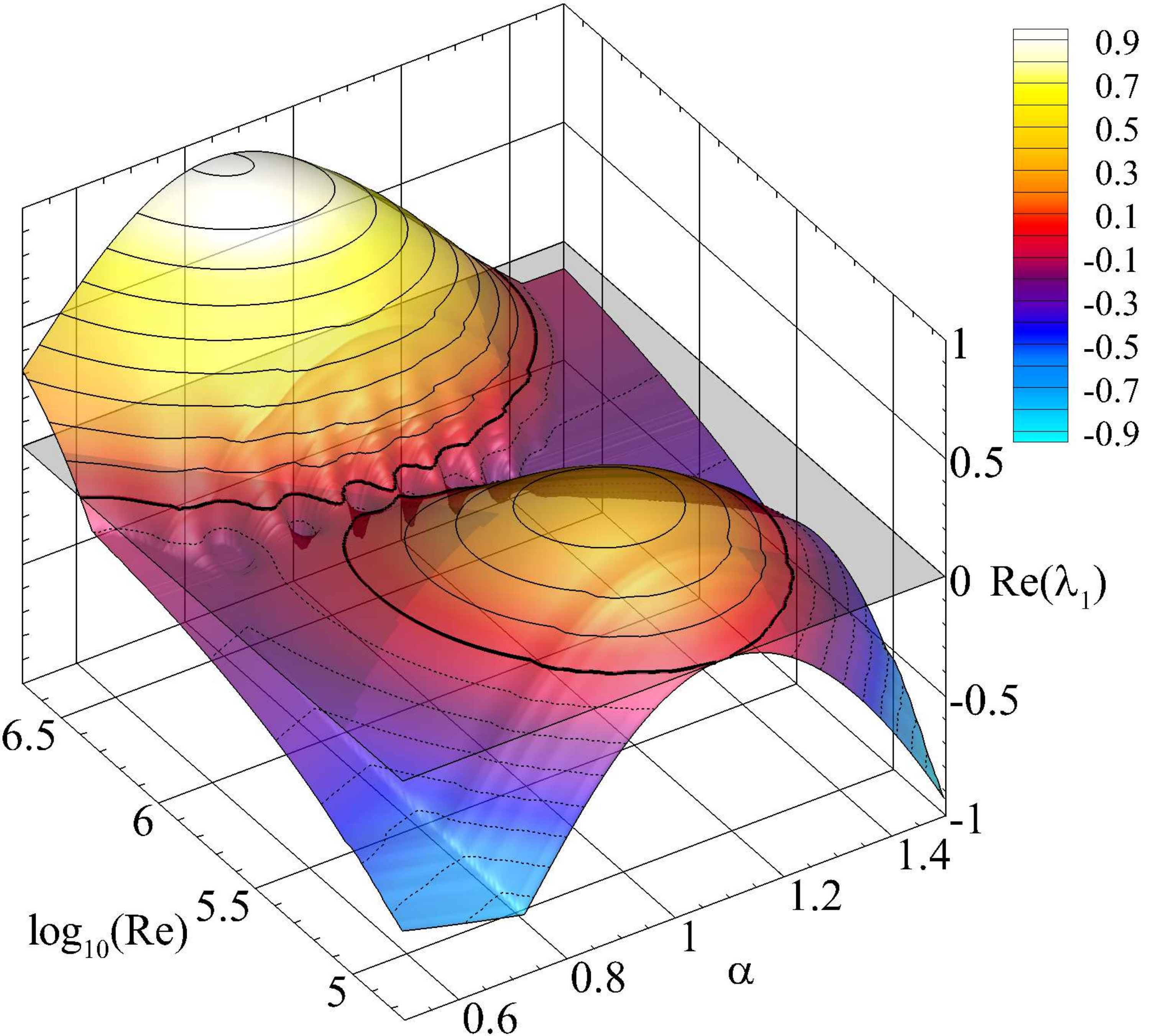}} &
\makecell{\\  \vspace{10mm} }
 & \makecell{\includegraphics[width=0.458\textwidth]{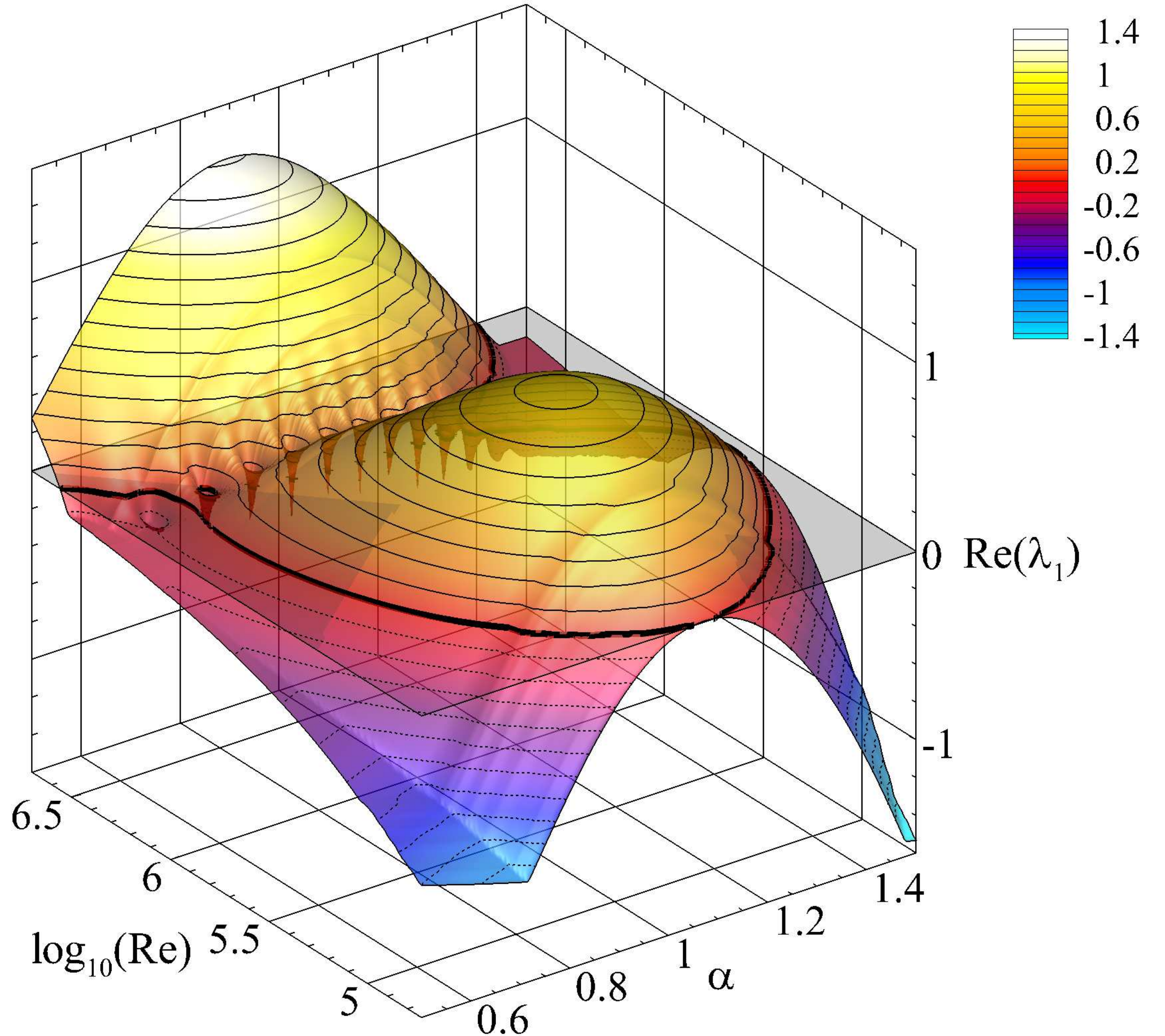}} \\
\end{tabular}
\addtolength{\tabcolsep}{+2pt}
\addtolength{\extrarowheight}{+10pt}
\end{center}
    \caption{Exponential growth rate as a function of $\alpha$ and $\Rey$ ($\ReyCrit$ through $5\times10^6$). At low $\Rey$, only the \TSL\ mode is unstable. After this mode restabilizes, multiple modes are excited, separated by sharp valleys. These modes have negative growth for some $\Rey$ at $\Sr=1.12\times10^{-2}$, but have positive growth at $\Sr=8\times10^{-3}$. Solid lines denote positive growth; dotted lines negative. Zero growth is emphasized with a thick black line on a gray intersecting plane. }
    \label{fig:Sr112008}
\end{figure}

Further considering $\Gamma=10$ and $H=10$, neutral (zero net growth) curves at several $\Sr$ are presented in \fig\ \ref{fig:neut_curves}. The $\Sr=1$ neutral curve is indistinguishable from that of the steady base flow \citep{Camobreco2020transition}. With decreasing $\Sr$, the critical Reynolds number rapidly increases and the neutral curve broadens, see \fig\ \ref{fig:neut_curves}(a). At $\Sr = 1.8\times10^{-2}$, just to the right of the discontinuity, waviness in the neutral curve reflects the excitation of multiple modes, as shown in \fig\ \ref{fig:Sr1718}(b). At $\Sr=1.748\times10^{-2}$, just to the left of the discontinuity, the \TSL\ mode is first destabilized. The increasing isolation of the oscillating boundary layers quickly restabilizes the flow, resulting in a very small instability pocket. Moving to \fig\ \ref{fig:neut_curves}(b), with slight decreases in $\Sr$, the (\TSL\ mode's) instability pocket rapidly occupies more of the wave number space, and the pocket terminates before it reaches the broader, pulsatile part of the neutral curve at $\Sr=1.12\times10^{-2}$ (the left-most point of the dashed curve in \fig\ \ref{fig:vary_Sr_G10}). With a slight drop to $\Sr=1.1\times10^{-2}$, the two curves meet, with a small throat allowing a path through wave number space with increasing $\Rey$ that always attains positive growth. At $\Sr= 1.12\times10^{-2}$, also shown in \fig\ \ref{fig:Sr112008}(a), the \TSL\ mode (the first local maximum) initially peaks, and then falls away with increasing $\Rey$. A small band of Reynolds numbers fail to produce net growth (along the line of six depressions in the wave number space). Increasing $\Rey$, multiple pulsatile modes become excited from the baseline spectrum, and become unstable. At $\Sr=1.1\times10^{-2}$, the rising pulsatile modes outpace the falling \TSL\ mode, so that at least one mode always maintains positive growth, see \fig\ \ref{fig:Sr112008}(b).

At $\Sr=10^{-2}$, three stable pockets are observed, see \fig\ \ref{fig:neut_curves}(c). At lower $\Sr$ the growth rates of the \TSL\ mode decrease more rapidly, leaving only pulsatile modes in control of the neutral stability behavior. Because these modes are excited in narrow resonant peaks in wave number space, stable regions can be present between the peaks. Thus, at lower $\Sr$, multiple stable pockets surrounded by unstable conditions form. Further reduction in $\Sr$ produces more resonant peaks, and more interleaved stable pockets, as shown at $\Sr = 8\times10^{-3}$ in \fig\ \ref{fig:Sr112008}(b). Further reducing $\Sr$, for large $H$ and $\Rey$, reaches the limit of the capability of the timestepper to cleanly resolve the entire neutral curves. By $\Sr=10^{-3}$, the part of the neutral curve able to be computed is approaching that of the steady base flow \citep{Camobreco2020transition}.

\begin{figure}
\begin{center}
\addtolength{\extrarowheight}{-10pt}
\addtolength{\tabcolsep}{-2pt}
\begin{tabular}{ llll }
\footnotesize{(a)} & \footnotesize{\hspace{4mm} $\Sr=1$}  &  \footnotesize{(b)} & \footnotesize{\hspace{4mm} $\Sr=10^{-2}$}  \\
\makecell{\vspace{10mm}  \\  \vspace{20mm} \rotatebox{90}{\footnotesize{$\rrs$}}} 
& \makecell{\includegraphics[width=0.458\textwidth]{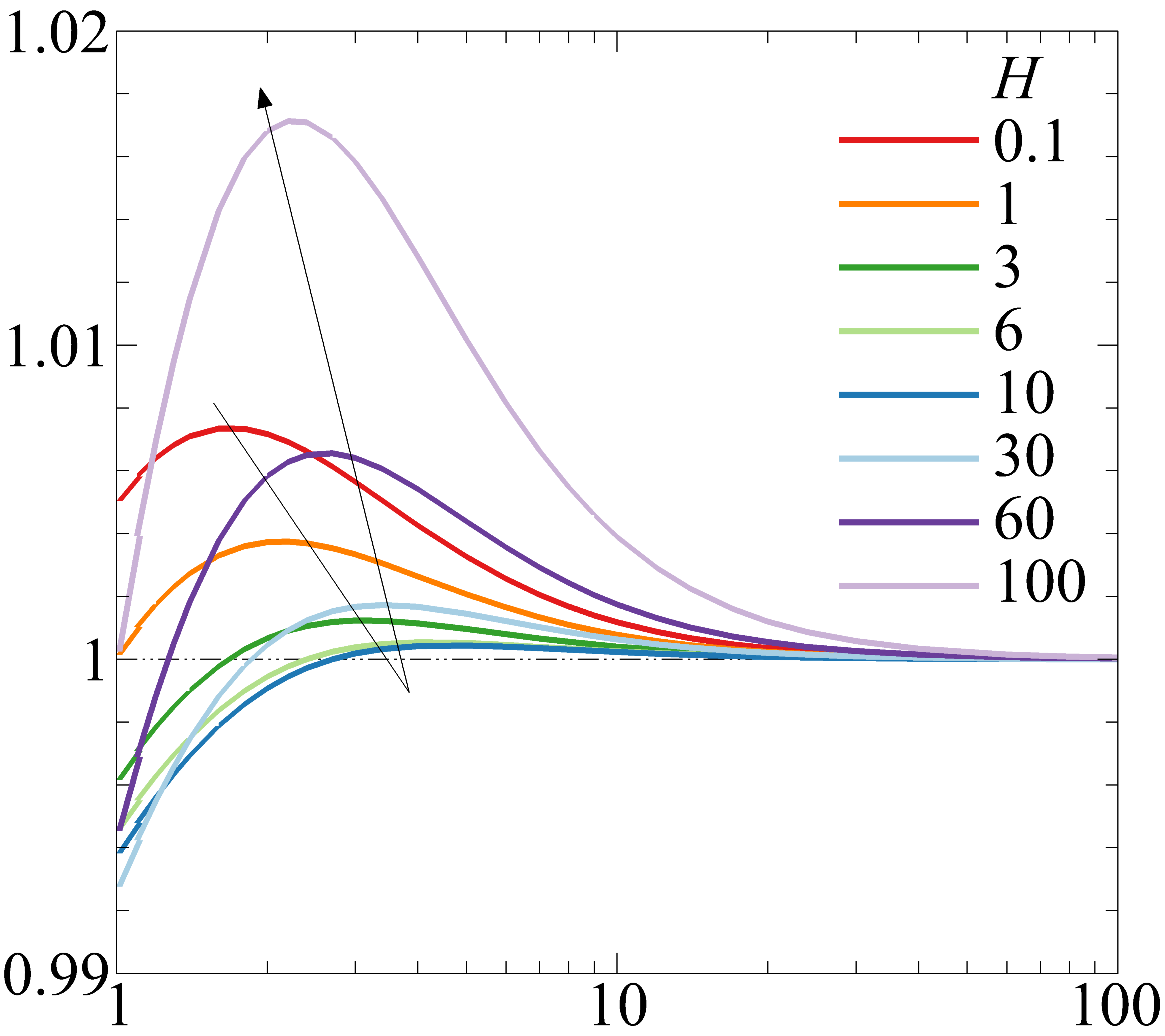}} &
\makecell{\vspace{10mm} \\  \vspace{20mm} \rotatebox{90}{\footnotesize{$\rrs$}}}
 & \makecell{\includegraphics[width=0.458\textwidth]{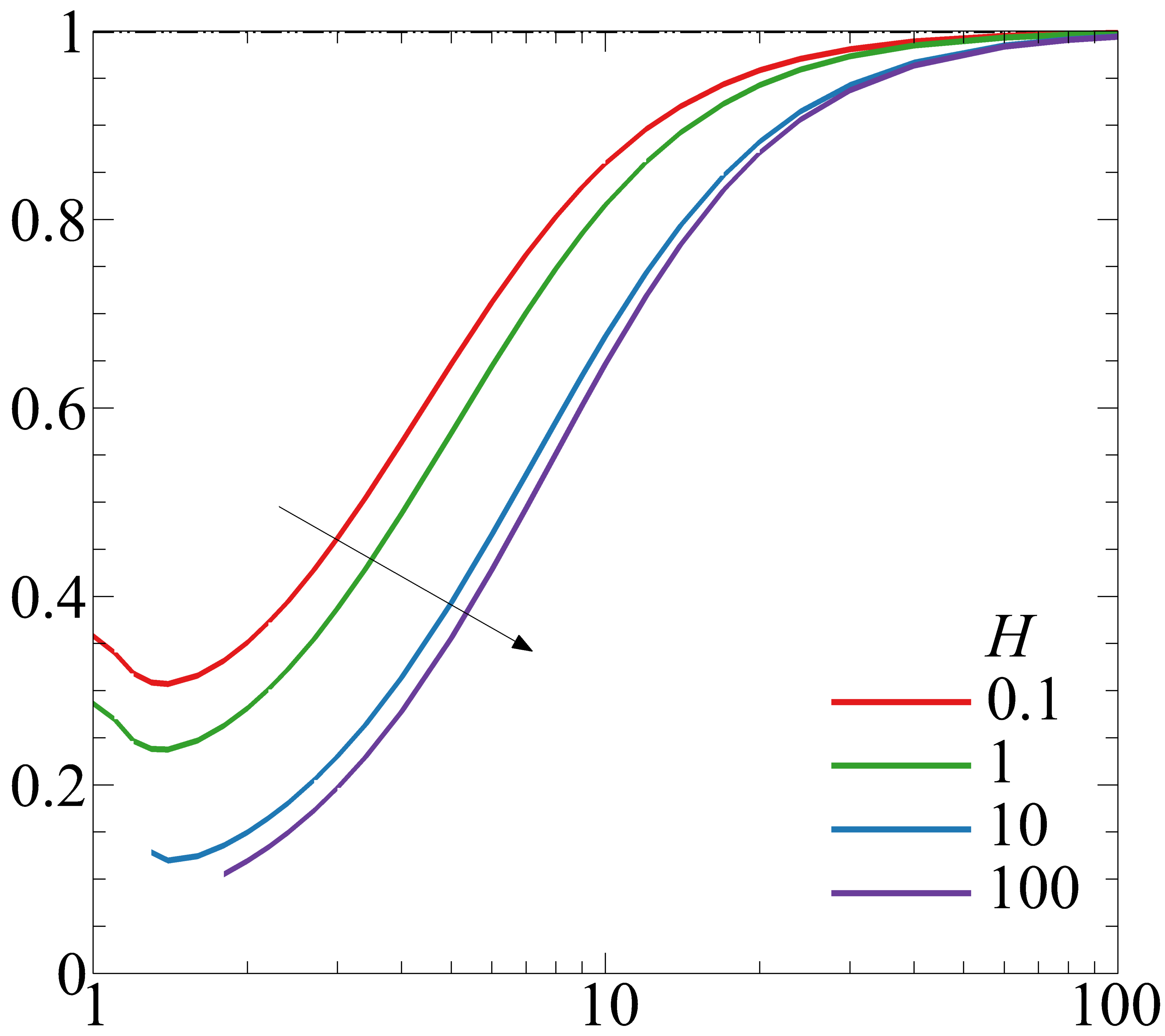}} \\
 & \hspace{36mm} \footnotesize{$\Gamma$} & & \hspace{36mm} \footnotesize{$\Gamma$} \\
\end{tabular}
\addtolength{\tabcolsep}{+2pt}
\addtolength{\extrarowheight}{+10pt}
\end{center}
    \caption{Variation in $\rrs$ as a function of $\Gamma\geq 1$ at $\Sr=1$ and $\Sr=10^{-2}$, curves of constant $H$ (arrows indicate increasing $H$). Small $\Sr$ and $\Gamma$ present significant potential for destabilization.}
    \label{fig:vary_G}
\end{figure}

The influence of $\Gamma$ is now considered. Over $1\leq\Gamma\leq 100$, different effects on $\rrs$ are observed at $\Sr = 1$, \fig\ \ref{fig:vary_G}(a), and at $\Sr = 10^{-2}$, \fig\ \ref{fig:vary_G}(b). As $\Sr=1$, close to the steady limit, $\rrs$ remains near unity. At small $H$, only stabilization is observed for all $\Gamma \geq 1$. With increasing $H$, a slight destabilization can be observed with increasing $H$, up to $H \approx 10$. Further increasing $H$ induces restabilization. This echoes the $\Sr$ variation, where the local minimum shifts to smaller $\Sr$ for $H \lesssim 10$, and shifts back to larger $\Sr$ for $H\gtrsim10$. At higher $H$, $H$ offsets $\Sr$, so the results for the steady base flow are only recovered at increasingly large $\Sr$.  On the other hand, at $\Sr = 10^{-2}$ in \fig\ \ref{fig:vary_G}(b), $\rrs$ is far from unity, and the effect of varying the Reynolds number on the base flow must again be considered. At smaller $\Rey$, the oscillating boundary layers are much thicker, with prominent inflection points well-placed to promote intracyclic growth. This part of the base flow becomes increasingly dominant with decreasing $\Gamma$, favoring the destabilization of the \TSL\ mode. Given that $(\Sr\Rey)^2 \gg H^2$, $\ReyCrit$ depends far more on the pulsatile process and only weakly on $H$, until $\Sr\Rey$ becomes small. However, the $\ReyCrit$ for the steady base flow strongly depends on $H$, so $\rrs$ reduces with increasing $H$. $\rrs$ continues to decrease up to $\Gamma \gtrsim 1$ for $H\leq 10$, matching well with the conclusion of Ref. \citep{Thomas2011linear} that the maximum reduction in $\ReyCrit$ occurs near unity amplitude ratio. At higher $H$, the magnitude of intracylic growth eventually limited computations (to $\Gamma > 1$).  At $H=100$, $\Sr = 10^{-2}$ no local minimum is observed for $\Gamma \geq 1$. However, these results still indicate that for $H \leq 100$ and $\Gamma \geq 1$, a $70$ to $90$\% reduction in the critical Reynolds number is possible with the addition of pulsatility. They further support that the percentage reduction in $\ReyCrit$ improves with increasing $H$. The mode defining this local minimum, even at small $\Gamma$, still appears to be directly related to the \TSL\ mode (as there were no sharp changes in the dominant $\alpha$ through the entire $\Sr-\Gamma-\Rey$ space). 


\begin{table}
\begin{center}
\begin{tabular}{ ccc cc cc ccc } 
\hline
$H$  &  $\Rey_\mathrm{crit,s}$ & $\alpha_\mathrm{crit,s}$ & $\Gamma$ & $\Sr$ & $\ReyCrit/(1+1/\Gamma)$ & $\alpha$ & $\rrs$ & $\als$ & \textbf{\% Reduction}   \\
\hline
$10^{-7}$ & 5772.22 & 1.02055  & 1.29 & $7.8\times10^{-3}$ & 1773.29 & 1.3812 & 0.3072  & 1.3534 & $\mathbf{69.28}$\\
0.01      & 5808.04 & 1.01991  & 1.29 & $7.8\times10^{-3}$ & 1777.58 & 1.3804 & 0.3061  & 1.3535 & $\mathbf{69.39}$\\
0.1       & 6136.85 & 1.01435  & 1.29 & $7.8\times10^{-3}$ & 1816.18 & 1.3823 & 0.2959  & 1.3628 & $\mathbf{70.41}$\\
0.3       & 6908.55 & 1.00291  & 1.27 & $7.6\times10^{-3}$ & 1902.79 & 1.3857 & 0.2754  & 1.3816 & $\mathbf{72.46}$\\
1         & 10033.2 & 0.97163 & 1.24 & $7.2\times10^{-3}$ & 2215.87 & 1.3980 & 0.2209  & 1.4388 & $\mathbf{77.91}$\\
3         & 21792.6 & 0.93194 & 1.19 & $6.3\times10^{-3}$ & 3185.90 & 1.4343 & 0.1462  & 1.5391 & $\mathbf{85.38}$\\
10        & 72436.8 & 0.96833 & 1.19 & $5.6\times10^{-3}$ & 7050    & 1.59    & 0.0973  & 1.6420 & $\mathbf{90.27}$\\
\hline
\end{tabular}
\caption{Optimisation of the pulsation (optimising $\Gamma$, $\Sr$ and $\alpha$) for the greatest reduction in the rescaled critical Reynolds number relative to the steady result. This is achieved at $\Gamma$ just above unity, and pulsation frequencies similar to those of the local minimum for the \TSL\ mode, \figs\ \ref{fig:vary_Sr_G100}(a) and \ref{fig:vary_Sr_G10}(a). Importantly, the percentage reduction improves with increasing $H$, with over an order of magnitude reduction in critical Reynolds number for $H \geq 10$.}
\label{tab:tab_3}
\end{center}
\end{table}


Given the results of \fig\ \ref{fig:vary_G}(b), it is worth considering the maximum reduction in $\rrs$ that can be obtained via optimisation of the pulsation over $10^{-4}<\Sr<1$ and $1<\Gamma<\infty$. These have been tabulated for increasing $H$ in \tbl\ \ref{tab:tab_3}. These optimized pulsations truly highlight how effective pulsatility can be in destabilizing a Q2D channel flow, both at hydrodynamic conditions, with a $69.3$\% reduction at $H=10^{-7}$, all the way up to a $90.3$\% reduction at $H=10$. Still larger percentage reductions are predicted at higher $H$, as $\rrs$ consistently decreases with increasing $H$.


\subsection{Intracylcic behavior}\label{sec:lin_res2}

This section is focused on processes taking place within each cycle that are obscured in the net growth quantifications. All results in this section are at $\ReyCrit$.

\begin{figure}
\begin{center}
\addtolength{\extrarowheight}{-10pt}
\addtolength{\tabcolsep}{-2pt}
\begin{tabular}{ llllll }
\footnotesize{(a)} & \footnotesize{\hspace{4mm} $\Sr=10^{-3}$} &  &  \footnotesize{(b)} & \footnotesize{\hspace{4mm} $\Sr=10^{-2}$}  &  \\
\makecell{ \\  \vspace{12mm} \rotatebox{90}{\footnotesize{$\twonv$}}} & 
\makecell{\includegraphics[width=0.44\textwidth]{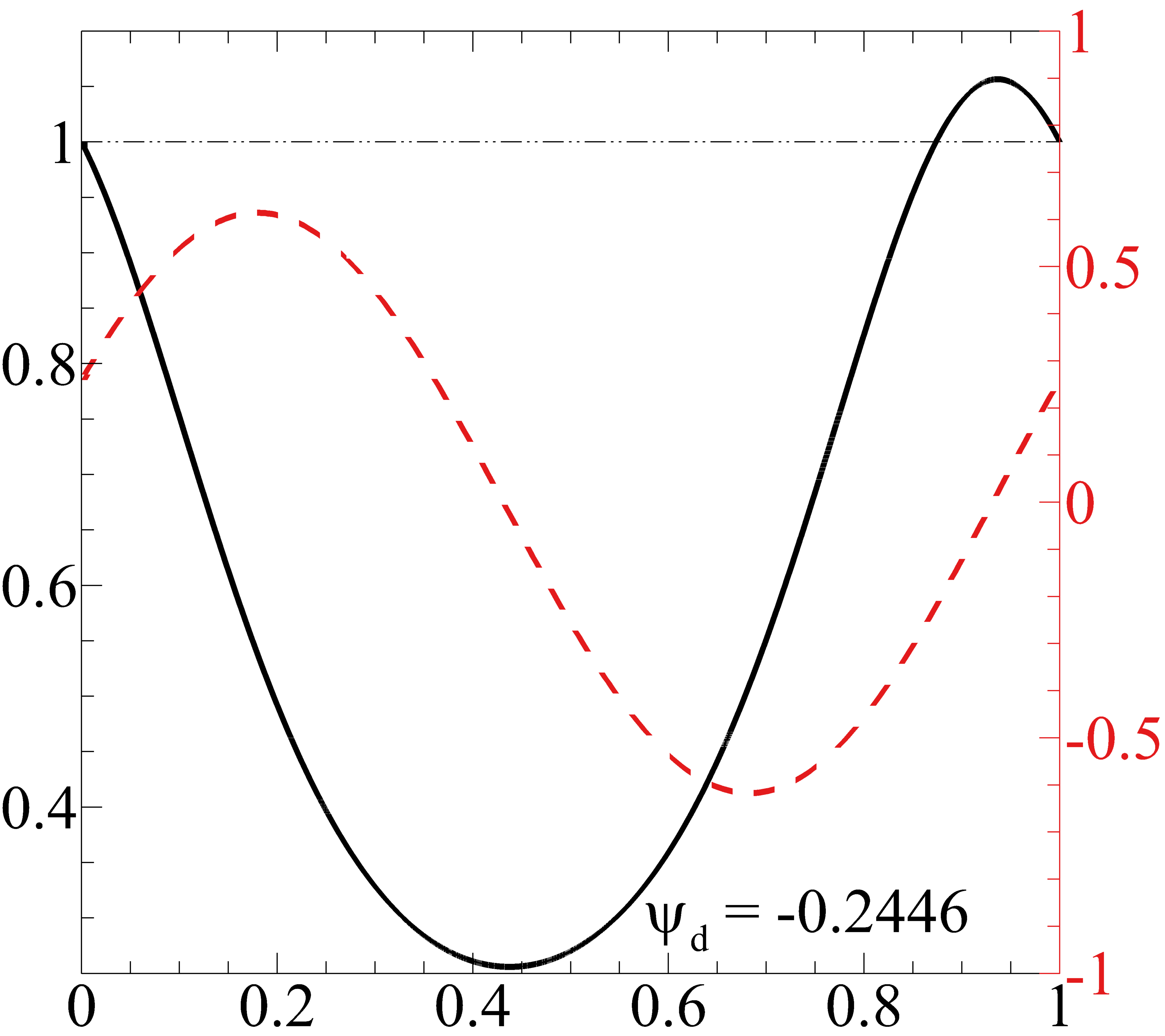}} &
\makecell{ \\  \vspace{11mm} \rotatebox{90}{\footnotesize{$\tmr{10^3\EU}$}}} &
\makecell{ \\  \vspace{12mm} \rotatebox{90}{\footnotesize{$\twonv$}}} &
\makecell{\includegraphics[width=0.44\textwidth]{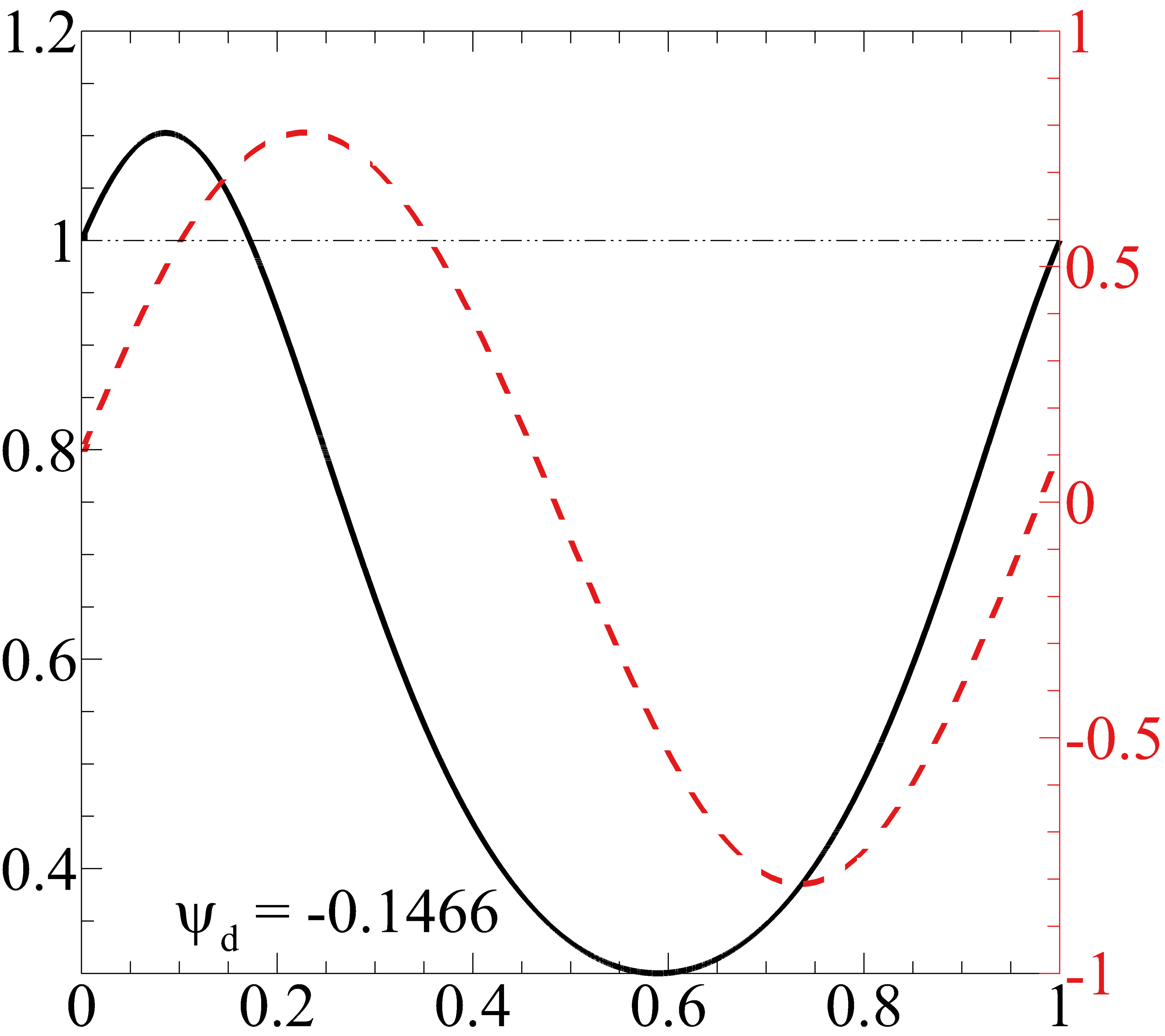}} &
\makecell{\\   \vspace{11mm} \rotatebox{90}{\footnotesize{$\tmr{10^4\EU}$}}}  \\
 & \hspace{34mm} \footnotesize{$\tpe$} & & & \hspace{34mm} \footnotesize{$\tpe$} &  \\
 \footnotesize{(c)} & \footnotesize{\hspace{4mm} $\Sr=10^{-1}$} &  &  \footnotesize{(d)} & \footnotesize{\hspace{4mm} $\Sr=1$} &   \\
 \makecell{ \\ \vspace{12mm} \rotatebox{90}{\footnotesize{$\twonv$}}} & 
 \makecell{\includegraphics[width=0.44\textwidth]{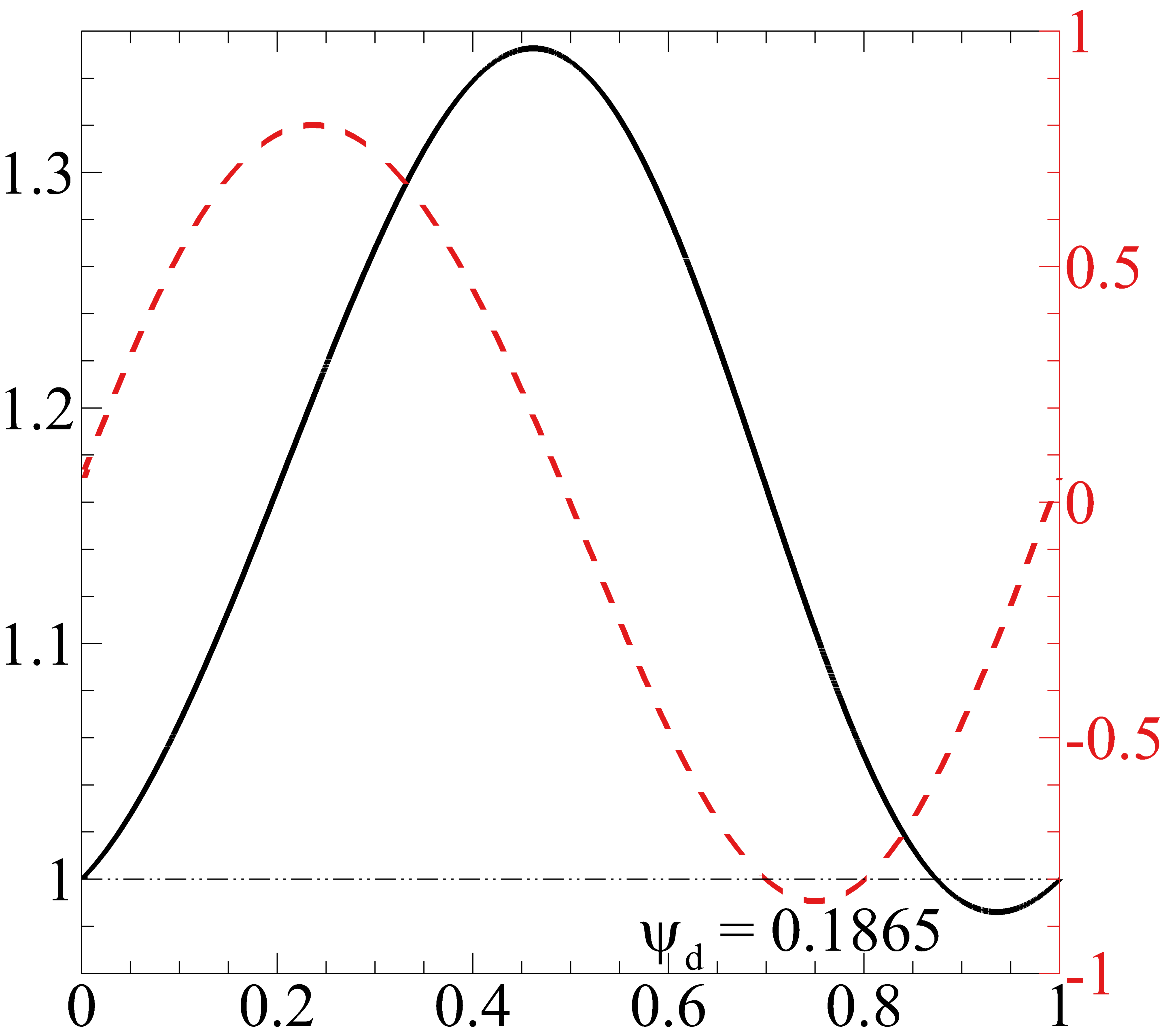}} &
\makecell{ \\  \vspace{11mm} \rotatebox{90}{\footnotesize{$\tmr{10^5\EU}$}}} &
\makecell{ \\  \vspace{12mm} \rotatebox{90}{\footnotesize{$\twonv$}}} &
\makecell{\includegraphics[width=0.44\textwidth]{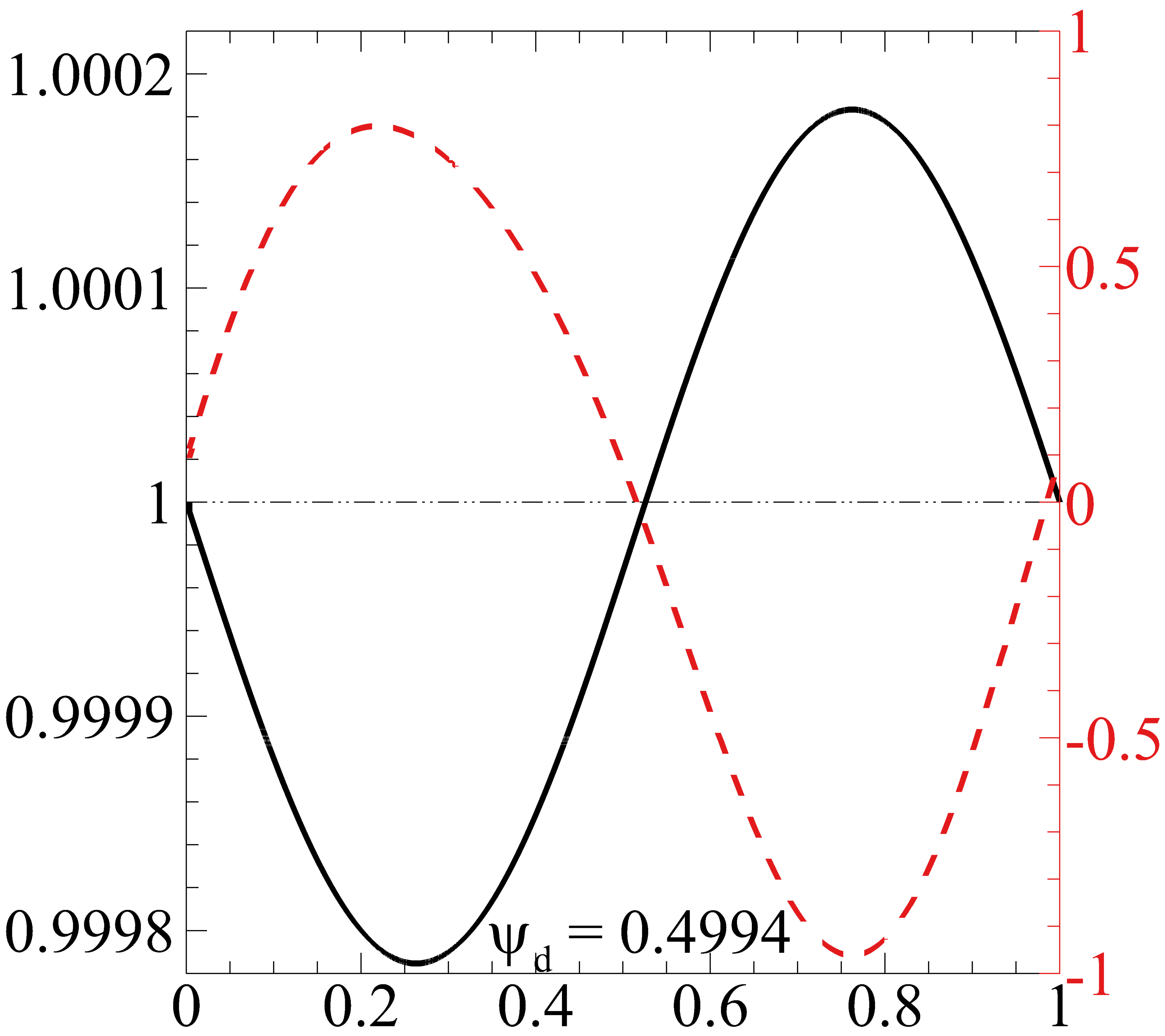}} &
\makecell{ \\  \vspace{11mm} \rotatebox{90}{\footnotesize{$\tmr{10^6\EU}$}}}  \\
 & \hspace{34mm} \footnotesize{$\tpe$} & & & \hspace{34mm} \footnotesize{$\tpe$} &  \\
\end{tabular}
\addtolength{\tabcolsep}{+2pt}
\addtolength{\extrarowheight}{+10pt}
\end{center}
    \caption{The perturbation norm (solid; black) and the base flow energy relative to the time mean (dashed; red) over one period at critical conditions at $H=100$, $\Gamma=100$, for various $\Sr$. The phase differences $\phased$ between the local minimums of each pair of curves are also annotated.}
    \label{fig:pert_base_norm100}
\end{figure}

The \TSL\ mode at $\Gamma=100$ and $H=100$ is considered first, in \fig\ \ref{fig:pert_base_norm100}, over a range of $\Sr$. The perturbation norm $\twonv$ is compared to $\EU(t) = \int U^2 \,\mathrm{d}y - \langle \int U^2 \,\mathrm{d}y \rangle_t$ (taking the value of the current base flow energy about the time mean solely to aid figure legibility). There are only simple, sinusoidal energy variations at these conditions, and perturbation energies remain order unity over the entire cycle (akin to the `cruising' regime). The key result is that the phase difference between the perturbation and base flow energy curves changes as $\Sr$ is varied. Measuring the phase difference $\phased$ of the local minimums of the perturbation and base flow energies appears most meaningful, and these values are annotated on \fig\ \ref{fig:pert_base_norm100}. The perturbation energy variation exhibits a lag to the base flow energy variation at $\Sr = 10^{-3}$, with $\phased = -0.2446$, and is closer to in phase by $\Sr = 10^{-2}$, $\phased = -0.1466$ (the optimal $\Sr$ is $1.5\times10^{-2}$ for minimising $\rrs$ at $\Gamma=100$). By $\Sr = 10^{-1}$, the perturbation energy leads the base flow energy  (positive $\phased$), and intracyclic growth in noticeably smaller. $\Sr = 1$ is close enough to the $\Sr \rightarrow \infty$ limit to produce negligible intracyclic growth. The minimum in $\rrs$ tends to occur when the perturbation and base flow energy growths are close to being in phase. Thus, selecting the optimal $\Sr$ to minimize $\rrs$ at a given $\Gamma$ (and $H$) amounts to tuning the frequency of the oscillating flow component, to ensure growth in the base flow and perturbation energies coincide.


\begin{figure}
\begin{center}
\addtolength{\extrarowheight}{-10pt}
\addtolength{\tabcolsep}{-2pt}
\begin{tabular}{ llllll }
\footnotesize{(a)} & \footnotesize{\hspace{4mm} $\Sr=10^{-3}$} &  &  \footnotesize{(b)} & \footnotesize{\hspace{4mm} $\Sr=10^{-2}$}  &  \\
\makecell{ \\  \vspace{12mm} \rotatebox{90}{\footnotesize{$\twonv$}}} & 
\makecell{\includegraphics[width=0.44\textwidth]{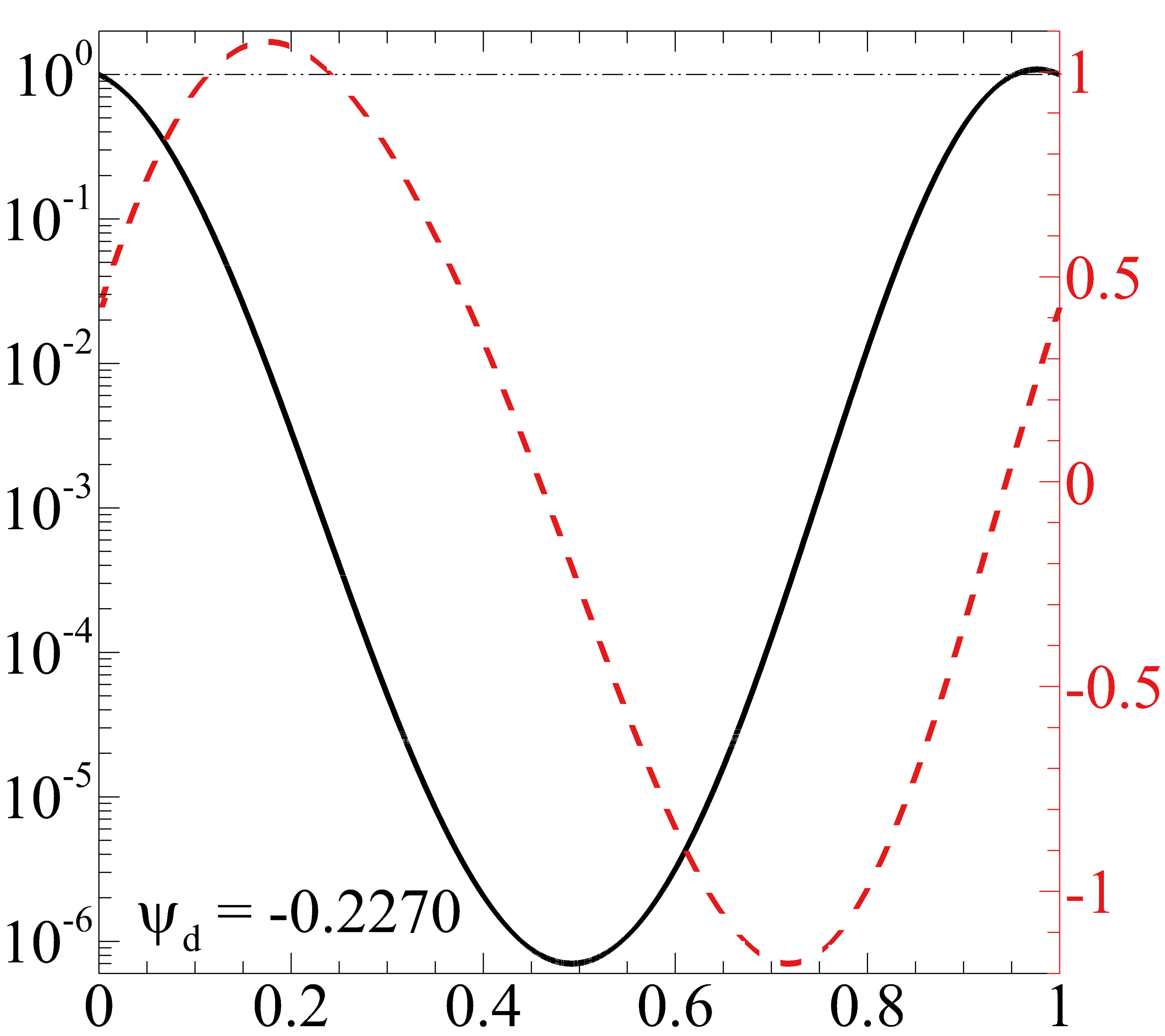}} &
\makecell{ \\  \vspace{11mm} \rotatebox{90}{\footnotesize{$\tmr{10^2\EU}$}}} &
\makecell{ \\  \vspace{12mm} \rotatebox{90}{\footnotesize{$\twonv$}}} &
\makecell{\includegraphics[width=0.44\textwidth]{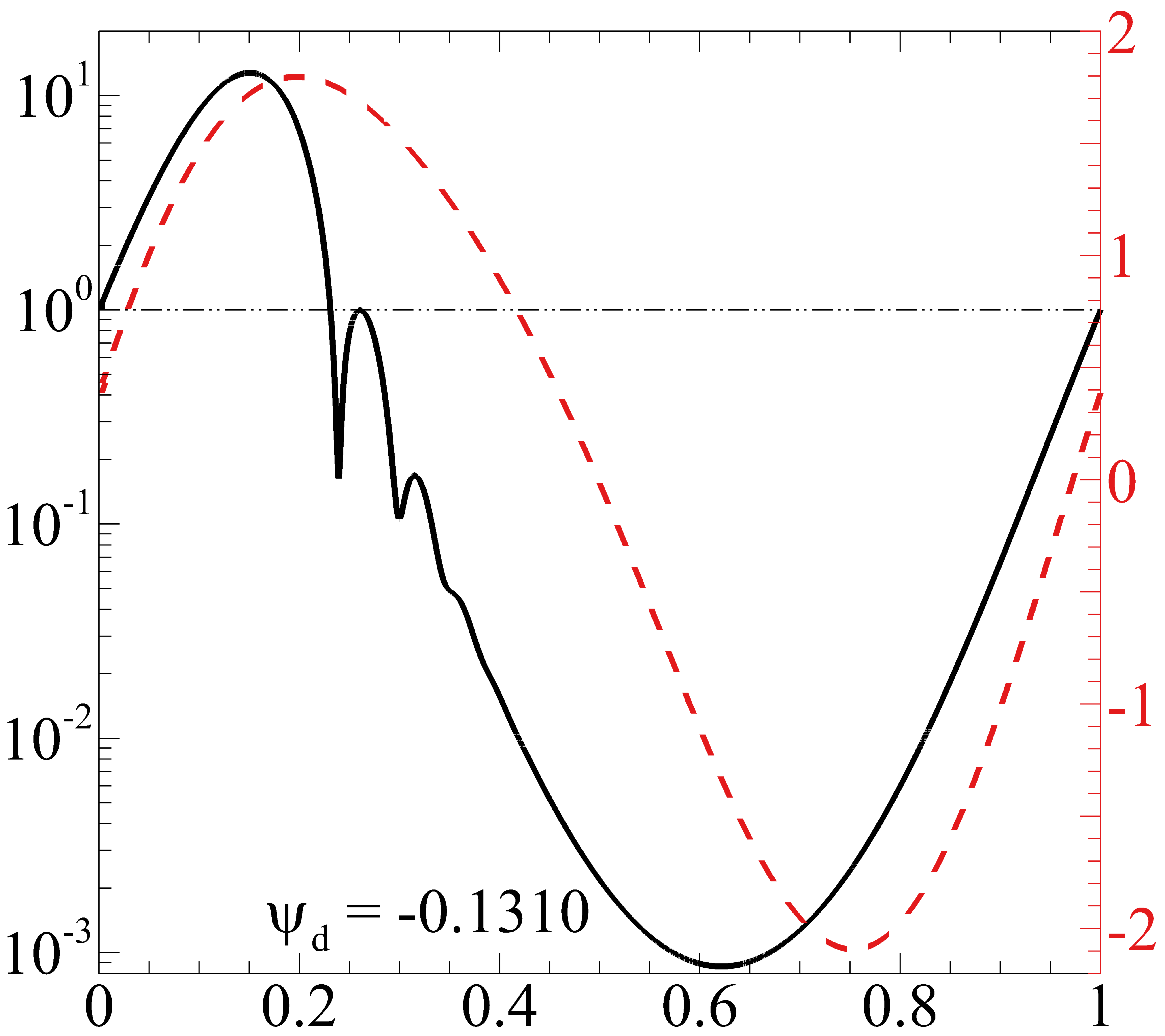}} &
\makecell{ \\  \vspace{11mm} \rotatebox{90}{\footnotesize{$\tmr{10^3\EU}$}}}  \\
 & \hspace{34mm} \footnotesize{$\tpe$} & & & \hspace{34mm} \footnotesize{$\tpe$} &  \\
 \footnotesize{(c)} & \footnotesize{\hspace{4mm} $\Sr=10^{-1}$} &  &  \footnotesize{(d)} & \footnotesize{\hspace{4mm} $\Sr=1$} &   \\
 \makecell{ \\ \vspace{12mm} \rotatebox{90}{\footnotesize{$\twonv$}}} &
\makecell{\includegraphics[width=0.44\textwidth]{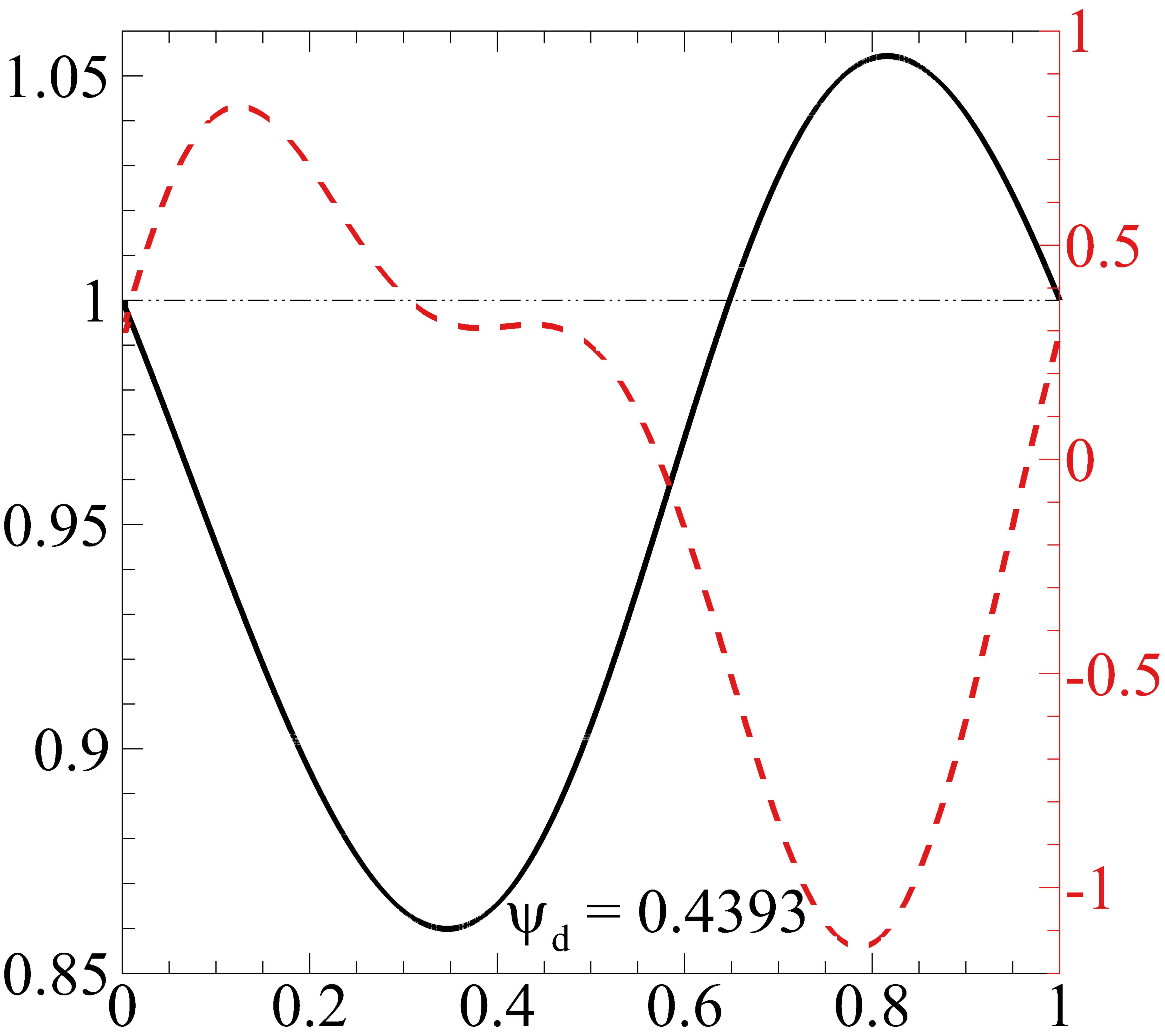}} &
\makecell{ \\  \vspace{11mm} \rotatebox{90}{\footnotesize{$\tmr{10^4\EU}$}}} &
\makecell{ \\  \vspace{12mm} \rotatebox{90}{\footnotesize{$\twonv$}}} &
\makecell{\includegraphics[width=0.44\textwidth]{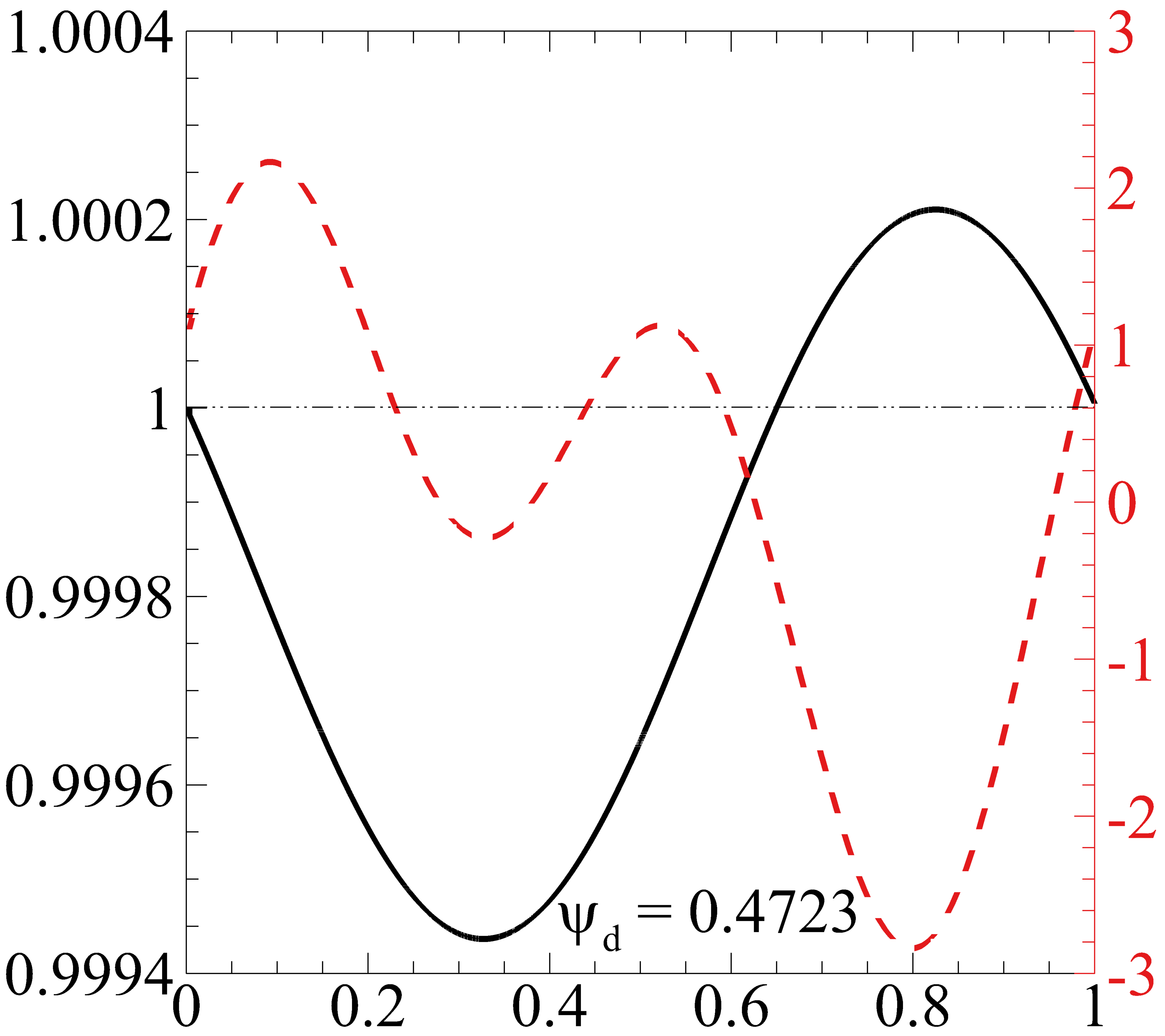}} &
\makecell{ \\  \vspace{11mm} \rotatebox{90}{\footnotesize{$\tmr{10^5\EU}$}}}  \\
 & \hspace{34mm} \footnotesize{$\tpe$} & & & \hspace{34mm} \footnotesize{$\tpe$} &  \\
\end{tabular}
\addtolength{\tabcolsep}{+2pt}
\addtolength{\extrarowheight}{+10pt}
\end{center}
    \caption{The perturbation norm (solid; black) and the base flow energy relative to the time mean (dashed; red) over one period at critical conditions at $H=10$, $\Gamma=10$, for various $\Sr$. The phase differences $\phased$ between the local minimums of each pair of curves are also annotated.}
    \label{fig:pert_base_norm10}
\end{figure}


The energy norms at $\Gamma = 10$, $H=10$ are displayed in \fig\ \ref{fig:pert_base_norm10}. At $\Sr = 10^{-3}$, \fig\ \ref{fig:pert_base_norm10}(a), toward the steady base flow limit, the variation of the perturbation is again a simple sinusoid, slightly lagging behind the base flow energy variation, as for $\Gamma=100$, \fig\ \ref{fig:pert_base_norm100}(a). However, at $\Gamma=10$, the increase in intracylcic growth with reducing $\Gamma$ can be clearly observed, eclipsing 6 orders of magnitude. Thus, at lower $\Gamma$ and $\Sr$, a behavior akin to the `ballistic' regime is reached. At $\Sr = 10^{-2}$, intracyclic growth remains large (the local minimum in $\rrs$ occurs at $\Sr = 9\times 10^{-3}$). An additional complexity in the form of a brief growth in perturbation energy (at $\tpe \approx 0.25$) occurs during the acceleration phase of the base flow, and is not detected at $\Sr < 9\times10^{-3}$. The additional growth incurred by the presence of inflection points is somewhat obscured by the lower $\ReyCrit$ at $\Sr=10^{-2}$. Increasing $\Sr$ to $10^{-1}$, the \TSL\ mode is no longer the least stable. At this $\Sr$ the intracyclic growth is relatively small, likely falling in the `cruising' regime, while by $\Sr=1$ the intracylcic growth again becomes trivial.

\begin{figure}
\begin{center}
\addtolength{\extrarowheight}{-10pt}
\addtolength{\tabcolsep}{-2pt}
\begin{tabular}{ llllll }
\footnotesize{(a)} & \footnotesize{\hspace{4mm} $H=100$, $\Gamma=100$, $\Sr=10^{-2}$} &  &  \footnotesize{(b)} & \footnotesize{\hspace{4mm} $H=10$, $\Gamma=10$, $\Sr=10^{-2}$}  &  \\
\makecell{ \\  \vspace{12mm} \rotatebox{90}{\footnotesize{$y$}}} & 
\makecell{\includegraphics[width=0.44\textwidth]{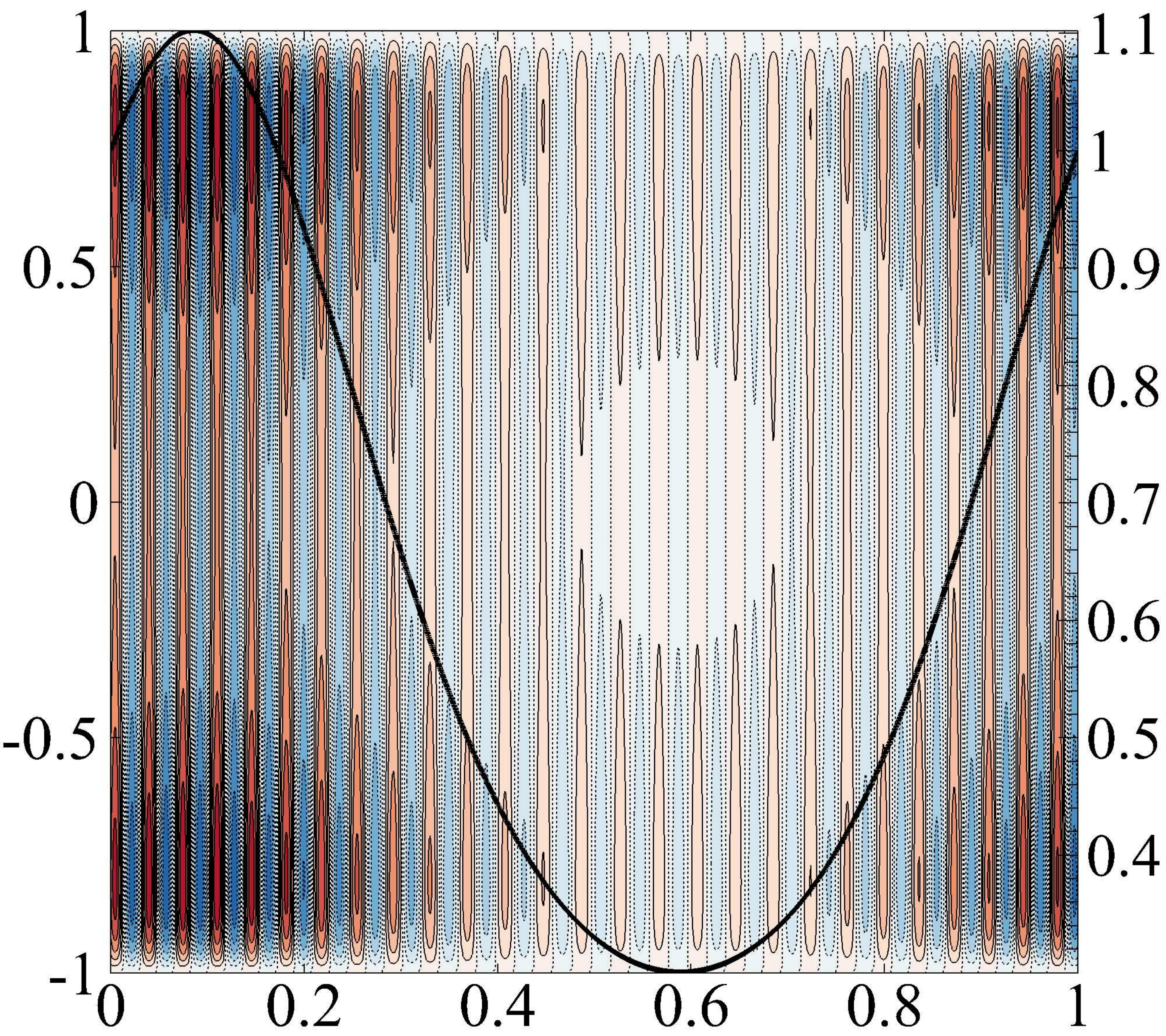}} &
\makecell{ \\  \vspace{11mm} \rotatebox{90}{\footnotesize{$\twonv$}}} &
\makecell{ \\  \vspace{12mm} \rotatebox{90}{\footnotesize{$y$}}} &
\makecell{\includegraphics[width=0.44\textwidth]{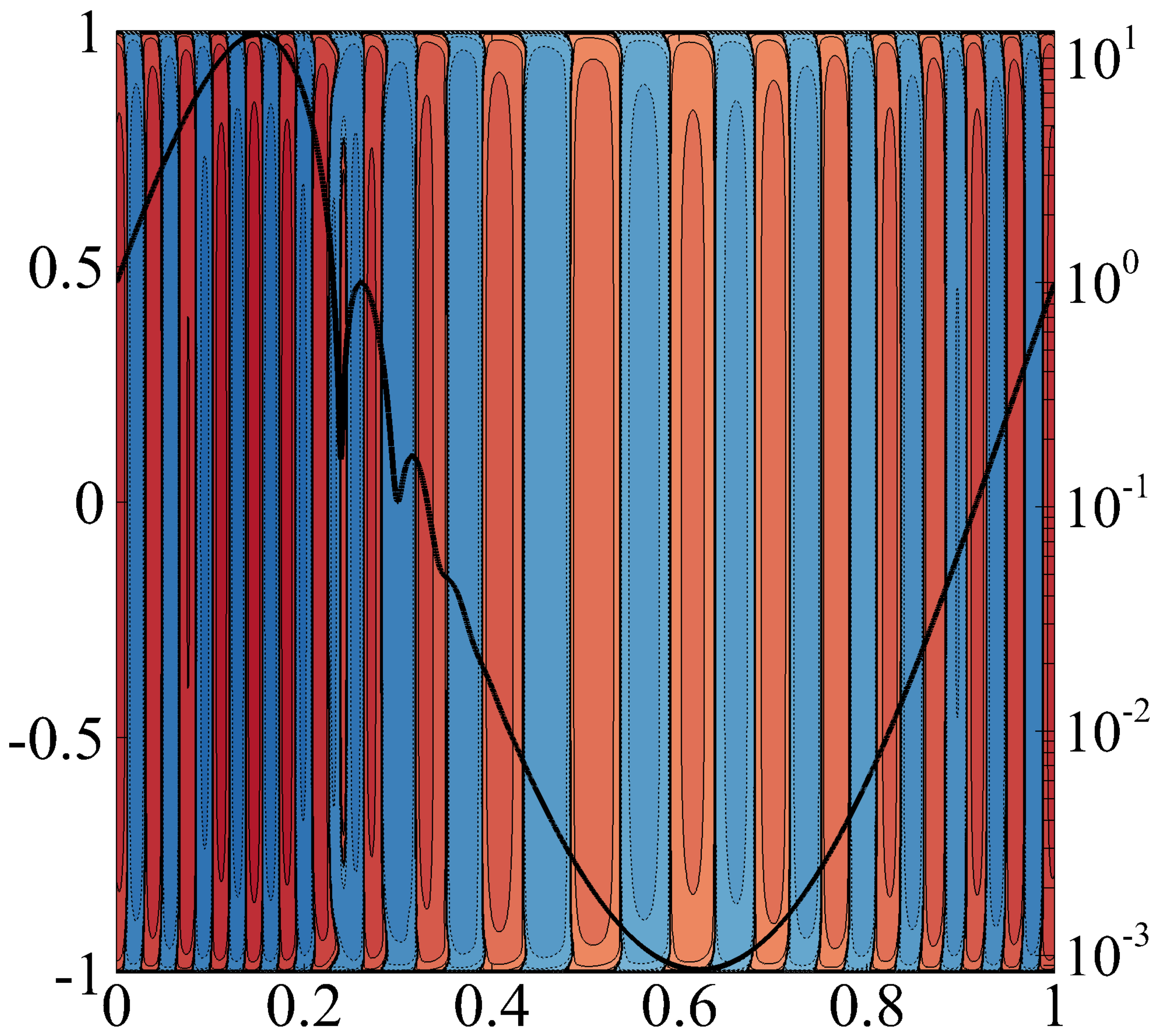}} &
\makecell{\\  \vspace{11mm} \rotatebox{90}{\footnotesize{$\twonv$}}}  \\
 & \hspace{34mm} \footnotesize{$\tpe$} & & & \hspace{34mm} \footnotesize{$\tpe$} &  \\
 \footnotesize{(c)} & \footnotesize{\hspace{4mm} $H=100$, $\Gamma=100$, $\Sr=10^{-1}$} &  &  \footnotesize{(d)} & \footnotesize{\hspace{4mm} $H=10$, $\Gamma=10$, $\Sr=10^{-1}$} &   \\
 \makecell{ \\  \vspace{12mm} \rotatebox{90}{\footnotesize{$y$}}} &
\makecell{\includegraphics[width=0.44\textwidth]{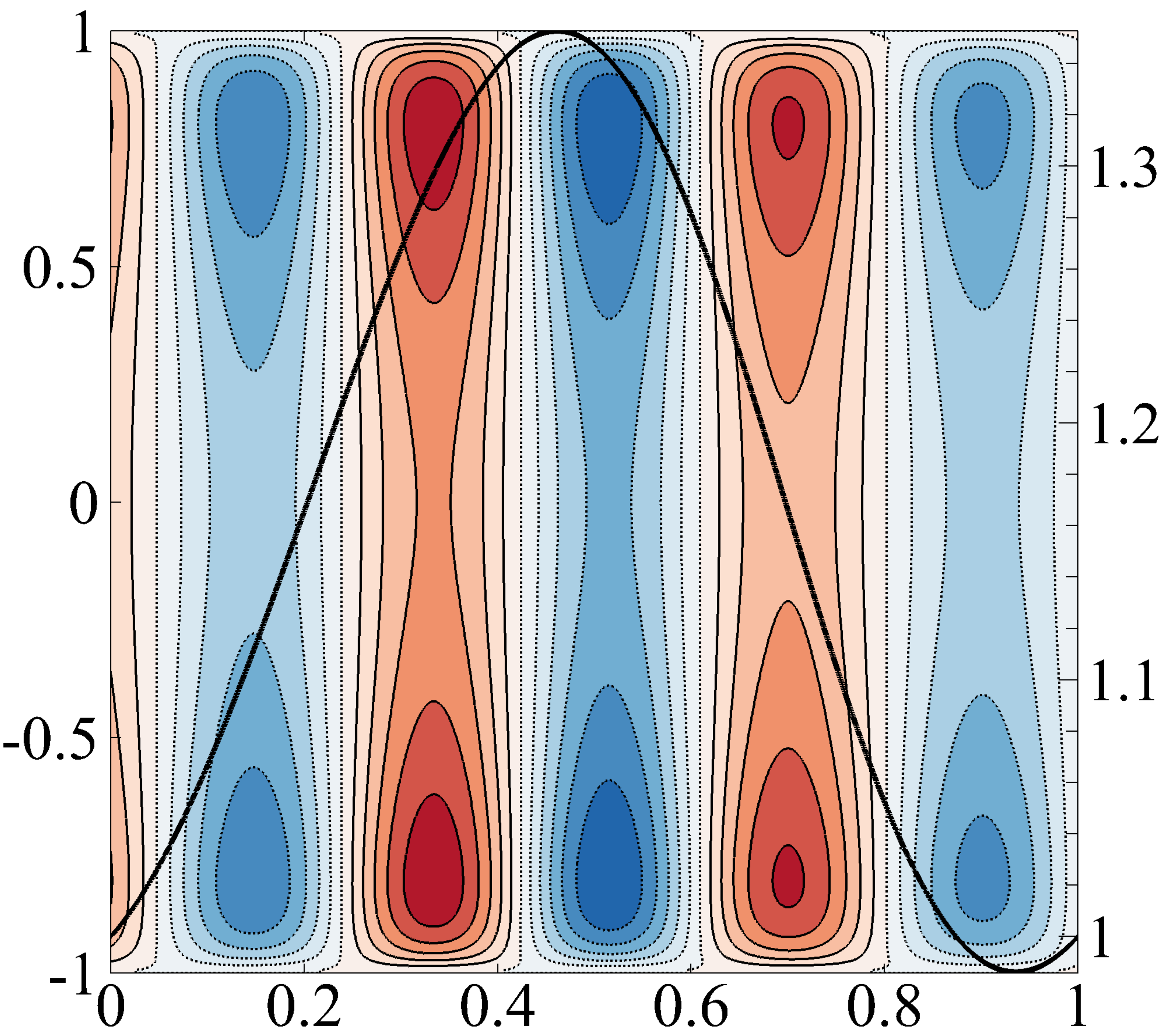}} &
\makecell{ \\  \vspace{11mm} \rotatebox{90}{\footnotesize{$\twonv$}}} &
\makecell{ \\  \vspace{12mm} \rotatebox{90}{\footnotesize{$y$}}} &
\makecell{\includegraphics[width=0.44\textwidth]{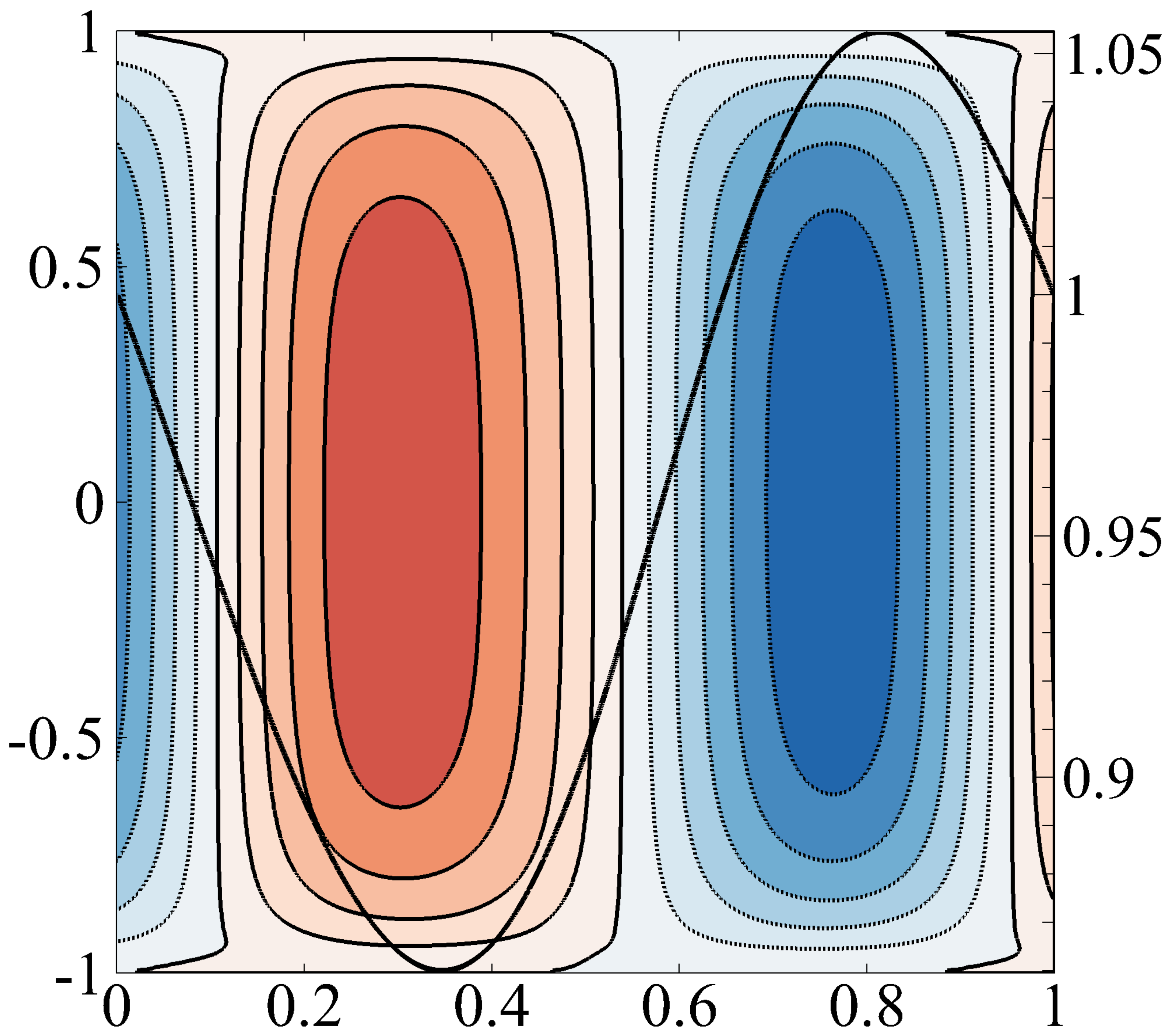}} &
\makecell{ \\  \vspace{11mm} \rotatebox{90}{\footnotesize{$\twonv$}}}  \\
 & \hspace{34mm} \footnotesize{$\tpe$} & & & \hspace{34mm} \footnotesize{$\tpe$} &  \\
  \footnotesize{(e)} & \footnotesize{\hspace{4mm} $H=100$, $\Gamma=100$, $\Sr=1$} &  &  \footnotesize{(f)} & \footnotesize{\hspace{4mm} $H=10$, $\Gamma=10$, $\Sr=1$} &   \\
 \makecell{ \\  \vspace{12mm} \rotatebox{90}{\footnotesize{$y$}}} &
\makecell{\includegraphics[width=0.44\textwidth]{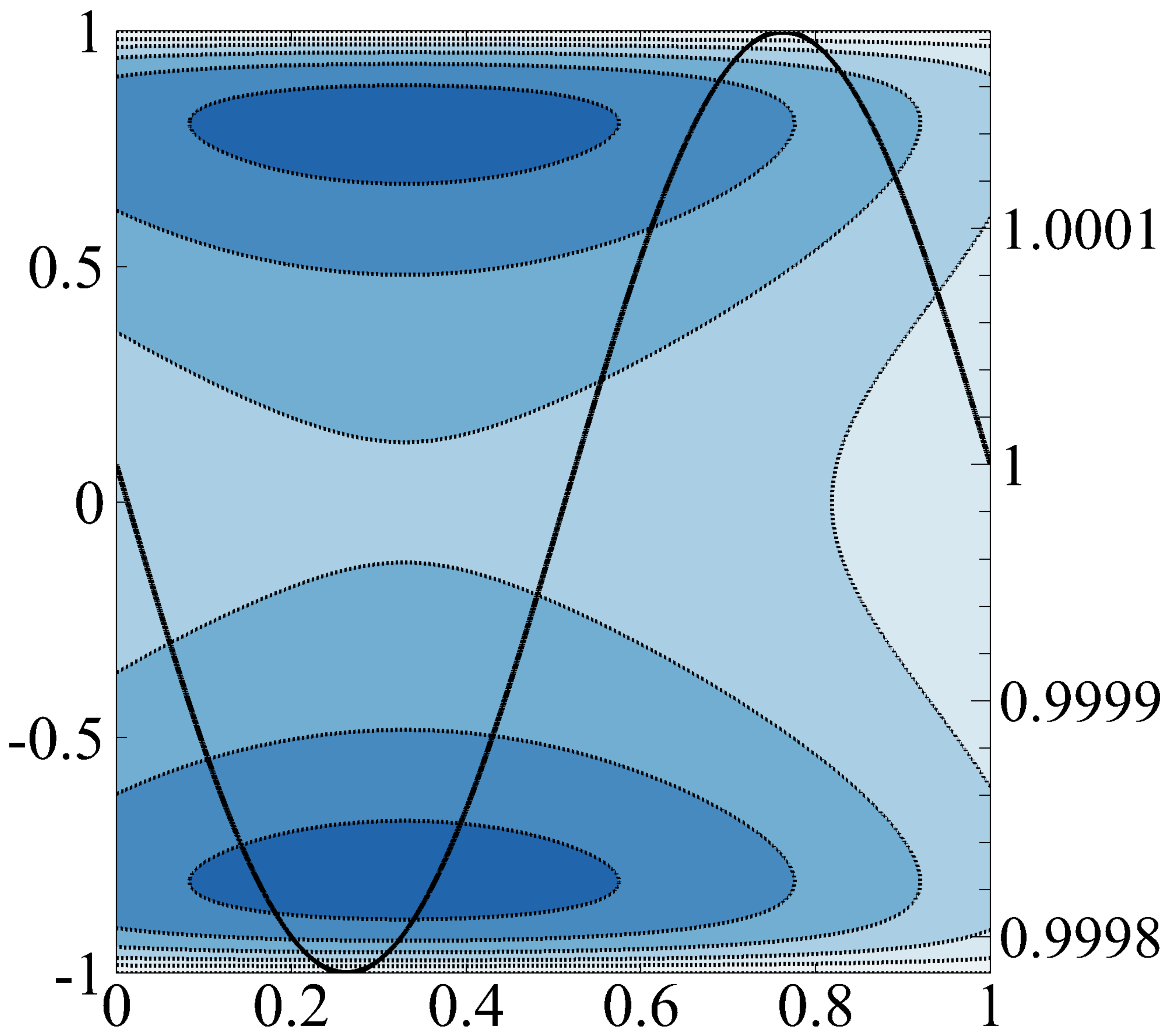}} &
\makecell{ \\  \vspace{11mm} \rotatebox{90}{\footnotesize{$\twonv$}}} &
\makecell{ \\  \vspace{12mm} \rotatebox{90}{\footnotesize{$y$}}} &
\makecell{\includegraphics[width=0.44\textwidth]{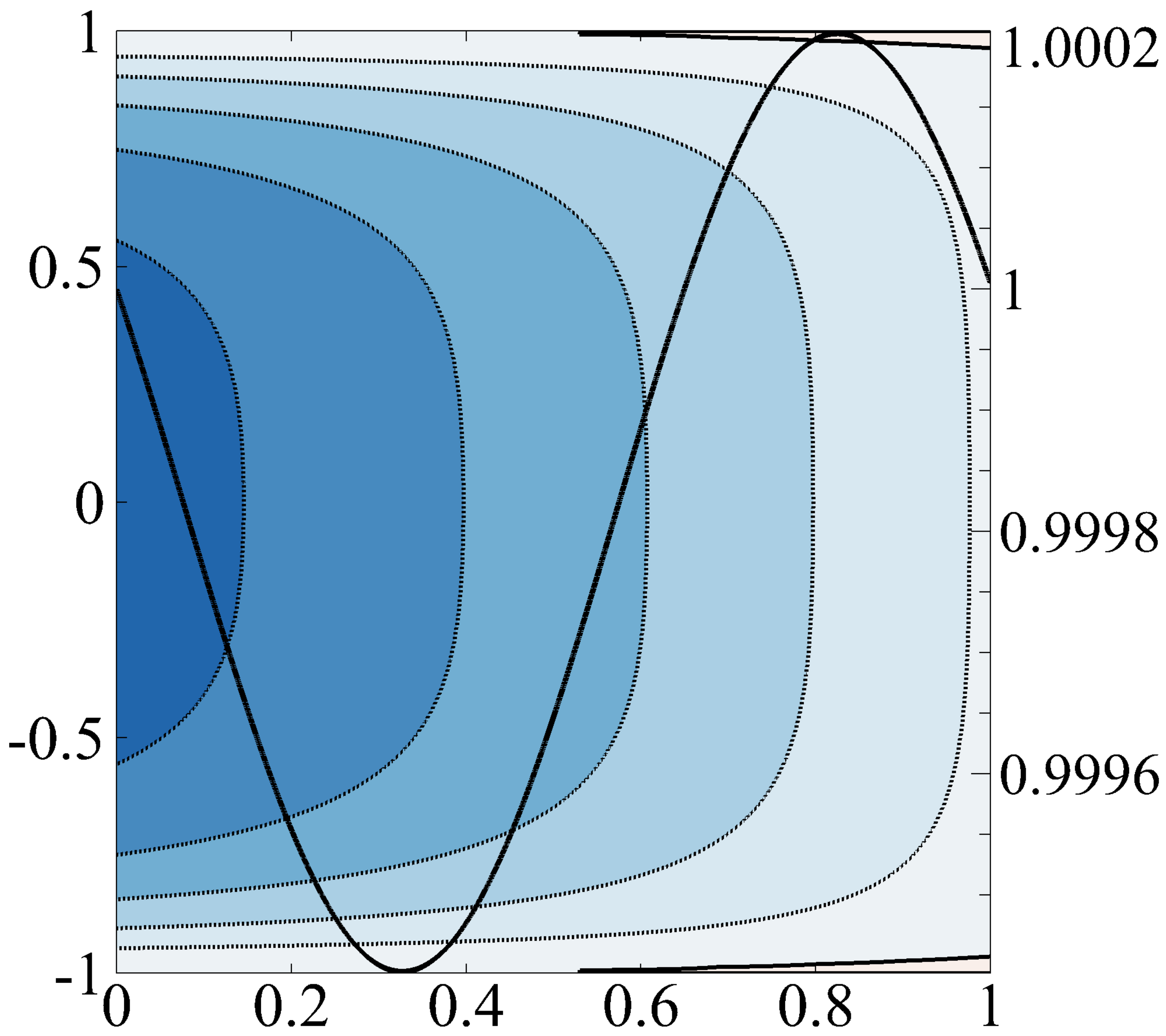}} &
\makecell{ \\  \vspace{11mm} \rotatebox{90}{\footnotesize{$\twonv$}}}  \\
 & \hspace{34mm} \footnotesize{$\tpe$} & & & \hspace{34mm} \footnotesize{$\tpe$} &  \\
\end{tabular}
\addtolength{\tabcolsep}{+2pt}
\addtolength{\extrarowheight}{+10pt}
\end{center}
    \caption{The linear evolution of the leading eigenvector $\tilde{v}(y,t)$ over one period. Linearly spaced contours between $\pm \max|\tilde{v}|$ are plotted, solid lines (red flooding) denote positive values, dotted lines (blue flooding) negative values, except for $H=10$, $\Sr=10^{-2}$, with logarithmically spaced contours between $-15$ and $15$. Perturbation norms $\twonv$ from \figs\ \ref{fig:pert_base_norm100} and \ref{fig:pert_base_norm10} are overlaid.}
    \label{fig:pert_xt_norm10}
\end{figure}

The linearized evolutions of the leading eigenvector are depicted over the period of the base flow in \fig\ \ref{fig:pert_xt_norm10}. At $\Gamma = 100$ the dominant mode is the \TSL\ mode for all $\Sr$, with a structure that does not observably change with time, as shown in the accompanying animation \cite{Supvideos2020}. The amplitude variations are also small; many repetitions of the wave are visible at lower $\Sr$ as the advection timescale is much smaller than the transient inertial timescale. Although the mode has a very similar appearance to that of a steady \TS\ wave, the additional isolation of the boundary layers means that the $H=100$ pulsatile mode resembles a $H=400$ steady mode \citep{Camobreco2020transition}. Once $H$ is reduced, separate \TS\ waves are no longer observed at each wall, but appear as a single conjoined structure. While at larger $\Sr$, the $H=10$, $\Gamma = 10$ mode structure still displays minimal time variation. Only at $\Sr=10^{-2}$ is significant unsteadiness observed, slightly towards the walls, and prominently during the disruption of the decay phase (at $\tpe \approx 0.25$). However, the general appearance of the structure as a conjoined \TS\ wave persists (this case is also animated \cite{Supvideos2020}). 


\begin{figure}
\begin{center}
\addtolength{\extrarowheight}{-10pt}
\addtolength{\tabcolsep}{-2pt}
\begin{tabular}{ llllll }
\footnotesize{(a)} & \footnotesize{\hspace{4mm} $\Sr=10^{-3}$} &  &  \footnotesize{(b)} & \footnotesize{\hspace{4mm} $\Sr=4\times10^{-3}$}  &  \\
\makecell{ \\  \vspace{12mm} \rotatebox{90}{\footnotesize{$\twonv$}}} & 
\makecell{\includegraphics[width=0.44\textwidth]{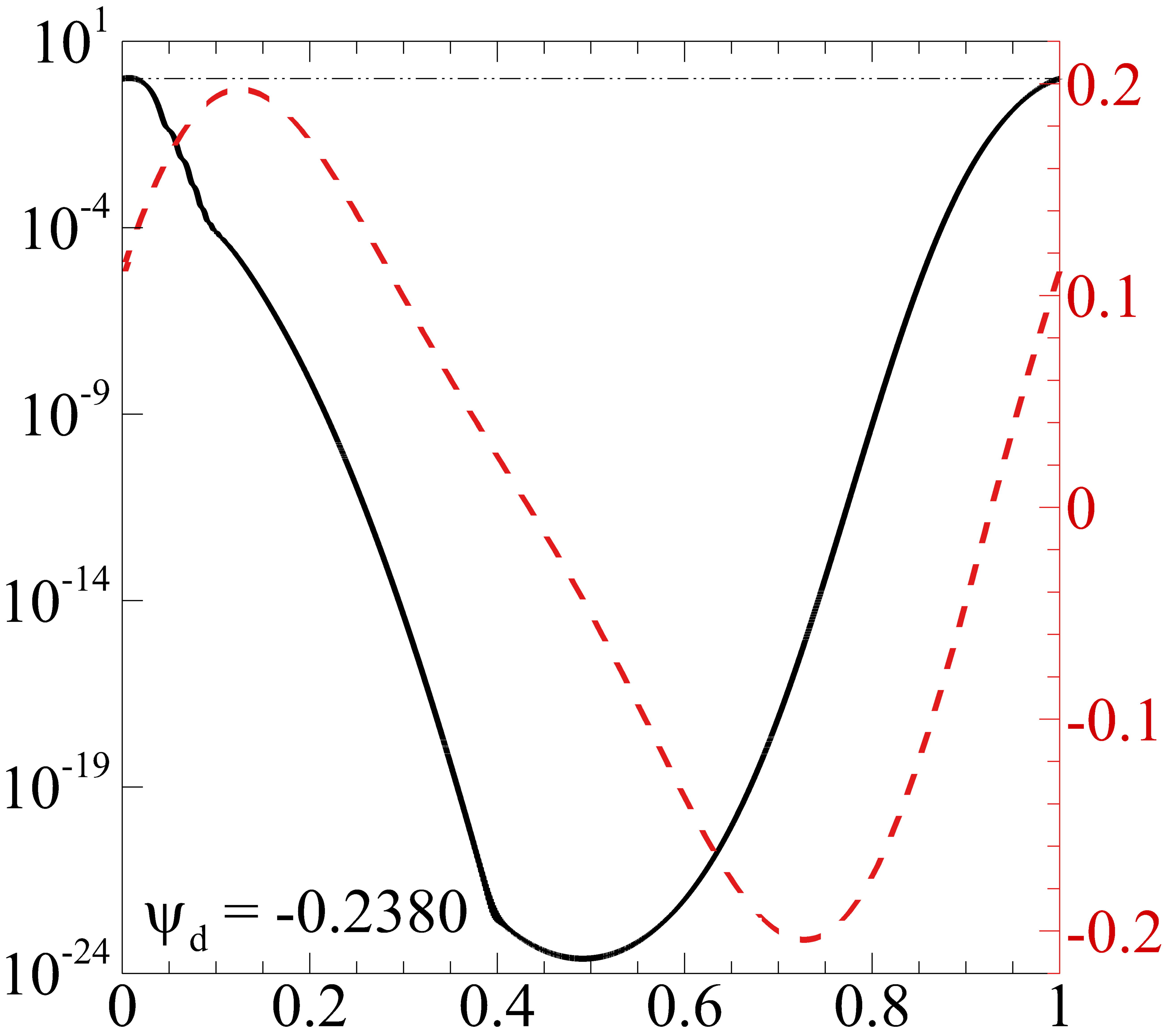}} &
\makecell{ \\  \vspace{11mm} \rotatebox{90}{\footnotesize{$\tmr{\EU}$}}} &
\makecell{ \\  \vspace{12mm} \rotatebox{90}{\footnotesize{$\twonv$}}} &
\makecell{\includegraphics[width=0.44\textwidth]{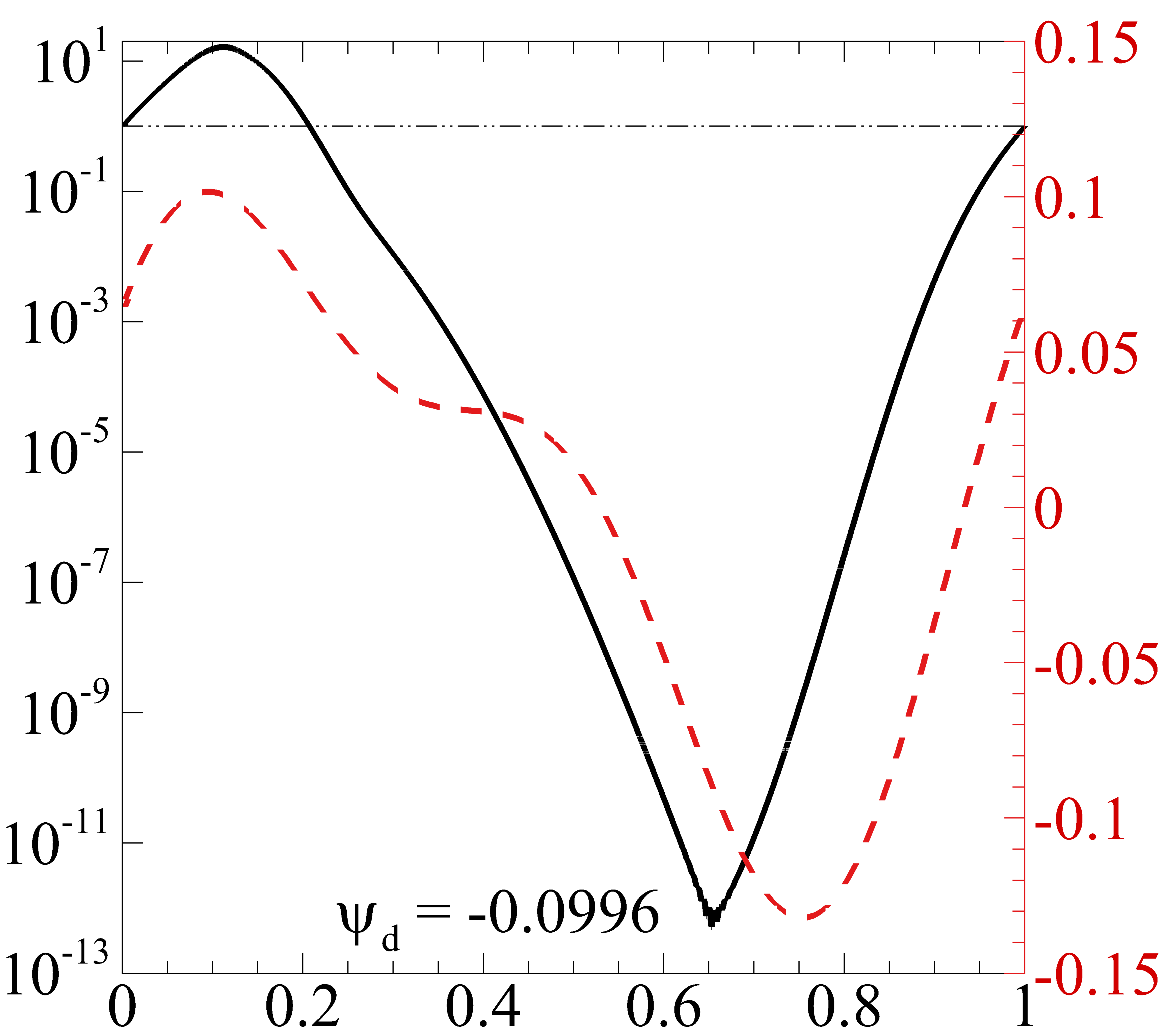}} &
\makecell{ \\  \vspace{11mm} \rotatebox{90}{\footnotesize{$\tmr{\EU}$}}}  \\
 & \hspace{34mm} \footnotesize{$\tpe$} & & & \hspace{34mm} \footnotesize{$\tpe$} &  \\
 \footnotesize{(c)} & \footnotesize{\hspace{4mm} $\Sr=7.2\times10^{-3}$} &  &  \footnotesize{(d)} & \footnotesize{\hspace{4mm} $\Sr=10^{-2}$} &   \\
 \makecell{ \\ \vspace{12mm} \rotatebox{90}{\footnotesize{$\twonv$}}} &
\makecell{\includegraphics[width=0.44\textwidth]{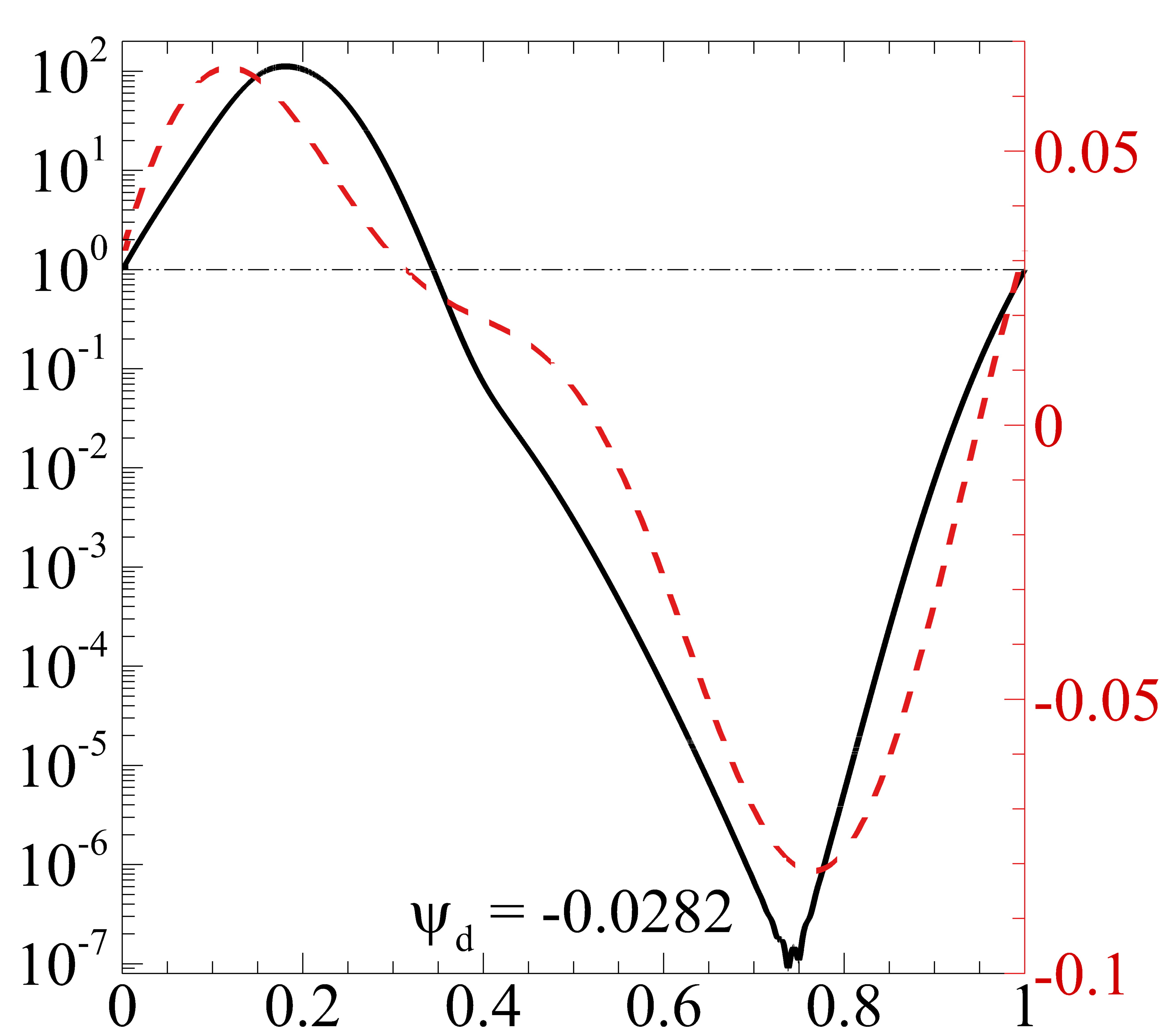}} &
\makecell{ \\  \vspace{11mm} \rotatebox{90}{\footnotesize{$\tmr{\EU}$}}} &
\makecell{ \\  \vspace{12mm} \rotatebox{90}{\footnotesize{$\twonv$}}} &
\makecell{\includegraphics[width=0.44\textwidth]{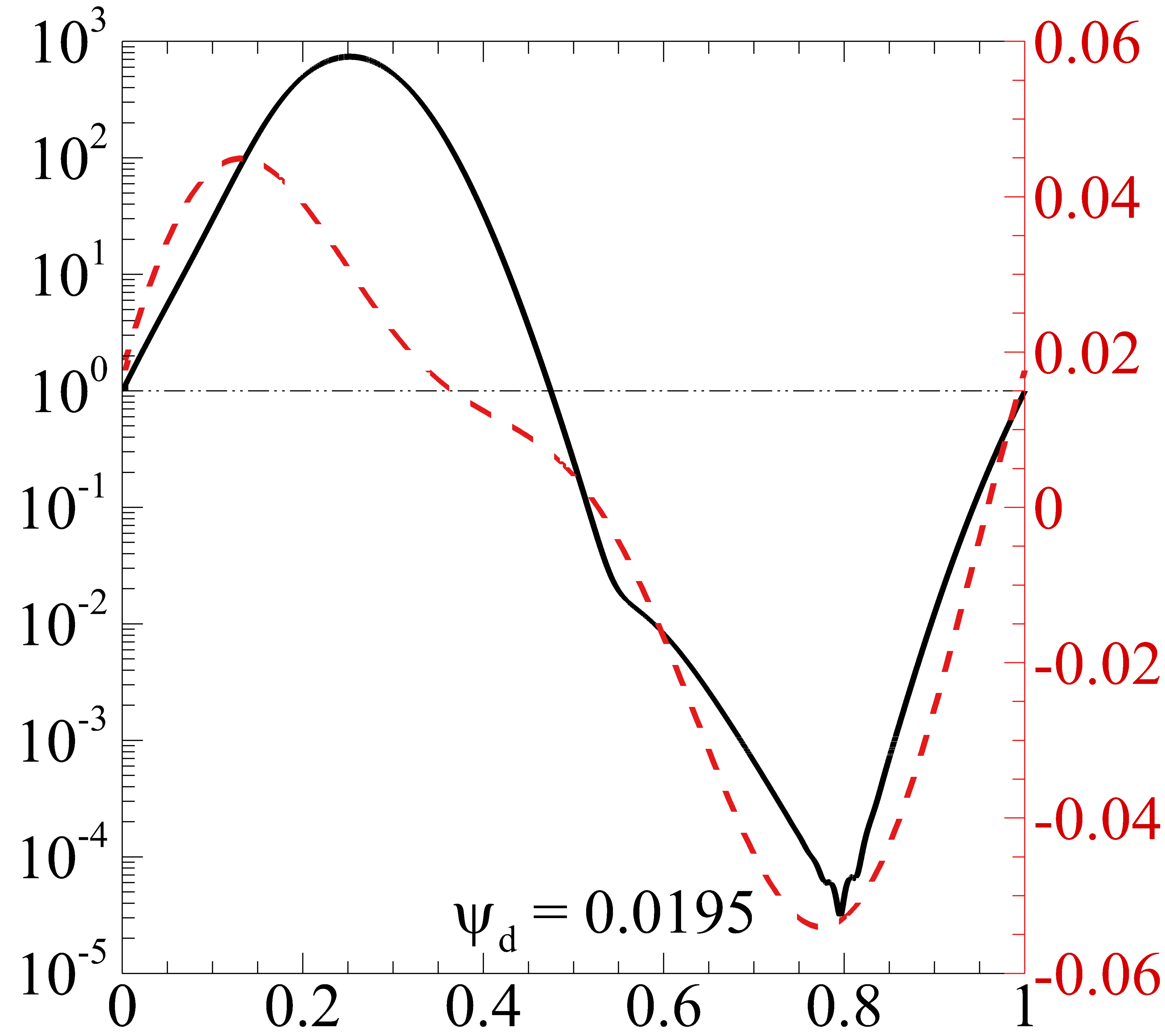}} &
\makecell{ \\  \vspace{11mm} \rotatebox{90}{\footnotesize{$\tmr{\EU}$}}}  \\
 & \hspace{34mm} \footnotesize{$\tpe$} & & & \hspace{34mm} \footnotesize{$\tpe$} &  \\
\end{tabular}
\addtolength{\tabcolsep}{+2pt}
\addtolength{\extrarowheight}{+10pt}
\end{center}
    \caption{The perturbation norm (solid; black) and the base flow energy relative to the time mean (dashed; red) over one period at critical conditions at $H=10$, $\Gamma=1.24$, for various $\Sr$. The $\Sr=7.2\times10^{-3}$ case represents the optimized pulsation for this $H$, recalling \tbl\ \ref{tab:tab_3}. The phase differences $\phased$ between the local minimums of each pair of curves are also annotated.}
    \label{fig:pert_base_norm1}
\end{figure}

Finally, at $H=1$, the optimized conditions ($\Gamma = 1.24$, $\Sr = 7.2 \times 10^{-3}$) and nearby $\Sr$ are considered, with the energy norms displayed in \fig\ \ref{fig:pert_base_norm1}. A smaller $\Gamma$ features staggering intracyclic growth, with almost 24 orders of magnitude of growth at $\Sr = 10^{-3}$. Similar to previous cases, at lower $\Sr$ the local minimum in perturbation energy significantly lags behind the minimum in the base flow energy, $\phased=-0.2380$. However, an additional feature at smaller $\Sr$ and $\Gamma$ is that the perturbation decay is more rapid, and almost plateaus at low energies (with neither a smooth transitioning from growth to decay or sharp bounce back up). At the slightly larger $\Sr = 4\times 10^{-3}$, the decay is not so rapid (decaying over $0.112<\tpe<0.653$ compared to $0.008<\tpe<0.491$),with a sharp bounce back to growth and a smaller lag in the locations of the local minima, $\phased = -0.0996$. At the optimized $\Sr = 7.2 \times 10^{-3}$, the decay rate of the perturbation is matched to the period of the base flow, the local minima in energy are close to coinciding ($\phased = -0.0282$), and so inflection points are maintained throughout the deceleration phase ($\rrs$ is then minimized). At larger $\Sr$, the perturbation energy leads the base flow energy ($\phased = 0.0195$), and the deceleration phase is not used to its full extent.

\begin{figure}
\begin{center}
\addtolength{\extrarowheight}{-10pt}
\addtolength{\tabcolsep}{-2pt}
\begin{tabular}{ ll ll }
\footnotesize{(a)} & \footnotesize{\hspace{3mm} $\tpe=0$, $\max(|\hat{v}|)=2.991\times10^{-1}$}  &
\footnotesize{(b)} & \footnotesize{\hspace{3mm} $\tpe=0.1$, $\max(|\hat{v}|)=8.215\times10^{0}$} \\
\makecell{ \\  \vspace{10mm} \rotatebox{90}{\footnotesize{$y$}}} & \makecell{\includegraphics[width=0.458\textwidth]{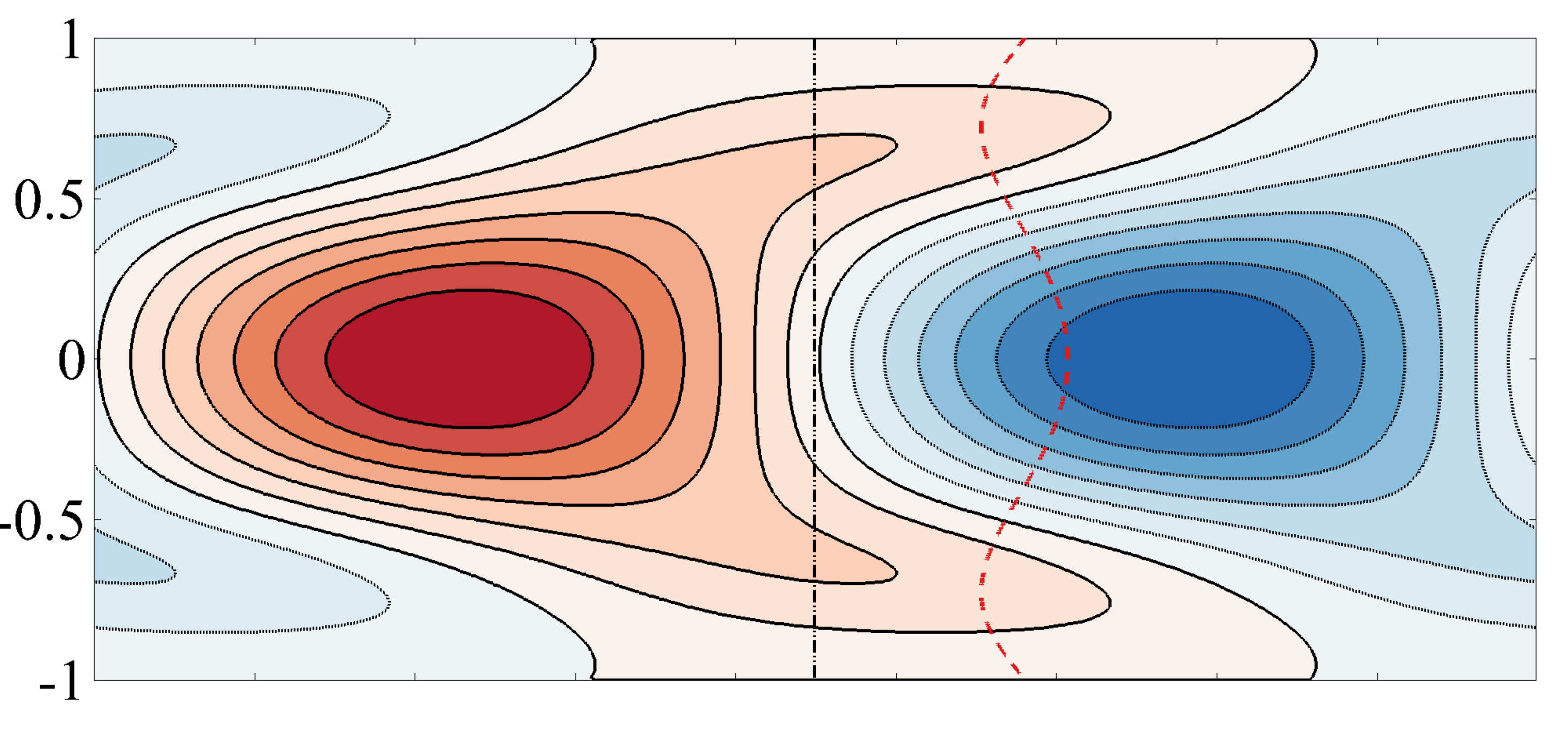}} & 
\makecell{ \\  \vspace{10mm} } & \makecell{\includegraphics[width=0.458\textwidth]{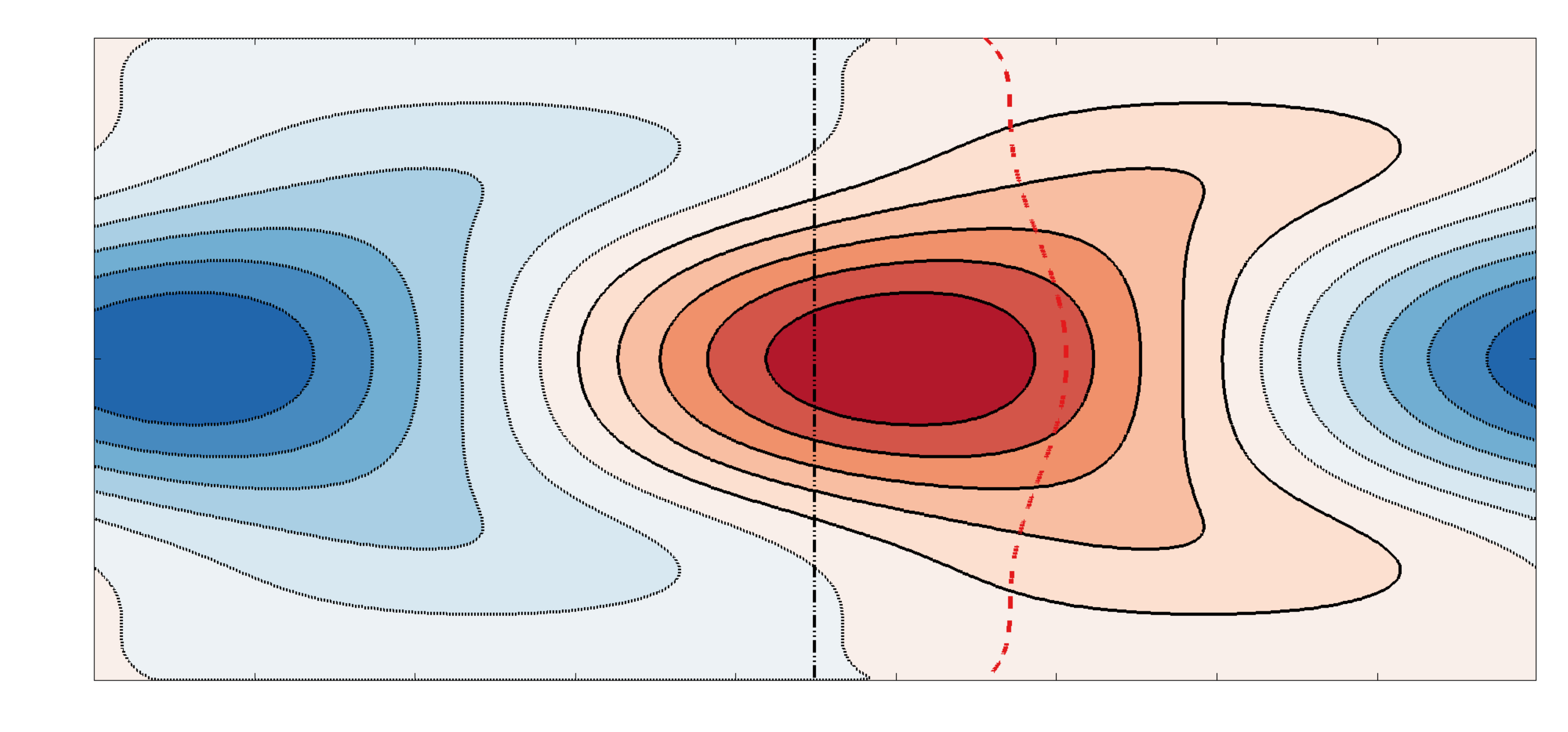}} \\
\footnotesize{(c)} & \footnotesize{\hspace{3mm} $\tpe=0.2$, $\max(|\hat{v}|)=3.273\times10^{1}$} &
\footnotesize{(d)} & \footnotesize{\hspace{3mm} $\tpe=0.3$, $\max(|\hat{v}|)=2.474\times10^{0}$} \\
\makecell{ \\  \vspace{10mm} \rotatebox{90}{\footnotesize{$y$}}}  & \makecell{\includegraphics[width=0.458\textwidth]{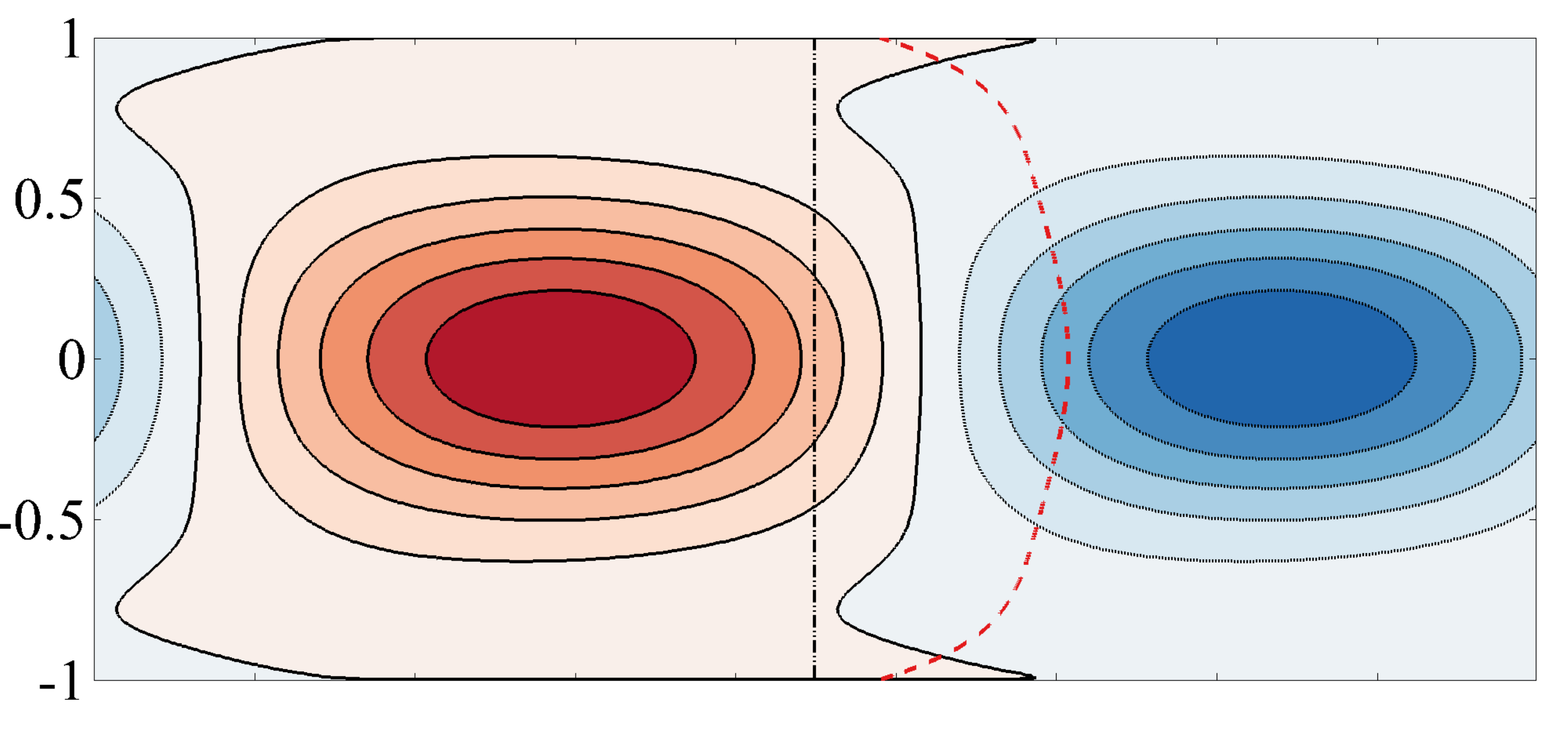}} & 
\makecell{ \\  \vspace{10mm} } & \makecell{\includegraphics[width=0.458\textwidth]{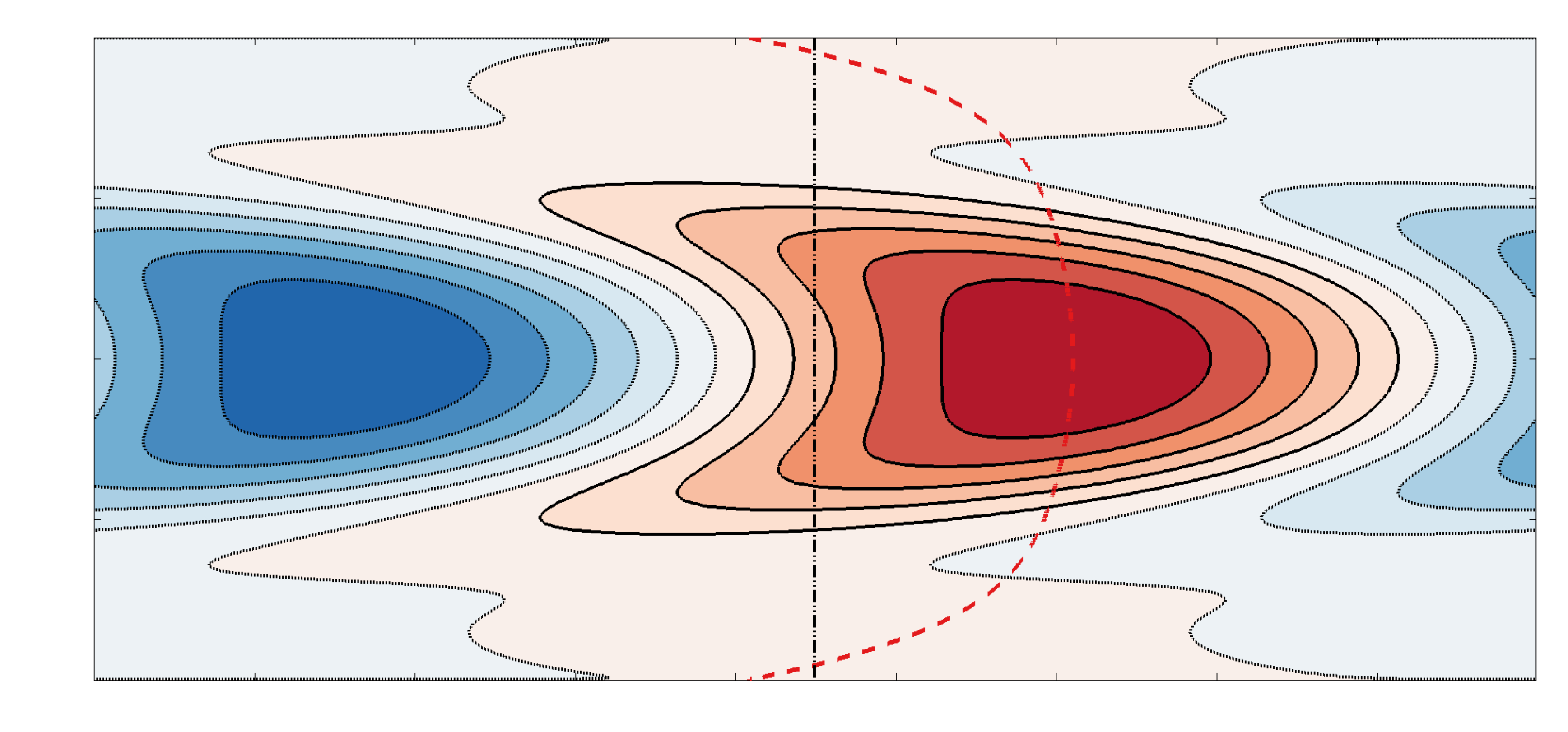}} \\
\footnotesize{(e)} & \footnotesize{\hspace{3mm} $\tpe=0.4$, $\max(|\hat{v}|)=2.879\times10^{-2}$}  &
\footnotesize{(f)} & \footnotesize{\hspace{3mm} $\tpe=0.7$, $\max(|\hat{v}|)=3.051\times10^{-7}$} \\
\makecell{ \\  \vspace{10mm} \rotatebox{90}{\footnotesize{$y$}}} & \makecell{\includegraphics[width=0.458\textwidth]{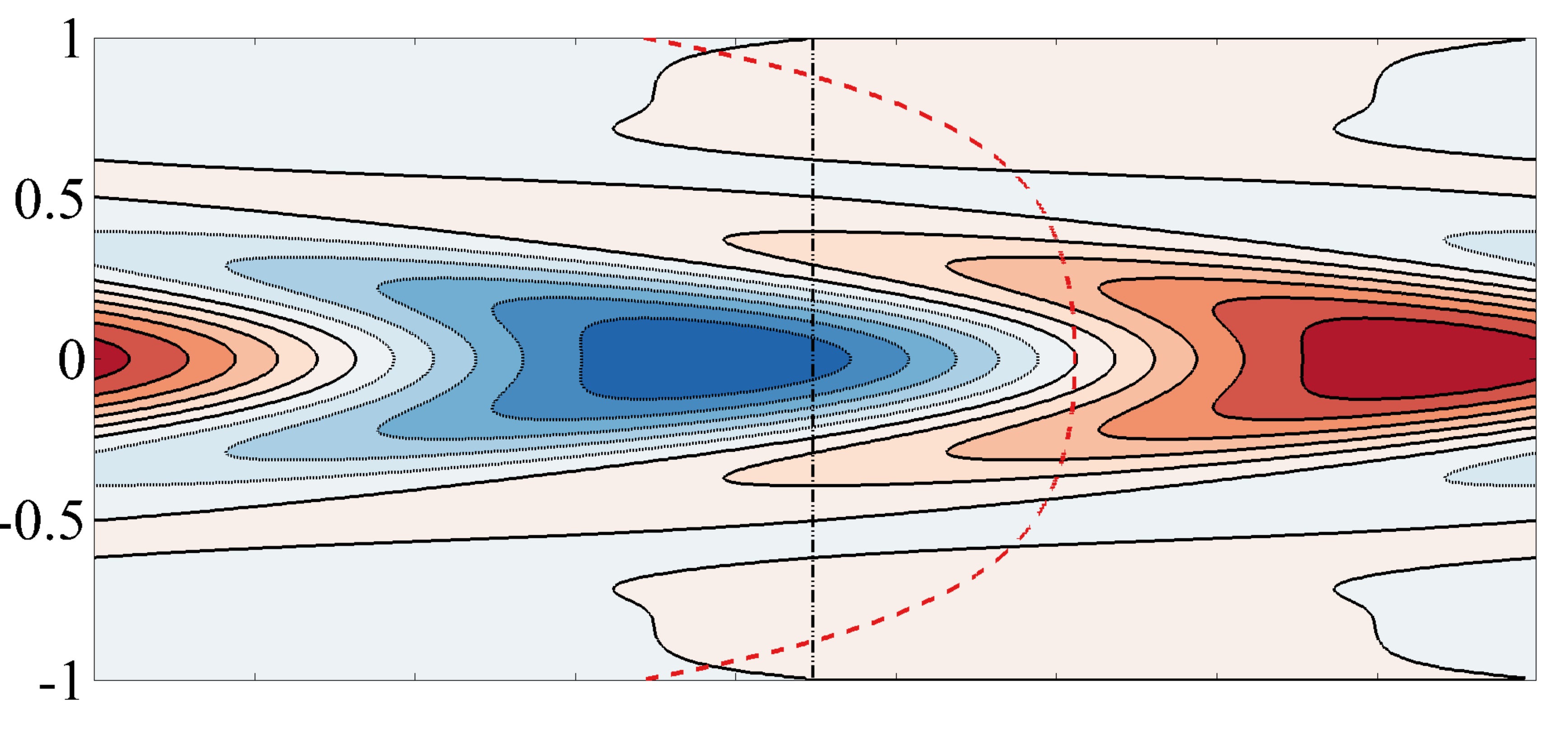}} & 
\makecell{ \\  \vspace{10mm} } & \makecell{\includegraphics[width=0.458\textwidth]{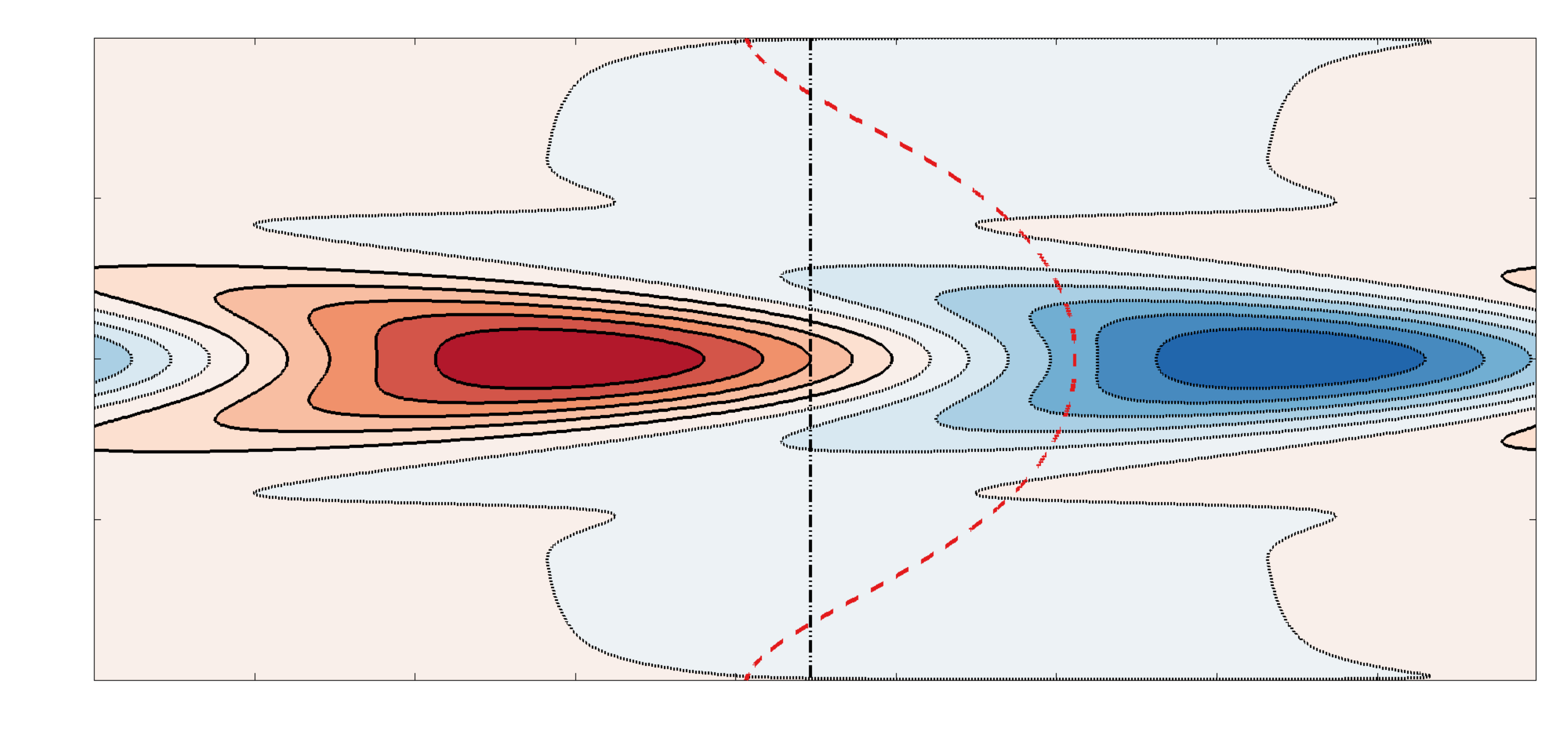}} \\
\footnotesize{(g)} & \footnotesize{\hspace{3mm} $\tpe=0.75$, $\max(|\hat{v}|)=2.954\times10^{-8}$} &
\footnotesize{(h)} & \footnotesize{\hspace{3mm} $\tpe=0.85$, $\max(|\hat{v}|)=6.752\times10^{-5}$} \\
\makecell{ \\  \vspace{10mm} \rotatebox{90}{\footnotesize{$y$}}}  & \makecell{\includegraphics[width=0.458\textwidth]{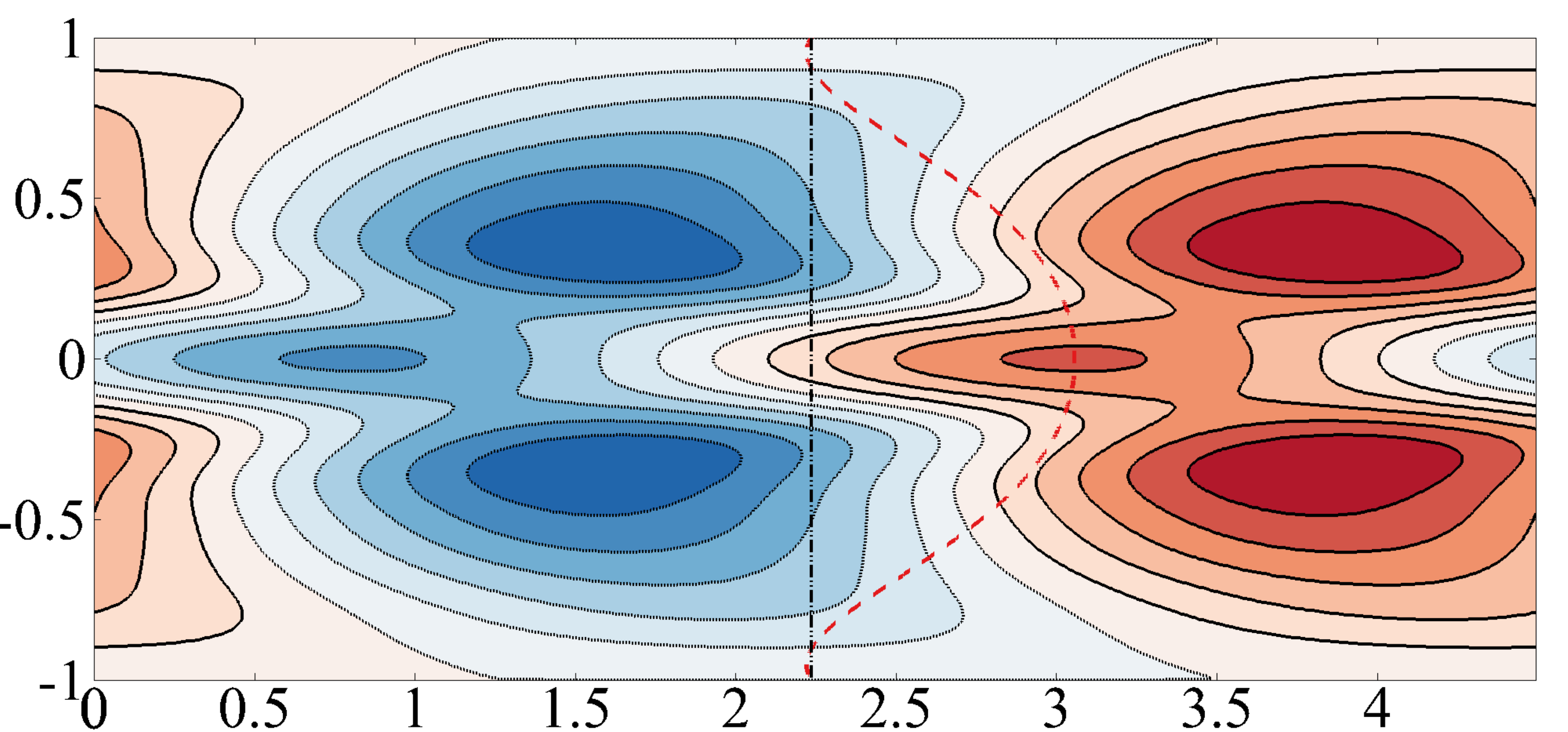}} & 
\makecell{ \\  \vspace{10mm} } & \makecell{\includegraphics[width=0.458\textwidth]{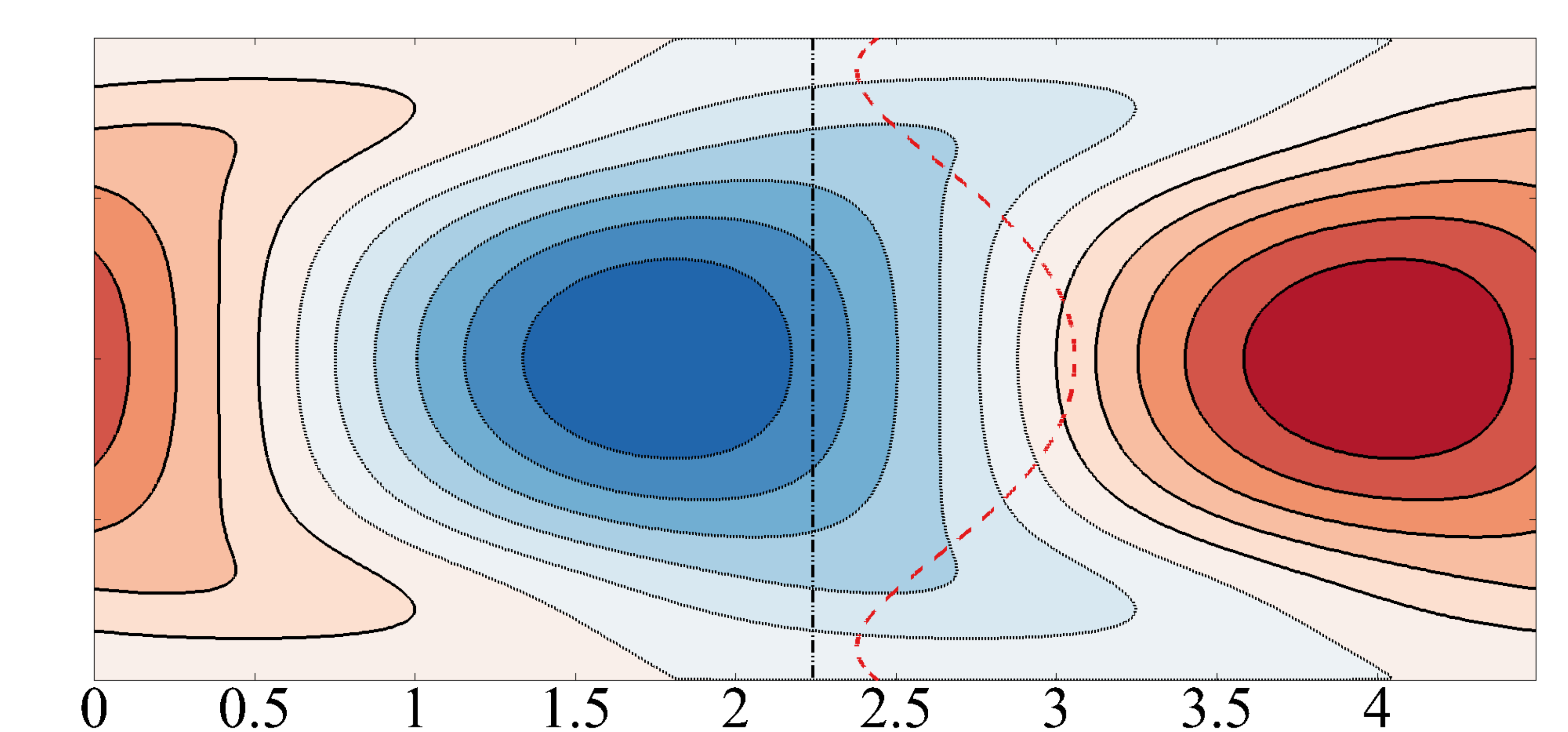}} \\
  & \hspace{38mm} \footnotesize{$x$} &  & \hspace{38mm} \footnotesize{$x$} \\
\end{tabular}
\addtolength{\tabcolsep}{+2pt}
\addtolength{\extrarowheight}{+10pt}
\end{center}
    \caption{Snapshots of the eigenvector expanded in the streamwise direction $\hat{v}=\tilde{v}(y,t)\exp(i\alpha x)$ through one cycle $\tpe \in [0,1]$ at $H=1$, $\Gamma=1.24$, $\Sr=7.2\times10^{-3}$. The base flow is overlaid (the black dashed line indicates zero base flow velocity). Red flooding positive; blue flooding negative.}
    \label{fig:snapshots_H1_helper}
\end{figure}

The evolution of the optimized perturbation at $H=1$ is shown in \fig\ \ref{fig:snapshots_H1_helper}, and in a supplementary animation \cite{Supvideos2020}. From $\tpe=0$, the perturbation is slowly growing, aided by the single large inflection points present in each half of the domain. As these become less pronounced, the `wings' of the perturbation are pulled in ($\tpe=0.2$). By this point, inflection points in the base flow have vanished, as the wall oscillation follows through to negative velocities, although a small amount of residual growth is maintained. The pull of the walls on the central structure sweeps the `wings' forward ($\tpe = 0.3$) as the base flow velocity in the central region is smaller than the velocities near the walls. The downstream pull of the walls acts to increasingly shear the structure, with perturbation decay until $\tpe =0.738$. The structure rapidly reorients to the wider forward `winged' structure just as inflection points reappear in the base flow, near $\tpe = 0.75$. As these inflection points become more pronounced rapid growth occurs, while the `wings' are swept further forward.


\section{Nonlinear analysis}\label{sec:nlin_all}
\subsection{Formulation and validation}\label{sec:nlin_for}

We now seek to investigate the nonlinear behavior of the optimized pulsations at various $H$. As a first step in investigating transitions to turbulence, the modal instabilities predicted in the preceding sections are targeted by the DNS. Although linear or nonlinear transiently growing disturbances may initiate bypass transition scenarios \cite{Kerswell2014optimization, Reshotko2001transient, Schmid2001stability, Trefethen1993hydrodynamic, Waleffe1995transition}, the modal instability seemed the natural starting point. Furthermore, if the modal instability has a large decay rate, linear transient growth mechanisms can be strongly compromised  \cite{Lozano2020cause}, as observed for cylinder wakes in particular \cite{Abdessemed2009transient}. Finally, previous work on steady Q2D transistions observed that only turbulence generated by a modal instability \cite{Camobreco2020role, Camobreco2020transition} was sustainable in wall-driven channel flows.

The direct numerical simulation of Eqs.~(\ref{eq:non_dim_m}) and (\ref{eq:non_dim_c}) is performed as follows. The initial field is solely the analytic solution from Sec.~\ref{sec:prob_set}, $\vect{u}=U(y,t=0)$. The initial phase did not prove relevant with either an initial seed of white noise, or no initial perturbation. The flow is driven by a constant pressure gradient, $\partial P/\partial x = \gamma_1 (\cosh(H^{1/2})/(\cosh(H^{1/2}-1))H/\Rey$, with the pressure decomposed into a linearly varying and fluctuating periodic component, as $p = P + p'$, respectively. Periodic boundary conditions, $\vect{u}(x=0)=\vect{u}(x=W)$ and $p'(x=0)=p'(x=W)$, are applied at the downstream and upstream boundaries. The domain length $W=2\pi/\alpha_\mathrm{max}$ is set to match the wave number that achieved maximal linear growth $\alpha_\mathrm{max}$. Synchronous lateral wall movement generates the oscillating flow component, with boundary conditions $U(y \pm 1,t) = \gamma_2 \cos(t)$. 
 
Simulations are performed with an in-house spectral element solver, employing a third order backward differencing scheme, with operator splitting, for time integration. High-order Neumann pressure boundary conditions are imposed on the oscillating walls to maintain third order time accuracy \citep{Karniadakis1991high}. The Cartesian domain is discretized with quadrilateral elements over which Gauss--Legendre--Lobatto nodes are placed. The mesh design is identical to that of Ref.~\citep{Camobreco2020transition}. The wall-normal resolution was unchanged, although the streamwise resolution was doubled. Elements are otherwise uniformly distributed in both streamwise and transverse directions, ensuring perturbations remain well resolved during all phases of their growth.  The solver, incorporating the SM82 friction term, has been previously introduced and validated \citep{Cassels2016heat, Cassels2019from3D, Hussam2012optimal, Sheard2009cylinders}.

\begin{figure}
\begin{center}
\addtolength{\extrarowheight}{-10pt}
\addtolength{\tabcolsep}{-2pt}
\begin{tabular}{ llll }
\makecell{\footnotesize{(a)} \vspace{19.5mm}  \\  \vspace{24mm} \rotatebox{90}{\footnotesize{$\int |\hat{v}| \mathrm{d}\Omega,\, \twonv$}}} 
& \makecell{\includegraphics[width=0.458\textwidth]{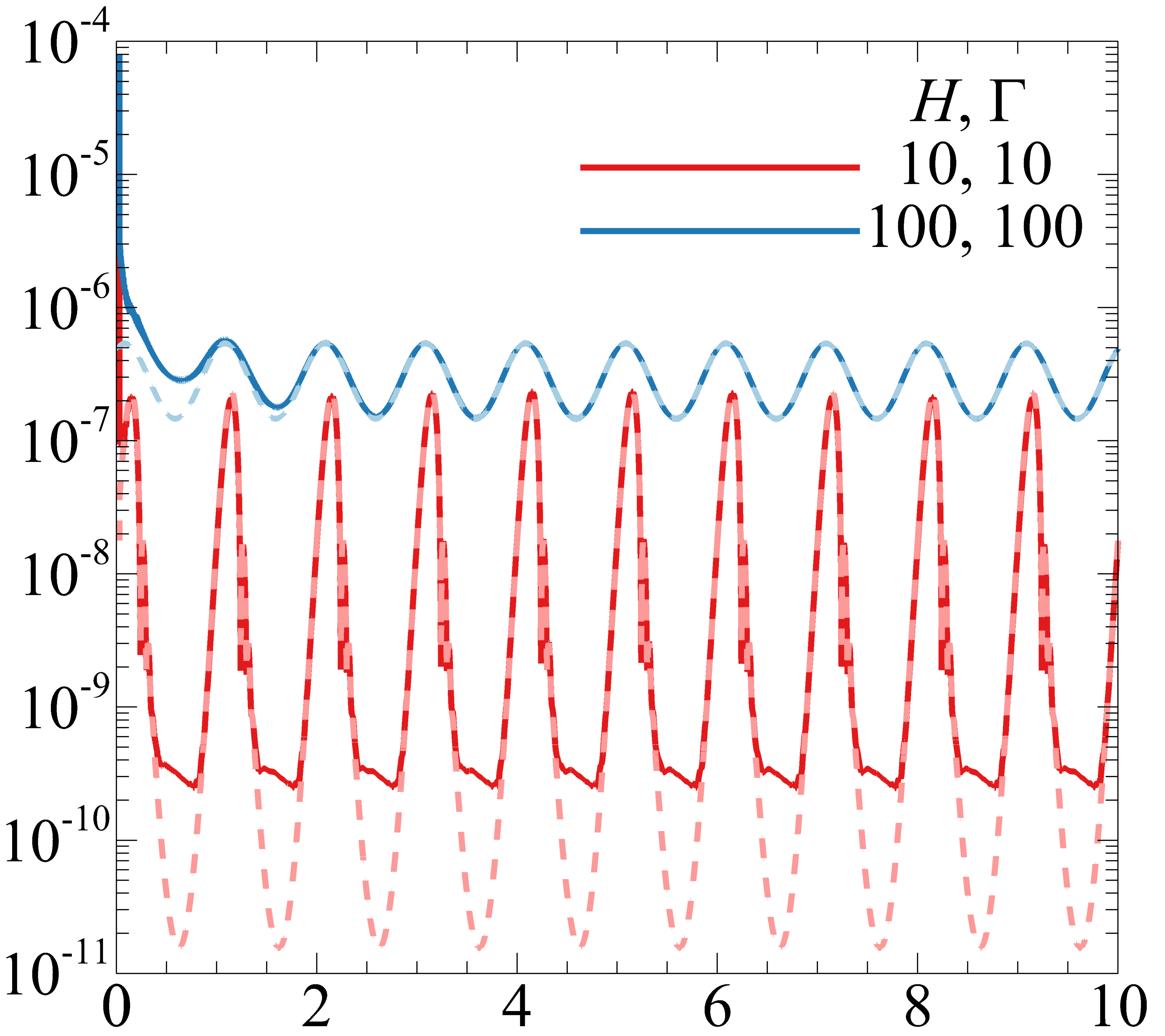}} &
\makecell{\footnotesize{(b)} \vspace{26.5mm} \\  \vspace{33mm} \rotatebox{90}{\footnotesize{$\Euv$}}}
 & \makecell{\includegraphics[width=0.458\textwidth]{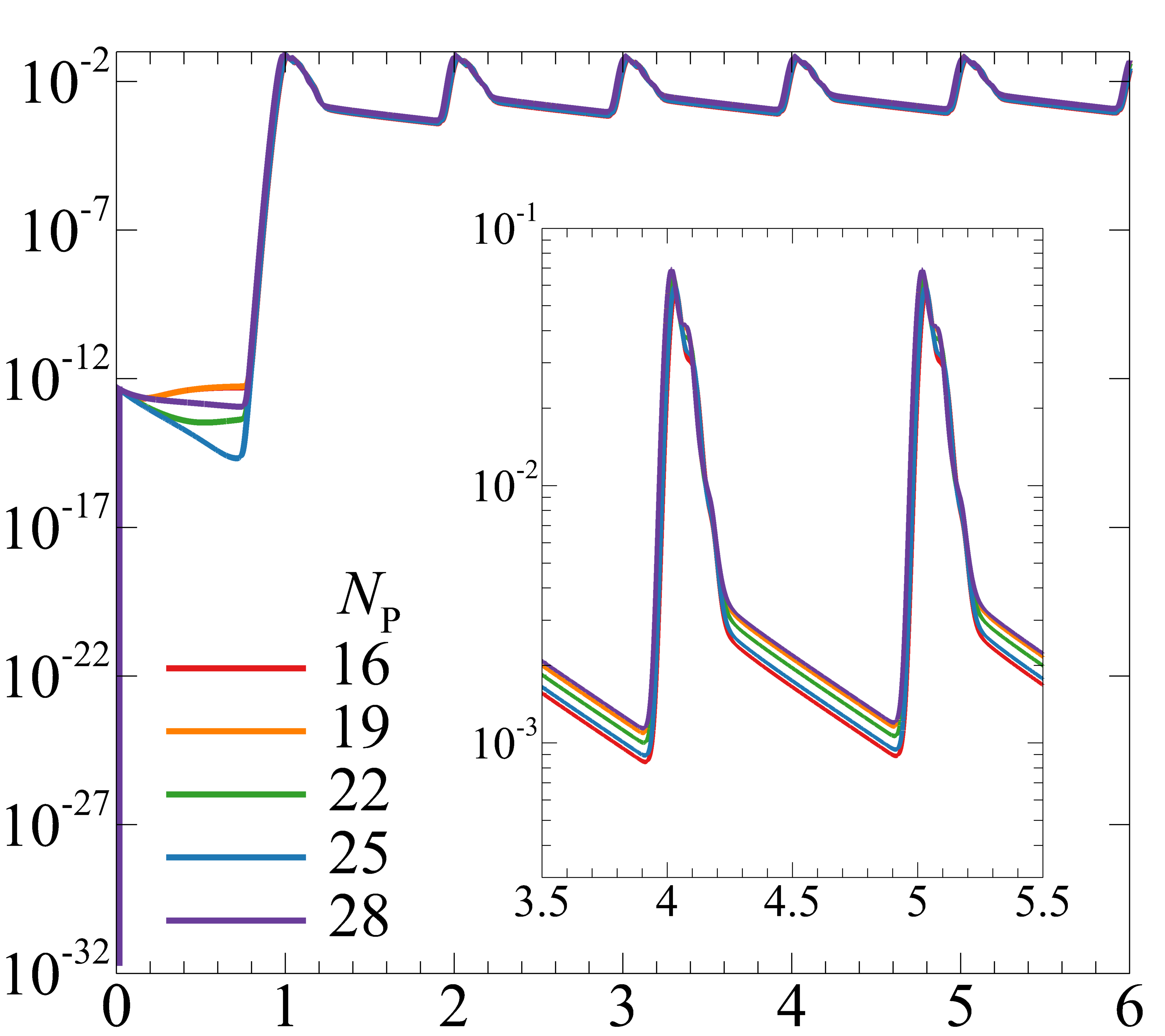}} \\
 & \hspace{38mm} \footnotesize{$\tpe$} & & \hspace{38mm} \footnotesize{$\tpe$} \\
\end{tabular}
\addtolength{\tabcolsep}{+2pt}
\addtolength{\extrarowheight}{+10pt}
\end{center}
    \caption{Resolution testing at critical conditions. (a) Comparison of nonlinear DNS (in-house solver; solid lines) and linearized timestepper (dashed lines) at $\Sr = 10^{-2}$. An initial perturbation of white noise with $E_0=10^{-6}$ was applied to the DNS. (b) Nonlinear DNS with no initial perturbation of the $H=1$ optimized pulsation ($\Gamma=1.24$, $\Sr = 7.2\times10^{-3}$), varying polynomial order. }
    \label{fig:val_DNS}
\end{figure}

Further validation, depicted in \fig\ \ref{fig:val_DNS}(a), is a comparison between the nonlinear time evolution in primitive variables (the in-house solver, referred to as DNS in future) and the linearized evolution with the timestepper, introduced earlier. These are both computed using the $\ReyCrit$ and $\alphaCrit$ from the Floquet method, at $H=10$, $\Gamma=10$ and $H=100$, $\Gamma=100$, both at $\Sr=10^{-2}$ (cases discussed in Sec.~\ref{sec:lin_res2}). Initial seeds of white noise have specified initial energy $E_{0}(t=0)=\int\hat{u}^2+\hat{v}^2\,\dUP\Omega/\int U^2(t=0)\,\dUP\Omega$, where $\Omega$ represents the computational domain. Linearity is ensured with $E_0 = 10^{-6}$. The DNS settles after a short period of decay, and then attains excellent agreement with the intracyclic growth curves from the linearized timestepper, both in magnitude, and dynamics over the cycle. The only difference is that for the $\Gamma=10$ case, at small perturbation amplitudes (near $10^{-10}$) the nonlinear evolution cuts out, and remains at roughly constant energy until the deceleration phase of the base flow to begin growth again, while the linearized evolution continues on a smooth decay-growth trajectory. 

The resolution requirements are assessed by varying the polynomial order $\Np$ of the spectral elements. \Fig\ \ref{fig:val_DNS}(b) depicts simulations, with no initiating perturbation, driven by the optimized pulsation at $H=1$, at critical conditions. Excluding the initial growth, which is always resolution `dependent', the agreement in the intracylic growth stages is excellent (see box out). The slight differences predominantly originate from the initial growth stage, translating the curves with respect to one another. $\Np=19$ was deemed sufficient for the pulsatile problem, as for the steady base flow problem \citep{Camobreco2020transition}.

Fourier analysis is also performed in the nonlinear simulations, exploiting the streamwise periodicity of the domain. The absolute values of the Fourier coefficients $c_\kappa = \lvert (1/\Nf)\sum_{n=0}^{n=\Nf-1}\hat{f}(x_n) e^{-2\pi i \kappa n/\Nf}\rvert$ were obtained using the discrete Fourier transform in MATLAB, where $x_n$ represents the $n$'th $x$-location linearly spaced between $x_0=0$ and $x_{\Nf}=W$. $\hat{f}$ may be $\hat{u}$, $\hat{v}$, $\hat{\omega}_z = \partial \hat{v}/ \partial x - \partial \hat{u} /\partial {y}$ or $\hat{u}^2+\hat{v}^2$, depending on the property of interest. In the $y$-direction, either a mean Fourier coefficient $\meanfoco$ is obtained by averaging the coefficients obtained at 21 $y$-values, and taking $\Nf=10000$. Alternately, considering 912 $y$-values, and taking $\Nf=380$, all except the $j$'th (and $\Nf-j$'th) Fourier coefficients were set to zero, $c_{\kappa,\neg j}=0$, and the inverse discrete Fourier transform $\hat{f}_j= \sum_{\kappa=0}^{\kappa=\Nf-1}c_{\kappa,\neg j} e^{2\pi i \kappa n/\Nf}$ computed. After isolating the $j$'th mode in the physical domain $\hat{f}_j$, an assessment of the degree of symmetry within that mode was determined by computing $\hat{f}_{\mathrm{s},j} = (\sum_{m=0}^{m=\Ny}[\hat{f}_j(y_m)-\hat{f}_j(-y_m)]^2)^{1/2}$, where $y_m$ represents the $m$'th $y$-location linearly spaced between $y_0=-1$ and $y_{\Ny}=0-1/(\Ny-1)$, and taking $\Ny = 912/2$. Thus, a purely symmetric mode has $\hat{f}_{\mathrm{s},j}=0$ as $\hat{f}_j(y_m)$=$\hat{f}_j(-y_m)$ for all $y_m$.


\subsection{Critical conditions}\label{sec:nlin_cres}

\begin{figure}
\begin{center}
\addtolength{\extrarowheight}{-10pt}
\addtolength{\tabcolsep}{-2pt}
\begin{tabular}{ llll }
\makecell{\footnotesize{(a)} \vspace{23mm}  \\  \vspace{27mm} \rotatebox{90}{\footnotesize{$\Ev$, $\twonv^2$}}} 
& \makecell{\includegraphics[width=0.458\textwidth]{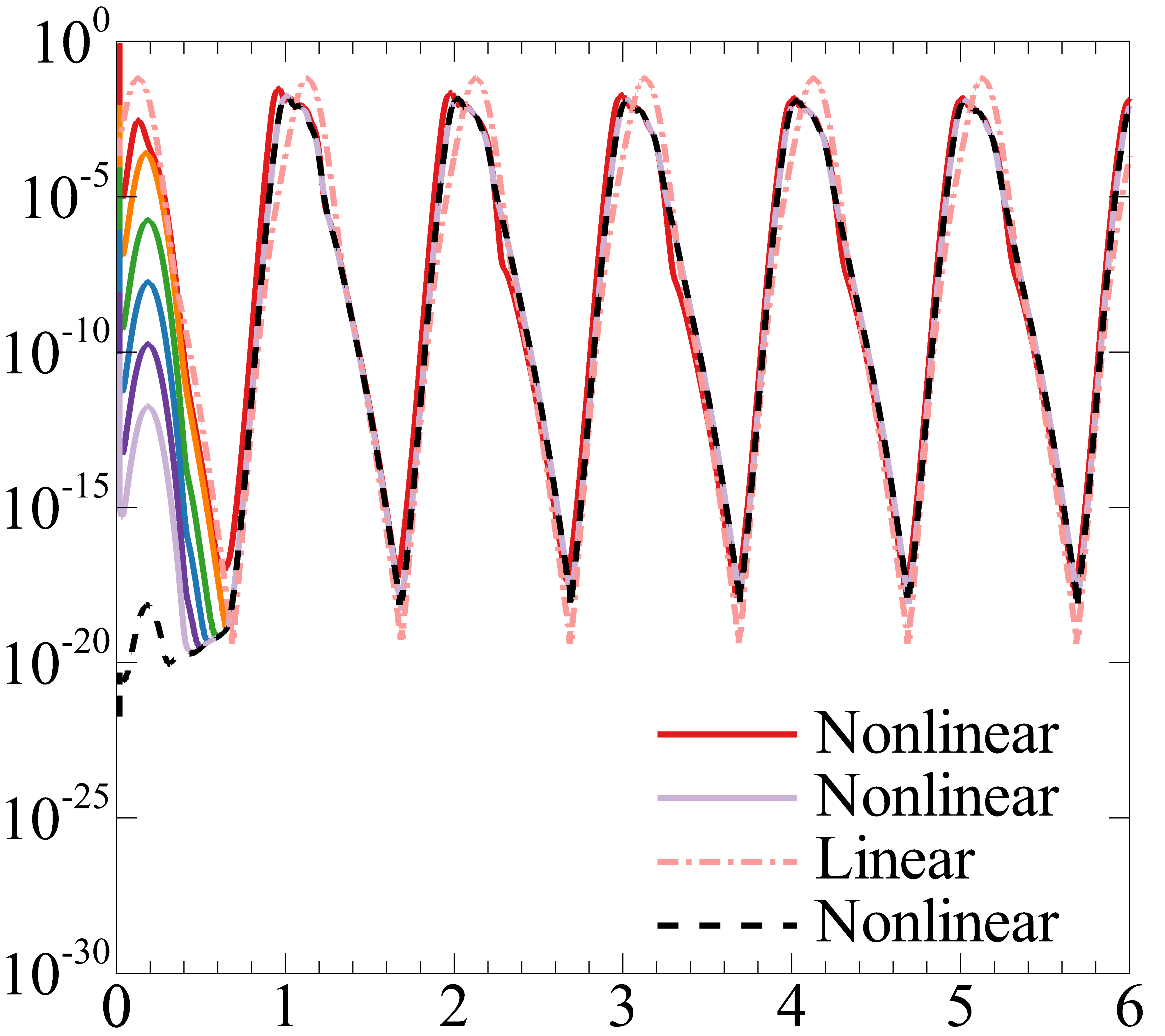}} &
\makecell{\footnotesize{(b)} \vspace{26.5mm} \\  \vspace{33mm} \rotatebox{90}{\footnotesize{$\Euv$}}}
 & \makecell{\includegraphics[width=0.458\textwidth]{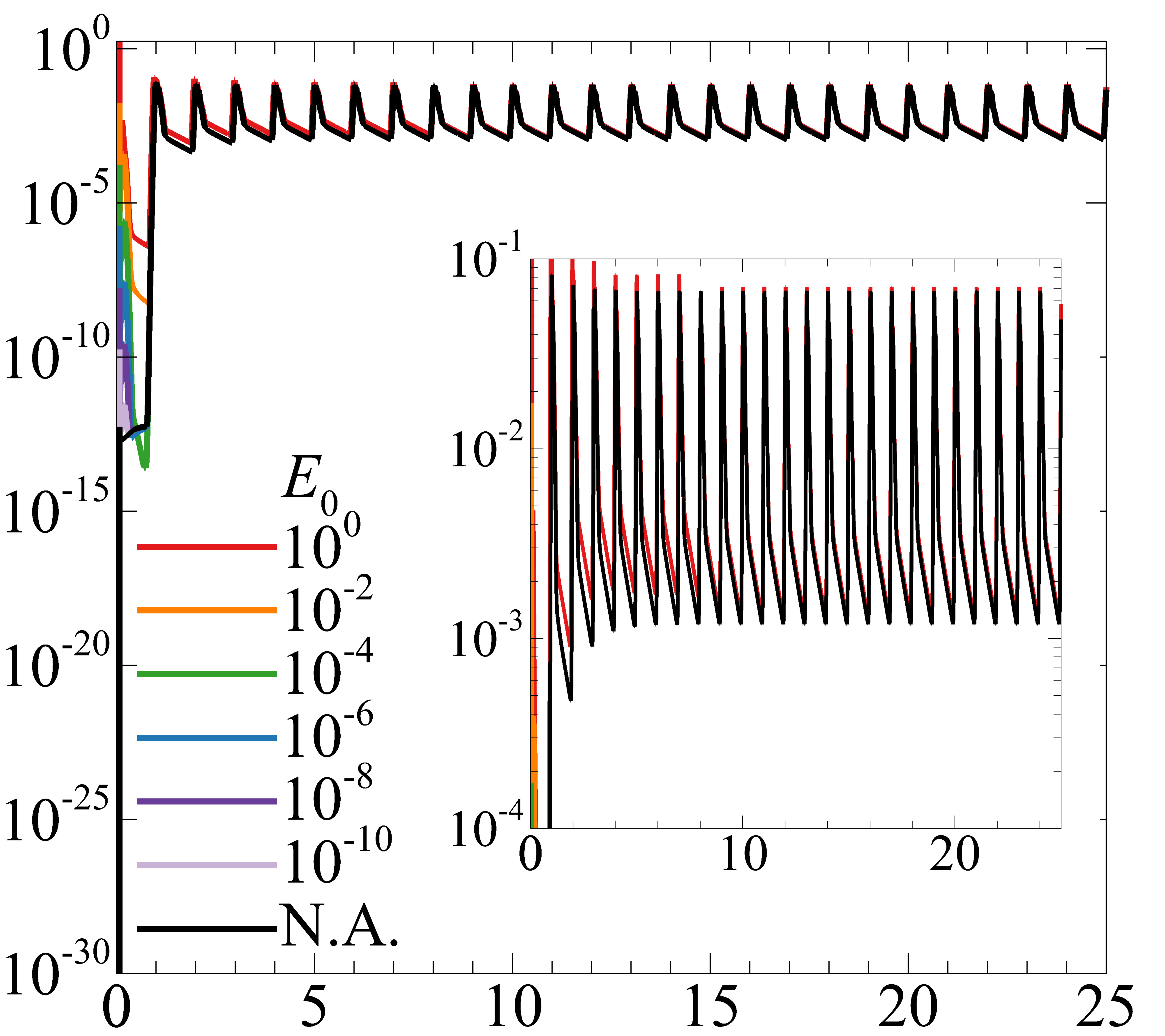}} \\
 & \hspace{38mm} \footnotesize{$\tpe$} & & \hspace{38mm} \footnotesize{$\tpe$} \\
\end{tabular}
\addtolength{\tabcolsep}{+2pt}
\addtolength{\extrarowheight}{+10pt}
\end{center}
    \caption{Effect of varying $E_0$, between $10^0$ and $10^{-10}$, on nonlinear evolution, compared to a case without an initial perturbation (black dashed line in (a) and solid line in (b)) and a case linearly evolved (pink dot-dashed line), for the optimized pulsation at $H=1$, $\Gamma = 1.24$, $\Sr = 7.2\times10^{-3}$. (a) $\Ev = \int \hat{v}^2 \mathrm{d} \Omega$. (b) $\Euv = \int \hat{u}^2 + \hat{v}^2 \mathrm{d} \Omega$.}.
    \label{fig:varyEz}
\end{figure}

\begin{figure}
\begin{center}
\addtolength{\extrarowheight}{-10pt}
\addtolength{\tabcolsep}{-2pt}
\begin{tabular}{ llll }
\makecell{\footnotesize{(a)} \vspace{26.5mm}  \\  \vspace{33.5mm} \rotatebox{90}{\footnotesize{$\Ev$}}} 
& \makecell{\includegraphics[width=0.458\textwidth]{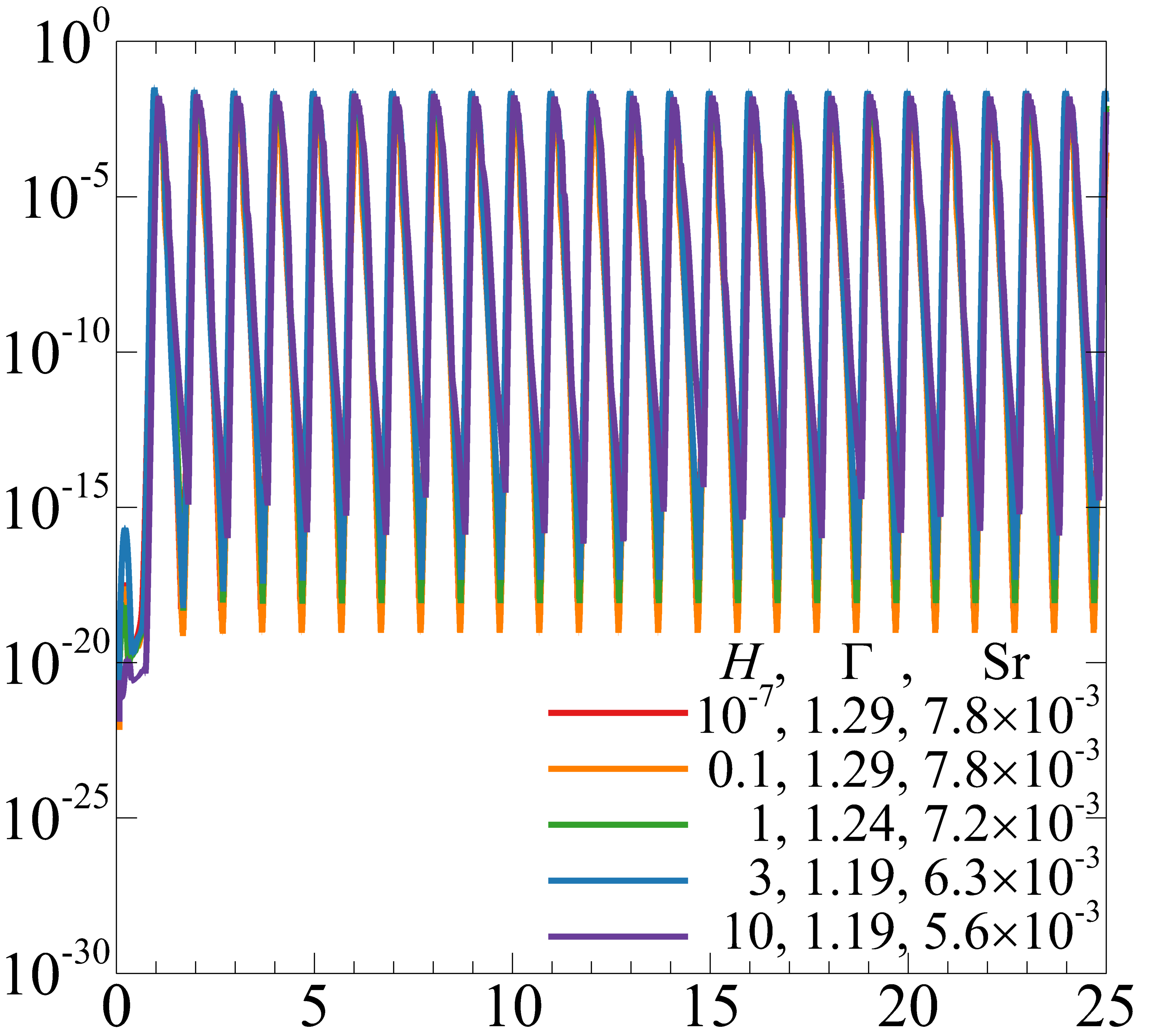}} &
\makecell{\footnotesize{(b)} \vspace{27mm} \\  \vspace{33.5mm} \rotatebox{90}{\footnotesize{$\Euv$}}}
 & \makecell{\includegraphics[width=0.458\textwidth]{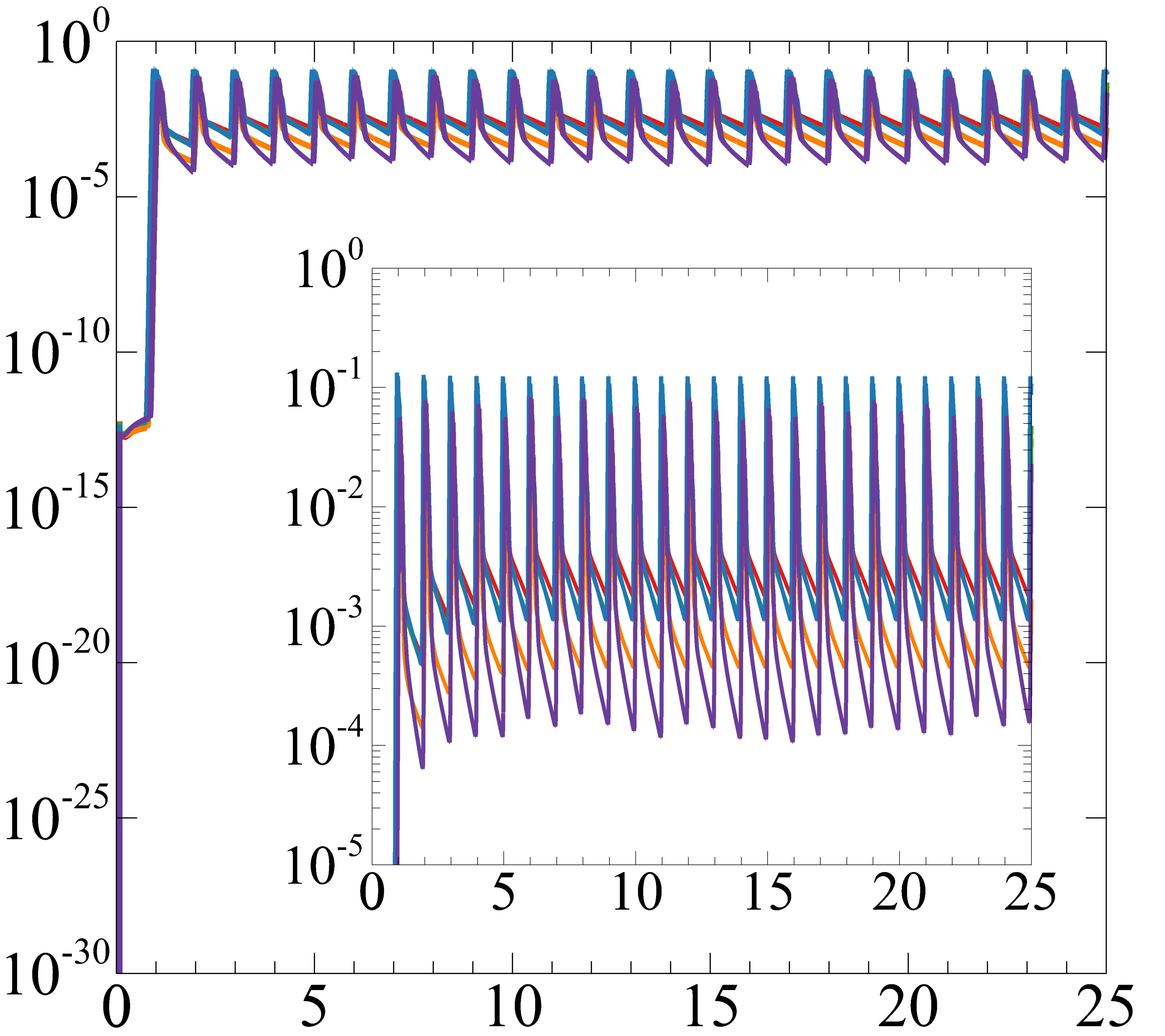}} \\
 & \hspace{38mm} \footnotesize{$\tpe$} & & \hspace{38mm} \footnotesize{$\tpe$} \\
\end{tabular}
\addtolength{\tabcolsep}{+2pt}
\addtolength{\extrarowheight}{+10pt}
\end{center}
    \caption{Nonlinear evolutions of the optimized pulsations, at various $H$, from \tbl\ \ref{tab:tab_3}. (a) $\Ev = \int \hat{v}^2 \mathrm{d} \Omega$. (b) $\Euv = \int \hat{u}^2 + \hat{v}^2 \mathrm{d} \Omega$. The ultimate result of the nonlinear evolutions is no net growth at $\ReyCrit$.}
    \label{fig:nogro_varyH}
\end{figure}

This section focuses solely on the minimum $\rrs$ conditions of \tbl\ \ref{tab:tab_3}, at $\ReyCrit$. The first factor is the role of the initial perturbation. Comparing a simulation without an initiating perturbation (\eg\ `numerical noise'), and simulations initiated with white noise of specified magnitude, \fig\ \ref{fig:varyEz}, yields two key results. The first is that all the initial energy trajectories collapse to the numerical noise result within the first period of evolution, except $\Ezero=1$ (slightly offset). For $\Ezero<1$ the perturbation energy decays no further than for the case initiated from numerical noise, and plateaus until the next deceleration phase of the base flow. Once this occurs, all energies grow in unison. As the $\Gamma$, $\Sr$ optima are within the `ballistic' regime, they decay to linearly small energies every period \cite{Pier2017linear}. Hence, unless a transition to turbulence occurs in the first period of the base flow, the initial energy has no influence on subsequent cycles. The second key result is that the linear and nonlinear evolutions compared via $\Ev = \int \hat{v}^2 \mathrm{d}\Omega$ are similar, see \fig\ \ref{fig:varyEz}(a), while they are not via $\Euv = \int \hat{u}^2+\hat{v}^2 \mathrm{d}\Omega$, \fig\ \ref{fig:varyEz}(b). In the second period of the base flow, the nonlinear intracyclic decay is largely truncated. After another period, the nonlinear case saturates to relatively constant energy maxima and minima (\fig\ \ref{fig:varyEz}(b) inset). Previous works \cite{Camobreco2020role, Camobreco2020transition} have shown that growth in $\hat{v}$ is stored in streamwise independent structures, $\hat{u}$, in nonlinear modal and nonmodal growth scenarios of steady quasi-two-dimensional base flows. A similar process occurs here, as is further discussed shortly. 

The lack of nonlinear net growth at the critical conditions for the remaining cases in \tbl\ \ref{tab:tab_3} is depicted in \fig\ \ref{fig:nogro_varyH}, again without specifying an initial perturbation. At higher $H$, nonlinear intracyclic growth was smaller than expected (linearly, intracyclic growth increased with increasing $H$ at $\Rey = \ReyCrit$). However, the final result of no net growth is still maintained, as expected at $\ReyCrit$. The only slight difference is that at higher $H$, and thereby larger $\Rey$, the maximum and minimum energies reached are becoming inconsistent (see box-out). In the linear solver, such inconsistencies would eventually limit the accurate computation of $\ReyCrit$.

\subsection{Supercritical conditions}\label{sec:nlin_sres}

\begin{figure}
\begin{center}
\addtolength{\extrarowheight}{-10pt}
\addtolength{\tabcolsep}{-2pt}
\begin{tabular}{ llll }
\makecell{\footnotesize{(a)} \vspace{26.5mm}  \\  \vspace{33.5mm} \rotatebox{90}{\footnotesize{$\Ev$}}} 
& \makecell{\includegraphics[width=0.458\textwidth]{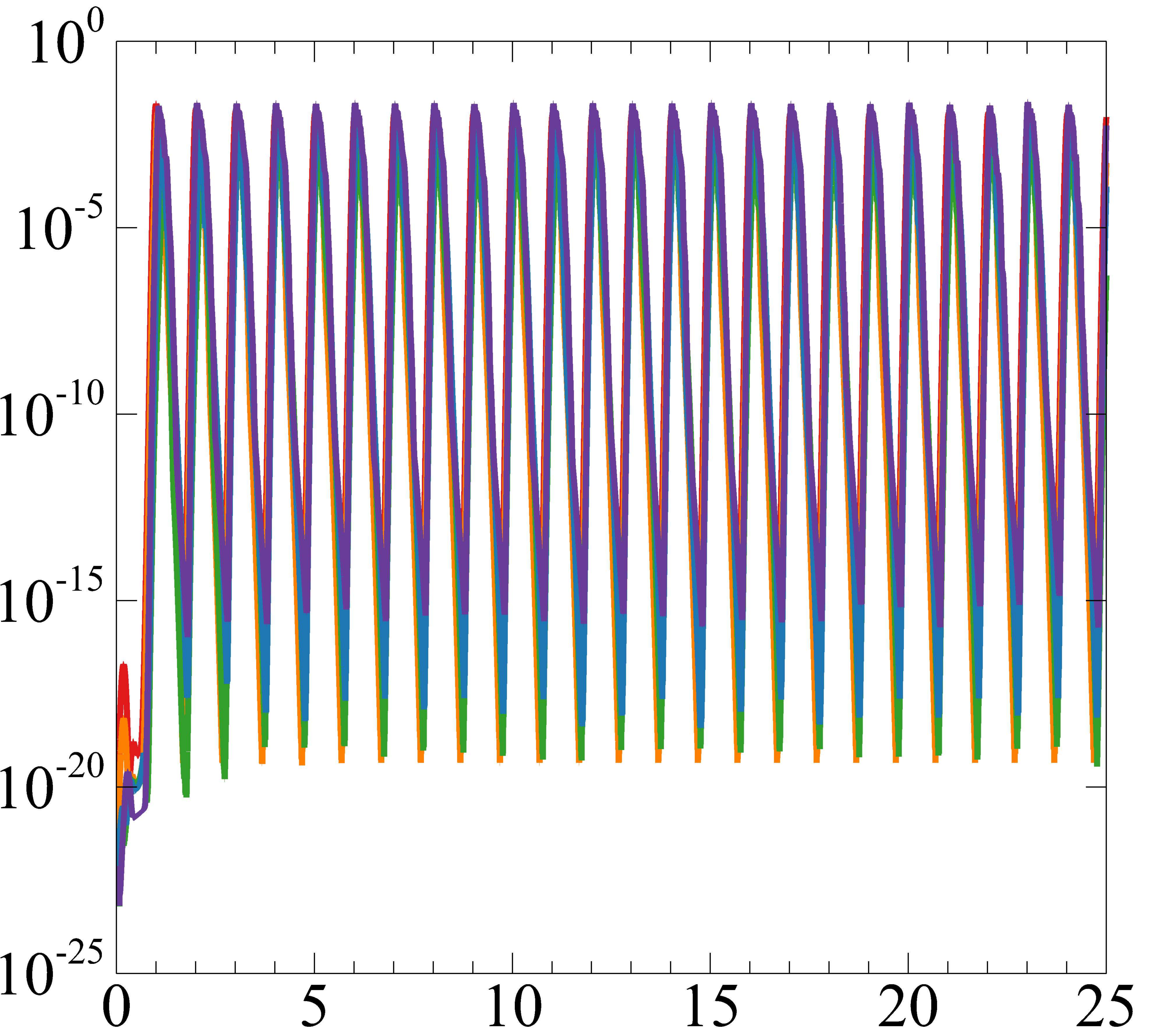}} &
\makecell{\footnotesize{(b)} \vspace{27mm} \\  \vspace{33.5mm} \rotatebox{90}{\footnotesize{$\Euv$}}}
 & \makecell{\includegraphics[width=0.458\textwidth]{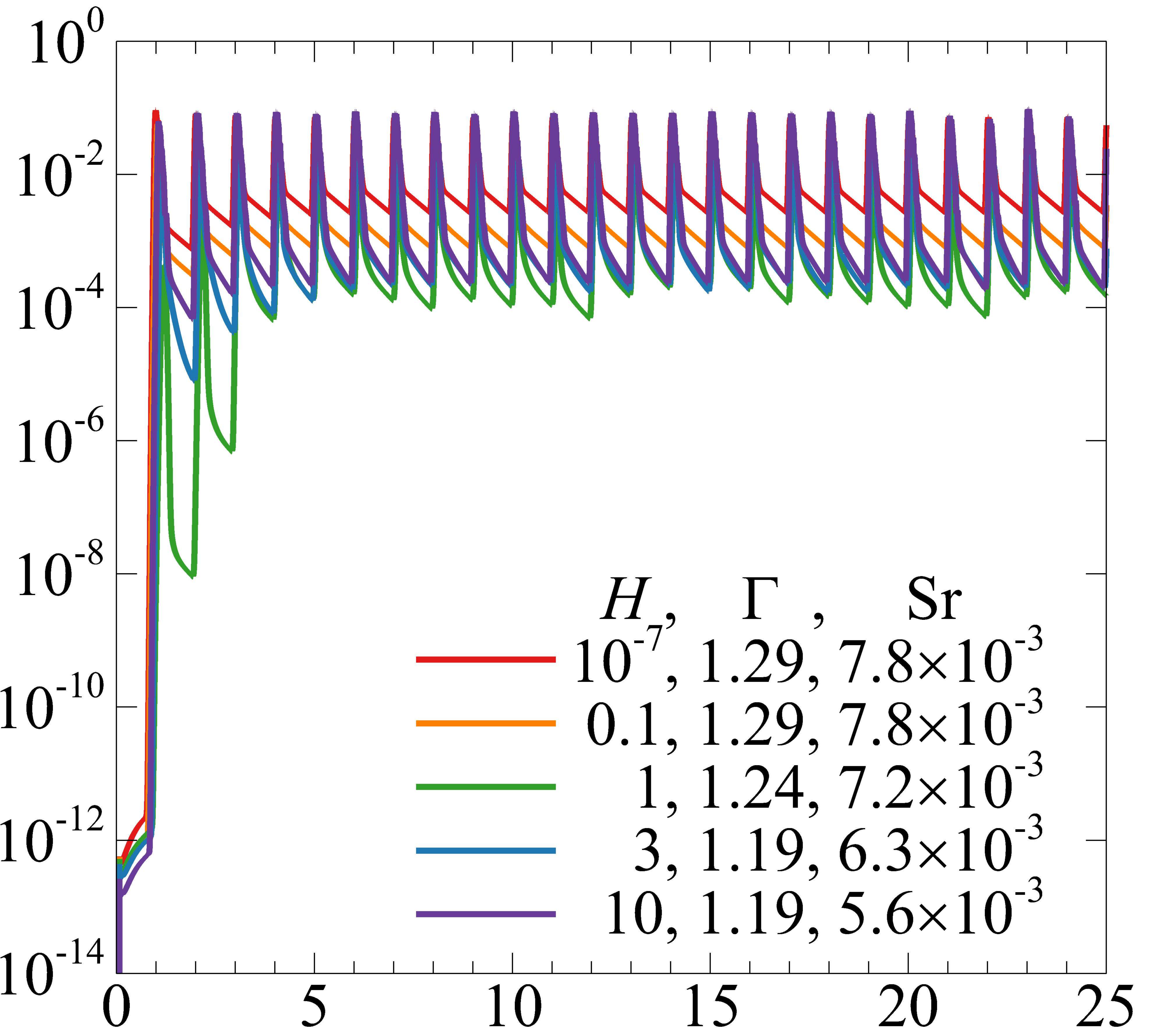}} \\
 & \hspace{38mm} \footnotesize{$\tpe$} & & \hspace{38mm} \footnotesize{$\tpe$} \\
\end{tabular}
\addtolength{\tabcolsep}{+2pt}
\addtolength{\extrarowheight}{+10pt}
\end{center}
    \caption{Nonlinear evolutions of the optimized $\Sr$ and $\Gamma$ for minimum $\rrs$, for various $H$, at $\Rey/\ReyCrit=1.1$. (a) $\Ev = \int \hat{v}^2 \mathrm{d} \Omega$. (b) $\Euv = \int \hat{u}^2 + \hat{v}^2 \mathrm{d} \Omega$. These results are very similar to those at $\Rey/\ReyCrit=1$ (\fig\ \ref{fig:nogro_varyH}) in spite of the fact that linearly, exponential growth is predicted.}
    \label{fig:nogro11_varyH}
\end{figure}

Supercritical Reynolds numbers are briefly considered, again without specifying an initial perturbation. As the base flow is Reynolds number dependent, only a 10\% and a 20\% increase (not shown) in the Reynolds number were attempted, for the values of $\Gamma$ and $\Sr$ that minimize $\rrs$ for $H \leq 10$. The overall behaviors at $\Rey/\ReyCrit=1$ (\fig\ \ref{fig:nogro_varyH}) and $\Rey/\ReyCrit=1.1$ (\fig\ \ref{fig:nogro11_varyH}) are virtually identical, even though exponential growth is predicted linearly at $\Rey/\ReyCrit=1.1$. Nonlinearly, the intracyclic growth in the first period is large enough to reach nonlinear amplitudes, which quickly modulates the base flow, resulting in the no net growth behavior. However, turbulence is not observed at these supercritical conditions, with only some chaotic behavior following the symmetry breaking of the linear mode. The severity of the decay in the acceleration phase may be the main factor preventing the transition to turbulence. However, the magnitude of $H$ and $\Rey$ could be a factor, since $H<3$ are unable to trigger turbulence for the case of a steady  base flow at the equivalent $\Rey/\ReyCrit$ ratio \cite{Camobreco2020transition}. Although higher $H$ were able to trigger turbulence in the classical duct flow, the magnitude of the Reynolds numbers were larger for the steady base flow, as optimising for minimum $\rrs$ results in an order of magnitude reduction in $\ReyCrit$. 


\subsection{Role of streamwise and wall-normal velocity components}\label{sec:str_wal}

\begin{figure}
\begin{center}
\addtolength{\extrarowheight}{-10pt}
\addtolength{\tabcolsep}{-2pt}
\begin{tabular}{ ll ll }
\footnotesize{(a)} & \footnotesize{\hspace{3mm} $\tpe=1.61$, $\max(|\hat{v}|)=3.319\times10^{-8}$}  &
\footnotesize{(b)} & \footnotesize{\hspace{3mm} $\tpe=1.7$, $\max(|\hat{v}|)=1.348\times10^{-9}$} \\
\makecell{ \\  \vspace{10mm} \rotatebox{90}{\footnotesize{$y$}}} & \makecell{\includegraphics[width=0.458\textwidth]{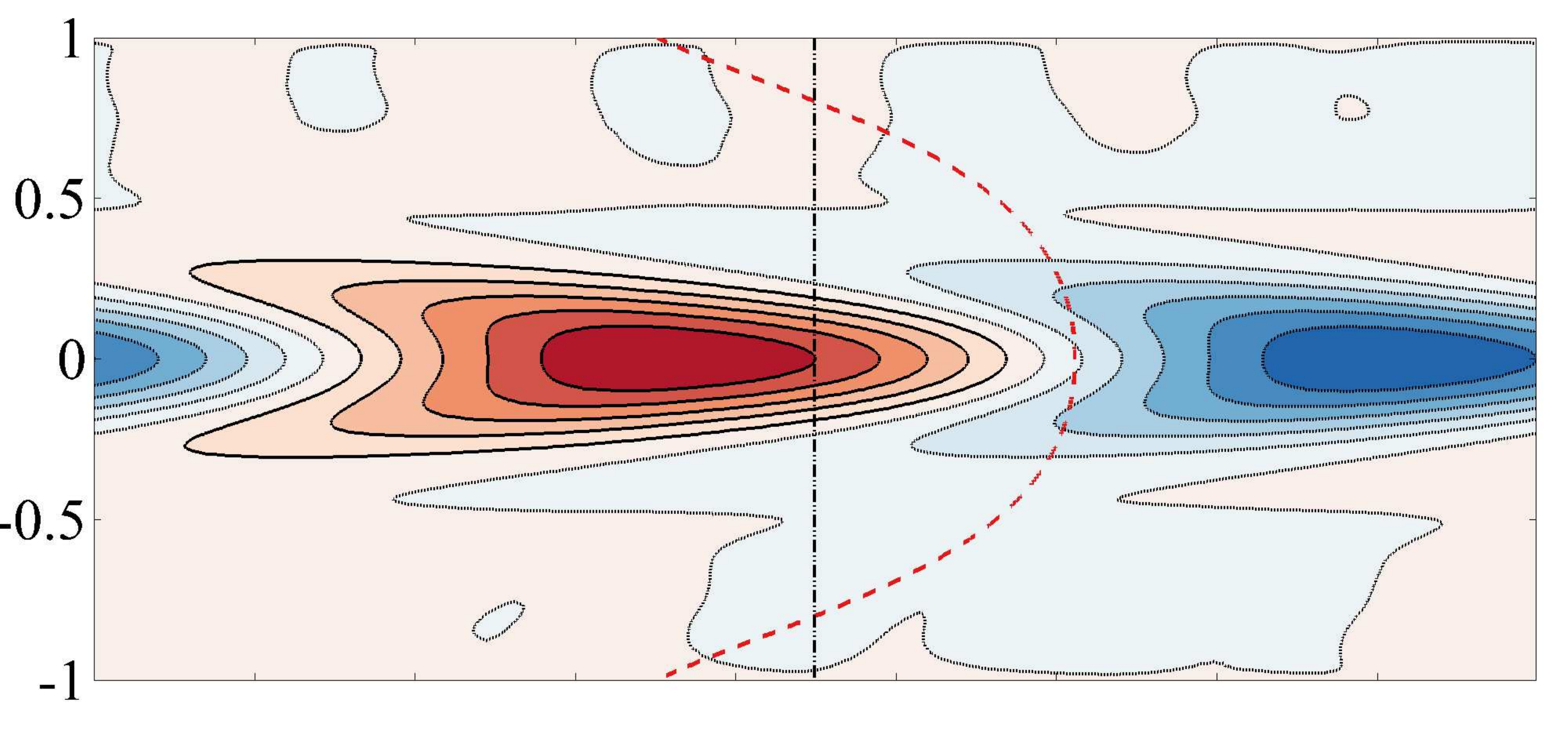}} & 
\makecell{ \\  \vspace{10mm} } & \makecell{\includegraphics[width=0.458\textwidth]{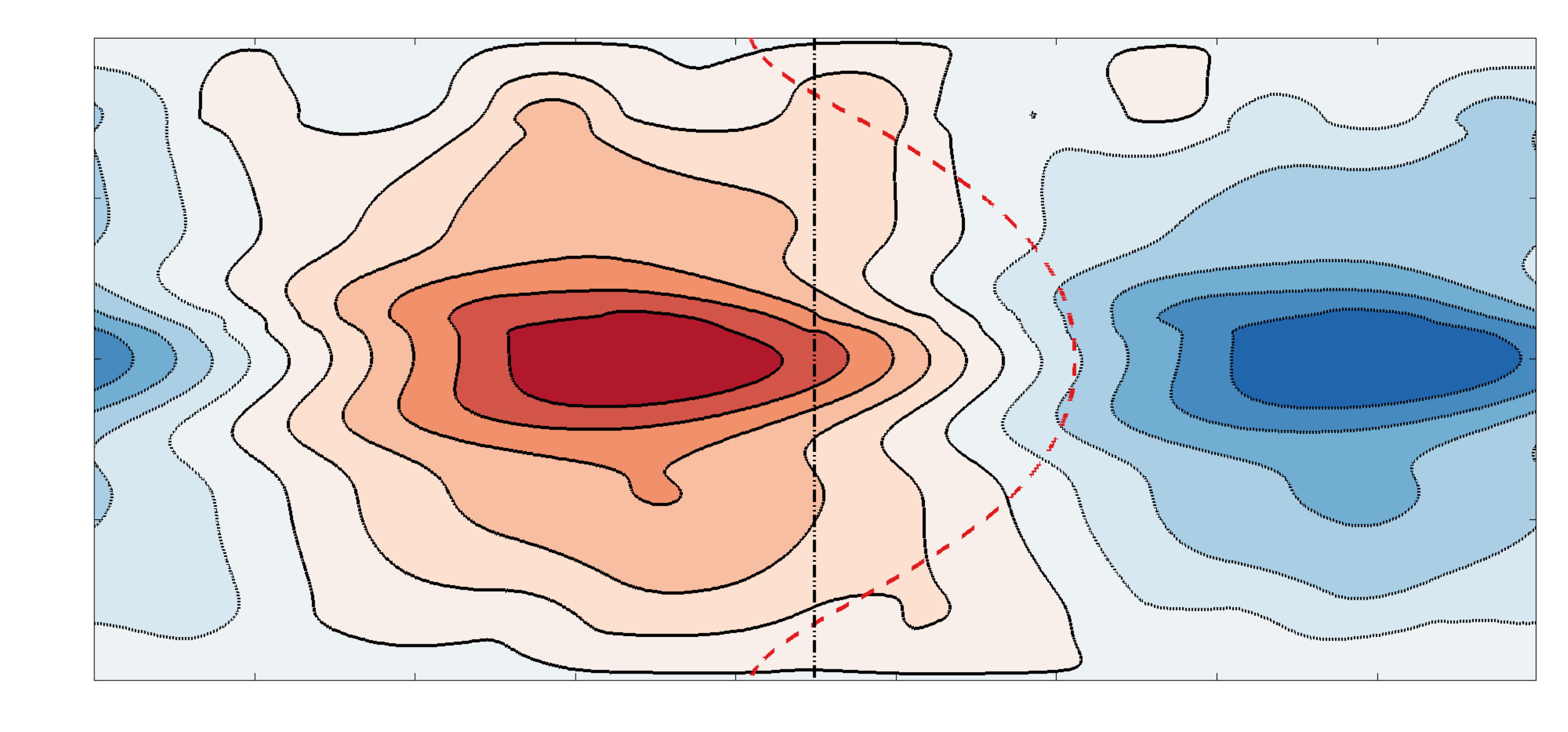}} \\
\footnotesize{(c)} & \footnotesize{\hspace{3mm} $\tpe=1.75$, $\max(|\hat{v}|)=1.646\times10^{-8}$} &
\footnotesize{(d)} & \footnotesize{\hspace{3mm} $\tpe=1.925$, $\max(|\hat{v}|)=3.491\times10^{-3}$} \\
\makecell{ \\  \vspace{10mm} \rotatebox{90}{\footnotesize{$y$}}}  & \makecell{\includegraphics[width=0.458\textwidth]{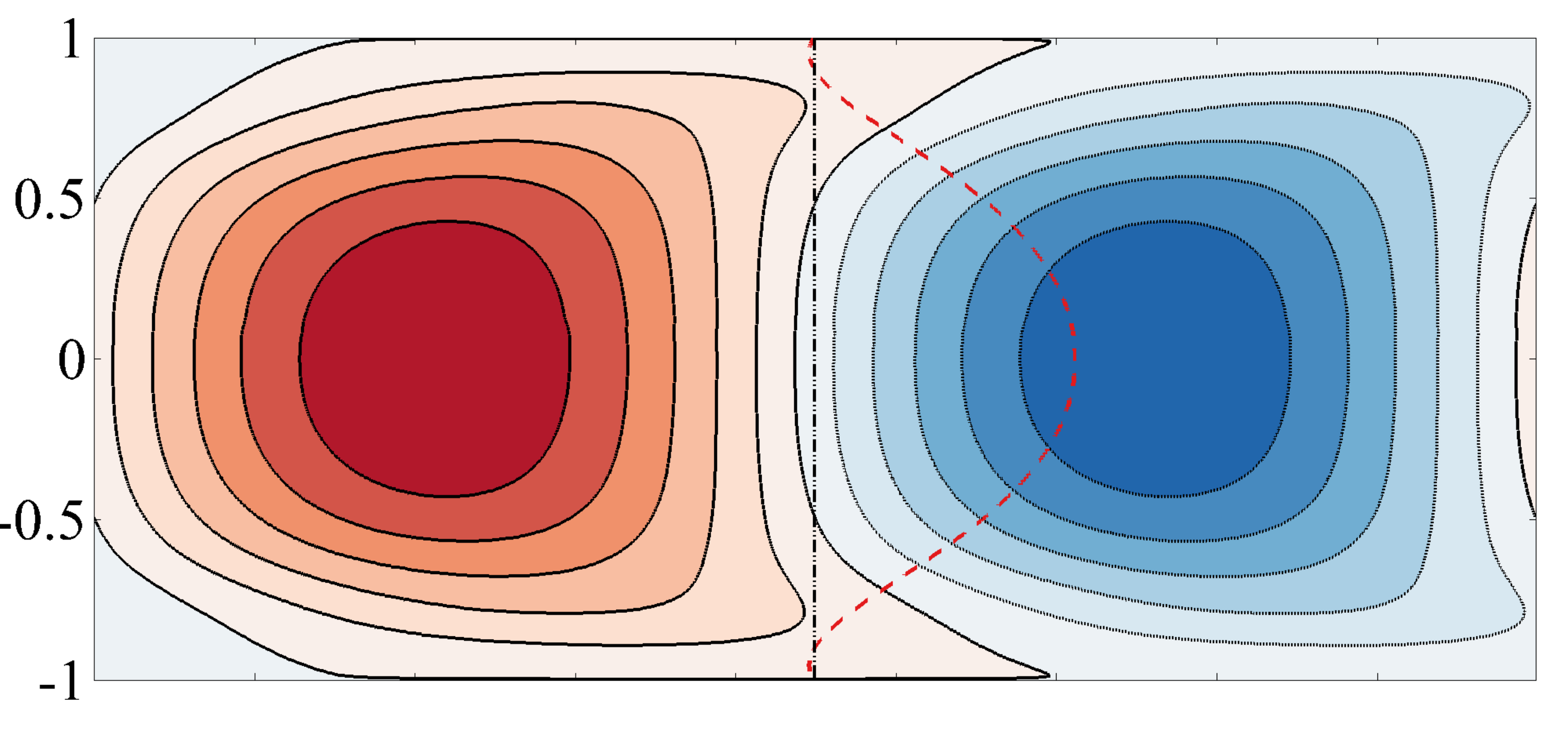}} & 
\makecell{ \\  \vspace{10mm} } & \makecell{\includegraphics[width=0.458\textwidth]{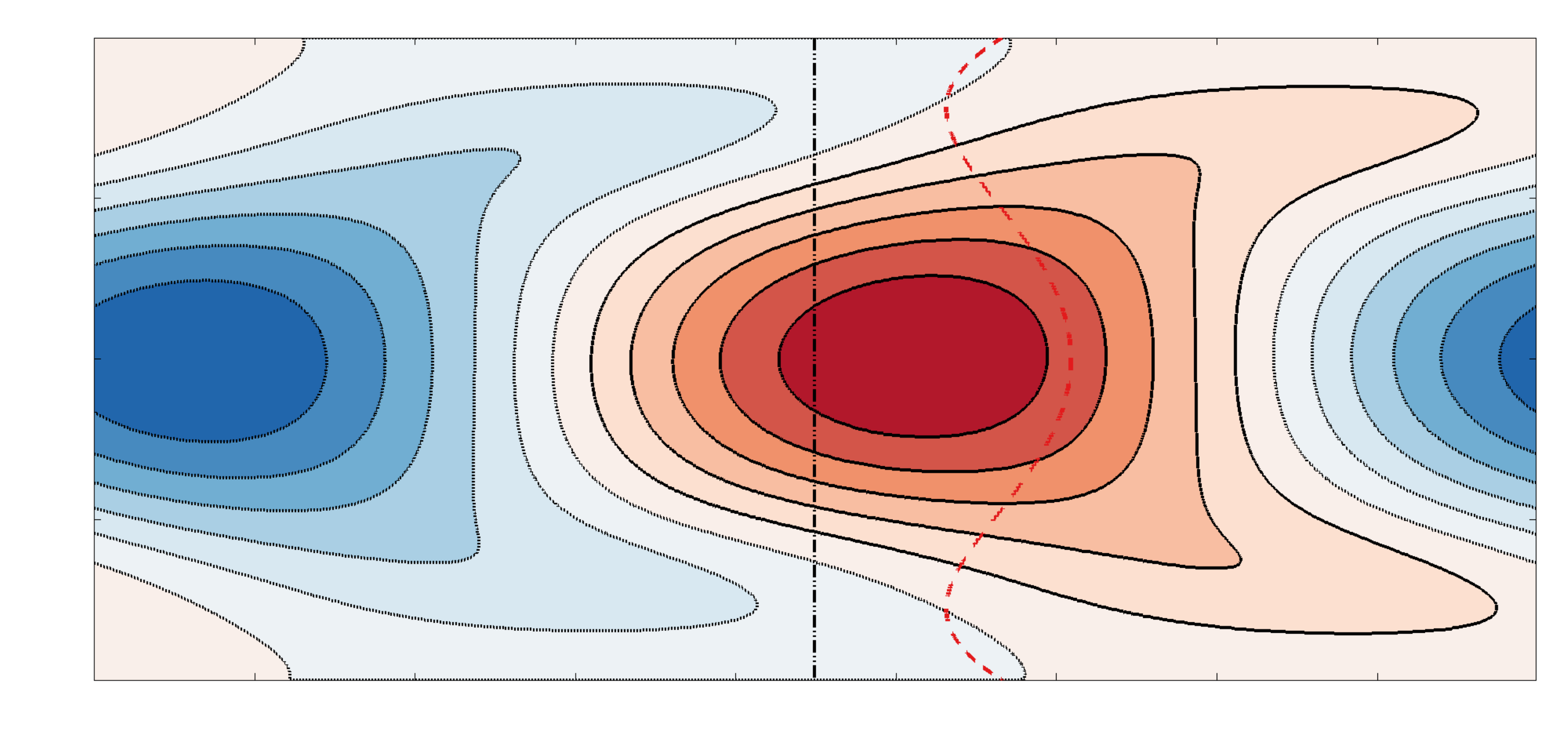}} \\
\footnotesize{(e)} & \footnotesize{\hspace{3mm} $\tpe=1.965$, $\max(|\hat{v}|)=2.310\times10^{-2}$}  &
\footnotesize{(f)} & \footnotesize{\hspace{3mm} $\tpe=2$, $\max(|\hat{v}|)=6.188\times10^{-2}$} \\
\makecell{ \\  \vspace{10mm} \rotatebox{90}{\footnotesize{$y$}}} & \makecell{\includegraphics[width=0.458\textwidth]{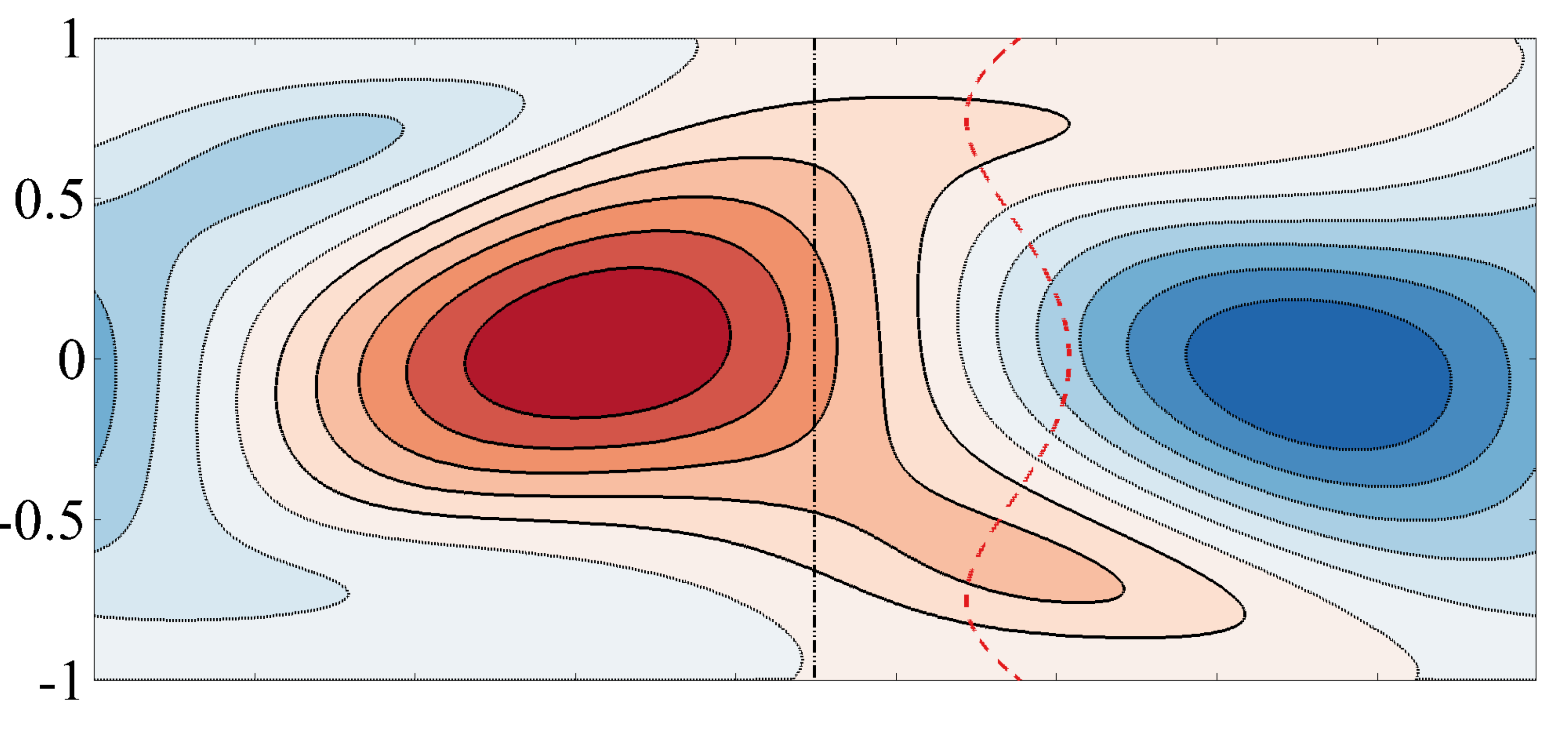}} & 
\makecell{ \\  \vspace{10mm} } & \makecell{\includegraphics[width=0.458\textwidth]{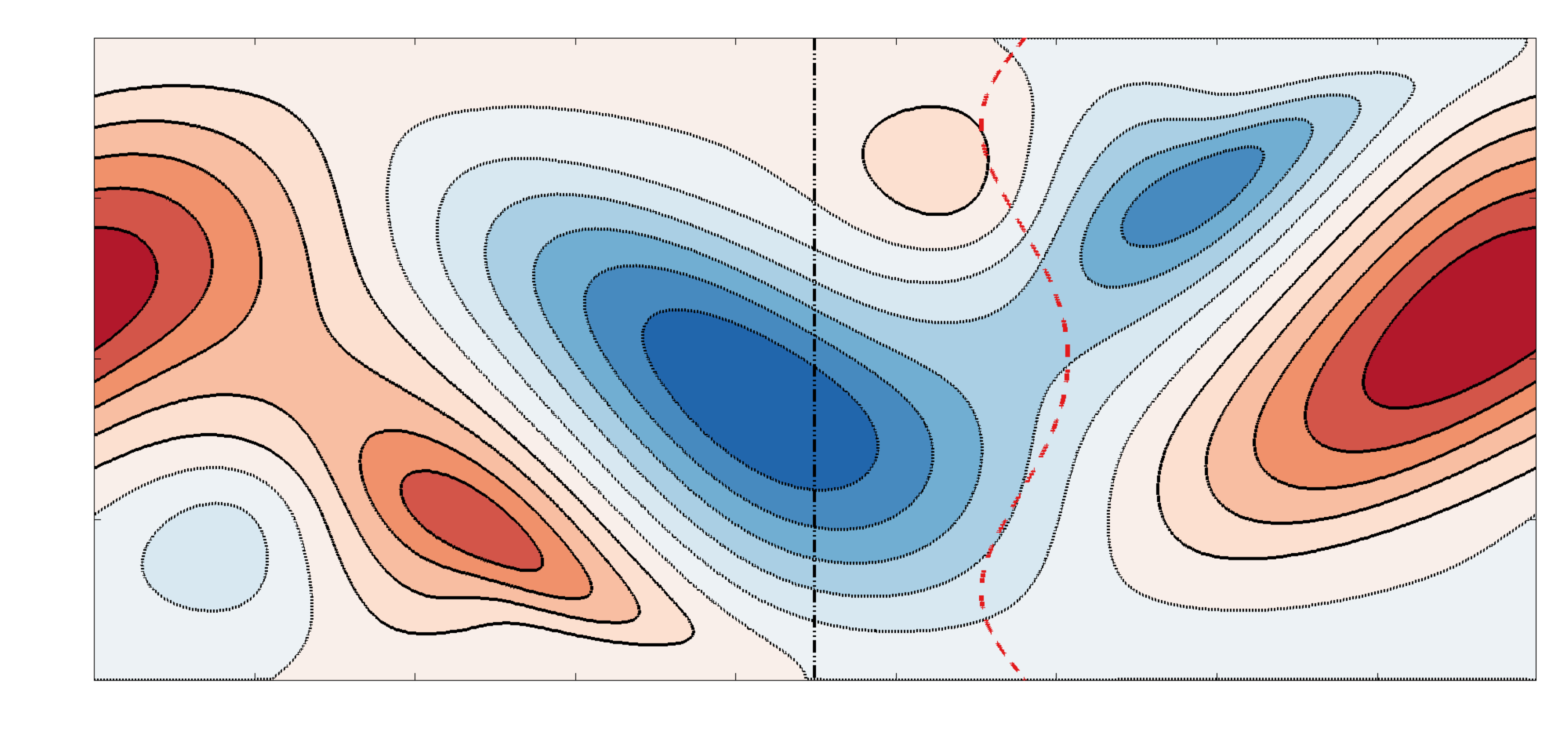}} \\
\footnotesize{(g)} & \footnotesize{\hspace{3mm} $\tpe=2.155$, $\max(|\hat{v}|)=3.278\times10^{-2}$} &
\footnotesize{(h)} & \footnotesize{\hspace{3mm} $\tpe=2.22$, $\max(|\hat{v}|)=7.450\times10^{-3}$} \\
\makecell{ \\  \vspace{10mm} \rotatebox{90}{\footnotesize{$y$}}}  & \makecell{\includegraphics[width=0.458\textwidth]{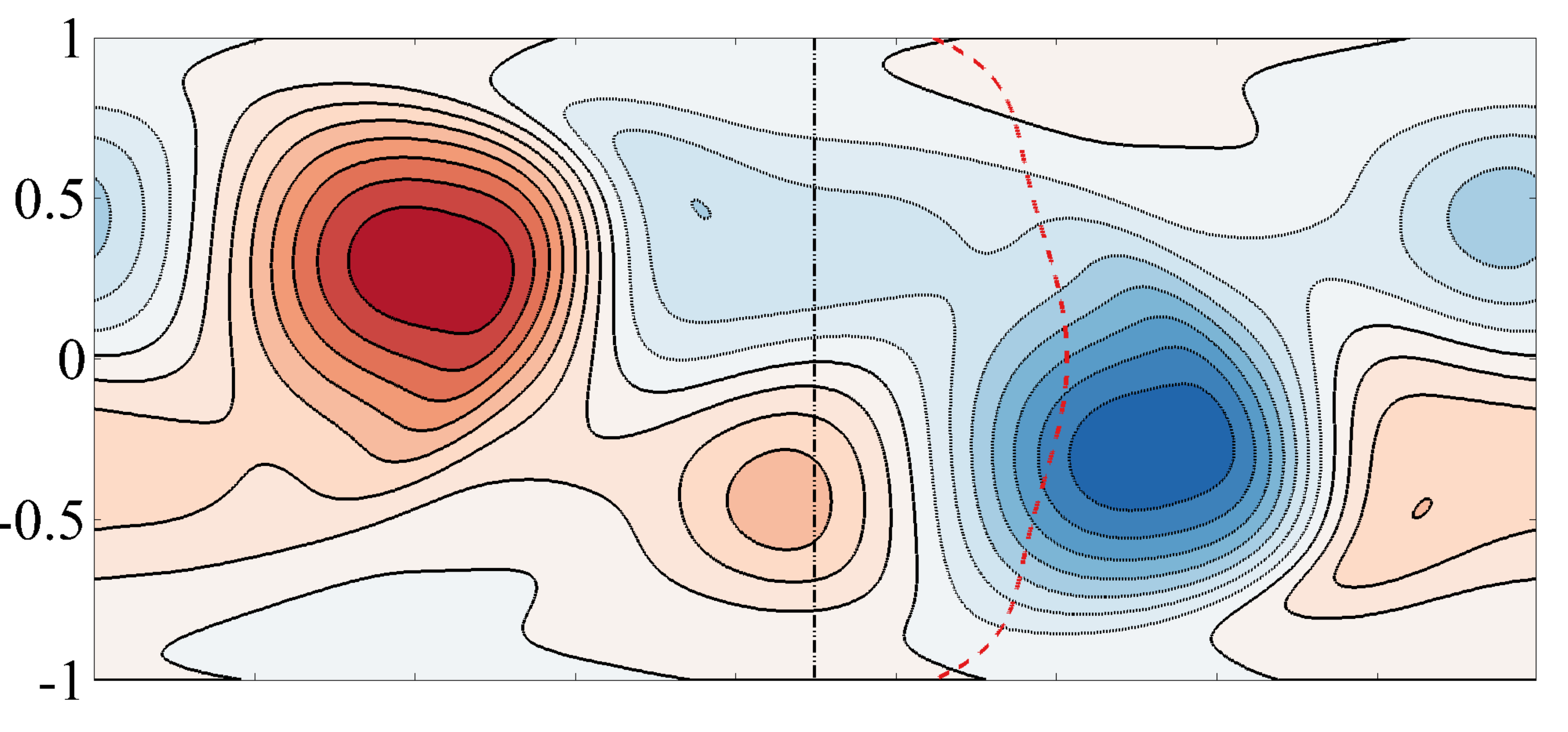}} & 
\makecell{ \\  \vspace{10mm} } & \makecell{\includegraphics[width=0.458\textwidth]{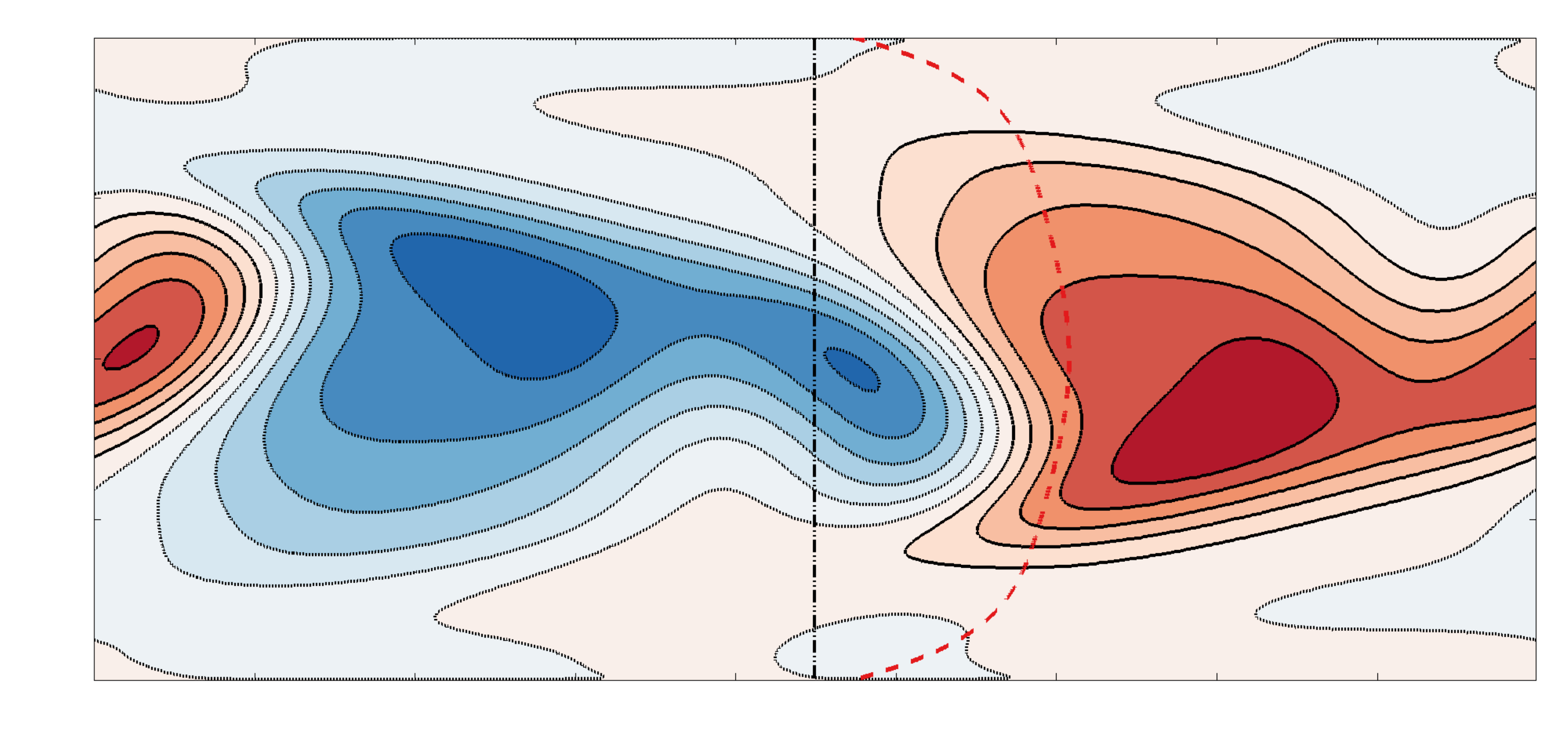}} \\
\footnotesize{(i)} & \footnotesize{\hspace{3mm} $\tpe=2.3$, $\max(|\hat{v}|)=8.864\times10^{-4}$} &
\footnotesize{(j)} & \footnotesize{\hspace{3mm} $\tpe=2.4$, $\max(|\hat{v}|)=4.572\times10^{-5}$} \\
\makecell{ \\  \vspace{10mm} \rotatebox{90}{\footnotesize{$y$}}}  & \makecell{\includegraphics[width=0.458\textwidth]{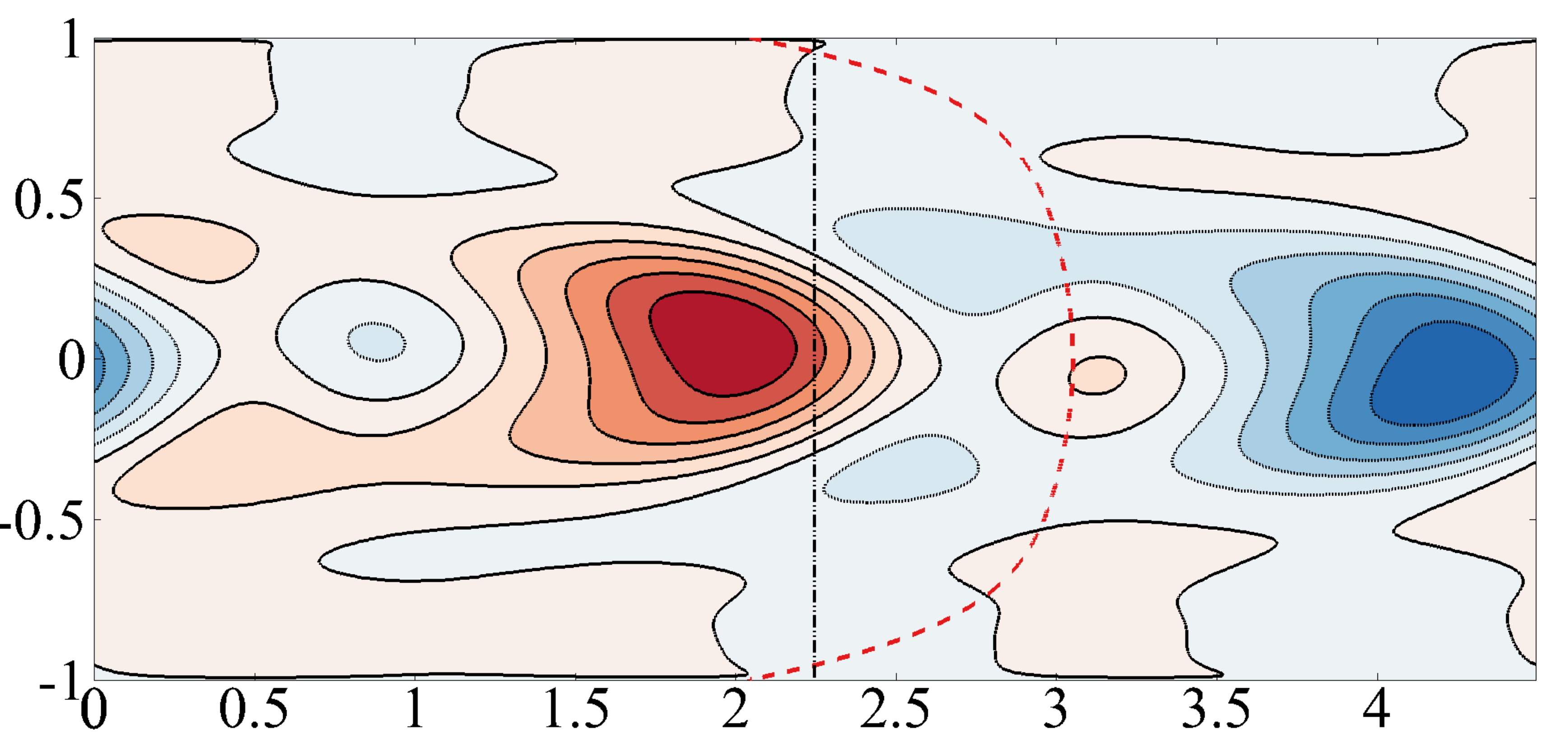}} & 
\makecell{ \\  \vspace{10mm} } & \makecell{\includegraphics[width=0.458\textwidth]{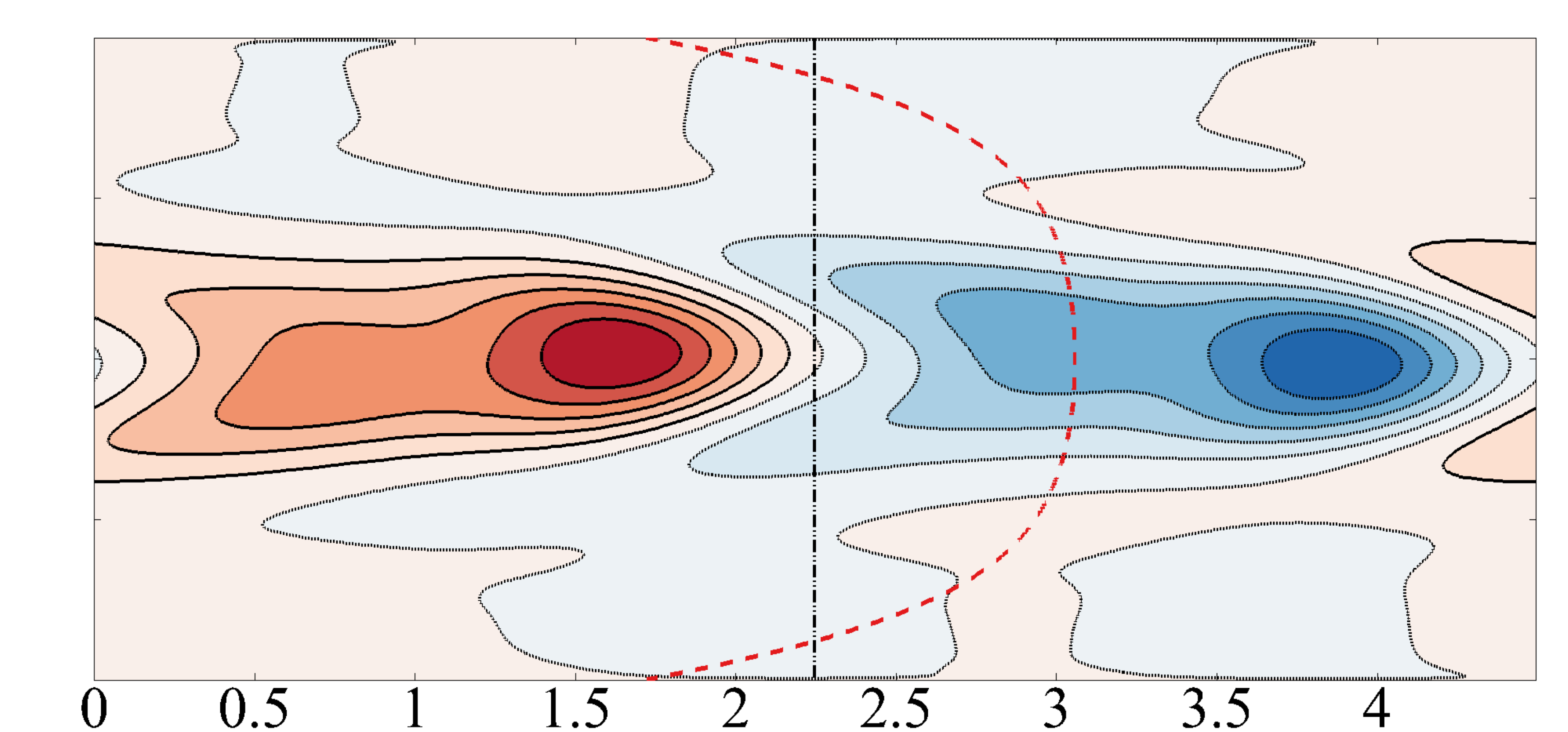}} \\
  & \hspace{38mm} \footnotesize{$x$} &  & \hspace{38mm} \footnotesize{$x$} \\
\end{tabular}
\addtolength{\tabcolsep}{+2pt}
\addtolength{\extrarowheight}{+10pt}
\end{center}
    \caption{Nonlinear evolution of $\hat{v}$-velocity perturbation contours at $H=1$, $\Gamma=1.24$, $\Sr=7.2\times10^{-3}$ through one cycle $\tpe \in [1.5,2.5]$. The base flow is overlaid (the black dashed line indicates zero base flow velocity). Red flooding positive; blue flooding negative.}
    \label{fig:snaps_nonlinear_v}
\end{figure}
Two aspects of the nonlinear evolution are considered in more detail. The first is the slight difference between the linear and nonlinear growth in $\hat{v}$, observed in \fig\ \ref{fig:varyEz}(a). Snapshots of the $\hat{v}$-velocity from the DNS are depicted in \fig\ \ref{fig:snaps_nonlinear_v} over $\tpe \in [1.5,2.5]$; the linear case at the same conditions was shown in \fig\ \ref{fig:snapshots_H1_helper}, over $\tpe \in [0,1]$. An animation comparing these cases is also provided \cite{Supvideos2020}. When at small energies at $\tpe=1.61$, the highly sheared structure along the centreline of the nonlinear case has a very similar appearance to its linear counterpart (around $\tpe = 1.7$). However, some higher wave number effects are still visible near the walls in the nonlinear case even at these small energies. The reformation of the nonlinear structure, as it spreads over the duct ($\tpe = 1.7$-$1.75$) and as the `wings' pull forward ($\tpe=1.925$), when inflection points form in the base flow, are also very similar to the linear case. However, past $\tpe \approx 1.925$, the linear growth rate slightly diminishes, while the nonlinear growth rate remains higher, again recalling \fig\ \ref{fig:varyEz}(a). This is related to nonlinearity inducing a symmetry breaking of the linear mode, from around $\tpe=1.965$, with the region of positive $\hat{v}$-velocity structure tilting downward, and the region of negative velocity tilting upward. Eventually, secondary structures separate from each core before the structures eventually break apart around $\tpe=2.155$. From $\tpe=2.22$ through to $\tpe=2.5$, the decay induced by the downstream pull of the walls creates a single highly sheared structure along the centreline, as for the linear case. 

\begin{figure}
\begin{center}
\addtolength{\extrarowheight}{-10pt}
\addtolength{\tabcolsep}{-2pt}
\begin{tabular}{ ll ll }
\footnotesize{(a)} & \footnotesize{\hspace{3mm} $\tpe=0.22$, $\max(|\hat{u}|)=2.722\times10^{-7}$}  &
\footnotesize{(b)} & \footnotesize{\hspace{3mm} $\tpe=0.76$, $\max(|\hat{u}|)=3.679\times10^{-7}$} \\
\makecell{ \\  \vspace{10mm} \rotatebox{90}{\footnotesize{$y$}}} & \makecell{\includegraphics[width=0.458\textwidth]{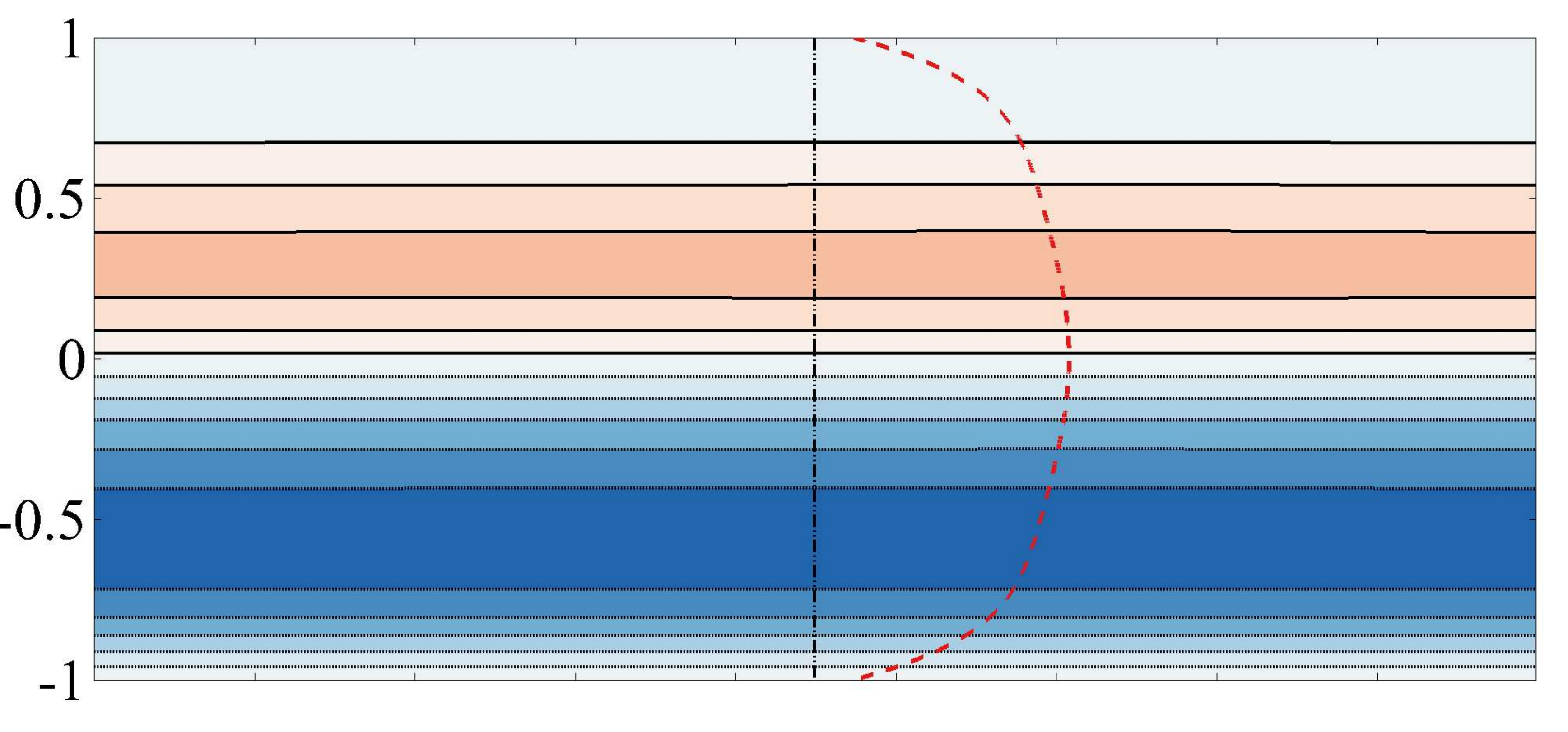}} & 
\makecell{ \\  \vspace{10mm} } & \makecell{\includegraphics[width=0.458\textwidth]{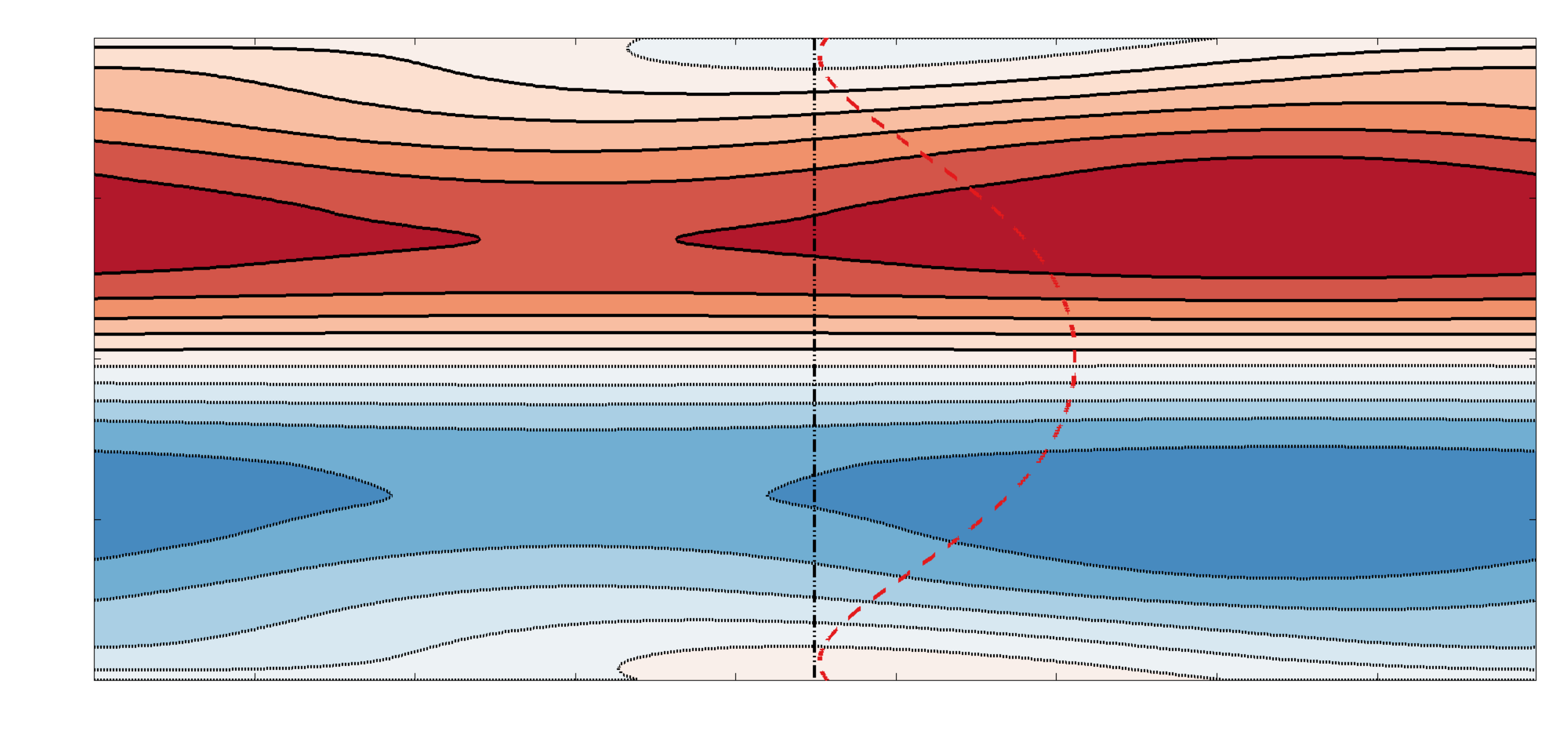}} \\
\footnotesize{(c)} & \footnotesize{\hspace{3mm} $\tpe=0.925$, $\max(|\hat{u}|)=1.010\times10^{-2}$} &
\footnotesize{(d)} & \footnotesize{\hspace{3mm} $\tpe=0.965$, $\max(|\hat{u}|)=7.503\times10^{-2}$} \\
\makecell{ \\  \vspace{10mm} \rotatebox{90}{\footnotesize{$y$}}}  & \makecell{\includegraphics[width=0.458\textwidth]{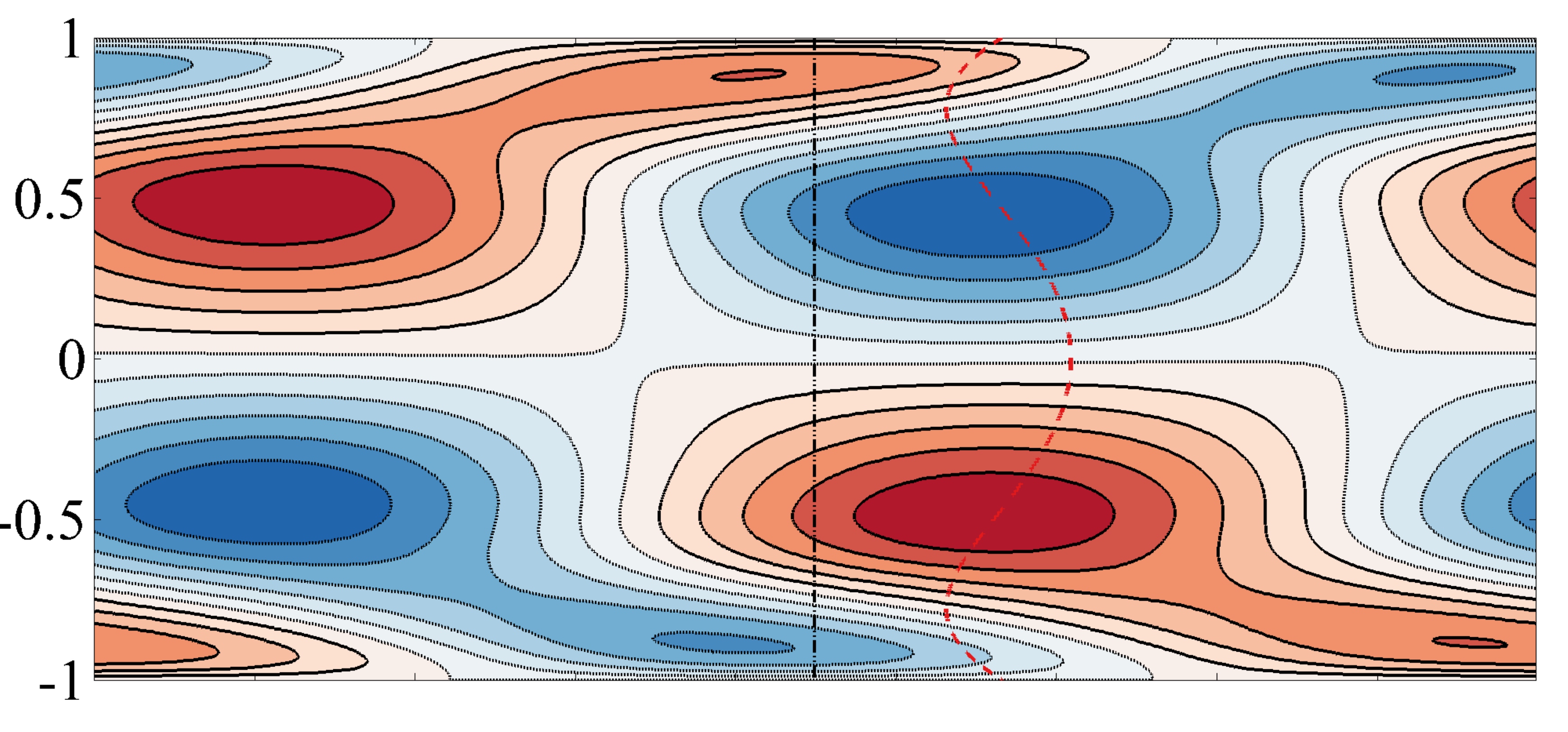}} & 
\makecell{ \\  \vspace{10mm} } & \makecell{\includegraphics[width=0.458\textwidth]{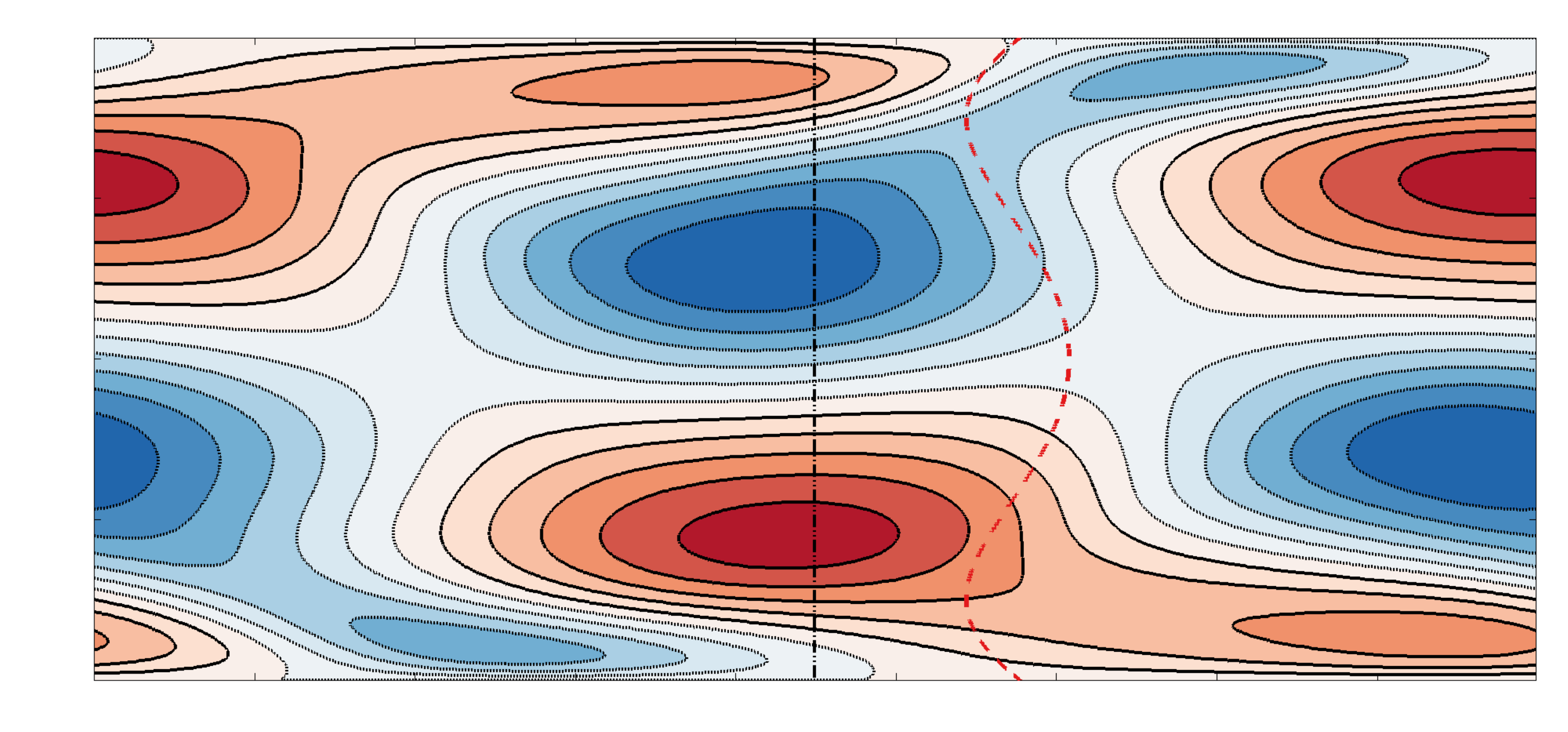}} \\
\footnotesize{(e)} & \footnotesize{\hspace{3mm} $\tpe=1.155$, $\max(|\hat{u}|)=5.001\times10^{-2}$}  &
\footnotesize{(f)} & \footnotesize{\hspace{3mm} $\tpe=1.3$,   $\max(|\hat{u}|)=1.943\times10^{-2}$} \\
\makecell{ \\  \vspace{10mm} \rotatebox{90}{\footnotesize{$y$}}} & \makecell{\includegraphics[width=0.458\textwidth]{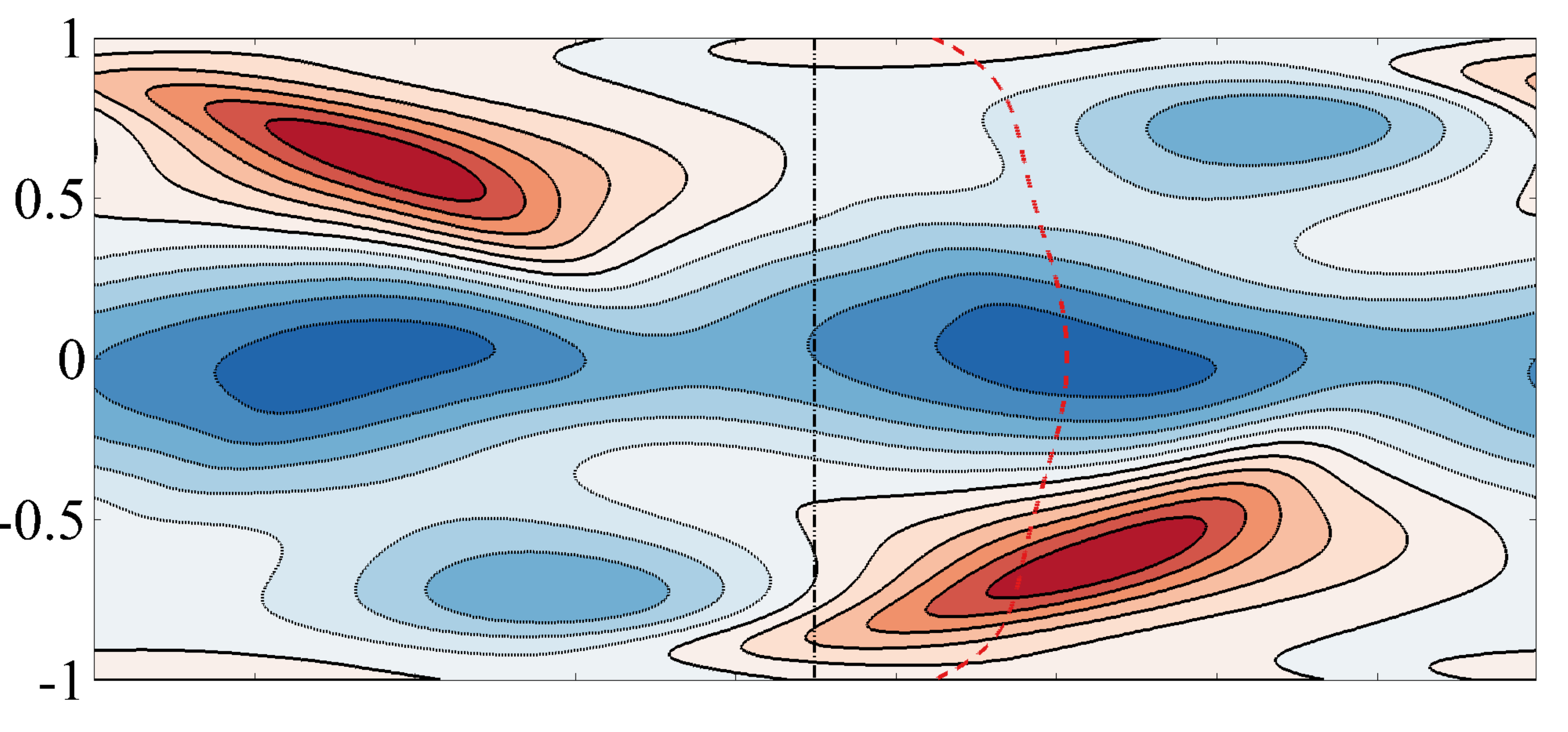}} & 
\makecell{ \\  \vspace{10mm} } & \makecell{\includegraphics[width=0.458\textwidth]{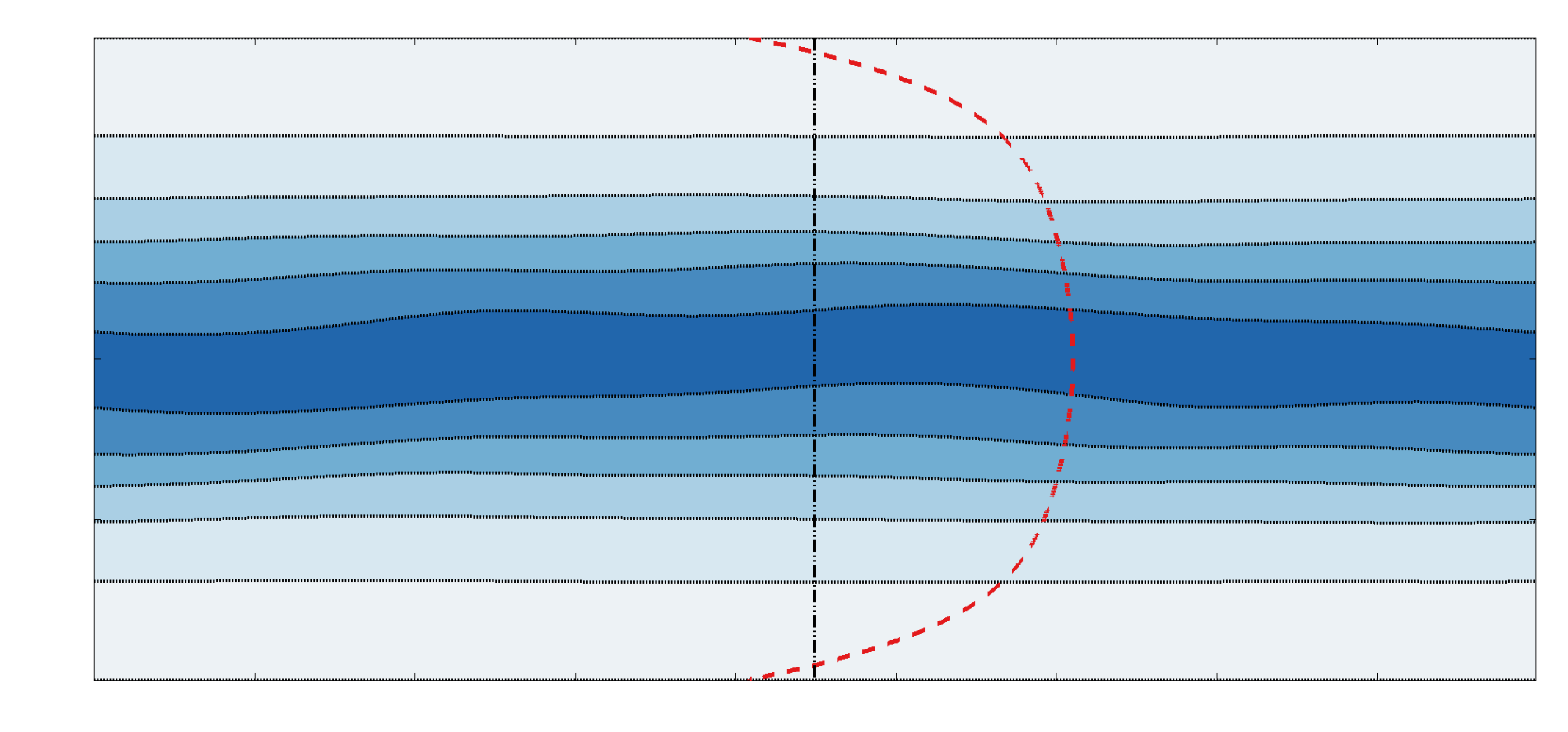}} \\
\footnotesize{(g)} & \footnotesize{\hspace{3mm} $\tpe=1.4$, $\max(|\hat{u}|)=1.599\times10^{-2}$} &
\footnotesize{(h)} & \footnotesize{\hspace{3mm} $\tpe=1.9$, $\max(|\hat{u}|)=9.752\times10^{-3}$} \\
\makecell{ \\  \vspace{10mm} \rotatebox{90}{\footnotesize{$y$}}}  & \makecell{\includegraphics[width=0.458\textwidth]{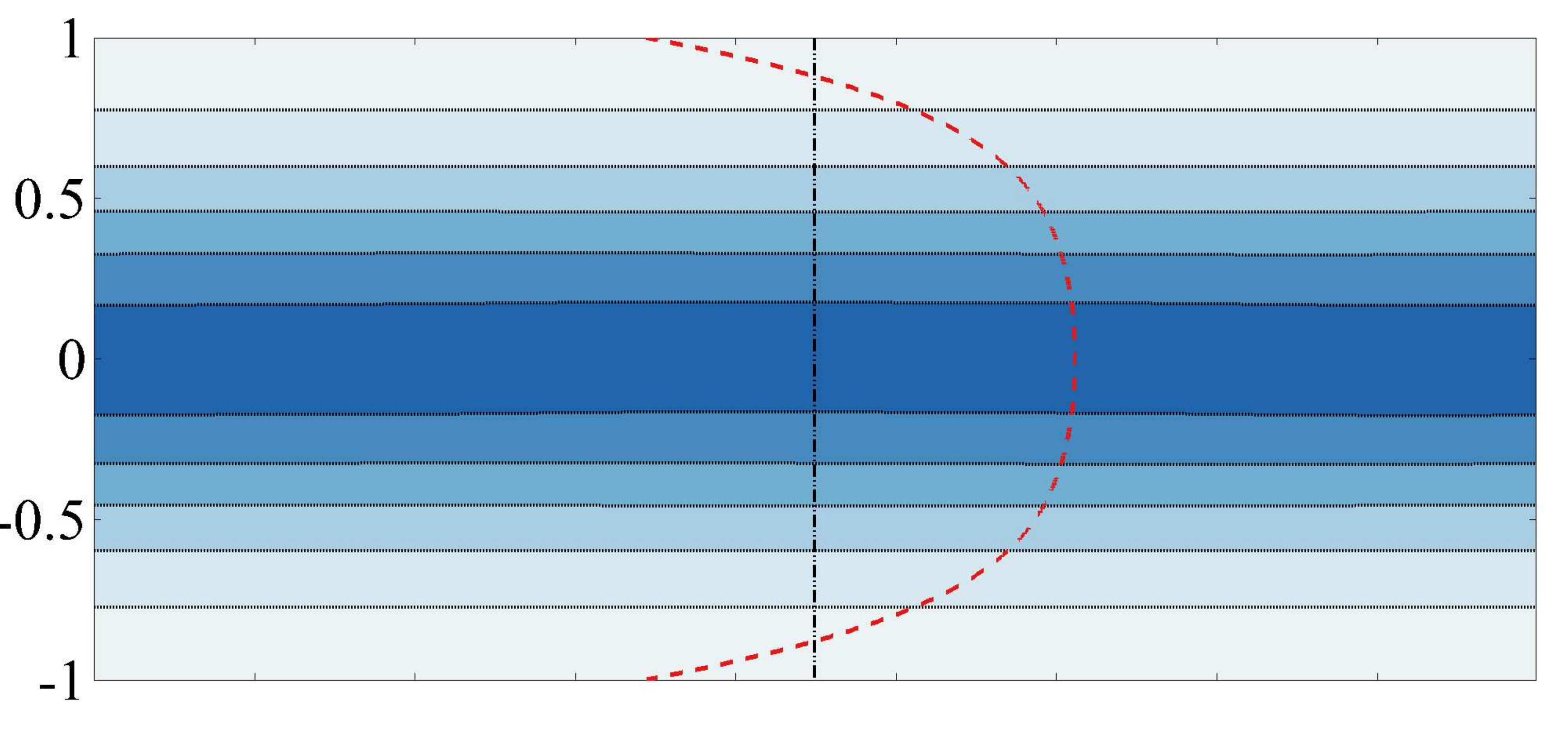}} & 
\makecell{ \\  \vspace{10mm} } & \makecell{\includegraphics[width=0.458\textwidth]{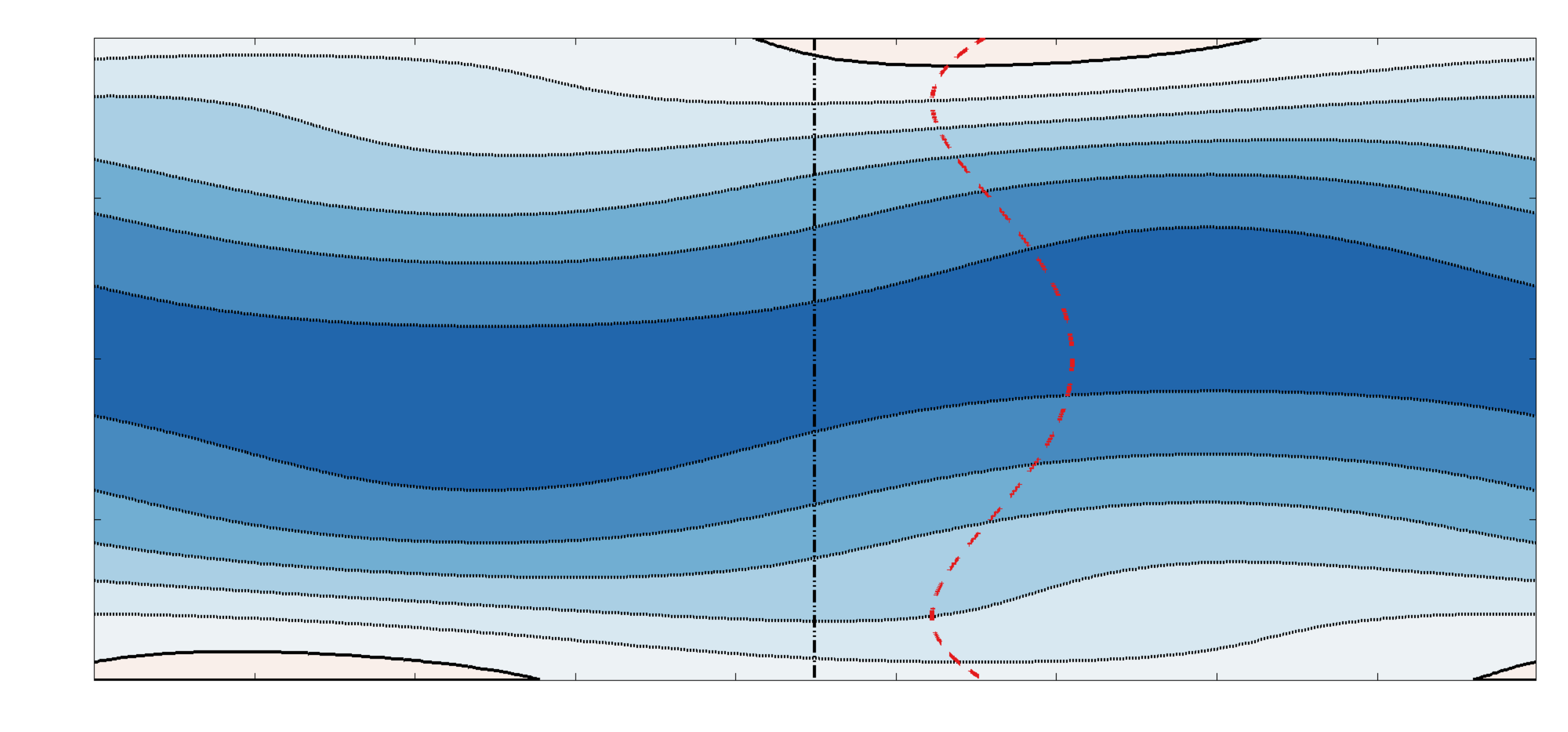}} \\
\footnotesize{(i)} & \footnotesize{\hspace{3mm} $\tpe=1.925$, $\max(|\hat{u}|)=1.428\times10^{-2}$} &
\footnotesize{(j)} & \footnotesize{\hspace{3mm} $\tpe=1.965$, $\max(|\hat{u}|)=5.821\times10^{-2}$} \\
\makecell{ \\  \vspace{10mm} \rotatebox{90}{\footnotesize{$y$}}}  & \makecell{\includegraphics[width=0.458\textwidth]{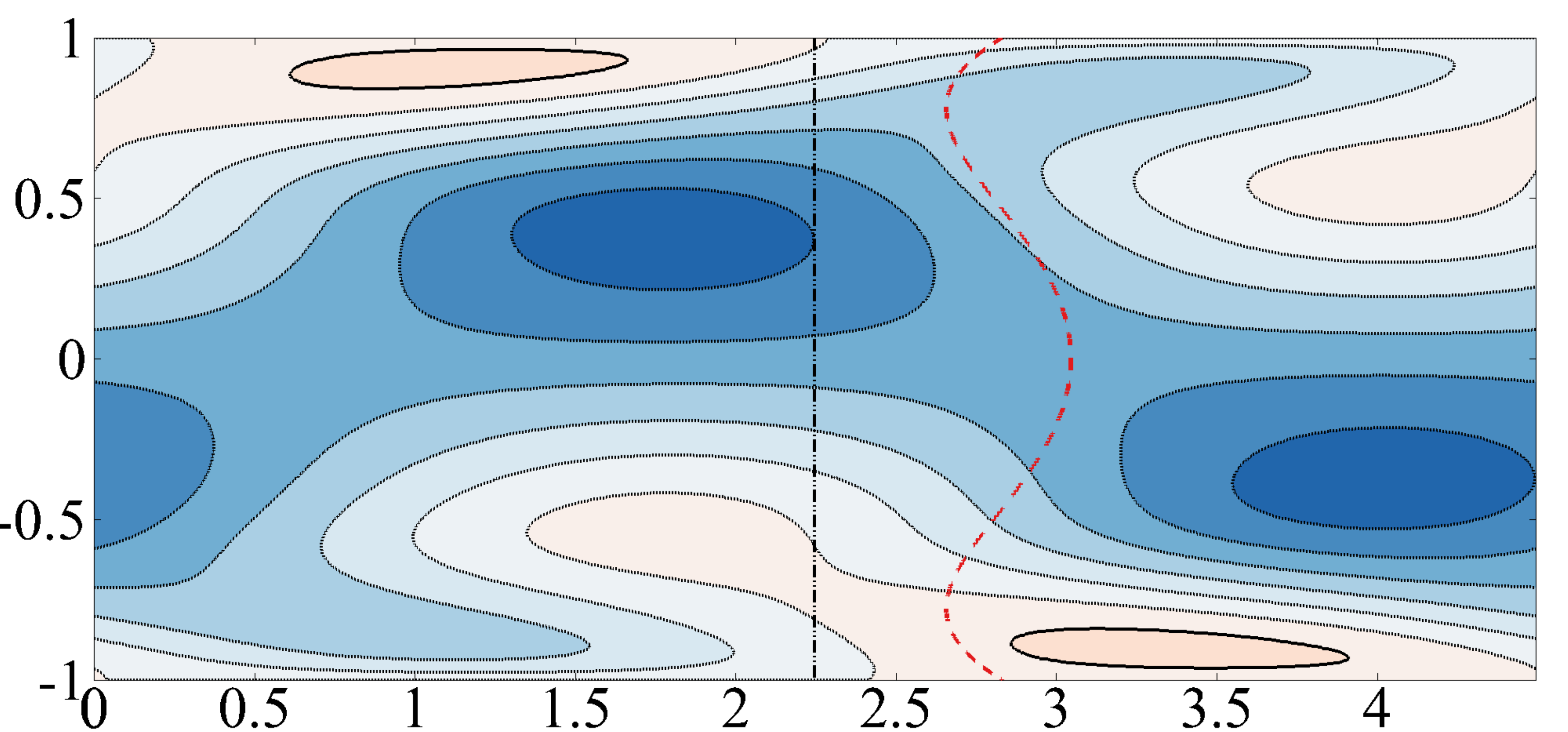}} & 
\makecell{ \\  \vspace{10mm} } & \makecell{\includegraphics[width=0.458\textwidth]{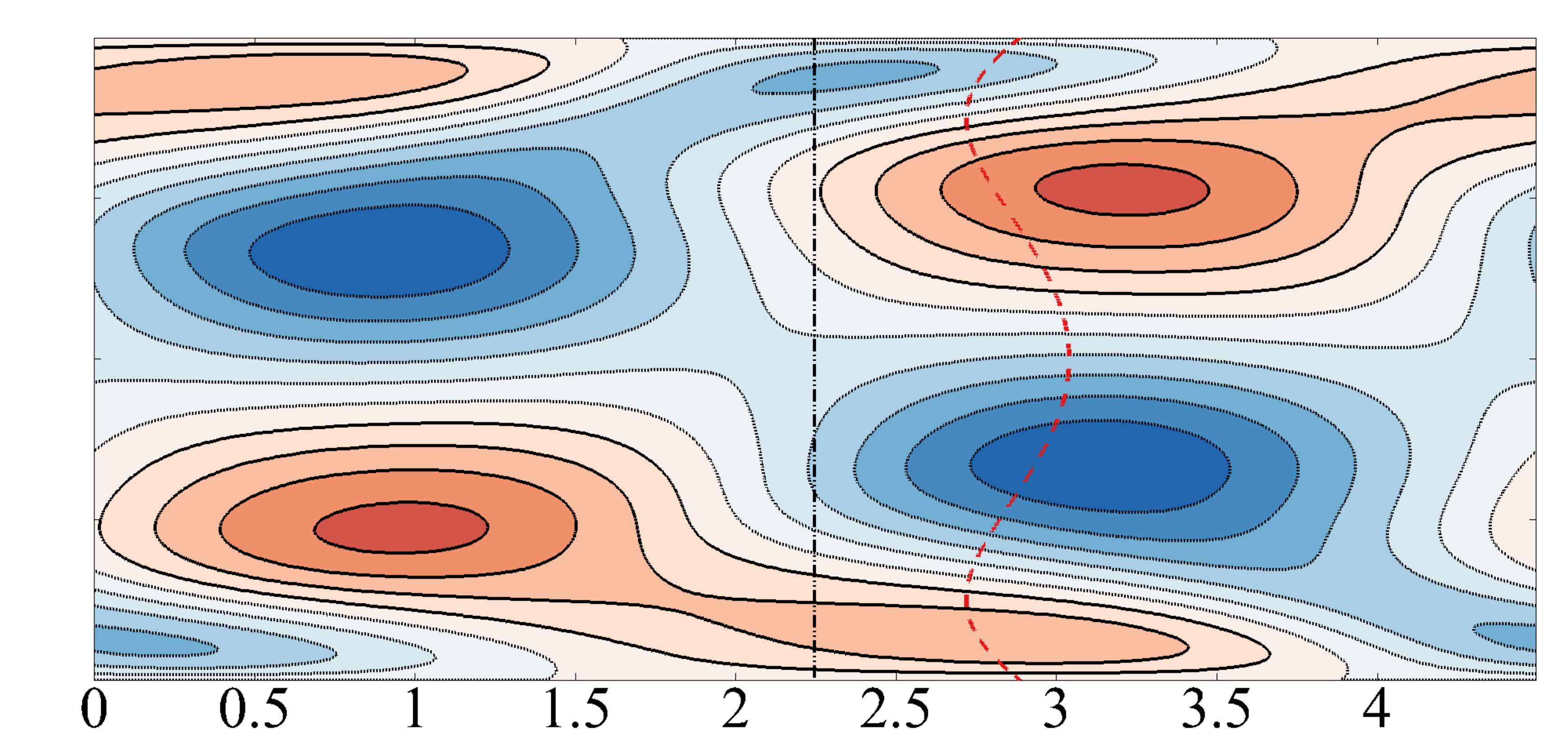}} \\
  & \hspace{38mm} \footnotesize{$x$} &  & \hspace{38mm} \footnotesize{$x$} \\
\end{tabular}
\addtolength{\tabcolsep}{+2pt}
\addtolength{\extrarowheight}{+10pt}
\end{center}
    \caption{Nonlinear evolution of $\hat{u}$-velocity perturbation contours at $H=1$, $\Gamma=1.24$, $\Sr=7.2\times10^{-3}$ through two cycles $\tpe \in [0,2]$. The base flow is overlaid (the black dashed line indicates zero base flow velocity). Red flooding positive; blue flooding negative.}
    \label{fig:snaps_nonlinear_u}
\end{figure}

The second aspect of the nonlinear evolution is the limited decay of $\Euv = \int \hat{u}^2+\hat{v}^2 \mathrm{d}\Omega$, of only 3 orders of magnitude, compared to the 18 or so orders of magnitude of decay in $\Ev = \int \hat{v}^2 \mathrm{d}\Omega$ (\fig\ \ref{fig:varyEz} or \ref{fig:nogro_varyH}). Snapshots of the $\hat{u}$-velocity from the DNS are shown in \fig\ \ref{fig:snaps_nonlinear_u} over the first two periods. An animation comparing the linear and nonlinear $\hat{u}$-velocity is also provided as supplementary material \cite{Supvideos2020}. The $\hat{u}$ perturbation is intially close to symmetric (see animation), with a central positive streamwise sheet of velocity, bounded by two negative sheets at each wall. The negative sheet of velocity near the bottom wall intensifies, and expands to fill the lower half of the duct, while pushing the positive sheet of velocity into the upper half of the duct, at $\tpe=0.22$ (the sheet of negative velocity near the top wall almost vanishing). By $\tpe=0.6$, the $\hat{u}$ perturbation is close to purely antisymmetric. However, opposite signed velocity near the walls begins encroaching on the streamwise sheets around the time when inflection points form in the base flow. This generates the linear mode observable at $\tpe=0.925$. At $\tpe=0.965$ the symmetry breaking observed in $\hat{v}$ is also observed in $\hat{u}$, disrupting the linear mode. This disruption eventually eliminates the positive velocity structures, leaving a wavy sheet of negative velocity, at $\tpe=1.3$. Throughout the acceleration phase of the base flow the sheet smooths out until it is streamwise invariant. This now symmetric sheet of negative velocity stores a large amount of perturbation energy, that produces a relatively large minimum $\hat{u}$-velocity. This sheet acts as a modulation to the base flow, and is highly persistent. Similar behaviors are observed in steady duct flows \cite{Camobreco2020transition}. Throughout the linear growth stage, the linear perturbation is able to form over the negative sheet, between $\tpe=1.9$ to $\tpe = 1.965$, before nonlinearity again breaks symmetry in the linear mode past $\tpe=1.965$.

\subsection{Symmetry breaking}\label{sec:sym_break}


\begin{figure}
\begin{center}
\addtolength{\extrarowheight}{-10pt}
\addtolength{\tabcolsep}{-2pt}
\begin{tabular}{ llll }
\makecell{\footnotesize{(a)} \vspace{26.5mm}  \\  \vspace{33mm} \rotatebox{90}{\footnotesize{$\vsk$}}} 
& \makecell{\includegraphics[width=0.458\textwidth]{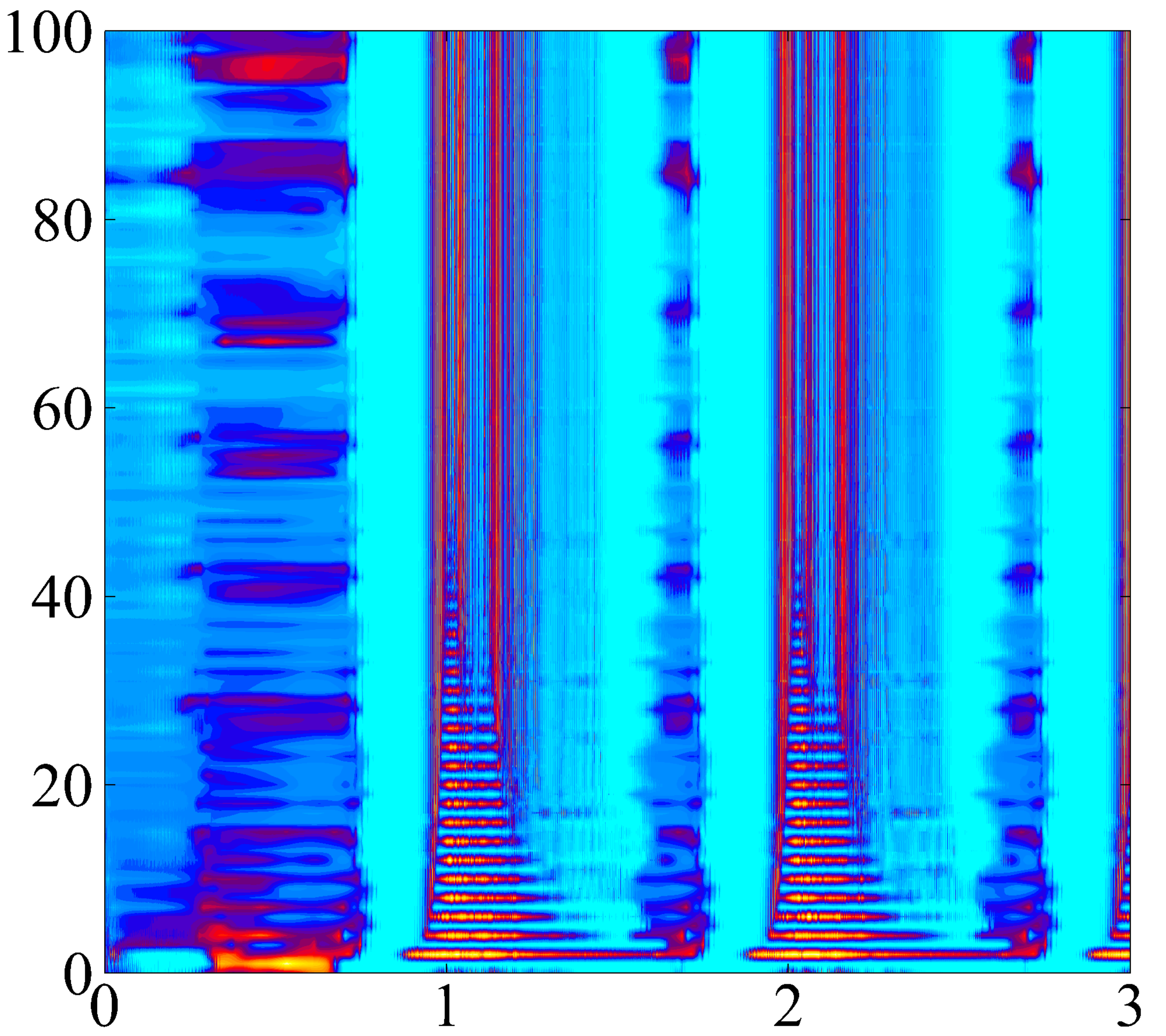}} &
\makecell{\footnotesize{(b)} \vspace{26.5mm} \\  \vspace{33mm} \rotatebox{90}{\footnotesize{$\usk$}}}
 & \makecell{\includegraphics[width=0.458\textwidth]{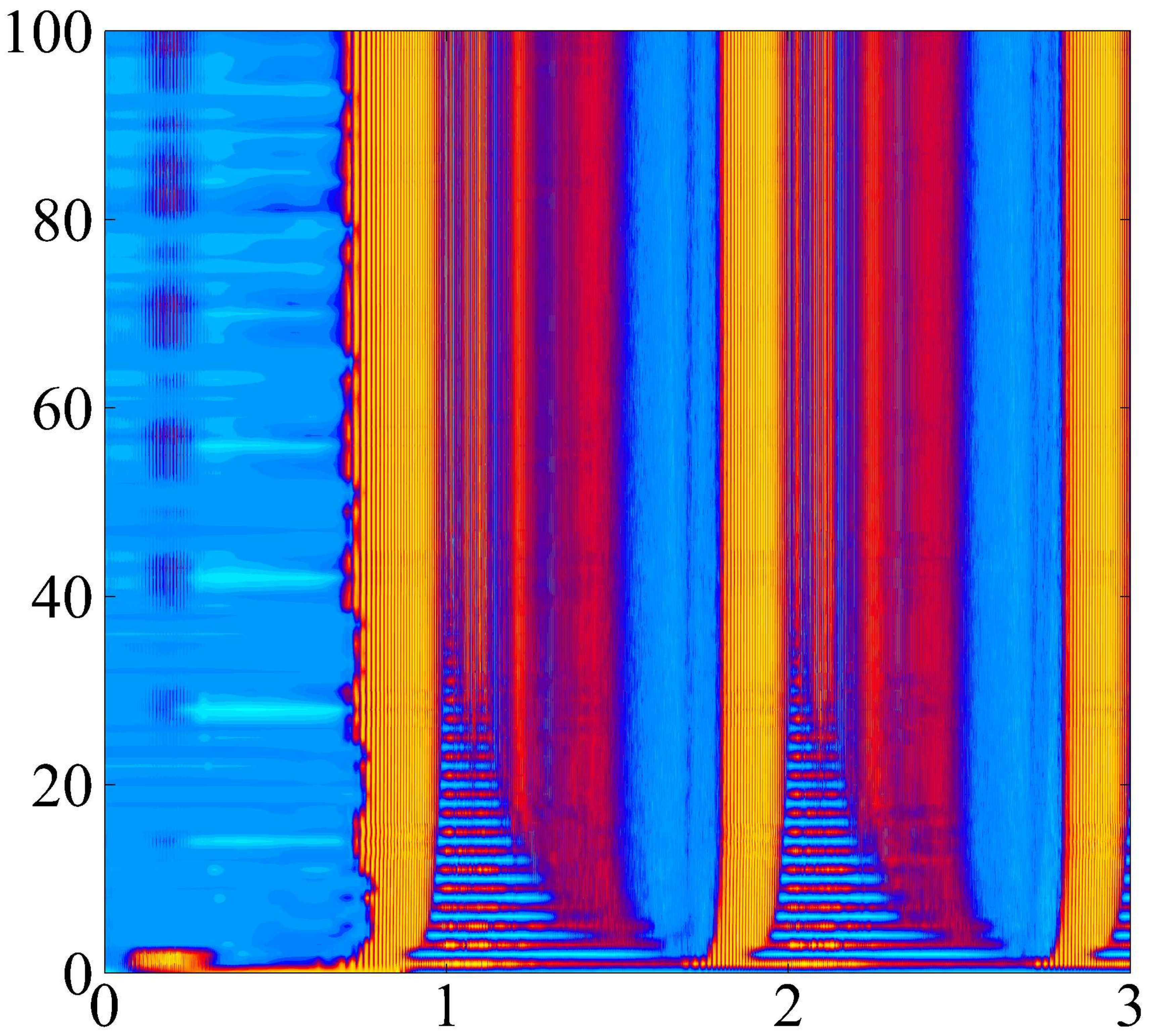}} \\
 & \hspace{38mm} \footnotesize{$\tpe$} & & \hspace{38mm} \footnotesize{$\tpe$} \\
 \makecell{\footnotesize{(c)} \vspace{26.5mm}  \\  \vspace{33mm} \rotatebox{90}{\footnotesize{$\csk$}}} 
& \makecell{\includegraphics[width=0.458\textwidth]{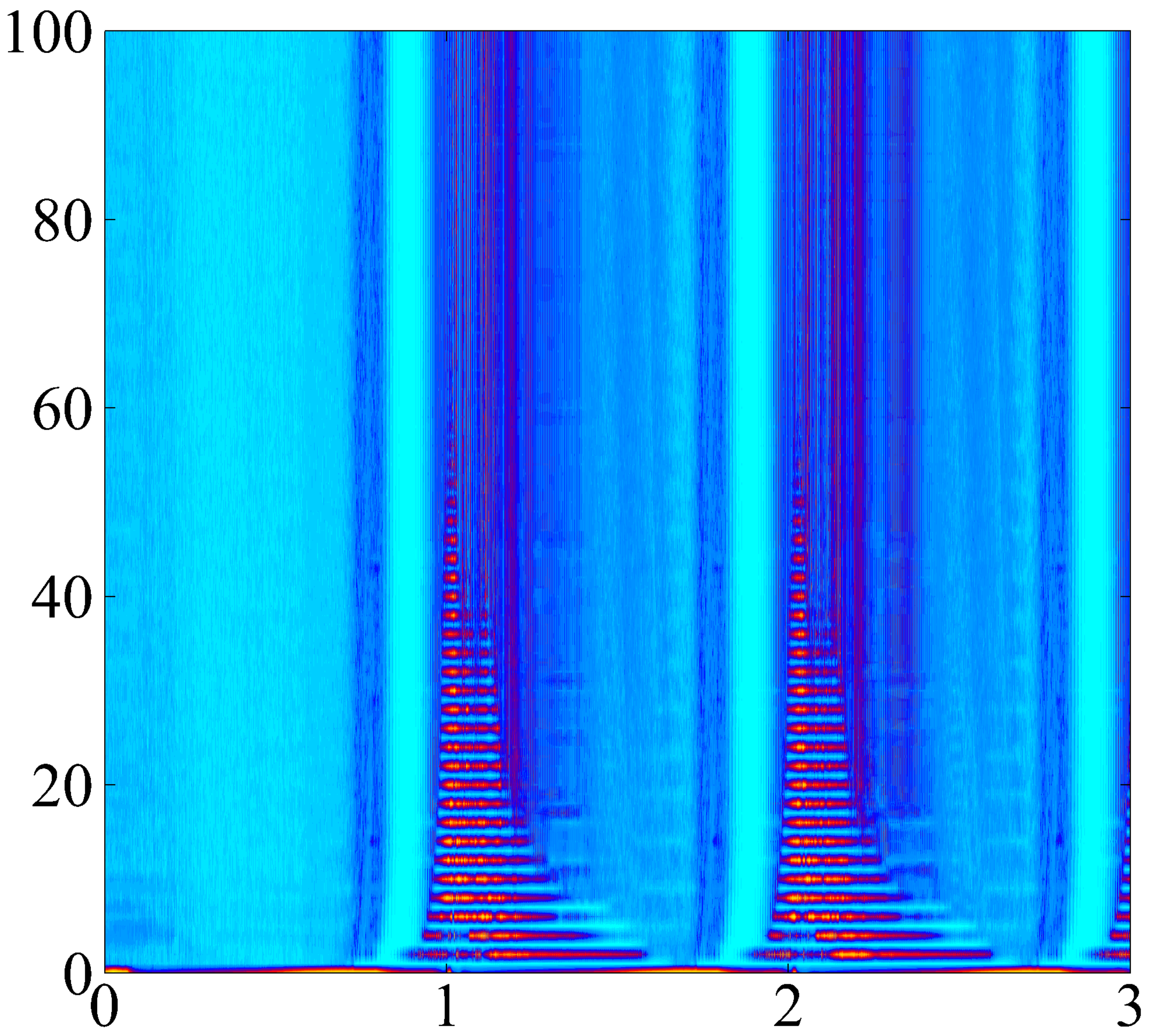}} &
\makecell{\footnotesize{(d)} \vspace{26.5mm} \\  \vspace{33mm} \rotatebox{90}{\footnotesize{$\meanfoco$, $\Euv$}}}
 & \makecell{\includegraphics[width=0.458\textwidth]{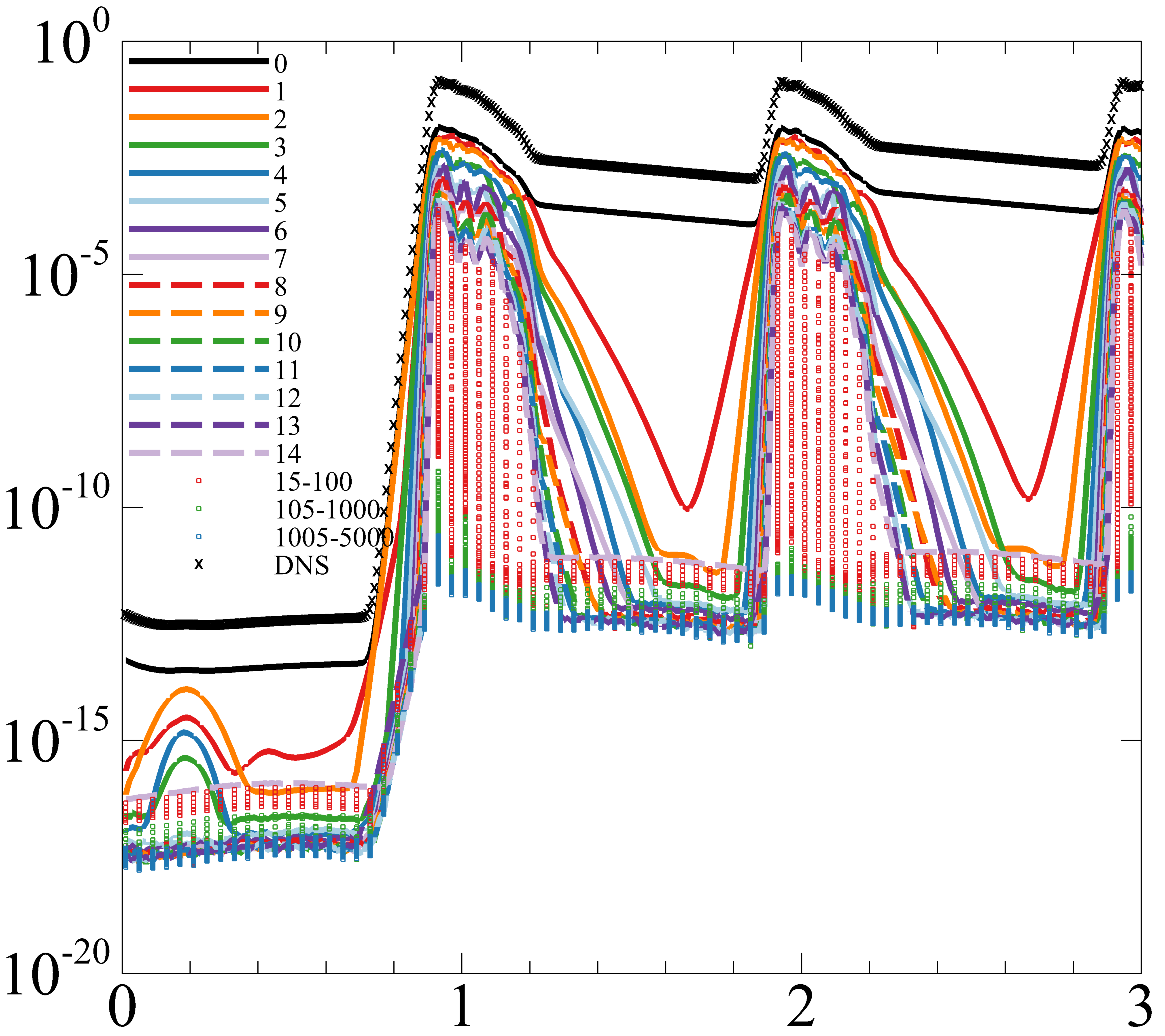}} \\
 & \hspace{38mm} \footnotesize{$\tpe$} & & \hspace{38mm} \footnotesize{$\tpe$} \\
\end{tabular}
\addtolength{\tabcolsep}{+2pt}
\addtolength{\extrarowheight}{+10pt}
\end{center}
    \caption{A measure of the symmetry in the zeroth through one-hundredth isolated streamwise Fourier modes. (a) Wall-normal velocity perturbation. (b) Streamwise velocity perturbation. (c) In-plane vorticity perturbation. Small values of the symmetry measure indicate the mode is almost symmetric (light blue), while large vales indicate the mode is almost antisymmetric (orange/yellow). (d) The $y$-averaged Fourier coefficient for each mode, based on $\hat{f} = \hat{u}^2+\hat{v}^2$, compared to the DNS measure $\Euv = \int \hat{u}^2 + \hat{v}^2 \mathrm{d} \Omega$. Note that for modes $100 < \kappa \leq 5000$ only every fifth $\kappa$ is plotted.}
    \label{fig:sym_measure}
\end{figure}

The symmetry breaking process was further analysed by measuring the degree of symmetry, separately for each mode $j$, via $\hat{f}_{\mathrm{s},j} = (\sum_{m=0}^{m=\Ny}[\hat{f}_j(y_m)-\hat{f}_j(-y_m)]^2)^{1/2}$. This is depicted for $\hat{v}$, $\hat{u}$ and $\hat{\omega}_z$ in \figs\ \ref{fig:sym_measure}(a) through (c), while a measure of the $y$-averaged energy in each mode is provided in \fig\ \ref{fig:sym_measure}(d). The key result is that when the nonlinear DNS had a similar appearance and growth rate to the linear simulation (e.g.~from $\tpe \approx 0.75+q$ to $\tpe \approx 0.95+q$, for $q=0$, $1$, $2$), every resolved $\hat{v}$ mode ($\kappa = 0$ through $100$) is close to purely symmetric, \fig\ \ref{fig:sym_measure}(a). Once symmetry breaking occurs,  at $\tpe\approx0.965$, every odd $\hat{v}$ mode (first, third, etc.) becomes antisymmetric. See also see the vorticity measure, \fig\ \ref{fig:sym_measure}(c), for the first 50 or 60 modes. Thus, the symmetry breaking does not appear to be connected to any asymmetry introduced by numerical noise in the initial perturbation, as every mode becomes symmetric through the preceding linear phase. The measure of symmetry in $\hat{u}$ is effectively the photo negative of $\hat{v}$ (if $\hat{v}$ is almost symmetric, $\hat{u}$ is almost antisymmetric). The exception is the zeroth mode, which remains symmetric after the first period. The zeroth mode stores a large amount of perturbation energy, \fig\ \ref{fig:sym_measure}(d), and decays very slowly compared to the higher modes. Hence, the DNS measure of the perturbation energy $\Euv$ closely resembles the energy in the zeroth mode. As a final note, although a large number of modes become appreciably energized, the floor of the energy in the highest modes (after the base flow modulation occurs) is not clearly raised, and no distinct inertial subrange forms (not shown). Hence, as turbulence is not observed, it cannot initiate the symmetry breaking. However, exactly how nonlinearity induces the symmetry breaking remains unknown.


\section{Conclusions} \label{sec:conc}

This work numerically investigates the stability of a pulsatile quasi-two-dimensional duct flows, motivated by their relevance to the cooling conduits of magnetic confinement fusion reactors. The linear stability over a large $\Rey$, $H$, $\Sr$, $\Gamma$ parameter space was analysed, to both determine the pulsation optimized for the greatest reduction in $\ReyCrit$, and more generally to understand the role of transient inertial forces in unsteady MHD duct flows. At large amplitude ratios ($\Gamma=100$, near the conditions of a steady base flow) the effect of varying $\Sr$ was clearest. Increasing $\Sr$ lead to both more prominent inflection points, acting to reduce $\ReyCrit$, and thinner oscillating boundary layers, acting to increase $\ReyCrit$. Although more prominent inflection points generated additional growth during the deceleration of the base flow, the effective length of the deceleration phase increases with decreasing $\Sr$.  Thus, by tuning $\Sr$ (for a given $H$, $\Gamma$) the minimum $\ReyCrit$ is reached as the perturbation and base flow energy variations fall in phase, so long as inflection points remain prominent. Furthermore, the percentage reduction in $\ReyCrit$ always improved with increasing $H$, when free to adjust $\Sr$. This observation, that pulsatility was still effective at destabilizing the flow in (or toward) fusion relevant regimes, satisfies the first question the paper put forward.

At intermediate amplitude ratios ($\Gamma=10$), the addition of the oscillating flow component lead to large changes in $\ReyCrit$ compared to the steady base flow. At these amplitude ratios the effect of $\Rey$ on the base flow becomes important. Increasing $\Rey$ reduces the oscillating boundary layer thickness and restabilizes the flow for a small range of frequencies. Although the base flow became more stable with increasing $\Rey$, a large enough $\Rey$ was eventually reached to destabilize other instability modes (different to the \TSL\ mode).

At smaller, near unity amplitude ratios (equal steady and oscillating base flow maxima) the largest advancements in $\ReyCrit$ over the steady value were observed. At $H=10^{-7}$, an almost $70$\% reduction in $\ReyCrit$ was attained, while by $H=10$, there was over an order of magnitude reduction ($90.3$\%). These improvements were attained at $\Sr$ of order $10^{-3}$, a region of the parameter space more than amenable to both SM82 modelling, and fusion relevant applications. Particularly in the latter case, a low frequency driving force would be far simpler to engineer than a high frequency oscillation. These results answer the second and third questions put forth in the paper.

At these conditions, the onset of turbulence was not observed in nonlinear DNS. Within the first oscillation period the intracyclic growth was able to propel an initial perturbation of numerical noise to nonlinear amplitudes. This modulated the base flow, by generating a sheet of negative velocity along the duct centreline. Although this modulated base flow had no effect on the growth of the wall-normal velocity perturbation, it was able to saturate the exponential growth at supercritical Reynolds numbers. Although turbulence was not triggered, the nonlinear growth was still a promising result. However, without a wider nonlinear investigtion of the parameter space, the capability for $\ReyCrit$ reductions to translate to reductions in the $\Rey$ at which turbulence is observed (the fourth question put forward), remains partially unresolved. At nonlinear amplitudes, a symmetry breaking process was observed within each cycle. The ensuing chaotic flow may naturally improve mixing, improving cooling conduit performance, without the severe increases in frictional losses accompaning a turbulent flow \cite{Kuhnen2018destabilizing}. This is an avenue for future work. 


Finally, the capability for the optimized pulsations to nonlinearly modulate the base flow within one cycle favors linear transient growth as a strong contender for enabling bypass transitions to turbulence. This is a key area of future research, as if the flow is transiently driven over a partial oscillation cycle (and steadily driven thereafter), turbulence may be rapidly triggered. A caveat to such a method is that it is the continually driven time periodic base flow which yields eigenvalues with positive growth rates at greatly reduced Reynolds numbers. Without such an underlying base flow, the leading eigenvalues may be strongly negative, and severely limit any transient growth, as for cylinder wake flows \cite{Abdessemed2009transient}. This may be particularly problematic if large amounts of regenerative transient growth are the key to sustaining turbulent states \cite{Budanur2020upper,Lozano2020cause}, a point that also requires further investigation.

Overall, the large reductions in $\ReyCrit$, occurring in a viable region of the parameter space, form too promising a direction to cease investigating. The first steps to this are to assess the heat transfer characteristics of the pulsatile base flow, which may naturally be more efficient than the steady equivalent, and investigating linear transient optimals. Other than linear transient growth, the use of pulsatility in concert with one of the various Q2D vortex promoters \citep{Cassels2016heat, Hussam2011dynamics, Hussam2012enhancing, Buhler1996instabilities, Hamid2016combining, Hamid2016heat, Cuevas2006flow} could aid in sustaining turbulence. Past the Q2D setup, the full 3D duct flow could be tackled. In particular, the interaction between the Stokes and Hartmann layers could result in new avenues to reach turbulence. The reduced constriction of the full 3D domain may also aid in sustaining turbulence. Note that for fusion applications, oscillatory wall motion is not viable. Therefore, in the context of a 3D domain, oscillatory pressure gradients are more relevant (note that the fully nonlinear wall- and pressure-driven flows are only equivalent in the 2D averaged equations). Lastly, with a broader scope, even electrically conducting walls could be investigated. Although less prevalent in self-cooled designs \cite{Smolentsev2008characterization}, the larger shear present in boundary layers forming on conducting walls provides conditions more susceptible to transitions to turbulence, and larger turbulent fluctuations \cite{Burr2000turbulent}. The interactions between flow pulsatility and electrically conducting walls could yield many new insights.

\begin{acknowledgments}
C.J.C.\ receives support from the Australian Government Research Training Program (RTP). This research was supported by the Australian Government via the Australian Research Council (Discovery Grant DP180102647), the National Computational Infrastructure (NCI) and Pawsey Supercomputing Centre (PSC), by Monash University via the MonARCH HPC cluster and by the Royal Society under the International Exchange Scheme between the UK and Australia (Grant E170034).
\end{acknowledgments}
%


%

\end{document}